%% file: main_v1.0.tex
\documentclass[%
reprint,
 amsmath,amssymb,
 aps,
longbibliography,  %MP
rmp,author-year,
]{revtex4-1}
\usepackage{graphicx}% Include figure files
\usepackage{dcolumn}% Align table columns on decimal point
\usepackage{bm}% bold math
\usepackage{xcolor}
\usepackage[breaklinks,colorlinks,citecolor=blue,urlcolor=blue,linkcolor=blue,linktoc=none]{hyperref}
\usepackage{dsfont}
\usepackage{upgreek}
\usepackage{diagbox}
\usepackage{babel}
\usepackage{fontenc}
\usepackage{inputenc}
\usepackage{bbold}
\usepackage{array}
\usepackage{epsfig}
\usepackage{amsbsy}
\usepackage{soul}
\usepackage{xfrac}
\usepackage{natbib}
\usepackage{multirow}
\usepackage[mediumqspace,mediumspace,amssymb,Gray]{SIunits}
\usepackage{mathrsfs}
\usepackage{tikz}
\usepackage{dcolumn}

\begin{document}

\preprint{APS/123-QED}

\title{Elimination of angular dependence in the quantum three-body problem made easy}

\author{Anjan Sadhukhan}
\affiliation{Department of Applied Chemistry, National Yang Ming Chiao Tung University, 
1001 University Road, Hsinchu, 300, Taiwan}
\author{Grzegorz Pestka}
\affiliation{Faculty of Humanities, Social and Information Technology Sciences, The Mazovian University in Płock, 
Pl. Dabrowskiego 2, 09-402 Płock, Poland}
\author{Rafał Podeszwa}
\affiliation{Institute of Chemistry, University of Silesia, ul. Szkolna 9, 40-006 Katowice, Poland}
\author{Henryk A. Witek}
\email[Contact author: ]{hwitek@nycu.edu.tw}
\affiliation{Department of Applied Chemistry and Institute of Molecular Science, National Yang Ming Chiao Tung University, 1001 University Road, Hsinchu, 300, Taiwan}
\date{\today}
\begin{abstract}

We present a systematic account of the separation of the angular degrees of freedom from the nonrelativistic Schr\"{o}dinger equation for a three-body quantum system with arbitrary masses, charges, total angular momentum, and parity. The resulting reduced Schr\"{o}dinger equation (RSE) for the partial-wave components, expressed as functions solely of the interparticle distances, is reported in a compact matrix operator form. The remnants of the angular dependence, essential for the hermiticity of the RSE and consequently the stability of variational computations, appear in the RSE formalism as additional angular factors, derived by expanding minimal bipolar harmonics into a basis of Wigner functions \texorpdfstring{$\mathcal{D}$}{D}. We validate the final form of the RSE by computing accurate energy levels for helium states using an explicitly correlated Hylleraas-type basis. This work serves as a self-contained reference for the RSE formulation, consolidating elements previously scattered throughout the literature, thereby offering a convenient foundation for further analytical and numerical studies of general three-body quantum systems.

\end{abstract}

\maketitle

\tableofcontents

\section{Introduction}\label{intro}
The separation of variables in the nonrelativistic quantum mechanical treatment of the three-body problem has received considerable attention in the literature \cite{Breit1930, Hughes1930, hirschfelder1935, DattaMajumdar1952, Jackson1954, Kemeny1964, Majumdar1964, Proriol1967, Iwai1987, BHATIA1964, BHATIA1965, Mukherjee1994, Pont1995, Hsiang1997, Ma1999, Ma2000, Gu2001, Gu2001a, MEREMIANIN2003, Hsiang2007, Chi2007,Pestka2008}. The elimination of the center of mass reduces the 9D Schr\"{o}dinger equation (SE) to a 6D partial differential equation (PDE) describing two quasiparticles in a central field. This problem can be further transformed into a set of 3D PDEs by eliminating the rotational degrees of freedom (the Euler angles $\alpha,\beta,\gamma$), as implied by the rotational invariance of the Hamiltonian. This is trivial for states with total angular momentum $L=0$, in which case the 6D wave function $\Psi(\bm{r}_{1},\bm{r}_{2})$ of the two quasiparticles simply reduces to $\psi(r_{1},r_{2},r_{12})$, a Hylleraas function of the interparticle distances $r_1=\left| \bm{r}_{1} \right|$, $r_2=\left| \bm{r}_{2} \right|$, and $r_{12}=\left| \bm{r}_{1} - \bm{r}_{2} \right|$ \cite{Hylleraas1928, Hylleraas1929}. For higher values of $L$, the procedure is considerably more involved: $\Psi^{LM\uppi}(\bm{r}_{1},\bm{r}_{2})$, corresponding to definite angular momentum $L$, its projection $M$ on the laboratory $z$ axis, and parity $\uppi$, can be represented as
\begin{equation}\label{wfa1}
    \Psi^{LM\uppi}(\bm{r}_{1},\bm{r}_{2}) \!=\! \sum_{l=d}^{L}\psi_{l}^{L\uppi}(r_{1},r_{2},r_{12}) \,\,\mathscr{Y}_{l}^{LM\uppi}(\bm{r}_{1},\bm{r}_{2})
\end{equation}
where $\mathscr{Y}_{l}^{LM\uppi}(\bm{r}_{1},\bm{r}_{2})$ with $l=d,\ldots,L$ are angular gen\-er\-a\-tors spanning the $\left(L,M,\uppi\right)$-invariant subspaces; $\psi_{l}^{L\uppi}(r_{1},r_{2},r_{12})$ are the corresponding $M$-independent\footnote{For details, see Sec.~\ref{ss:psiexp}.} reduced coefficients depending solely on the interparticle distances; and $d=0$ for states with natural parity ($\uppi= \text{n}$) and $d=1$ for states with unnatural parity ($\uppi= \text{u}$). Note that the distinction between natural parity states ($\text{S}^\text{e}$, $\text{P}^\text{o}$, $\text{D}^\text{e}$, $\text{F}^\text{o}$, $\ldots$)   and unnatural parity states ($\text{P}^\text{e}$, $\text{D}^\text{o}$, $\text{F}^\text{e}$, $\text{G}^\text{o}$, $\ldots$) seems more appropriate here than the usual distinction between even parity states ($\text{S}^\text{e}$, $\text{P}^\text{e}$, $\text{D}^\text{e}$, $\text{F}^\text{e}$, $\ldots$) and odd parity states ($\text{P}^\text{o}$, $\text{D}^\text{o}$, $\text{F}^\text{o}$, $\text{G}^\text{o}$, $\ldots$).  The action of the Hamiltonian on the wave function $\Psi^{LM\uppi}(\bm{r}_{1},\bm{r}_{2})$ and the subsequent elimination of~the angular components produce a system of coupled 3D PDEs for the $L+1-d$ reduced coefficients $\psi_{l}^{L\uppi}(r_{1},r_{2},r_{12})$. The resulting system of PDEs---representing states with arbitrary angular momentum $L$ and parity $\uppi$---is hereafter referred to as the reduced SE (RSE). 

RSEs were first derived for several distinct values of $L$: $L=0$ ($\text{S}^\text{e}$ states) \cite{Hylleraas1928,Hylleraas1929}, $L=1$ ($\text{P}^\text{o}$ and $\text{P}^\text{e}$ states) \cite{Breit1930}, and $L=2$ ($\text{D}^\text{e}$ and $\text{D}^\text{o}$ states) \cite{Schwartz1961}, before a closed-form RSE for arbitrary $L$ and $\uppi$ was obtained \cite{DattaMajumdar1952,BHATIA1964,BHATIA1965,Kalotas1965} using parity-adapted linear combinations of Wigner functions $\mathcal{D}^{MK}_L(\alpha,\beta,\gamma)$ as angular generators. The re\-sult\-ing equations have a rather complicated algebraic form; this is particularly evident in the weak variational form of the RSE derived by \textcite{Mukherjee1994, Mukherjee1995}.
Much simpler expressions emerge when  minimal bipolar harmonics (MBHs) $\Omega_{l}^{LM\uppi}\equiv\Omega_{l}^{LM\uppi}(\bm{r}_{1},\bm{r}_{2})$ are selected as angular generators   \cite{Manakov1998,King1967,Breit1930,Schwartz1961,Drake1978,Frolov1996,MEREMIANIN2003,Drake1987,Drake1990}, since they preserve the partial angular momenta $l_1=l$ and $l_2=L-l+d$ of the individual quasiparticles, thereby simplifying the Hamiltonian's action. The RSE in the MBH basis for general values of $L$ and $\uppi$ was derived in the limit $m_3\rightarrow \infty$ [Eq.~(34) of \textcite{Jackson1954}, Eqs.~(9)--(12) of \textcite{Pont1995}, and Eq.~(22) of \textcite{Bottcher1994}], as well as for finite masses of all particles [Eq.~(26) of \textcite{Efros1986} and Eq.~(67) of \textcite{MEREMIANIN2003}]. Several other publications \cite{Drake1978, Frolov1996, Gu2001, Harris2004} computed the action of the kinetic energy operator on the wavefunction $\Psi^{LM\uppi}(\bm{r}_{1},\bm{r}_{2})$ in the form given by Eq.~(\ref{wfa1}) with MBHs as the angular generators, but without explicit formulation of the corresponding RSE. 

While the separation of angular degrees of freedom in the basis of parity-adapted Wigner functions $\mathcal{D}^{MK}_L(\alpha,\beta,\gamma)$ was thoroughly explored in the excellent exposition by \textcite{BHATIA1964, BHATIA1965}, an analogous elementary treatment using minimal bipolar harmonics is unavailable, with the existing body of results scattered across many studies. The seminal work of \textcite{Breit1930} is limited to states with $L=1$. Similarly, the thorough and essentially complete work of \textcite{Pont1995} relies on the infinite-mass approximation for one of the particles. The comprehensive account of \textcite{MEREMIANIN2003} relies heavily on the irreducible tensor approach, and since the attention of the authors is primarily on the analysis of gauge singularities in the RSE for $N$ particles, the details of the derivations are kept to a minimum. The RSE formulated by \textcite{Efros1986} [see Eq.~(26) of his work] emerges as a byproduct of the deriving matrix elements in three-body calculations; consequently, the explicit mathematical steps are largely omitted. 

Motivated by this gap in the literature, we present here a complete, self-contained construction of the RSE in the MBH basis, applicable to a general Coulomb system of three quantum particles with arbitrary masses, charges, angular momentum quantum numbers $L$ and $M$, and parity $\uppi$. This approach casts the resulting equations in a matrix-operator form, providing a particularly transparent exposition of the algebraic structure of the RSE. By preserving the partial angular momenta of the constituent particles, the formalism enables a clear interpretation of the results in the usual language of one-particle excitations typical of quantum chemical approaches, including configuration-interaction descriptions of two-electron atomic systems.

This paper serves several primary objectives:
\begin{enumerate}
    \item Establishing a complete, mathematically rigorous, pedagogically-oriented, and self-contained ref\-er\-ence for the derivation of the RSE in the MBH basis, in the spirit of the exposition by \textcite{BHATIA1964, BHATIA1965}, who used the basis of parity-adapted Wigner functions $\mathcal{D}^{MK}_L(\alpha,\beta,\gamma)$. 
    \item Casting the RSE into a matrix-operator form readily applicable to variational computations of energy levels and wave functions for a general Coulomb system of three quantum particles with arbitrary masses, charges, angular momentum quantum numbers $L$ and $M$, and parity $\uppi$.
    \item Deriving an analytical expression that expands MBHs in the basis of Wigner functions $\mathcal{D}^{MK}_L(\alpha,\beta,\gamma)$, thereby providing a methodological tool to translate between the present work and the analogous formalism developed by \textcite{BHATIA1964, BHATIA1965}.
    \item Expressing in elementary algebraic form the angular integral originally introduced by Drake \cite{Drake1978, Calais1962} in terms of Wigner 3-$j$ and 6-$j$ symbols and given explicitly in Eq.~(15) of \textcite{Frolov1996}.
    \item Numerically verifying the correctness of the derived equations by computing accurate nonrelativistic energy levels for selected states of helium and comparing them to existing literature values.
    \item Providing an algebraically simple form of the RSE, which constitutes a natural starting point for future attempts of analytical solutions to the Schr\"{o}dinger equation for three quantum particles with non-vanishing angular momentum, following the treatment of the $\text{S}^e$ states of helium initiated by \textcite{Fock1954}, \textcite{Hylleraas1960}, and \textcite{Pluvinage1955}, and continued by others \cite{Bartlett1955, Frost1964a, Frost1964b, Frost1964c, White1970, Haftel1983, Davis1982, Davis1982a, Davis1983, Davis1983a, Davis1983b, Abbott1984, Gottschalk1985, Abbott1986, Abbott1987, Gottschalk1987a, Gottschalk1987b, McIsaac1987, Liverts2010, Liverts2013, Liverts2018, Langner2022}. 
    \item Organizing, summarizing, and reviewing the existing formulas, derivations, and results scattered across many studies.
\end{enumerate} 
 
The paper is organized as follows. Sec.~\ref{sec:prel} provides the mathematical and quantum-mechanical preliminaries required to eliminate the rotational degrees of freedom in the nonrelativistic treatment of the three-body problem. Section~\ref{method} details the derivation of the RSE, illustrated by the applications of the resulting formalism for several low values of $L$. To validate the derived formalism, Sec.~\ref{numerical} presents accurate numerical evaluations of nonrelativistic energy levels for selected low-lying states of helium with both natural and unnatural parity. Finally, Sec.~\ref{conc} provides concluding remarks. Readers interested in the numerical and analytical treatment of the exact wave functions of three-body systems may skip the preliminaries and derivation details and proceed directly to Sec.~\ref{method}, where Eqs.~(\ref{hamtme}) and~(\ref{rrse}) constitute the required formal basis for those tasks.

\section{Preliminaries} \label{sec:prel}
\subsection{Center of mass separation}
The Schr\"{o}dinger equation for three quantum particles with masses $m_{i}$ and charges $q_{i}$ ($i=1,2,3$) interacting via pairwise Coulomb potentials is given in atomic units in the laboratory-fixed ($\overline{x}\,\overline{y}\,\overline{z}$) reference frame (Fig.~\ref{fig:comparison}) as
\begin{equation}\label{selab}
    \left[\sum_{i=1}^{3}\frac{-\overline{\Delta}_{i}}{2\,m_{i}} + \sum_{j>i=1}^{3} \frac{q_{i}q_{j}}{\overline{r}_{ij}}-\varepsilon \right]\Phi(\overline{\bm{r}}_{1}, \overline{\bm{r}}_{2}, \overline{\bm{r}}_{3}) = 0
\end{equation}
where $\overline{\bm{r}}_i=\left[\overline{x}_i,\overline{y}_i,\overline{z}_i\right]^\mathsf{T}$ is the position of particle $i$ in the ($\overline{x}\,\overline{y}\,\overline{z}$) frame,  $\overline{r}_{ij}=\left|\overline{\bm{r}}_{i}-\overline{\bm{r}}_{j}\right|$, and $\overline{\Delta}_{i}=\partial_{\,\overline{x}_{i}\overline{x}_{i}} + \partial_{\,\overline{y}_{i}\overline{y}_{i}} + \partial_{\,\overline{z}_{i}\overline{z}_{i}}$. Introducing a new set of coordinates ($\bm{R},\bm{r}_{1},\bm{r}_{2}$) with $\bm{r}_{i}=\left[x_i, y_i, z_i\right]^\mathsf{T}\!=\overline{\bm{r}}_{i}-\overline{\bm{r}}_{3}$ ($i=1,2$), $\bm{R} =\left[R_{x}, R_{y}, R_{z}\right]^\mathsf{T} \!=\!\frac{m_{1}\overline{\bm{r}}_{1}+m_{2}\overline{\bm{r}}_{2}+m_{3}\overline{\bm{r}}_{3}}{M}$, and $M\!=\!m_1\!+\!m_2\!+\!m_3$ transforms the SE given in Eq.~\eqref{selab} into a new form
\begin{widetext}
    \begin{equation}\label{secm0}
        \left[\vphantom{\frac{1}{m_{3}}}\right.\underbrace{\vphantom{\frac{1}{m_{3}}}-\frac{1}{2M}\Delta_{R}}_{\hat{\mathcal{T}}_R} \, \underbrace{-\frac{1}{2\mu_{1}}\Delta_{1}-\frac{1}{2\mu_{2}}\Delta_{2}-\frac{1}{m_{3}}\nabla_{1}\cdot\nabla_{2}+\frac{q_{1}q_{2}}{r_{12}}+\frac{q_{1}q_{3}}{r_{1}}+\frac{q_{2}q_{3}}{r_{2}}}_{\hat{\mathcal{H}}} \left.\vphantom{\frac{1}{m_{3}}}\right]~\Phi(\bm{R},\bm{r}_{1},\bm{r}_{2}) = \varepsilon~\Phi(\bm{R},\bm{r}_{1},\bm{r}_{2})
    \end{equation}
\end{widetext}
where $\mu_{i}=\frac{m_{i\vphantom{y}}m_{3}}{m_{i}+m_{3}}$ are the (reduced) masses of the quasiparticles $i=1,2$ located at positions $\bm{r}_{1}$ and $\bm{r}_{2}$, respectively, while $r_1=\left| \bm{r}_{1} \right|$, $r_2=\left| \bm{r}_{2} \right|$, and  $r_{12}=\left| \bm{r}_{1} - \bm{r}_{2} \right|$. The quasiparticle associated with $\bm{R}$ is interpreted as the center of mass (CM) with mass $M$ and no charge (and, hence, no potential energy). The additive separability of the operators $\hat{\mathcal{T}}_{R}$ and $\hat{\mathcal{H}}$ in Eq.~\eqref{secm0} allows for a product ansatz
\begin{equation}
    \Phi(\bm{R},\bm{r}_{1},\bm{r}_{2}) = \Psi_{CM}(\bm{R})~\Psi(\bm{r}_{1},\bm{r}_{2})
\end{equation}
which splits the SE in Eq.~\eqref{secm0} into two RSEs coupled by a separation constant $E_{CM}$
\begin{equation}\label{secm1}
    \hat{\mathcal{T}}_R~\Psi_{CM}(\bm{R}) = E_{CM}~\Psi_{CM}(\bm{R})
\end{equation}
\begin{equation}\label{secm2}
    \hat{\mathcal{H}}~\Psi(\bm{r}_{1},\bm{r}_{2}) = \underbrace{\left( \varepsilon - E_{CM} \right)}_{E}~\Psi(\bm{r}_{1},\bm{r}_{2})
\end{equation}
The solution to Eq.~\eqref{secm1} is trivial [see, for example, Eq.~(17.8) of \textcite{Landau1977ch3}] and is not discussed further here. The kinetic energy operator within $\hat{\mathcal{H}}$ in Eq.~\eqref{secm2} consists of the mass-polarization operator $-\frac{1}{m_{3}}\nabla_{1}\cdot\nabla_{2}=-\frac{1}{m_{3}}\left(\partial_{x_{1} x_{2}}+\partial_{y_{1} y_{2}} + \partial_{z_{1} z_{2}}\right)$ in addition to the usual two Laplacian terms $\Delta_{i}=\left(\partial_{x_{i} x_{i}} + \partial_{y_{i} y_{i}} + \partial_{z_{i} z_{i}}\right)$ for $i=1,2$.

\begin{figure}[h]
    \centering
    \hspace{-0.08\columnwidth}
    \begin{minipage}[b]{0.492\columnwidth}
        \centering
        \scalebox{0.49}{\input{lab_frame1}}

        (a)
    \end{minipage}
    \hfill
    \begin{minipage}[b]{0.492\columnwidth}
        \centering
        \scalebox{0.49}{\input{lab_frame2}}

        (b)
    \end{minipage}
    \caption{(a) Graphical illustration of the position $\bm{R}$ of the center of mass (CM) and the positions $\overline{\bm{r}}_1$, $\overline{\bm{r}}_2$, and $\overline{\bm{r}}_3$ of the three original particles with respect to the origin of the laboratory-fixed ($\overline{x}\,\overline{y}\,\overline{z}$) reference frame. (b) Graphical illustration of the positions $\bm{r}_{1}$ and $\bm{r}_{2}$ of the two emerging quasiparticles with respect to the new laboratory-fixed $(x\,y\,z)$ reference frame, obtained from ($\overline{x}\,\overline{y}\,\overline{z}$) via translation by $\overline{\bm{r}}_3$.}
    \label{fig:comparison}
\end{figure}

\subsection{Angular momentum operators}
Based on the classical definition of angular momentum $\bm{l}_{i}=\bm{r}_{i}\times\bm{p}_{i}$, the total angular momentum operator $\hat{\bm{L}}_{xyz}=\hspace{1pt}\hat{\hspace{-1pt}\bm{l}}_{1}+\hspace{1pt}\hat{\hspace{-1pt}\bm{l}}_{2}=\left[L_{x},L_{y},L_{z}\right]^\mathsf{T}$ of a system of two quasiparticles in the translated laboratory-fixed ($x\,y\,z$) reference frame (see Fig.~\ref{fig:comparison}) is defined as
    \begin{equation}\label{amocc}
    \hat{\bm{L}}_{xyz} = \underbrace{-\textbf{\textit{i}}\!\left[\begin{array}{ccc}
         y_{1}\partial_{z_{1}}&\!\!\!-\!\!\!&z_{1}\partial_{y_{1}}   \\
         z_{1}\partial_{x_{1}}&\!\!\!-\!\!\!&x_{1}\partial_{z_{1}}  \\
         x_{1}\partial_{y_{1}}&\!\!\!-\!\!\!&y_{1}\partial_{x_{1}} 
    \end{array}\right]}_{\hat{\hspace{4pt}{\bm{l}}_{1}}}
    \underbrace{\!-\textbf{\textit{i}}\!\left[\begin{array}{ccc}
         y_{2}\partial_{z_{2}}&\!\!\!-\!\!\!&z_{2}\partial_{y_{2}}  \\
         z_{2}\partial_{x_{2}}&\!\!\!-\!\!\!&x_{2}\partial_{z_{2}} \\
         x_{2}\partial_{y_{2}}&\!\!\!-\!\!\!&y_{2}\partial_{x_{2}}
    \end{array}\right]}_{\hat{\hspace{4pt}{\bm{l}}_{2}}}
\end{equation}
where $\textbf{\textit{i}}^2=-1$. Direct evaluation shows that $\hat{L}^{2}$ (defined as $\hat{\bm{L}}_{xyz}\cdot\hat{\bm{L}}_{xyz}$) as well as 
all the components of $\hat{\bm{L}}_{xyz}$ commute with the Hamiltonian $\hat{\mathcal{H}}$ in Eq.~\eqref{secm2}. This fact, together with $[\hat{L}^{2}, \hat{L}_{z}]=0$, ensures that $\Psi(\bm{r}_{1}, \bm{r}_{2})$ can be defined as a simultaneous eigenfunction of $\hat{\mathcal{H}}$, $\hat{L}^{2}$, and $\hat{L}_{z}$. These eigenproperties of $\Psi(\bm{r}_{1}, \bm{r}_{2})$ allow us to separate the rotational and internal degrees of freedom. To this end, we consider the $\left(L,M\right)$-invariant subspaces, labeled by the quantum numbers $L$ and $M$ corresponding to the operators $\hat{L}^2$ and $\hat{L}_{z}$, respectively. For each such subspace, we choose a suitable angular generator basis. The Wigner functions $\mathcal{D}^{MK}_L(\alpha,\beta,\gamma)$ and the MBHs $\Omega_{l}^{LM\uppi}(\theta_{1},\phi_{1},\theta_2,\phi_2)$ are customary choices for the angular basis. To discuss the properties of $\Omega_{l}^{LM\uppi}$ and $\mathcal{D}^{MK}_L$ needed for further derivations, we first express angular momentum operators in bispherical coordinates $(r_{1},\theta_{1},\phi_{1},r_{2},\theta_{2},\phi_{2})$ and in (internal and Euler-angle) coordinates $(r_{1},r_{2},\theta,\alpha,\beta,\gamma)$.

The operator $\hat{\bm{L}}_{xyz}$ defined in Eq.~(\ref{amocc}) can be expressed in bispherical coordinates  $(r_{1},\theta_{1},\phi_{1},r_{2},\theta_{2},\phi_{2})$ following the usual relation between Cartesian and spherical coordinates
\begin{equation}\label{ctspc}
    \left[ x,y,z\right]^\mathsf{T} = \left[ r \sin\theta \cos\phi, r \sin\theta \sin\phi, r \cos\theta\right]^\mathsf{T}    
\end{equation}
yielding
    \begin{equation} \label{eq:Lbisph}
        \hat{\bm{L}}_{xyz}\! =\!\!\left[\begin{array}{r}
        \!\hat{L}_x\! \\ \!\hat{L}_y\! \\ \!\hat{L}_z\! \end{array}\right] \!\!= \textbf{\textit{i}} \sum_{i=1}^2 \!\left[\begin{array}{r}
        \!+ \sin\phi_{i}\,\partial_{\theta_{i}} \!+\frac{\cos\theta_{i}\cos{\phi_i}}{\sin\theta_{i}}\,\partial_{\phi_{i}} \!\!  \\
        \!- \cos\phi_{i}\,\partial_{\theta_{i}} \!+\frac{\cos\theta_{i}\sin{\phi_i}}{\sin\theta_{i}}\,\partial_{\phi_{i}} \!\!   \\
        -\,\partial_{\phi_{i}} \!\!
    \end{array}\right]
    \end{equation}
The bispherical representation of the operator $\hat{L}^{2}=\hat{\bm{L}}_{xyz}\cdot\hat{\bm{L}}_{xyz}$  follows directly from Eq.~\eqref{eq:Lbisph}
\begin{eqnarray}\label{eq:L2bisph}
        \hspace{-15pt}\hat{L}^2 &=& \hat{l}_{1}^{\hspace{1pt}2}\!\! +\! \hat{l}_{2}^{\hspace{1pt}2} -{\scriptstyle 2\sin\left(\phi_{1}\!-\!\phi_{2}\right)}\!\!\left(\!\tfrac{\cos\theta_{2}}{\sin\theta_{2}}\partial_{\theta_{1}\phi_{2}}\!\!-\!\tfrac{\cos\theta_{1}}{\sin\theta_{1}}\partial_{\theta_{2}\phi_{1}}\!\!\right)\!\nonumber \\[-3pt]
        && \hspace{-25pt}-2\!\left(\!1\!+\! \tfrac{\cos\theta_{1}\cos\theta_{2}}{\sin\theta_{1}\sin\theta_{2}} \,  { \scriptstyle \cos\left(\phi_{1}\!-\!\phi_{2}\right)}\!\right) \partial_{\phi_{1}\phi_{2}}\!\!-{\scriptstyle 2\cos\left(\phi_{1}\!-\!\phi_{2}\right)}\,\partial_{\theta_{1}\theta_{2}}
\end{eqnarray}
where 
\begin{equation} \label{eq:l2op}
   \hat{l}_{i}^{\hspace{1pt}2} = -\partial_{\theta_{i}\theta_{i}} - \tfrac{\cos{\theta_{i}}}{\sin\theta_{i}}\,\partial_{\theta_{i}}-\tfrac{1}{\left(\sin\theta_{i}\right)^2}\,\partial_{\phi_{i}\phi_{i}} 
\end{equation}
is the standard squared angular momentum operator for a single quasiparticle $i$.

\begin{figure}[htb]
    \centering
    \scalebox{0.8}{\input{body_frame}}
    \caption{The orientation of the body-fixed $(\hspace{-0.8pt}XY\hspace{-1pt}Z)$ reference frame is determined by two non-axial vectors $\bm{r}_{1}$ and $\bm{r}_{2}$ lying in the  $XY$ plane. The origin of the frame is located at the position of the third particle. The $X$ axis bisects the angle $\theta = \arccos(\frac{\bm{r}_{1}\!\cdot\bm{r}_{2}}{r_{1}r_{2}})$, while the $Z$ axis coincides with the vector $\bm{r}_{1}\times\bm{r}_{2}$.}
    \label{fig:body_frame}
\end{figure}
Analogous expressions in coordinates $(r_{1},r_{2},\theta,\alpha,\beta,\gamma)$ can be constructed by introducing the so-called body-fixed  $(\hspace{-0.8pt}XY\hspace{-1pt}Z)$ frame description of the two quasiparticles using three shape variables $\left(r_{1},r_{2},\theta\right)$ (for details, see Fig.~\ref{fig:body_frame}) and relating it to the laboratory-fixed $(x\,y\,z)$ frame by three Euler angles $\left(\alpha,\beta,\gamma\right)$. The positions of the quasiparticles in the  $(\hspace{-0.8pt}XY\hspace{-1pt}Z)$ frame are chosen here to be
\begin{equation} \label{bodyfixed}
    \left[ \begin{array}{c} X_{i} \\ Y_{i} \\ Z_{i} \end{array}\right] = \left[ \begin{array}{c} \phantom{(-1)^i~ }r_{i} ~\cos\left(\frac{\theta}{2}\right) \\[2pt] (-1)^i~ r_{i}~ \sin\left(\frac{\theta}{2}\right)  \\[2pt] \phantom{(-1)^i~ }0 ~\phantom{\sin\left(\frac{\theta}{2}\right)}\end{array}\right]\!\!, \,\,i=1,2    
\end{equation}
The $(\hspace{-0.8pt}XY\hspace{-1pt}Z)$ frame can be reoriented to match the $(x\,y\,z)$ frame through three consecutive Euler rotations by proper Euler angles $\left(\alpha,\beta,\gamma\right)$ associated with the following sequence of rotations
\begin{equation}\label{eulerrotation0}
    XY\!Z\! \xrightarrow{\left(Z\phantom{^{'}}\!\!,\gamma\right)} X^{'}\!Y^{'}\!Z^{'} \xrightarrow{\left(Y^{'}\!\!,\beta\right)} X^{''}\!Y^{''}\!Z^{''} \xrightarrow{\left(Z^{''}\!\!,\alpha\right)} x\,y\,z
\end{equation}
where the notation $\left(W,\vartheta\right)$ denotes a counterclockwise rotation by the angle $\vartheta$ about the axis $W$ and where $X^{'}\!Y^{'}\!Z^{'}$ and $X^{''}\!Y^{''}\!Z^{''}$ denote two intermediate coordinate frames in the transition from $(\hspace{-0.8pt}XY\hspace{-1pt}Z)$ to $\left(x\,y\,z\right)$ \cite{Biedenharn1984}. The general theory of Euler angle rotations, which restricts the values of Euler angles to $\alpha\in \left[0,2\pi\right]$, $\beta\in \left[0,\pi\right]$, and $\gamma\in \left[0,2\pi\right]$ \cite{wigner2013group}, allows us to relate the coordinates $\left[X,Y,Z\right]^\mathsf{T}$ of an arbitrary point in the $(\hspace{-0.8pt}XY\hspace{-1pt}Z)$ frame to the coordinates $\left[x,y,z\right]^\mathsf{T}$ in the $\left(x\,y\,z\right)$ frame via the following relation originating from Eq.~\eqref{eulerrotation0}
\begin{equation}\label{eulerrotation1}
    \left[\begin{array}{c} x \\ y \\ z \end{array}\right] = \bm{R}_{\texttt{z}}^{\alpha}\,\bm{R}_{\texttt{y}}^{\beta}\,\bm{R}_{\texttt{z}}^{\gamma}\left[\begin{array}{c} X \\ Y \\ Z \end{array}\right]
\end{equation}
where 
\begin{equation}\label{rmatrix}
    \bm{R}_{\texttt{z}}^{\vartheta} \!=\!\! \left[\!\!\begin{array}{ccc}
       \cos{\vartheta}  & \!\!-\!\sin{\vartheta} & \!0\\
       \sin{\vartheta} & \!\!+\!\cos{\vartheta} & \!0\\
       0 & 0 & \!1
    \end{array}\!\!\right]\!\text{ and }\bm{R}_{\texttt{y}}^{\vartheta}\! =\!\! \left[\!\!\begin{array}{ccc}
       \!+\!\cos{\vartheta}  & \!0\! & \sin{\vartheta}\\
       0 & \!1\! & 0\\
       \!-\!\sin{\vartheta}  & \!0\! & \cos{\vartheta}
    \end{array}\!\!\right]
\end{equation}
Combining Eq.~\eqref{eulerrotation1} with Eq.~\eqref{bodyfixed} yields the fol\-low\-ing transformation relation between coordinates $\left(x_{1},y_{1},z_{1},x_{2},y_{2},z_{2}\right)$ and $(r_{1},r_{2},\theta,\alpha,\beta,\gamma)$
\begin{equation}\label{ctetrans}
    \left[\begin{array}{c}
         x_{i} \\
         y_{i} \\
         z_{i} 
    \end{array}\right] = r_{i}\!\!\left[\begin{array}{l}
         +\cos\gamma_{i}\cos\beta\cos\alpha-\sin\gamma_{i}\sin\alpha  \\
         +\cos\gamma_{i}\cos\beta\sin\alpha+\sin\gamma_{i}\cos\alpha \\
         -\cos\gamma_{i}\sin\beta
    \end{array}\right]
\end{equation}
where $\gamma_{i}=\gamma+(-1)^{i}\,\frac{\theta}{2}$ and $i=1,2$. 

The explicit form of the operator $\hat{\bm{L}}_{xyz}$ in Eq.~\eqref{amocc}, expressed in coordinates $(r_{1},r_{2},\theta,\alpha,\beta,\gamma)$ using the transformation relations given by Eq.~\eqref{ctetrans}, is
\begin{equation}\label{lineuler}
    \hat{\bm{L}}_{xyz} = -\textbf{\textit{i}}\!\left[\begin{array}{c}
         -\frac{\cos\alpha\cos\beta}{\sin\beta}\,\partial_{\alpha}-\sin\alpha\,\partial_{\beta}+\frac{\cos\alpha}{\sin\beta}\,\partial_{\gamma}  \\
         -\frac{\sin\alpha\cos\beta}{\sin\beta}\,\partial_{\alpha}+\cos\alpha\,\partial_{\beta}+\frac{\sin\alpha}{\sin\beta}\,\partial_{\gamma} \\
         \phantom{-\frac{\sin\alpha\cos\beta}{\sin\beta}\,}\partial_{\alpha}\phantom{+\cos\alpha\,\partial_{\beta}+\frac{\sin\alpha}{\sin\beta}\,\partial_{\gamma} }
    \end{array}\right]
\end{equation}
and the corresponding form of the operator $\hat{L}^2$  is
\begin{equation}\label{l2eeler}
    \hat{L}^2 \!=\! \tfrac{-1}{\vphantom{{}^{\dagger}}\left(\sin\beta\right)^2}\!\left(\partial_{\alpha\alpha}\!\!+\!\partial_{\gamma\gamma}\!\!-\!2\cos\beta\,\partial_{\alpha\gamma}\right)-\tfrac{\cos\beta}{\sin\beta}\,\partial_{\beta}-\partial_{\beta\beta}
\end{equation}
For the forthcoming considerations, it is advantageous to express the components of the angular momentum operator $\hat{\bm{L}}_{xyz}$ in the body-fixed $(\hspace{-0.8pt}XY\hspace{-1pt}Z)$ frame by implicitly defining  a new operator $\hat{\bm{L}}_{XYZ}=\left[\hat{L}_{X},\hat{L}_{Y},\hat{L}_{Z}\right]^\mathsf{T}$ as
\begin{equation}\label{leularbf}
    \left[\!\!\begin{array}{c}
         \hat{L}_{x}  \\
         \hat{L}_{y} \\
         \hat{L}_{z}
    \end{array}\!\!\right] = \bm{R}_{\texttt{z}}^{\alpha}\,\bm{R}_{\texttt{y}}^{\beta}\,\bm{R}_{\texttt{z}}^{\gamma}\left[\!\!\begin{array}{c}
         \hat{L}_{X}  \\
         \hat{L}_{Y} \\
         \hat{L}_{Z}
    \end{array}\!\!\right]
\end{equation}
in analogy with Eq.~(\ref{eulerrotation1}). Using the orthogonality of the operator $\bm{R}_{\texttt{z}}^{\alpha}\,\bm{R}_{\texttt{y}}^{\beta}\,\bm{R}_{\texttt{z}}^{\gamma}$, i.e., the fact that $\left[\bm{R}_{\texttt{z}}^{\alpha}\,\bm{R}_{\texttt{y}}^{\beta}\,\bm{R}_{\texttt{z}}^{\gamma}\right]^{-1}=\bm{R}_{\texttt{z}}^{-\gamma}\,\bm{R}_{\texttt{y}}^{-\beta}\,\bm{R}_{\texttt{z}}^{-\alpha}$, it is a straightforward exercise to show that $\hat{\bm{L}}_{XYZ}$ is given by
\begin{equation}\label{LXYZeuler}
    \hat{\bm{L}}_{XYZ}= -\textbf{\textit{i}}\!\left[\!\begin{array}{l}
    +\frac{\cos\gamma\cos\beta}{\sin\beta}\,\partial_{\gamma}+\sin\gamma\,\partial_{\beta}-\frac{\cos\gamma}{\sin\beta}\,\partial_{\alpha}  \\
    -\frac{\sin\gamma\cos\beta}{\sin\beta}\,\partial_{\gamma}+\cos\gamma\,\partial_{\beta}+\frac{\sin\gamma}{\sin\beta}\,\partial_{\alpha} \\
    \phantom{-\frac{\sin\gamma\cos\beta}{\sin\beta}\,}\partial_{\gamma}
    \end{array}\!\right]\!\!
\end{equation}
Direct computation shows that $\hat{\bm{L}}_{XYZ}\!\cdot\!\hat{\bm{L}}_{XYZ}\!=\!\hat{L}^2\!=\!\hat{\bm{L}}_{xyz}\!\cdot\!\hat{\bm{L}}_{xyz}$, with the explicit form of $\hat{L}^2$ given by Eq.~(\ref{l2eeler}). The operators $\hat{L}_z$ and $\hat{L}_Z$ can be considered as the pro\-jec\-tions of the total angular momentum onto the lab\-o\-ra\-to\-ry-fixed $z$ axis and the body-fixed $Z$ axis, respectively. 

The commutation relations for the components of $\hat{\bm{L}}_{xyz}$ and $\hat{\bm{L}}_{XYZ}$ are given by $\left[\hat{L}_{v},\hat{L}_{V}\right] = 0$ for $V\in \{X,Y,Z\}$ and $v \in \{x,y,z\}$, in addition to the usual commutation relations $\left[\hat{L}^2,\hat{L}_{V}\right] =\left[\hat{L}^2,\hat{L}_{v}\right] =0$ and
\begin{equation*}\begin{array}{lll}
     \left[\hat{L}_{x},\hat{L}_{y}\right] \hspace{-3pt}=\hspace{-2pt} \textbf{\textit{i}}\hat{L}_{z}\text{, }&
    \hspace{-5pt}\left[\hat{L}_{y},\hat{L}_{z}\right] \hspace{-3pt}=\hspace{-2pt} \textbf{\textit{i}}\hat{L}_{x}\text{, }&
    \hspace{-5pt}\left[\hat{L}_{z},\hat{L}_{x}\right] \hspace{-3pt}=\hspace{-2pt} \textbf{\textit{i}}\hat{L}_{y}\\[5pt]
    \left[\hat{L}_{X},\hat{L}_{Y}\right] \hspace{-3pt}=\hspace{-2pt} -\textbf{\textit{i}}\hat{L}_{Z}\text{, }&
    \hspace{-5pt}\left[\hat{L}_{Y},\hat{L}_{Z}\right] \hspace{-3pt}=\hspace{-2pt} -\textbf{\textit{i}}\hat{L}_{X}\text{, }&
    \hspace{-5pt}\left[\hat{L}_{Z},\hat{L}_{X}\right] \hspace{-3pt}=\hspace{-2pt} -\textbf{\textit{i}}\hat{L}_{Y}\end{array}
\end{equation*}
as expected\footnote{Note, however, the anomalous factor $-1$ appearing for the components of $\hat{\bm{L}}_{XYZ}$, which is responsible for the unusual sign convention in Eq.~(\ref{ladderbodyeuler}) for $\hat{L}^{\raisebox{-0.1pt}{$\scriptstyle\pm$}}= \hat{L}_{X}\mp\textbf{\textit{i}}\hat{L}_{Y}$.}  for the components of angular momentum. The fact that $\hat{L}^2$, $\hat{L}_z$, and $\hat{L}_Z$ form a set of simultaneously commuting operators will be used in the next section to construct the Wigner functions $\mathcal{D}^{MK}_L(\alpha,\beta,\gamma)$ spanning the $\left(L,M\right)$-invariant subspaces. The construction proc\-ess becomes particularly simple if we introduce the fol\-low\-ing four ladder operators
\begin{equation}\label{ladderlabeuler}
    \hat{L}_{\pm} = \hat{L}_{x}\pm\textbf{\textit{i}}\hat{L}_{y} = e^{\pm\textbf{\textit{i}}\alpha}\!\left(\!\textbf{\textit{i}}\,\tfrac{-\partial_{\gamma}\!+\cos\hspace{-1pt}\beta \, \partial_{\alpha}}{\sin\hspace{-1pt}\beta}\pm\partial_{\beta}\!\right)
\end{equation}
\begin{equation}\label{ladderbodyeuler}
    \hat{L}^{\raisebox{-0.1pt}{$\scriptstyle\pm$}} = \hat{L}_{X}\mp\textbf{\textit{i}}\hat{L}_{Y} = e^{\pm\textbf{\textit{i}}\gamma}\!\left(\!\textbf{\textit{i}}\,\tfrac{\partial_{\alpha}\!-\cos\hspace{-1pt}\beta \, \partial_{\gamma}}{\sin\hspace{-1pt}\beta}\mp\partial_{\beta}\!\right)
\end{equation}

For completeness, we also provide below the explicit representation of the operator $\hat{\bm{L}}_{XYZ}$ in bispherical coordinates $(r_{1},\theta_{1},\phi_{1},r_{2},\theta_{2},\phi_{2})$. The connection between  $(r_{1},r_{2},\theta,\alpha,\beta,\gamma)$
and $(r_{1},\theta_{1},\phi_{1},r_{2},\theta_{2},\phi_{2})$ implicitly originating from Eqs.~(\ref{ctspc}) and~(\ref{ctetrans}) is given by
\begin{equation}\label{bpctoeulr}
\begin{aligned}
    \cos{\theta_{i}} &= -\sin{\beta}\cos{\gamma_{i}} \\
    \sin{\theta_{i}} &= [1-\sin^2{\beta}\cos^2{\gamma_{i}}]^{\frac{1}{2}} \\
    \cos{\phi_{i}} &= \tfrac{\cos{\gamma_{i}}\cos{\beta}\cos{\alpha}-\sin{\gamma_{i}}\sin{\alpha}}{\sin{\theta_{i}}} \\
    \sin{\phi_{i}} &= \tfrac{\cos{\gamma_{i}}\cos{\beta}\sin{\alpha}+\sin{\gamma_{i}}\cos{\alpha}}{\sin{\theta_{i}}}
\end{aligned}
\end{equation}
where $\gamma_{i}=\gamma+(-1)^{i}\,\frac{\theta}{2}$ for $i=1,2$. Furthermore, we consider the scalar
\begin{equation}
    \label{eq:r1dotr2} \bm{r_{1}}\!\cdot\bm{r_{2}} = r_{1}r_{2} \cos \theta 
\end{equation}
where
\begin{equation}
    \label{eq:costheta} \cos \theta \hspace{3pt} =  \cos{\theta_{1}}\cos{\theta_{2}}\!+\sin{\theta_{1}}\sin{\theta_{2}}\cos{\left(\phi_{2}\!-\!\phi_{1}\right)}
\end{equation}
and the pseudovector $\bm{Q}=\bm{r_{1}}\times\bm{r_{2}}$ 
\begin{equation} \label{eq:pseudvQ}
   \bm{Q} = r_{1}r_{2}\!\left[\!\!\begin{array}{l}
         \sin{\theta_{1}}\sin{\phi_{1}}\cos{\theta_{2}}\!-\!\sin{\theta_{2}}\sin{\phi_{2}}\cos{\theta_{1}} \\
         \sin{\theta_{2}}\cos{\phi_{2}}\cos{\theta_{1}}\!-\!\sin{\theta_{1}}\cos{\phi_{1}}\cos{\theta_{2}} \\
         \sin{\theta_{1}}\sin{\theta_{2}}\sin{\left(\phi_{2}-\phi_{1}\right)}
    \end{array}\!\!\!\right] 
\end{equation}
with components $\bm{Q}=\left[Q_{x},Q_{y},Q_{z}\right]^\mathsf{T}$ and magnitude
\begin{eqnarray} \label{eq:Qmagn}
    Q=\vert \bm{Q}\vert&=&r_{1}r_{2} \sin \theta = r_{1}r_{2} \sqrt{1-\cos^2 \theta }
\end{eqnarray}
With these definitions, the inversion of the relations in Eq.~(\ref{ctetrans}) yields
\begin{equation}\label{eq:eutocart}
\begin{aligned}
    \cos{\theta} &= \tfrac{\vphantom{|}\bm{r_{1}}\!\text{$\cdot$}\bm{r_{2}}}{r_{1}r_{2}} & \sin{\theta} &= \tfrac{\vphantom{|}Q}{r_{1}r_{2}} \\
    \cos{\tfrac{\theta}{2}} &= \sqrt{\tfrac{1}{2}\!\!+\!\!\tfrac{\vphantom{|}\bm{r_{1}}\!\text{$\cdot$}\bm{r_{2}}}{2\,r_{1}r_{2}}} & \sin{\tfrac{\theta}{2}} &= \sqrt{\tfrac{1}{2}\!\!-\!\!\tfrac{\vphantom{|}\bm{r_{1}}\!\text{$\cdot$}\bm{r_{2}}}{2\,r_{1}r_{2}}} \\
    \cos{\alpha} &= \tfrac{\vphantom{\big|}Q_{x}}{\sqrt{Q_{x}^2+Q_{y}^2}} & \sin{\alpha} &= \tfrac{\vphantom{\big|}Q_{y}}{\sqrt{Q_{x}^2+Q_{y}^2}} \\
    \cos{\beta} &= \tfrac{\vphantom{|}Q_{z}}{Q} & \sin{\beta} &= \tfrac{\sqrt{Q_{x}^2+Q_{y}^2}}{Q} \\
    \cos{\gamma} &= \tfrac{\frac{-Q\,\left(\frac{z_{\scalebox{0.4}{1}}\vphantom{y_y}}{r_{\scalebox{0.4}{1}}\vphantom{y_y}}+\frac{z_{\scalebox{0.4}{2}}\vphantom{y_y}}{r_{\scalebox{0.4}{2}}\vphantom{y_y}}\right)}{\sqrt{Q_{x}^2+Q_{y}^2}}}{2\,\sqrt{\frac{1}{2}+\frac{\vphantom{y_y}\bm{r_{1}}\!\text{$\cdot$}\bm{r_{2}}}{2\,r_{1}r_{2}}}}\hspace{10pt} & \sin{\gamma} &= \tfrac{\frac{-Q\,\left(\frac{z_{\scalebox{0.4}{1}}\vphantom{y_y}}{r_{\scalebox{0.4}{1}}\vphantom{y_y}}-\frac{z_{\scalebox{0.4}{2}}\vphantom{y_y}}{r_{\scalebox{0.4}{2}}\vphantom{y_y}}\right)}{\sqrt{Q_{x}^2+Q_{y}^2}}}{2\,\sqrt{\frac{1}{2}-\frac{\vphantom{y_y}\bm{r_{1}}\!\text{$\cdot$}\bm{r_{2}}}{2\,r_{1}r_{2}}}}
\end{aligned}
\end{equation}
Using these formulas together with Eqs.~(\ref{leularbf}) and~(\ref{eq:Lbisph}) produces, after a lengthy sequence of algebraic manipulations, the following expressions for the components of $\hat{\bm{L}}_{XYZ}$ in bispherical coordinates $(r_{1},\theta_{1},\phi_{1},r_{2},\theta_{2},\phi_{2})$ 
\begin{eqnarray}
    \hat{L}_{X} &=& \frac{-\textbf{\textit{i}}}{2\cos{\frac{\theta}{2}}}\left[\vphantom{\tfrac{\cos{\theta_{2}}}{\sin{\theta_{1}}}} \sin{\left(\phi_{2}\!-\!\phi_{1}\right)}
    \left(\sin{\theta_{2}}\partial_{\theta_{1}}\!\!-\sin{\theta_{1}}\partial_{\theta_{2}}\right)\right.\nonumber\\
    &&\left. +\tfrac{\cos{\theta_{2\vphantom{y}}}-\cos{\theta_{1}}\cos{\theta}}{\sin^{2^{\vphantom{a}}}{\theta_{1}}}\partial_{\phi_{1}}\!\!+\!\tfrac{\cos{\theta_{1}}-\cos{\theta_{2\vphantom{y}}}\cos{\theta}}{\sin^{2^{\vphantom{a}}}{\theta_{2}}}\partial_{\phi_{2}}\right] \label{eq:lXsph}\\
    \hat{L}_{Y} &=& \frac{-\textbf{\textit{i}}}{2\sin{\frac{\theta}{2}}}\left[\vphantom{\tfrac{\cos{\theta_{2}}}{\sin{\theta_{1}}}} \sin{\left(\phi_{2}\!-\!\phi_{1}\right)}
    \left(\sin{\theta_{2}}\partial_{\theta_{1}}\!\!+\sin{\theta_{1}}\partial_{\theta_{2}}\right)\right.\nonumber\\
    &&\left. +\tfrac{\cos{\theta_{2\vphantom{y}}}-\cos{\theta_{1}}\cos{\theta}}{\sin^{2^{\vphantom{a}}}{\theta_{1}}}\partial_{\phi_{1}}\!\!-\!\tfrac{\cos{\theta_{1}}-\cos{\theta_{2\vphantom{y}}}\cos{\theta}}{\sin^{2^{\vphantom{a}}}{\theta_{2}}}\partial_{\phi_{2}}\right]\label{eq:lYsph}\\
    \hspace{-10pt}\hat{L}_{Z} &=& \frac{-\textbf{\textit{i}}}{\sin{\theta}} \!\left[\sin{\left(\phi_{2}\!\!-\!\phi_{1}\right)} \left( \tfrac{\sin{\theta_{2\vphantom{y}}}}{\sin{\theta_1}}\partial_{\phi_1} \!\!+ \tfrac{\sin{\theta_{1\vphantom{y}}}}{\sin{\theta_2}}\partial_{\phi_2}\!\right)\right.\nonumber\\
    &&\left. -\tfrac{\cos{\theta_{2\vphantom{y}}}-\cos{\theta_1}\cos{\theta}}{\sin{\theta_1}}\partial_{\theta_1}\!\!
    +\!\tfrac{\cos{\theta_{1\vphantom{y}}}-\cos{\theta_2}\cos{\theta}}{\sin{\theta_2}}\partial_{\theta_2}\vphantom{\tfrac{\sin{\theta_2}}{\sin{\theta_2}}}\right] \label{eq:lZsph}
\end{eqnarray}
where $\cos \theta$ is given by Eq.~(\ref{eq:costheta}), $\sin \theta$ by Eq.~(\ref{eq:Qmagn}), and $\sin \frac{\theta}{2}$ and $\cos \frac{\theta}{2}$ by Eq.~(\ref{eq:eutocart}).

The expressions for $\hat{\bm{L}}_{xyz}$ and $\hat{\bm{L}}_{XYZ}$ in bispherical coordinates $(r_{1},\theta_{1},\phi_{1},r_{2},\theta_{2},\phi_{2})$, given in Eq.~(\ref{eq:Lbisph}) and Eqs.~(\ref{eq:lXsph})--(\ref{eq:lZsph}), respectively, can be used to construct the corresponding ladder operators $\hat{L}_{\pm}$ and $\hat{L}^{\raisebox{-0.1pt}{$\scriptstyle\pm$}}$, yielding
\begin{eqnarray}\label{ladderlabbsph}
    \hat{L}_{\pm} &=& \hat{L}_{x}\pm \textbf{\textit{i}}\hat{L}_{y} = \sum_{i=1}^{2} e^{\pm\textbf{\textit{i}}\phi_{i}} \left(\frac{\textbf{\textit{i}}\cos\theta_{i}}{\sin\theta_{i}}\partial_{\phi_{i}}\pm\partial_{\theta_{i}}\right) \\
    \label{ladderbodybsph} \hat{\!L}^{\pm} & = & \hat{L}_{X}\mp\textbf{\textit{i}}\hat{L}_{Y} =\\
    &\pm &
    \frac{e^{\pm\textbf{\textit{i}}\frac{\theta}{2}}}{\sin{\theta}} \left(\!
    \sin{\left(\phi_{1}\!-\!\phi_{2}\right)}
    \sin{\theta_{2}}\partial_{\theta_{1}} \!\!-\! \tfrac{\cos{\theta_{2\vphantom{y}}}-\cos{\theta_{1}}\cos{\theta}}{\sin^{2^{\vphantom{a}}}{\theta_{1}}}\partial_{\phi_{1}}\!\right)\nonumber\\
    &\pm & \frac{e^{\mp\textbf{\textit{i}}\frac{\theta}{2}}}{\sin{\theta}}\left(\! \sin{\left(\phi_{1}\!-\!\phi_{2}\right)}
    \sin{\theta_{1}}\partial_{\theta_{2}} \!\!+\! \tfrac{\cos{\theta_{1}}-\cos{\theta_{2\vphantom{y}}}\cos{\theta}}{\sin^{2^{\vphantom{a}}}{\theta_{2}}}\partial_{\phi_{2}}\!\right)\nonumber
\end{eqnarray}
The ladder operators $\hat{L}_{\pm}$ commute with the Hamiltonian $\hat{\mathcal{H}}$ in Eq.~\eqref{secm2} because an analogous property also holds for the components of $\hat{\bm{L}}_{xyz}$. This property of $\hat{L}_{\pm}$ is extensively used in Sec.~\ref{actHwf}, where we compute the action of the Hamiltonian $\hat{\mathcal{H}}$ on a general wave function $\Psi^{LM\uppi}(\bm{r}_{1},\bm{r}_{2})$. Note, however, that this property does not hold for the body-fixed ladder operators $\hat{L}^{\raisebox{-0.1pt}{$\scriptstyle\pm$}}$. 

\subsection{Wigner functions \texorpdfstring{$\mathcal{D}^{MK}_{L}(\alpha,\beta,\gamma)$}{D} }
%\subsection{Wigner functions \texorpdfstring{$\mathcal{D}$}{D}}
The Wigner functions $\mathcal{D}^{MK}_{L}(\alpha,\beta,\gamma)$ \cite{wigner2013group, BHATIA1964} are the simultaneous eigenfunctions of the angular momentum operators $\hat{L}^2$, $\hat{L}_{z}$, and $\hat{L}_{Z}$, as given by Eqs.~\eqref{l2eeler}, \eqref{lineuler}, and~\eqref{LXYZeuler}, respectively. They are labeled using three quantum numbers ($L$, $M$, and $K$), which correspond to these three commuting operators.  The corresponding eigenvalue equations are
\begin{equation}\label{L2LzLZdmkl}
    \begin{aligned}
    \hat{L}^{2}\,\mathcal{D}^{MK}_{L}(\alpha,\beta,\gamma) &= L(L+1)\,\mathcal{D}^{MK}_{L}(\alpha,\beta,\gamma)\\
    \hat{L}_{z}\,\mathcal{D}^{MK}_{L}(\alpha,\beta,\gamma) &= M\,\mathcal{D}^{MK}_{L}(\alpha,\beta,\gamma)\\
    \hat{L}_{Z}\,\mathcal{D}^{MK}_{L}(\alpha,\beta,\gamma) &= K\,\mathcal{D}^{MK}_{L}(\alpha,\beta,\gamma)
    \end{aligned}
\end{equation}
Since $\hat{L}_{Z}$ corresponds to rotation about the body-fixed $Z$ axis, the eigenvalue $K$ can be interpreted as the projection of the total angular momentum onto the body-fixed $Z$ axis. Similarly,  $M$ can be interpreted as the projection of the total angular momentum onto the laboratory-fixed $z$ axis. Consequently, the quantum numbers $M$ and $K$ each take on $2L+1$ distinct integer values: $M,K\in\{-L,\ldots,L\}$.

The solutions to Eqs.~\eqref{L2LzLZdmkl} can be obtained in the usual way using separation of variables and the Frobenius method. The physically meaningful solutions $\mathcal{D}^{MK}_{L}(\alpha,\beta,\gamma)$ can be expressed as
\begin{eqnarray}\label{wignerd}
    \mathcal{D}^{MK}_{L}(\alpha,\beta,\gamma)&=&\mathscr{N}^{MK}_{L}  \left(\tfrac{1+\cos{\beta}}{2}\right) ^ {\frac{\vert  K + M \vert}{2} }\left(\tfrac{1-\cos{\beta}}{2}\right) ^ {\frac{\vert  K - M \vert}{2} }\nonumber\\
    &&\hspace{-40pt}e^{\textbf{\textit{i}} M \alpha} e^{\textbf{\textit{i}} K \gamma} {}_2F_1\left[\!\begin{array}{c}
         {\scriptstyle -L+\lambda,\, L+1+\lambda} \\
         {\scriptstyle{1+\left|K+M\right|}}
    \end{array};\tfrac{1+\cos\beta}{2}\right]
\end{eqnarray}
where $\lambda = \max\left(\left|K\right|\!,\!\left|M\right|\right) = \frac{\left|K+M\right|+\left|K-M\right|}{2}$, and where ${}_2F_1$ denotes the Gauss hypergeometric function \cite{slater1966}
\begin{equation}\label{ghgf}
    {}_2F_1\left[\!\begin{array}{c}
     a,b  \\
     c
\end{array};x\right]=\sum_{k=0}^{\infty}\frac{(a)_{k}(b)_{k}}{(c)_{k}}\frac{x^k}{k!}
\end{equation}
with 
\begin{equation} \label{eq:poch}
    (\alpha)_k=\alpha\,(\alpha\hspace{-1pt}+\!1)\cdot\! \ldots\!\cdot(\alpha\hspace{-1pt}+\!k\!-\!1) 
\end{equation}
denoting the Pochhammer symbol \cite{slater1966}. Since $-L+\lambda$  is always a non-positive integer, the function ${}_2F_1$ in Eq.~\eqref{wignerd} terminates and reduces to a polynomial of degree $L-\lambda$ in the variable $\cos\beta$. The normalization constant 
\begin{eqnarray}\label{ncwignerd}
    \mathscr{N}^{MK}_{L} &=& \tfrac{\scriptstyle (-1)^{\frac{\left|K+M\right|+K-M}{2}}}{2\pi}\left[\tfrac{\left(2L+1\right)}{2}\right]^{\frac{1}{2}}\nonumber\\
    &&\!\left[\!\!\left(\!\!\!\begin{array}{c}
         \scriptstyle L+\frac{\left|K+M\right|+K-M}{2}  \\
         \scriptstyle \left|K+M\right| 
    \end{array}\!\!\!\right)\!\!\!\left(\!\!\!\begin{array}{c}
          \scriptstyle {L+\frac{\left|K+M\right|-K+M}{2}}\\
          \scriptstyle \left|K+M\right|
    \end{array}\!\!\!\right)\!\!\right]^{\frac{1}{2}}
\end{eqnarray}
orthonormalizes the functions $\mathcal{D}^{MK}_{L}\equiv \mathcal{D}^{MK}_{L}(\alpha,\beta,\gamma)$ 
\begin{equation}\label{oncwd}
    \left\langle\mathcal{D}^{M^{'}K^{'}}_{L^{'}} \right|\left.\!\vphantom{\mathcal{D}^{M^{'}K^{'}}_{L^{'}}}\mathcal{D}^{MK}_{L} \right\rangle = \delta_{LL^{'}}\delta_{MM^{'}}\delta_{KK^{'}}
\end{equation}
with respect to the following integral
\begin{equation}\label{dtau}
    \left\langle f\vphantom{g} \right|\left.\!g\vphantom{f} \right\rangle =\!\!\int\limits_{0}^{2\pi}\hspace{-4pt} d\alpha\! \int\limits_{0}^{\pi}\hspace{-3pt}\sin{\beta}\,d\beta \hspace{-3pt}\int\limits_{0}^{2\pi}\hspace{-3pt} d\gamma \,\, (f^\ast g)
\end{equation}
where $f^\ast$ denotes the complex conjugate of $f$.

Alternatively, the set of Wigner functions $\mathcal{D}_{L}^{MK}(\alpha,\beta,\gamma)$ corresponding to a definite value of $L$ can be generated recursively using the ladder operators defined in Eqs.~\eqref{ladderlabeuler} and~\eqref{ladderbodyeuler} from the maximal Wigner function $\mathcal{D}_{L}^{LL}(\alpha,\beta,\gamma)$, which is given by the particularly simple expression
\begin{equation} \label{eq:DLLL}
    \mathcal{D}_{L}^{LL}(\alpha,\beta,\gamma) \!=\! (-1)^{\!L}\, \tfrac{\sqrt{L+\frac{1}{2}}}{2\pi}\,e^{\textbf{\textit{i}}L\left(\alpha+\gamma\right)}\!\left(\tfrac{1+\cos\beta}{2}\right)^{\!\!L}
\end{equation}
The action of the ladder operators $\hat{L}_{\pm}$ and $\hat{L}^{\raisebox{-0.1pt}{$\scriptstyle\pm$}}$, originally defined by Eqs.~(\ref{ladderlabeuler}) and~(\ref{ladderbodyeuler}), on $\mathcal{D}^{MK}_{L}$ is
\begin{equation} \label{eq:Lpmdown}
    \hat{L}_{\pm}~\mathcal{D}^{MK}_{L} = \sqrt{L\left(L+1\right)-M\left(M\pm1\right)}~\mathcal{D}^{M\pm1,K}_{L}
\end{equation}
\begin{equation} \label{eq:Lpmup}
    \hat{L}^{\raisebox{-0.1pt}{$\scriptstyle\pm$}}~\mathcal{D}^{MK}_{L} = \sqrt{L\left(L+1\right)-K\left(K\pm1\right)}~\mathcal{D}^{M,K\pm1}_{L}
\end{equation}
These relations, enabled by the choice of the phase factor $(-1)^{\frac{\left|K+M\right|+K-M}{2}}$ in Eq.~\eqref{ncwignerd}, yield
\begin{equation} \label{eq:DLMKalt}
    \mathcal{D}_{L}^{MK} = b_{MK}^{-1} ~\underbrace{\hat{L}_{-}\ldots \hat{L}_{-}}_{L-M~\text{times}}\,\underbrace{\vphantom{\hat{L}_{-}}\hat{L}^{-}\ldots \hat{L}^{-}}_{L-K~\text{times}}~\mathcal{D}_{L}^{LL}
\end{equation}
where the numerical factor $b_{MK}$ is given by
\begin{eqnarray} \label{eq:bMK}
    b_{MK} &=&  {\textstyle \prod\limits_{\mu=M+1}^{L}}\sqrt{L\left(L+1\right)-\mu\left(\mu-1\right)}\\ &&{\textstyle \prod\limits_{\kappa=K+1}^{L}}\sqrt{L\left(L+1\right)-\kappa\left(\kappa-1\right)}\nonumber
\end{eqnarray}
The choice of the maximal Wigner function $\mathcal{D}_{L}^{LL}(\alpha,\beta,\gamma)$ in Eqs.~(\ref{eq:DLLL}) and~(\ref{eq:DLMKalt}) and the operators $\hat{L}_{-}$ and $\hat{L}^{-}$ in Eq.~(\ref{eq:DLMKalt}) is somewhat arbitrary; one could equally well start the construction process, for example, from the maximal Wigner function 
\begin{equation} \label{eq:DLL-L}
    \mathcal{D}_{L}^{L,-L}(\alpha,\beta,\gamma) \!=\! (-1)^{\!L}\, \tfrac{\sqrt{L+\frac{1}{2}}}{2\pi}\,e^{\textbf{\textit{i}}L\left(\alpha-\gamma\right)}\!\left(\tfrac{1-\cos\beta}{2}\right)^{\!\!L}
\end{equation}
and the operators $\hat{L}_{-}$ and $\hat{L}^{+}$.

\subsection{Parity and antisymmetry of \texorpdfstring{$\mathcal{D}_{L}^{MK}(\alpha,\beta,\gamma)$}{D} }
%\subsection{Parity and antisymmetry of Wigner functions \texorpdfstring{$\mathcal{D}$}{D}}
The parity operation corresponds to the simultaneous inversion of the coordinates of the quasiparticles. The inversion operator $\hat{\uppi}: \left[x,y,z\right]^\mathsf{T} \mapsto \left[-x,-y,-z\right]^\mathsf{T}$ has the following effect on $\bm{r}_{1}$ and $\bm{r}_{2}$ and functions thereof
\begin{equation}\label{invop}
\begin{aligned}
    \hspace{1pt}\hat{\uppi}\left(\bm{r}_{1}\right)&\rightarrow -\bm{r}_{1},~~~  \hspace{1pt}\hat{\uppi}\left(\bm{r}_{1}\!\cdot\bm{r}_{2}\right)~\rightarrow \bm{r}_{1}\!\cdot\bm{r}_{2}\\
    \hspace{1pt}\hat{\uppi}\left(\bm{r}_{2}\right)&\rightarrow -\bm{r}_{2},~~~  \hspace{1pt}\hat{\uppi}\left(\bm{r}_{1}\!\!\times\!\bm{r}_{2}\right)\rightarrow \bm{r}_{1}\!\times\bm{r}_{2}
\end{aligned}
\end{equation}
Consequently, the behavior of the sine and cosine functions of the Euler angles under the inversion operation, as derived using Eqs.~\eqref{eq:eutocart}, is given by
\begin{equation}
    \begin{aligned}
        \hspace{1pt}\hat{\uppi}\left(\sin{\alpha}\right) &\rightarrow +\sin{\alpha},~~~~ \hspace{1pt}\hat{\uppi}\left(\cos{\alpha}\right) \rightarrow +\cos{\alpha}\\
        \hspace{1pt}\hat{\uppi}\left(\sin{\beta}\right) &\rightarrow +\sin{\beta},~~~~ \hspace{1pt}\hat{\uppi}\left(\cos{\beta}\right) \rightarrow +\cos{\beta}\\
        \hspace{1pt}\hat{\uppi}\left(\sin{\gamma}\right) &\rightarrow -\sin{\gamma},~~~~ \hspace{1pt}\hat{\uppi}\left(\cos{\gamma}\right) \rightarrow -\cos{\gamma}
    \end{aligned}
\end{equation}
This suggests the following transformation properties of the Euler angles under inversion
\begin{equation}
    \hspace{1pt}\hat{\uppi}\left(\alpha\right)\rightarrow\alpha,~~~ \hspace{1pt}\hat{\uppi}\left(\beta\right)\rightarrow\beta, ~~~\hspace{1pt}\hat{\uppi}\left(\gamma\right)\rightarrow\gamma+\pi
\end{equation}
These results determine the behavior of the Wigner functions $\mathcal{D}_{L}^{MK}(\alpha,\beta,\gamma)$ under inversion as
\begin{equation} \label{eq:wignerdpar}
  \hspace{1pt}\hat{\uppi}\left(\mathcal{D}_{L}^{MK}(\alpha,\beta,\gamma)\right) \,\,\rightarrow \,\, \left(-1\right)^{K}\mathcal{D}_{L}^{MK}(\alpha,\beta,\gamma) 
\end{equation}
because, according to Eq.~(\ref{wignerd}),  $\mathcal{D}_{L}^{MK}(\alpha,\beta,\gamma+\pi) = e^{\textbf{\textit{i}} K \pi}\mathcal{D}_{L}^{MK}(\alpha,\beta,\gamma)$. This signifies that the Wigner functions $\mathcal{D}_{L}^{MK}(\alpha,\beta,\gamma)$ possess even space parity when $K$ is even, and odd space parity when $K$ is odd.

In the case where two out of three particles (say, 1 and 2) are identical, the system possesses an additional discrete transformation property: antisymmetry (symmetry) under the permutation  $\hat{P}_{{12}}$ of the identical fermions (bosons). The effect of the permutation operator $\hat{P}_{{12}}$ on $\bm{r}_{1}$ and $\bm{r}_{2}$ and functions thereof
\begin{equation}\label{permutop}
\begin{aligned}
    \hat{P}_{{12}}\left(\bm{r}_{1}\right)&\rightarrow \bm{r}_{2},~~~  \hat{P}_{{12}}\left(\bm{r}_{1}\!\cdot\bm{r}_{2}\right)~\rightarrow \phantom{-}\bm{r}_{1}\!\cdot\bm{r}_{2}\\
    \hat{P}_{{12}}\left(\bm{r}_{2}\right)&\rightarrow \bm{r}_{1},~~~  \hat{P}_{{12}}\left(\bm{r}_{1}\!\!\times\!\bm{r}_{2}\right)\rightarrow -\bm{r}_{1}\!\times\bm{r}_{2}
\end{aligned}
\end{equation}
stipulates, according to Eqs.~\eqref{eq:eutocart}, that
\begin{equation} \label{eq:sincosperm}
    \begin{aligned}
        \hat{P}_{{12}}\left(\sin{\alpha}\right) &\rightarrow -\sin{\alpha},~~~~ \hat{P}_{{12}}\left(\cos{\alpha}\right) \rightarrow -\cos{\alpha}\\
        \hat{P}_{{12}}\left(\sin{\beta}\right) &\rightarrow +\sin{\beta},~~~~ \hat{P}_{{12}}\left(\cos{\beta}\right) \rightarrow -\cos{\beta}\\
        \hat{P}_{{12}}\left(\sin{\gamma}\right) &\rightarrow -\sin{\gamma},~~~~ \hat{P}_{{12}}\left(\cos{\gamma}\right) \rightarrow +\cos{\gamma}
    \end{aligned}
\end{equation}
which implies the following transformation properties of the Euler angles $\alpha$, $\beta$, and $\gamma$ under permutation 
\begin{equation} \label{eq:abgperm}
    \hat{P}_{{12}}\!\left(\alpha\right)\!\rightarrow\alpha\!+\!\pi,\,\,\,\hat{P}_{{12}}\!\left(\beta\right)\!\rightarrow\pi\! -\! \beta, \,\,\,\hat{P}_{{12}}\!\left(\gamma\right)\!\rightarrow2\pi\!-\!\gamma
\end{equation}
Following \textcite{wigner2013group} (p.~216) and  \textcite{Edmonds1974} (Sec.~4.2), it can be established that
\begin{equation} \label{eq:wignerdperm}
  \hat{P}_{{12}}\!\left(\mathcal{D}_{L}^{MK}(\alpha,\beta,\gamma)\right) \,\,\rightarrow \,\, \left(-1\right)^{L}\mathcal{D}_{L}^{M,-K}(\alpha,\beta,\gamma) 
\end{equation}
An elementary demonstration of this fact can also be obtained from the action of the operator $\hat{P}_{{12}}$ on the Wigner function $\mathcal{D}_{L}^{MK}$ given by Eq.~(\ref{eq:DLMKalt}), and on
the operators $\hat{L}_{\pm}$ and $\hat{L}^{\raisebox{-0.1pt}{$\scriptstyle\pm$}}$ given by Eqs.~(\ref{ladderlabeuler}) and~(\ref{ladderbodyeuler})
\begin{equation} \label{eq:Lpmperm}
    \hat{P}_{{12}}\!\left(\hat{L}^{\raisebox{-0.1pt}{$\scriptstyle\pm$}}\right)\!\rightarrow \hat{L}^{\mp},\hspace{20pt}\hat{P}_{{12}}\!\left(\hat{L}_{\pm}\right)\!\rightarrow \hat{L}_{\pm}
\end{equation}
This is easily established by taking into account Eq.~(\ref{eq:sincosperm}) and the relations $\hat{P}_{{12}}\!\left(\partial_\alpha\right)\!\rightarrow\partial_\alpha$, $\hat{P}_{{12}}\!\left(\partial_\beta\right)\!\rightarrow\!-\partial_\beta$, and  $\hat{P}_{{12}}\!\left(\partial_\gamma\right)\!\rightarrow\!-\partial_\gamma$, which follow naturally from Eq.~(\ref{eq:abgperm}). We have
\[
  \hat{P}_{{12}}\!\left(\mathcal{D}_{L}^{MK}\right) = b_{MK}^{-1} ~\underbrace{\hat{L}_{-}\ldots \hat{L}_{-}}_{L-M~\text{times}}\,\underbrace{\vphantom{\hat{L}_{-}}\hat{L}^{+}\ldots \hat{L}^{+}}_{L-K~\text{times}}~\hat{P}_{{12}}\!\left(\mathcal{D}_{L}^{LL}\right) 
\]
It  follows from Eqs.~(\ref{eq:DLLL}),~(\ref{eq:sincosperm}), and~(\ref{eq:DLL-L}) that
\[
\hat{P}_{{12}}\!\left(\mathcal{D}_{L}^{LL}\right) = (-1)^{\!L}~ \mathcal{D}_{L}^{L,-L}  
\]
The proof of Eq.~(\ref{eq:wignerdperm}) follows immediately from the definition of $b_{MK}$ in Eq.~(\ref{eq:bMK}) and the usual ladder algebra for the operators $\hat{L}_\pm$ and $\hat{L}^\pm$ generated by Eqs.~(\ref{eq:Lpmup}) and~(\ref{eq:Lpmdown}), if one uses the fact that 
\[
\prod_{\kappa=K+1}^{L}\hspace{-7pt}\sqrt{L\left(L\!+\!1\right)\!-\!\kappa\left(\kappa\!-\!1\right)} = \hspace{-5pt}\prod^{-K-1}_{\kappa=-L}\hspace{-5pt}\sqrt{L\left(L\!+\!1\right)\!-\!\kappa\left(\kappa\!+\!1\right)}
\] 

\subsection{Minimal bipolar harmonics \texorpdfstring{$\Omega_{l}^{LM\uppi}(\theta_{1},\phi_{1},\theta_2,\phi_2)$}{ Omega}  }

Consider two quasiparticles with individual angular momenta $l_1$ and $l_2$. They are usually described by two spherical harmonics (SHs) $Y_{m_1}^{l_1}(\theta_1,\phi_1)$ and $Y_{m_2}^{l_2}(\theta_2,\phi_2)$ defined implicitly by the eigenequations
\begin{eqnarray}
  \hat{l}_{i}^{\hspace{1pt}2} ~Y_{m_i}^{l_i}(\theta_i,\phi_i) & = & l_{i}(l_{i}\hspace{-2pt}+\hspace{-2pt}1) ~Y_{m_i}^{l_i}(\theta_i,\phi_i) \\
  \hat{l}_{iz} ~Y_{m_i}^{l_i}(\theta_i,\phi_i) & = & m_i ~Y_{m_i}^{l_i}(\theta_i,\phi_i)
\end{eqnarray}
where $\hat{l}_{i}^{\hspace{1pt}2}$ is given by Eq.~(\ref{eq:l2op}) and $\hat{l}_{iz}=-\textbf{\textit{i}}\partial_{\phi_{i}}$ is implied by Eqs.~(\ref{amocc}) and~(\ref{eq:Lbisph}). The explicit definition of the SHs is given by
\begin{eqnarray}\label{eq:sph}
    Y_{m}^{l}(\theta,\phi) &=& N_{lm}e^{\textbf{\textit{i}}m\phi}(\sin{\theta})^{\vert m\vert}\nonumber\\
    && {}_2F_1\left[\!\!\begin{array}{c}
         {\scriptstyle -l+\left|m\right|,l+\left|m \right|+1} \\
         {\scriptstyle{1+\left|m\right|}}
    \end{array}\!\!;\tfrac{1+\cos\theta}{2}\right]
\end{eqnarray}
where $N_{lm}$ is the normalization phase factor
\begin{eqnarray}
    N_{lm} = \frac{\left(-1\right)^{l+\frac{m-\vert m\vert }{2}}}{\vert m\vert ! ~2^{1+\vert m\vert}}~\sqrt{\frac{(2l+1)(l+\vert m\vert)!}{\pi~(l-\vert m\vert)!}}
\end{eqnarray}
Coupling the individual angular momenta $l_1$ and $l_2$ produces a total angular momentum $L$ that can take any integer value from $\left| l_1 - l_2 \right|$ to $l_1+l_2$. The resulting eigenfunctions $\Omega_{l_{1} \hspace{-0.7pt}l_{2}}^{L\hspace{-0.7pt}M}\equiv \Omega_{l_{1} \hspace{-0.7pt}l_{2}\vphantom{y}}^{L\hspace{-0.7pt}M} \hspace{-0.7pt}(\theta_{1}\hspace{-0.5pt},\phi_{1}\hspace{-0.7pt},\theta_{2},\phi_{2})$, usually referred to in the literature as bispherical harmonics (BHs), have the following eigenproperties 
\begin{alignat}{6}\label{eq: mbheve}
    &\hat{L}^{\hspace{-0.7pt}2}~ & \Omega_{l_{1} \hspace{-0.7pt}l_{2}}^{L\hspace{-0.7pt}M} &= L\hspace{-0.5pt}(L\hspace{-2.5pt}+\hspace{-3pt}1)\hspace{2pt}\Omega_{l_{1}  \hspace{-0.7pt}l_{2}}^{L\hspace{-0.7pt}M} \hspace{20pt}& \hat{L}_{z}\hspace{2pt} & \Omega_{l_{1} \hspace{-0.7pt}l_{2}}^{L\hspace{-0.7pt}M} &&= M\hspace{2pt}\Omega_{l_{1} \hspace{-0.7pt}l_{2}}^{L\hspace{-0.7pt}M} \\
    &\hat{l}_{i}^{\hspace{1pt}2}\hspace{3pt} & \Omega_{l_{1} \hspace{-0.7pt}l_{2}}^{L\hspace{-0.7pt}M} &= \hspace{1pt}l_{i}\hspace{-1pt}(l_{i}\hspace{-3pt}+\hspace{-3pt}1)\hspace{2pt}\Omega_{l_{1} \hspace{-0.7pt}l_{2}}^{L\hspace{-0.7pt}M} \hspace{20pt}& \hat{\uppi}\hspace{7pt} & \Omega_{l_{1} \hspace{-0.7pt}l_{2}}^{L\hspace{-0.7pt}M} &&= (-1)^{l_1\hspace{-1.4pt}+\hspace{-0.6pt}l_2}\hspace{2pt}\Omega_{l_{1} \hspace{-0.7pt}l_{2}}^{L\hspace{-0.7pt}M} \nonumber
\end{alignat}
These BHs are constructed from products of the individual SHs using the standard angular momentum coupling scheme
\begin{equation}\label{bsheqn0}
\Omega_{l_{1} \hspace{-0.7pt}l_{2}}^{L\hspace{-0.7pt}M} = 
\sum_{\mu}
C_{l_1,\mu,l_2,M\hspace{-1.5pt}-\hspace{-1pt}\mu}^{LM}~
Y_{\mu}^{l_1}\!(\theta_{1},\!\phi_{1}\hspace{-0.7pt}) ~
Y_{M\!-\!\mu}^{l_2}\hspace{-1.5pt}(\theta_{2},\!\phi_{2}\hspace{-0.5pt})
\end{equation}
where a compact form of the Clebsch-Gordan  coefficients $C_{l_1\!,m_1\!,l_2\hspace{-0.6pt},m_2}^{LM}$ was given by \textcite{Shimpuku1963} as
\begin{eqnarray}\label{eq:Ccoef}
    &&C_{l_1\!,m_1\!,l_2\hspace{-0.6pt},m_2}^{LM} = \sqrt{\tfrac{\left(\!\!\!\begin{array}{c}
         \scriptstyle 2l_{1}  \\
         \scriptstyle l_{1}+l_{2}-L
    \end{array}\!\!\!\right)\!\!\left(\!\!\!\begin{array}{c}
         \scriptstyle  2l_{2}  \\
         \scriptstyle l_{1}+l_{2}-L
    \end{array}\!\!\!\right)}{\left(\!\!\!\begin{array}{c}
         \scriptstyle l_{1}+l_{2}+L+1  \\
         \scriptstyle l_{1}+l_{2}-L
    \end{array}\!\!\!\right)\!\!\left(\!\!\!\begin{array}{c}
         \scriptstyle 2l_{1}  \\
         \scriptstyle l_{1}-m_{1}
    \end{array}\!\!\!\right)\!\!\left(\!\!\!\begin{array}{c}
         \scriptstyle 2l_{2}  \\
         \scriptstyle l_{2}-m_{2}
    \end{array}\!\!\!\right)\!\!\left(\!\!\!\begin{array}{c}
         \scriptstyle 2L  \\
         \scriptstyle L-M
    \end{array}\!\!\!\right)}}\nonumber\\
    &&\hspace{30pt}\hspace{-5pt}\sum_{\kappa=0}^{l_{1}\!-\hspace{-0.5pt}m_{1}}\hspace{-5pt}\left(-1\right)^{\kappa}\!\!\left(\!\!\!\begin{array}{c}
         \scriptstyle l_{1}+l_{2}-L  \\
         \scriptstyle \kappa
    \end{array}\!\!\!\right)\!\!\left(\!\!\!\begin{array}{c}
         \scriptstyle l_{1}-l_{2}+L  \\
         \scriptstyle l_{1}-m_{1}-\kappa
    \end{array}\!\!\!\right)\!\!\left(\!\!\!\begin{array}{c}
         \scriptstyle -l_{1}+l_{2}+L  \\
         \scriptstyle l_{2}+m_{2}-\kappa
    \end{array}\!\!\!\right) 
\end{eqnarray}
for $\left| l_1 - l_2 \right| \leq L \leq l_1+l_2$ and $m_1+m_2=M$, with $C_{l_1\!,m_1\!,l_2\hspace{-0.6pt},m_2}^{LM} = 0$ otherwise.

\medskip

For any  values of $L$, $M$, and $\uppi=(-1)^d$, the set 
\begin{equation}
    U^{L\hspace{-0.7pt}M\hspace{-0.7pt}\uppi} = \left\{ \Omega_{l_{1} \hspace{-0.7pt}l_{2}}^{L\hspace{-0.7pt}M\hspace{-0.7pt}\uppi}: l_1,l_2 \in \mathbb{N}_0  \right\}
\end{equation}
contains infinitely many elements. However, as first observed by \textcite{Schwartz1961} (see Appendix~I of his work) and \textcite{King1967}, and later demonstrated by \textcite{Manakov1996} [see Sec.~2.2 of~their work], the set $U^{L\hspace{-0.7pt}M\hspace{-0.7pt}\uppi}$ can be generated from its finite subset 
\begin{equation} \label{eq:Ugen}
    \mathscr{U}^{L\hspace{-0.7pt}M\hspace{-0.7pt}\uppi} = \left\{\Omega_{l}^{L\hspace{-0.7pt}M\hspace{-0.7pt}\uppi}: l=d,\ldots,L \right\} \subset U^{L\hspace{-0.7pt}M\hspace{-0.7pt}\uppi}
\end{equation}
consisting of exactly $L+1-d$ generators $\Omega_{l}^{L\hspace{-0.7pt}M\hspace{-0.7pt}\uppi} \equiv \Omega_{l}^{L\hspace{-0.7pt}M\hspace{-0.7pt}\uppi}  \hspace{-0.7pt}(\theta_{1}\hspace{-0.5pt},\phi_{1}\hspace{-0.7pt},\theta_{2},\phi_{2})$, which are explicitly defined in Eq.~(\ref{bsheqn1}) below. This subset of generators is referred to as the  minimal bipolar harmonics (MBHs) \cite{Schwartz1961,MEREMIANIN2003, King1967, Manakov1998} with quantum numbers $L$ and $M$ and parity $\uppi$. The generators $\Omega_{l}^{L\hspace{-0.7pt}M\hspace{-0.7pt}\uppi}$ span the $\left(L,M,\uppi\right)$-invariant subspaces, implying that every function $\Omega_{l_{1} \hspace{-0.7pt}l_{2}}^{L\hspace{-0.7pt}M\hspace{-0.7pt}\uppi}$ from $U^{L\hspace{-0.7pt}M\hspace{-0.7pt}\uppi}$ can be expressed as a linear combination of the generators $\Omega_{l}^{L\hspace{-0.7pt}M\hspace{-0.7pt}\uppi} $ with coefficients $b_l(\cos \theta)$ depending only on the internal shape angle $\theta$ defined implicitly by Eq.~(\ref{eq:costheta}) and depicted in Fig.~\ref{fig:body_frame}.  

The parameter $d$, introduced in Eq.~(\ref{wfa1}), characterizes the parity $\uppi$ of a given state. We have
\begin{equation} \label{eq:d}
    d = \left\{ \begin{array}{llr}
        0 & \text{for }  \,\uppi\!=\!\text{n}\!\equiv\!\phantom{-}(-1)^L &\text{(natural parity)} \\
        1 & \text{for } \,\uppi\!=\!\text{u}\!\equiv\!-(-1)^L &\text{(unnatural parity)} 
    \end{array}\right.
\end{equation}
Recall that the term ``natural'' signifies  that the parity $\uppi=(-1)^L$ of a many-particle state with angular mo\-men\-tum $L$ naturally agrees with the parity $(-1)^l$ characterizing a single-particle state with angular mo\-men\-tum $l$. Note that in the process of constructing a quantum angular theory of many particles, the designation of states with natural (n) and unnatural (u) parity leads to a more transparent way of presenting the final formulas than the usual designation of states with even (e) and odd (o) parity  commonly used in the literature on the subject.

For definite values of $L$, $M$, $\uppi$, and $l$, the explicit form of the MBH is given as a linear combination of products of two spherical harmonics
\begin{equation}\label{bsheqn1}
\Omega_{l}^{LM\uppi} \!=\hspace{-8pt} 
\sum_{\mu=\mu_{\text{min}}}^{\mu_{\text{max}}}\hspace{-6pt}
C_{l\hspace{-0.8pt},\mu\hspace{-0.8pt},L\hspace{-0.6pt}+\hspace{-0.6pt}d\hspace{-0.8pt}-\hspace{-0.8pt}l\hspace{-0.8pt},M\!-\!\mu}^{LM\uppi}\,
Y_{\mu}^{l}\hspace{-1pt}(\theta_{1}\hspace{-0.5pt},\hspace{-1pt}\phi_{1}\hspace{-0.7pt}) \,
Y_{M\!-\!\mu}^{L\hspace{-0.6pt}+\hspace{-0.6pt}d\hspace{-0.8pt}-\hspace{-0.8pt}l}\hspace{-1pt}(\theta_{2},\!\phi_{2}\hspace{-0.5pt})
\end{equation}
where $\mu_{\text{min}}=-\min(l,L+d-l-M)$ and $\mu_{\text{max}}=\min(l,L+d-l+M)$, and the parity-adapted Clebsch-Gordan coefficient $C_{l_1\!,m_1\!,l_2\hspace{-0.6pt},m_2}^{LM\uppi}$ can be obtained from Eq.~(\ref{eq:Ccoef}) in the compact form 
\begin{equation}
    C_{l_1\!,m_1\!,l_2\hspace{-0.6pt},m_2}^{LM\uppi} \!\!=\!\! \sum_{\kappa=0}^{d}\!\!\tfrac{(-1)^{\kappa}\!\!\sqrt{\!\!\left(\!\!\!\begin{array}{c}
         \scriptstyle 2l_{1}  \\
         \scriptstyle d
    \end{array}\!\!\!\right)\!\!\left(\!\!\!\begin{array}{c}
         \scriptstyle 2l_{2}  \\
         \scriptstyle d
    \end{array}\!\!\!\right)}\!\!\left(\!\!\!\begin{array}{c}
         \scriptstyle 2l_{1}-d  \\
         \scriptstyle l_{1}-m_{1}-\kappa 
    \end{array}\!\!\!\right)\!\!\left(\!\!\!\begin{array}{c}
         \scriptstyle 2l_{2}-d  \\
         \scriptstyle l_{2}+m_{2}-\kappa 
    \end{array}\!\!\!\right)}{\sqrt{\!\!\left(\!\!\!\begin{array}{c}
         \scriptstyle 2L+1+d  \\
         \scriptstyle d 
    \end{array}\!\!\!\right)\!\!\left(\!\!\!\begin{array}{c}
         \scriptstyle 2l_{1}  \\
         \scriptstyle l_{1}-m_{1}
    \end{array}\!\!\!\right)\!\!\left(\!\!\!\begin{array}{c}
         \scriptstyle 2l_{2}  \\
         \scriptstyle l_{2}-m_{2}
    \end{array}\!\!\!\right)\!\!\left(\!\!\!\begin{array}{c}
         \scriptstyle 2L  \\
         \scriptstyle L-M
    \end{array}\!\!\!\right)\!\!}}
\end{equation}
The MBHs defined in this way are mutually orthonormal 
\begin{eqnarray} \label{eq:omegaorth}
    \left\langle\Omega_{l'}^{L'M'\uppi'} \right|\left.\!\vphantom{\Omega_{l'}^{L'M'\pi}}\Omega_{l}^{LM\uppi} \right\rangle = \delta_{LL'}\delta_{MM'}\delta_{ll'}\delta_{\uppi'\uppi}
\end{eqnarray}
with respect to the inner product
\begin{equation}\label{bipolnormcondunp}
    \left\langle f\vphantom{g} \right|\left.\!g\vphantom{f} \right\rangle =\!\!\int\limits_{0}^{2\pi} \hspace{-3pt}d\phi_{1\!\!}\int\limits_{0}^{\pi}\hspace{-2.5pt} \sin{\theta_{1}}\,d\theta_{1}\!\!\int\limits_{0}^{2\pi} \hspace{-3pt}d\phi_{2}\!\!\int\limits_{0}^{\pi}\hspace{-2.5pt} \sin{\theta_{2}}\,d\theta_{2}\,\, (f^\ast g)
\end{equation}
An interested reader might find it useful to know that the MBHs \emph{are not}  mutually orthonormal 
\begin{equation}\label{alpikk2a1}
    \left\langle\hspace{-2pt}\Omega_{l'}^{L\hspace{-0.5pt}'M'\uppi'} \right|\left.\!\vphantom{\Omega_{l'}^{L'M'\pi}}\Omega_{l}^{LM\uppi} \hspace{-2pt}\right\rangle\hspace{-2pt}=\hspace{-1pt}\delta_{LL\hspace{-0.5pt}'}\delta_{MM'}\delta_{\uppi'\uppi}\hspace{1pt}\left\langle\hspace{-2pt}\vphantom{\Omega_{l'}^{L'M'\pi}}\Omega_{l\hspace{-0.3pt}'\hspace{-0.3pt}}^{LM\uppi}\right|\left.\vphantom{\Omega_{l'}^{L'M'\pi}}\Omega_{l}^{LM\uppi}\hspace{-2pt}\right\rangle
\end{equation}
with respect to the inner product given by Eq.~(\ref{dtau}), where $\left\langle\hspace{-2pt}\left.\Omega_{l\hspace{-0.3pt}'\hspace{-0.3pt}}^{LM\uppi}\right|\Omega_{l}^{LM\uppi}\right\rangle$ is given later in Eq.~(\ref{alpikk2a}).
This relation plays a crucial role in the elimination of angular dependence from the SE in Secs.~\ref{sec:eleangdep} and~\ref{sec:angint}.

A natural extension of the MBHs $\Omega_{l}^{LM\uppi}$ given by Eq.~(\ref{bsheqn1}) is the set of solid minimal bipolar harmonics (SMBHs) $\bm{\Omega}_{l}^{LM\uppi} \hspace{-2pt}\equiv \bm{\Omega}_{l}^{LM\uppi}(\bm{r}_{1},\bm{r}_{2})$, defined as 
\begin{equation}\label{nsbphgen}
    \bm{\Omega}_{l}^{LM\uppi} = r_{1}^l r_{2}^{L-l+d}~\Omega_{l}^{LM\uppi}(\theta_{1},\phi_{1},\theta_{2},\phi_{2})
\end{equation}
The relationship between the MBHs and the SMBHs is analogous to that between the SHs and the solid SHs. For example, the SMBHs are harmonic functions of the operators $\Delta_1$, $\Delta_2$, and $\nabla_1 \cdot \nabla_2$ [for details, see Eqs.~(17) and~(18) of \textcite{Efros1986} and also Eqs.~(\ref{ldoafi0}) and~(\ref{ldoaf30})], just as the solid SHs are harmonic functions of the Laplace operator $\Delta$.

\subsection{Properties of the minimal bipolar harmonics and the solid minimal bipolar harmonics}

Using $\Omega_{l}^{LM\uppi}$ as a shorthand notation for both the minimal bipolar harmonics $\Omega_{l}^{LM\uppi}(\theta_{1}\hspace{-0.5pt},\phi_{1}\hspace{-0.7pt},\theta_{2},\phi_{2})$ and the solid minimal bipolar harmonics $\bm{\Omega}_{l}^{LM\uppi}(\bm{r}_{1},\bm{r}_{2})$, it is straightforward to verify that both the MBHs and the SMBHs possess the following eigenvalue and symmetry properties
\begin{alignat}{2}\label{eq: mbheprop}
    &\hat{l}_{1}^{\hspace{1pt}2}\hspace{3pt} &\Omega_{l}^{L\hspace{-0.7pt}M\hspace{-0.7pt}\uppi} &= \hspace{1pt}l(l\hspace{-2.5pt}+\hspace{-3pt}1)\hspace{2pt}\Omega_{l}^{L\hspace{-0.7pt}M\hspace{-0.7pt}\uppi}  \nonumber \\
    &\hat{l}_{2}^{\hspace{1pt}2}\hspace{3pt} &\Omega_{l}^{L\hspace{-0.7pt}M\hspace{-0.7pt}\uppi} &= \hspace{1pt}\left(L\hspace{-2.2pt}-\hspace{-2.2pt}l\hspace{-2.2pt}+\hspace{-2pt}d\right)\hspace{-1pt}(L\hspace{-2.2pt}-\hspace{-2.2pt}l\hspace{-2.2pt}+\hspace{-2pt}d\hspace{-2.5pt}+\hspace{-3pt}1)\hspace{2pt}\Omega_{l}^{L\hspace{-0.7pt}M\hspace{-0.7pt}\uppi}   \nonumber\\
    &\hat{L}^{\hspace{-0.7pt}2}             &\Omega_{l}^{L\hspace{-0.7pt}M\hspace{-0.7pt}\uppi} &= L(L\hspace{-2.2pt}+\hspace{-3pt}1)\hspace{2pt}\Omega_{l}^{L\hspace{-0.7pt}M\hspace{-0.7pt}\uppi}\\  
    &\hat{L}_{z}\hspace{2pt}                &\Omega_{l}^{L\hspace{-0.7pt}M\hspace{-0.7pt}\uppi} &= M\hspace{2pt}\Omega_{l}^{L\hspace{-0.7pt}M\hspace{-0.7pt}\uppi} \nonumber \\
    &\hat{P}_{{12}}\hspace{1pt}             &\Omega_{l}^{L\hspace{-0.7pt}M\hspace{-0.7pt}\uppi} &= (-1)^d\hspace{1pt}\Omega_{L-l+d}^{L\hspace{-0.7pt}M\hspace{-0.7pt}\uppi} \nonumber\\
    &\hat{\uppi}\hspace{5pt}                &\Omega_{l}^{L\hspace{-0.7pt}M\hspace{-0.7pt}\uppi} &= (-1)^{L+d}\hspace{2pt}\Omega_{l}^{L\hspace{-0.7pt}M\hspace{-0.7pt}\uppi} \nonumber
\end{alignat}
For the MBHs, the results involving the angular momentum and parity operators follow from Eqs.~(\ref{eq: mbheve}) combined with the inclusion relation $\mathscr{U}^{L\hspace{-0.7pt}M\hspace{-0.7pt}\uppi}  \subset U^{L\hspace{-0.7pt}M\hspace{-0.7pt}\uppi}$. For the SMBHs, these relations follow from the analogous properties of the MBHs, combined with the fact that the angular momentum operators and $\hat{\uppi}$ commute with $r_1$ and $r_2$. The action of the exchange operator $\hat{P}_{{12}}$  on the MBHs and the SMBHs follows directly from their definitions in Eqs.~(\ref{bsheqn1}) and~(\ref{nsbphgen}), respectively. 

The action of the ladder operators $\hat{L}_{\pm}$ defined by Eq.~(\ref{ladderlabbsph}) on the MBHs and the SMBHs is given by 
\begin{equation} \label{eq:ladderM}
    \hat{L}_{\pm}~\Omega_{l}^{LM\uppi} = \sqrt{L\left(L+1\right)-M\left(M\pm1\right)}~\Omega_{l}^{L,M\hspace{-0.8pt}\pm\hspace{-0.7pt} 1,\hspace{0.5pt}\uppi}
\end{equation}
These relations follow from the commutation relations $[\hat{L}^2,\hat{L}_{\pm}]=0$, $[\hat{L}_z,\hat{L}_{\pm}]=\pm\hat{L}_{\pm}$, and $[\hat{L}_+,\hat{L}_-]=2\hat{L}_z$ obeyed by the  operators $\hat{L}^2$, $\hat{L}_z$, and $\hat{L}_{\pm}$, the normalization condition for the MBHs given by Eq.~(\ref{eq:omegaorth}), the standard algebra of angular momentum operators and their normalized eigenstates [cf.\ Eqs.~(2.3.16) and~(2.3.17) of \textcite{Edmonds1974}], and the Condon-Shortley phase convention implicitly woven into the definition of the Clebsch-Gordan  coefficients  \cite{Shimpuku1963}
in Eq.~(\ref{eq:Ccoef}).

Note that the ladder operators $\hat{L}^{\raisebox{-0.1pt}{$\scriptstyle\pm$}}$ defined in Eq.~(\ref{ladderbodybsph}) do not play any pronounced role in the theory of the MBHs and the SMBHs. This is mainly because the generators $\Omega_{l}^{LM\uppi}$ \emph{are not} eigenfunctions of the operator $\hat{L}_Z$ defined in Eq.~(\ref{eq:lZsph}), in contrast to the Wigner functions $\mathcal{D}_{L}^{M^{\vphantom{l}}K}(\alpha,\beta,\gamma)$. This implies that every $\Omega_{l}^{L\hspace{-0.7pt}M\hspace{-0.7pt}\uppi^{\vphantom{l}}}  \hspace{-0.7pt}(\theta_{1}\hspace{-0.5pt},\phi_{1}\hspace{-0.7pt},\theta_{2},\phi_{2})$ with $l\in\left\{d,\ldots,L \right\}$ corresponds to a linear combination of Wigner functions $\mathcal{D}_{L}^{M^{\vphantom{l}}K}(\alpha,\beta,\gamma)$ with $K\in \left\{ -L,\ldots,L \right\}$ and with coefficients that are functions of $\theta$; a detailed discussion of this dependence is given in the next paragraph. 

\subsection{Relation between Wigner functions \texorpdfstring{$\mathcal{D}_{L}^{MK}(\alpha,\beta,\gamma)$}{D} and minimal bipolar harmonics \texorpdfstring{$\Omega_{l}^{LM\uppi}(\theta_{1},\phi_{1},\theta_2,\phi_2)$}{Omega} }

Both sets of generators, 
$\left\{\mathcal{D}_{L}^{MK}: K=-L,\ldots,L\right\}$
and
$\left\{\Omega_{l}^{L\hspace{-0.7pt}M\hspace{-0.7pt}\text{n}}  \hspace{-0.7pt}: l=0,\ldots,L\right\} \cup \left\{\Omega_{l}^{L\hspace{-0.7pt}M\hspace{-0.7pt}\text{u}}  \hspace{-0.7pt}: l=1,\ldots,L\right\}$, span equivalent $(2L+1)$-dimensional, $(L,M)$-invariant subspaces com\-pris\-ing functions of both natural and unnatural parity with definite values of $L$ and $M$. Consequently, it is possible to express one set of generators as a linear combination of the other, and vice versa. In what follows, we need to express a general MBH $\Omega_{l}^{LM\uppi}(\theta_{1},\phi_{1},\theta_{2},\phi_{2})$  as a linear combination of the Wigner functions $\mathcal{D}_{L}^{MK}(\alpha,\beta,\gamma)$. Such an expansion can be written as
\begin{equation}\label{mbhwdr}
    \Omega_{l}^{LM\uppi}(\theta_{1},\phi_{1},\theta_{2},\phi_{2}) =\!\!\! \sum_{K=-L}^{L}\!\!\!\Lambda_{Kl}^{L\uppi}\!\left(\theta\right)\mathcal{D}_{L}^{MK}\!\!\left(\alpha,\beta,\gamma\right)
\end{equation}
where the expansion coefficients are defined as \begin{equation}\label{intlambda}
    \Lambda_{Kl}^{L\uppi}\left(\theta\right) = \left\langle\left.\mathcal{D}_{L}^{MK}\right|\Omega_{l}^{LM\uppi}\right\rangle
\end{equation}
using the inner product  $\langle \cdot| \cdot \rangle$ defined in Eq.~\eqref{dtau}. In Sec.~I of Supplemental Material \cite{supplemental}, we demonstrate that the coefficients $\Lambda_{Kl}^{L\uppi}\left(\theta\right)$ are solely functions of $\theta$. Note that, owing to the parity symmetry properties of the Wigner functions $\mathcal{D}_{L}^{MK}\!\!\left(\alpha,\beta,\gamma\right)$ [see Eq.~(\ref{eq:wignerdpar}) for details], the summation in Eq.~(\ref{mbhwdr}) involves only even values of $K$ for MBHs with even parity,  and only odd values of $K$ for MBHs with odd parity. This signifies that approximately half of the coefficients $\Lambda_{Kl}^{L\uppi}\left(\theta\right)$ in Eq.~(\ref{mbhwdr}) are identically zero. 

The coefficients $\Lambda_{Kl}^{L\uppi}\left(\theta\right)$ do  not  depend explicitly on $M$. This can be easily demonstrated by substituting the following identity originating from Eq.~\eqref{eq:ladderM}
\begin{equation}\label{eq:mlumbh}
    \Omega_{l}^{LM\uppi} = \tfrac{\overbrace{\scriptstyle{\hat{L}_{-}\ldots\hat{L}_{-}}}^{\text{$L
    \!\!-\!\!M\!$ times}}\,\overbrace{\scriptstyle{\hat{L}_{+}\ldots\hat{L}_{+}}}^{\text{$L
    \!\!-\!\!M\!$ times}}}{\prod\limits_{{\scriptscriptstyle \mu=M}}^{{\scriptscriptstyle L-1}}\!\!\left[L(L+1)-\mu(\mu+1)\right]} \,\Omega_{l}^{LM\uppi}
\end{equation}
into Eq.~(\ref{intlambda}) and applying the relation $\hat{L}_{-}^\dagger=\hat{L}_{+}$ [see Eq.~(\ref{ladderlabeuler})]. We have
\begin{eqnarray}
    \left\langle\left.\mathcal{D}_{L}^{MK} \right|\Omega_{l}^{LM\uppi}\right\rangle\nonumber \!&=&\!\left\langle\vphantom{\tfrac{\scriptstyle{\hat{L}_{+}\ldots\hat{L}_{+}}\mathcal{D}_{L_{\vphantom{y}}}^{MK}}{\prod\limits_{{\scriptscriptstyle \mu=M}}^{{\scriptscriptstyle L-1}}\sqrt{L(L+1)-\mu(\mu+1)}}}\mathcal{D}_{L}^{MK} \right| \left.\tfrac{\overbrace{\scriptstyle{\hat{L}_{-}\ldots\hat{L}_{-}}}^{\text{$L
    \!\!-\!\!M\!$ times}}\,\overbrace{\scriptstyle{\hat{L}_{+}\ldots\hat{L}_{+}}}^{\text{$L
    \!\!-\!\!M\!$ times}}\Omega_{l}^{LM\uppi}}{\prod\limits_{{\scriptscriptstyle \mu=M}}^{{\scriptscriptstyle L-1}}\!\!\left[L(L+1)-\mu(\mu+1)\right]}\right\rangle \\
    &&\hspace{-40pt}=
    \left\langle \vphantom{\tfrac{\scriptstyle{\hat{L}_{+}\ldots\hat{L}_{+}}\mathcal{D}_{L_{\vphantom{y}}}^{MK}}{\prod\sqrt{L(L+1)-\mu(\mu+1)}}} \right.
    \hspace{-5pt}
    \tfrac{\overbrace{\scriptstyle{\hat{L}_{+}\ldots\hat{L}_{+}}}^{L-M~\text{times}}\mathcal{D}_{L_{\vphantom{y}}}^{MK}}{\prod\limits_{{\scriptscriptstyle \mu=M}}^{{\scriptscriptstyle L-1}}\hspace{-1pt}\sqrt{L(L+\hspace{-1pt}1)-\mu(\mu+\hspace{-1pt}1)}}\hspace{-3pt}
    \left.\vphantom{\tfrac{\scriptstyle{\hat{L}_{+}\ldots\hat{L}_{+}}\mathcal{D}_{L_{\vphantom{y}}}^{MK}}{\prod\limits_{{\scriptscriptstyle \mu=M}}^{{\scriptscriptstyle L-1}}\hspace{-1pt}\sqrt{L(L+1)-\mu(\mu+1)}}}\right|
    \!\tfrac{\overbrace{\scriptstyle{\hat{L}_{+}\ldots\hat{L}_{+}}}^{L-M~\text{times}}\Omega_{l_{\vphantom{y}}}^{LM\uppi}}{\prod\limits_{{\scriptscriptstyle \mu=M}}^{{\scriptscriptstyle L-1}}\hspace{-1pt}\sqrt{L(L+\hspace{-1pt}1)-\mu(\mu+\hspace{-1pt}1)}}\hspace{-6pt}
    \left. \vphantom{\tfrac{\scriptstyle{\hat{L}_{+}\ldots\hat{L}_{+}}\mathcal{D}_{L_{\vphantom{y}}}^{MK}}{\prod\hspace{-1pt}\sqrt{L(L+1)-\mu(\mu+1)}}}\,\,\right\rangle \nonumber\\[5pt]
     &&\hspace{-40pt}=\left\langle\left.\mathcal{D}_{L}^{LK} \right|\Omega_{l}^{LL\uppi}\right\rangle =  \Lambda_{Kl}^{L\uppi}\left(\theta\right) \label{eq:Mind}
\end{eqnarray}
which confirms that $\Lambda_{Kl}^{L\uppi}\left(\theta\right)$ does not depend on the initial value of $M$, as claimed above.

The detailed derivation of the explicit form of $\Lambda_{Kl}^{L\uppi}\left(\theta\right)$ defined by Eq.~\eqref{intlambda} is presented in Sec.~I of Supplemental Material \cite{supplemental}. Here, we briefly note that the derivation is based on expressing $\Omega_{l}^{LL\uppi}$ [see Eqs.~\eqref{mbhsLL} and~\eqref{mbhsLL1}] in terms of the Euler angles $\left(\alpha, \beta, \gamma\right)$ using the relations provided by Eqs.~\eqref{bpctoeulr}. The integral is then evaluated by elementary integration over the Euler angles, followed by a series of algebraic transformations and simplifications. This process yields
\begin{eqnarray}\label{eq:Lambdapcs}  
   \Lambda_{Kl}^{L\uppi}(\theta) &=& \left\{ \begin{array}{ll}
      \overline{\Lambda}_{Kl}^{L\uppi}(\theta)  & \text{ when } L-K+ d \text{ is even} \\[5pt]
      0  & \text{ when } L-K+ d \text{ is odd}
   \end{array}\right. 
\end{eqnarray}
with a parity-universal compact formula
\begin{eqnarray}\label{Lambdapi}
    \overline{\Lambda}_{Kl}^{L\uppi}\hspace{-2pt}\left(\!\theta\hspace{-1pt}\right)\hspace{-2pt} &=& \hspace{-2pt}\tfrac{8\pi^{2}\mathscr{N}^{LK}_{L_{\vphantom{a}}} N_{l}^{L\uppi}}{2L+1+d}\hspace{-4pt}\sum_{\eta,\zeta=0}^{d}\hspace{-6pt}\sum_{\texttt{j}=0}^{\frac{L}{2}\hspace{-1pt}-\hspace{-1pt}\frac{K}{2}\hspace{-1pt}+\hspace{-1pt}\frac{d}{2}\hspace{-1pt}-\hspace{-0.5pt}\eta}\hspace{-6pt}\tfrac{(-1)^{\frac{L}{2}\hspace{-1pt}-\hspace{-1pt}\frac{K}{2}\hspace{-1pt}-\hspace{-1pt}\frac{d}{2}\hspace{-1pt}-\hspace{-0.5pt}\eta\hspace{-0.5pt}+\hspace{-0.5pt}\zeta}}{\left(\hspace{-4pt}\begin{array}{c}
        \scriptscriptstyle 2L+d \\[2pt]
        \scriptscriptstyle L+K+\eta 
    \end{array}\hspace{-4pt}\right)}\\
    &&\hspace{-10pt} \left(\hspace{-4pt}\begin{array}{c}
        \scriptstyle l-\zeta \\
        \scriptstyle \texttt{j} 
    \end{array}\hspace{-4pt}\right)\hspace{-4pt}\left(\hspace{-4pt}\begin{array}{c}
         \scriptstyle L-l+\zeta \\
         \scriptstyle \frac{L}{2}\hspace{-1pt}-\hspace{-1pt}\frac{K}{2}\hspace{-1pt}+\hspace{-1pt}\frac{d}{2}\hspace{-1pt}-\hspace{-0.5pt}\eta\hspace{-0.5pt}-\hspace{-1pt}\texttt{j} 
    \end{array}\hspace{-4pt}\right) e^{\textit{\textbf{i}}\hspace{1pt}\left(4\texttt{j}-2l+K+2\eta-(-1)^{\zeta}2\eta\right)\frac{\theta}{2}}\nonumber
\end{eqnarray}
where $\mathscr{N}^{LK}_{L}$ follows from Eq.~\eqref{ncwignerd} setting $M=L$, and $ N_{l}^{L\uppi}$ is given by Eq.~\eqref{nlpik}.
The explicit form of $\overline{\Lambda}_{Kl}^{L\uppi}\hspace{-2pt}\left(\!\theta\hspace{-1pt}\right)$ in Eq.~\eqref{Lambdapi} can further be simplified to
\begin{eqnarray}\label{Lambdan}
    \overline{\Lambda}_{Kl}^{L\text{n}}\hspace{-2pt}\left(\!\theta\hspace{-1pt}\right)\hspace{-2pt} &=& \hspace{-2pt}\left(\!-1\!\right)^{\frac{L-K}{2}}\hspace{-4pt} \left[\hspace{-2pt}\tfrac{\left(\!L+\!K\!\right)!\left(\!L-\!K\!\right)!}{2\left(\!2L+\!1\!\right)!}\tfrac{\left(\!\frac{3}{2}\!\right)_{\!l}\hspace{-2pt}\left(\!\frac{3}{2}\!\right)_{\!L-l}}{l!\left(L-l\right)!}\hspace{-2pt}\right]^{\frac{1}{2}}\hspace{-2pt} \\
    &&\hspace{0pt} \hspace{-2pt}\sum\limits_{\texttt{j}=0}^{\frac{L-K}{2}}\hspace{-2pt}\left(\hspace{-4pt}\begin{array}{c}
        \scriptstyle l \\
        \scriptstyle \texttt{j} 
    \end{array}\hspace{-4pt}\right)\hspace{-4pt}\left(\hspace{-4pt}\begin{array}{c}
         \scriptstyle L-l \\
         \scriptstyle \frac{L-K}{2}\!-\!\texttt{j} 
    \end{array}\hspace{-4pt}\right) e^{\textit{\textbf{i}}\hspace{1pt}\left(4\texttt{j}-2l+K\right)\frac{\theta}{2}}\nonumber
\end{eqnarray}
for natural parity, and
\begin{eqnarray}\label{Lambdau}
    \overline{\Lambda}_{Kl}^{L\text{u}}\hspace{-2pt}\left(\!\theta\hspace{-1pt}\right)\hspace{-2pt} &=& \hspace{-2pt}\left(\!-1\!\right)^{\frac{L-K+1}{2}}\hspace{-4pt} \left[\hspace{-2pt}\tfrac{\left(\!L+\!K\!\right)!\left(\!L-\!K\!\right)!}{\left(\!L+\!1 \!\right)\left(\!2L+\!1\!\right)!}\tfrac{\left(\!\frac{3}{2}\!\right)_{\!l}\hspace{-2pt}\left(\!\frac{3}{2}\!\right)_{\!L-l+1}}{\left(l-1\right)!\left(L-l\right)!}\hspace{-2pt}\right]^{\frac{1}{2}}\hspace{-2pt} \\
    &&\hspace{-20pt} \hspace{-6pt}\sum\limits_{\texttt{j}=0}^{\frac{L-K+1}{2}}\hspace{-6pt}\tfrac{2\texttt{j}\left(L+1\right)-l\left(L-K+1\right)}{2l\left(L-l+1\right)}\left(\hspace{-4pt}\begin{array}{c}
        \scriptstyle l \\
        \scriptstyle \texttt{j} 
    \end{array}\hspace{-4pt}\right)\hspace{-4pt}\left(\hspace{-4pt}\begin{array}{c}
         \scriptstyle L-l+1 \\
         \scriptstyle \frac{L-K+1}{2}\!-\!\texttt{j} 
    \end{array}\hspace{-4pt}\right) e^{\textit{\textbf{i}}\hspace{1pt}\left(4\texttt{j}-2l\hspace{-1pt}+\hspace{-1pt}K\right)\frac{\theta}{2}}\nonumber
\end{eqnarray}
for unnatural parity.  

Previous attempts to derive an analytical relationship between MBHs and Wigner functions $\mathcal{D}_{L}^{MK}(\alpha,\beta,\gamma)$ have not been entirely successful: 
\begin{itemize}
    \item \textcite{Pont1995} con\-sid\-ered an expansion analogous to Eq.~(\ref{mbhwdr}) and found a closed-form expression for their expansion coefficients, $C^{K}_{l_1,l_2}(\theta)$,  given in Eqs.~(A11)--(A14) of \textcite{Pont1995}. However, this formula yields inconsistencies when compared to values of $C^{K}_{l_1,l_2}(\theta)$ computed directly from Eqs.~(A22) and~(A23) of \textcite{Pont1995}. An analysis of this discrepancy is presented in Sec.~II of Supplemental Material \cite{supplemental}. 
    \item \textcite{Nikitin1985a,Nikitin1985b} derived a recurrence relation [see Eq.~(4.7) of \textcite{Nikitin1985a}] that allows for the determination of $\Lambda_{Kl}^{L\text{n}}(\theta)$ [denoted as $F^{LK}_{l_{1}l_{2}}(\theta)$ in \cite{Nikitin1985a}] for arbitrary values of $L$ and $K$, with the recurrence boundary conditions given by Eqs.~(4.1) and~(4.8) of \textcite{Nikitin1985a}. The resulting coefficients $F^{LK}_{l_{1}l_{2}}(\theta)$ can be used in conjunction with Eq.~(3.11) of \textcite{Nikitin1985a} in a manner analogous to Eq.~\eqref{mbhwdr}. However, two drawbacks of this approach are apparent: (i)~no general solution for the recurrence relation in Eq.~(4.7) of \textcite{Nikitin1985a} is offered, and (ii)~the computational effort required to determine all necessary coefficients $F^{LK}_{l_{1}l_{2}}(\theta)$ for $K=0,\ldots,L$ grows steeply with $L$. Despite these limitations, the method proposed by \textcite{Nikitin1985a} can be used to determine the connection between (general) bispherical harmonics (BHs) $\Omega_{l_{1} \hspace{-0.7pt}l_{2}}^{L\hspace{-0.7pt}M}\equiv \Omega_{l_{1} \hspace{-0.7pt}l_{2}\vphantom{y}}^{L\hspace{-0.7pt}M} \hspace{-0.7pt}(\theta_{1}\hspace{-0.5pt},\phi_{1}\hspace{-0.7pt},\theta_{2},\phi_{2})$, whereas the equations derived herein correspond only to a subset of these functions, namely the minimal bipolar harmonics (MBHs) $\Omega_{l}^{LM\uppi}(\theta_{1},\phi_{1},\theta_{2},\phi_{2})$. For this specific subset, the recursive approach of \textcite{Nikitin1985a} yields results consistent with Eqs.~(\ref{eq:Lambdapcs}) and~(\ref{Lambdan}), as verified by us across a set of test cases.
\end{itemize}

%\subsection{Specialized variants of MBHs  with \texorpdfstring{$M=L$}{M=L} }\label{sec:mbhmeql}
\subsection{Specialized variants of minimal bipolar harmonics with \texorpdfstring{$M=L$}{M=L}}\label{sec:mbhmeql}
    
We show in the next section that the derivation of the reduced Schr\"{o}dinger equations (RSEs) for the partial wave coefficients $\psi_{l}^{L\uppi}(r_{1},r_{2},r_{12})$ can be performed for an arbitrary $M$, always yielding the same set of RSEs. The choice of $M=L$ considerably simplifies the an\-a\-lyt\-i\-cal form of the MBHs and, consequently, also the derivation of the RSEs. For the ensuing discussion, it is advantageous to explicitly consider four variants of specialized MBHs with $M=L$, each possessing slightly different properties:

(1)~The usual minimal bipolar harmonics (MBHs) $\Omega_{l}^{L\hspace{-0.7pt}L\hspace{-0.7pt}\uppi} \equiv \Omega_{l}^{L\hspace{-0.7pt}L\hspace{-0.7pt}\uppi} \hspace{-0.7pt}(\theta_{1}\hspace{-0.5pt},\phi_{1}\hspace{-0.7pt},\theta_{2},\phi_{2})$, given by Eq.~(\ref{bsheqn1}) subject to the extra condition $M\!=\!L$. They are normalized according to Eq.~(\ref{eq:omegaorth}) and depend only on the angular variables. The specialization of $M\!=\!L$ allows us to express the MBHs in a particularly simple form 
\begin{eqnarray}\label{mbhsLL}
    \Omega_{l}^{L\hspace{-0.7pt}L\hspace{-0.7pt}\text{n}}  \hspace{-0.7pt} &\!=& N_{l}^{L\text{n}}\left(\sin{\theta_{1}} \, e^{\textbf{\textit{i}}\phi_{1}}\right)^{\!l}
    \left(\sin{\theta_{2}} \, e^{\textbf{\textit{i}}\phi_{2}}\right)^{\!L-l}\\
    \Omega_{l}^{L\hspace{-0.7pt}L\hspace{-0.7pt}\text{u}}  \hspace{-0.7pt} &\!=& N_{l}^{L\text{u}}\left[\left(\sin{\theta_{1}} \, e^{\textbf{\textit{i}}\phi_{1}}\right)^{\!l}
    \left(\sin{\theta_{2}} \, e^{\textbf{\textit{i}}\phi_{2}}\right)^{\!L-l}\cos{\theta_{2}}\right. \nonumber\\
    &-& \left.\!\cos{\theta_{1}}\left(\sin{\theta_{1}} \, e^{\textbf{\textit{i}}\phi_{1}}\right)^{\!l-1}
    \left(\sin{\theta_{2}} \, e^{\textbf{\textit{i}}\phi_{2}}\right)^{\!L-l+1}\right] \label{mbhsLL1}
\end{eqnarray}
with the normalizing phase factors $N_{l}^{L\uppi}$ given by
\begin{equation}\label{nlpik}
    N_{l}^{L\uppi} = \frac{(-1)^L}{4\pi}\sqrt{\frac{2^d\left(L+1-d\right)!\left(\frac{3}{2}\right)_{l}\left(\frac{3}{2}\right)_{L-l+d}}{(l-d)!(L-l)!(L+1)!}}
\end{equation}
where $(\alpha)_k$ denotes the Pochhammer symbol given by Eq.~(\ref{eq:poch}), and where $d$ is defined in Eq.~(\ref{eq:d}).

(2)~The unnormalized minimal bipolar harmonics (UMBHs) $\overline{\Omega}_{l}^{LL\uppi} \equiv \overline{\Omega}_{l}^{LL\uppi}(\theta_{1},\phi_{1},\theta_{2},\phi_{2})$, defined implicitly by the relation
\begin{equation}\label{nbshnp}
    \Omega_{l}^{L\hspace{-0.7pt}L\hspace{-0.7pt}\uppi}  \hspace{-0.7pt}(\theta_{1}\hspace{-0.5pt},\phi_{1}\hspace{-0.7pt},\theta_{2},\phi_{2}) = N_{l}^{L\uppi}~\overline{\Omega}_{l}^{LL\uppi}(\theta_{1},\phi_{1},\theta_{2},\phi_{2})
\end{equation}
and given explicitly by the following compact formulas \cite{Gu2001}
\begin{equation}\label{ubshnp}
    \overline{\Omega}_{l}^{LL\text{n}} = \left(\sin{\theta_{1}} \, e^{\textbf{\textit{i}}\phi_{1}}\right)^{l}
    \left(\sin{\theta_{2}} \, e^{\textbf{\textit{i}}\phi_{2}}\right)^{L-l}
\end{equation}
for natural parity $\uppi=\text{n}$, and 
\begin{eqnarray}\label{ubshup}
    \overline{\Omega}_{l}^{LL\text{u}} = \cos{\theta_{2}}~\overline{\Omega}_{l}^{LL\text{n}}-\cos{\theta_{1}}~\overline{\Omega}_{l-1}^{LL\text{n}}
\end{eqnarray}
for unnatural parity $\uppi=\text{u}$. 

(3)~The solid minimal bipolar harmonics (SMBHs) $\bm{\Omega}_{l}^{LL\uppi} \equiv \bm{\Omega}_{l}^{LL\uppi}(\bm{r}_{1},\bm{r}_{2})$, defined explicitly by
\begin{equation}\label{nsbph}
    \bm{\Omega}_{l}^{LL\uppi}(\bm{r}_{1},\bm{r}_{2}) = r_{1}^l r_{2}^{L-l+d}~\Omega_{l}^{LL\uppi}(\theta_{1},\phi_{1},\theta_{2},\phi_{2})
\end{equation}

(4)~The unnormalized solid minimal bipolar harmonics (USMBHs) $\overline{\bm{\Omega}}_{l}^{LL\uppi}\hspace{-3pt} \equiv \overline{\bm{\Omega}}_{l}^{LL\uppi}(\bm{r}_{1},\bm{r}_{2})$, defined implicitly by
\begin{eqnarray}\label{nsbshccnp}
    \bm{\Omega}_{l}^{LL\uppi} = N_{l}^{L\uppi}~\overline{\bm{\Omega}}_{l}^{LL\uppi}
\end{eqnarray}
where $N_{l}^{L\uppi}$ was previously given in Eq.~(\ref{nlpik}). These functions have an elegant representation \cite{Gu2001} in Cartesian coordinates $\left(x_{1},y_{1},z_{1},x_{2},y_{2},z_{2}\right)$
\begin{eqnarray}\label{usbshccnp}
    \overline{\bm{\Omega}}_{l}^{LL\text{n}} \!= \,\left(x_{1}+\textbf{\textit{i}}y_{1}\right)^{l} \left(x_{2}+\textbf{\textit{i}}y_{2}\right)^{L-l}
\end{eqnarray}
for natural parity $\uppi=\text{n}$, and 
\begin{eqnarray}\label{usbshccup}
    \overline{\bm{\Omega}}_{l}^{LL\text{u}} \!= \,z_{2} ~\overline{\bm{\Omega}}_{l}^{LL\text{n}}-z_{1} ~\overline{\bm{\Omega}}_{l-1}^{LL\text{n}}
\end{eqnarray}
for unnatural parity $\uppi=\text{u}$. Note that the USMBHs also possess the  following ladder-like property
\begin{eqnarray}
    (x_{1}+\textbf{\textit{i}}y_{1}) ~\overline{\bm{\Omega}}_{l}^{LL\uppi} \!&=& \,\overline{\bm{\Omega}}_{l+1}^{L+1,L\uppi} \label{sprel1} \\
    (x_{2}+\textbf{\textit{i}}y_{2}) ~\overline{\bm{\Omega}}_{l}^{LL\uppi} \!&=& \,\overline{\bm{\Omega}}_{l}^{L+1,L\uppi} \label{sprel2}    
\end{eqnarray}
which are valid for an arbitrary $l\in\left\{d,\ldots,L \right\}$ and are used extensively in the derivation of the results presented in the next paragraph.

Evaluating the action of the kinetic energy operator on $\Psi^{LL\uppi}(\bm{r}_{1},\bm{r}_{2})$  relies heavily on the following properties of the USMBHs 
\begin{eqnarray}
    \hspace{-15pt}\nabla_{1}\overline{\bm{\Omega}}_{l}^{LL\text{n}} \hspace{-7pt} &=&\hspace{-7pt}  \left[\!\!\begin{array}{c}
     1 \\ \textbf{\textit{i}} \\ 0 \end{array}\!\!\right]l ~\overline{\bm{\Omega}}_{l-1}^{L-1,L\text{n}}  \label{eq:nablanp1}\\
        \hspace{-15pt}
   \nabla_{2}\overline{\bm{\Omega}}_{l}^{LL\text{n}} \hspace{-7pt}&=&\hspace{-7pt} \left[\!\!\begin{array}{c}
      1 \\ \textbf{\textit{i}} \\ 0 \end{array}\!\!\right]\! \left(L\!-\!l\right) \overline{\bm{\Omega}}_{l}^{L-1,L\text{n}} \label{eq:nablanp2}\\
    \hspace{-15pt}\nabla_{1}\overline{\bm{\Omega}}_{l}^{LL\text{u}}\hspace{-7pt} &=&\hspace{-7pt} \left[\!\!\begin{array}{c}
         1  \\ \textbf{\textit{i}} \\ 0 \end{array}\!\!\right]\!\!\!\left(\!\!(\!l\hspace{-2.5pt}-\hspace{-2.5pt}1)\overline{\bm{\Omega}}_{l-1}^{L\hspace{-1pt}-\hspace{-1pt}1,L\text{u}}\!\!+\!z_{2}\overline{\bm{\Omega}}_{l-1}^{L\hspace{-1pt}-\hspace{-1pt}1,L\text{n}}\!\right)\!\!-\!\!\left[\!\!\begin{array}{c}
         0 \\ 0 \\ 1
    \end{array}\!\!\right]\!\!\overline{\bm{\Omega}}_{l-1}^{LL\text{n}}\label{eq:nabla1up}\\
    \hspace{-15pt}\nabla_{2}\overline{\bm{\Omega}}_{l}^{LL\text{u}} \hspace{-7pt}&=&\hspace{-7pt}\left[\!\!\begin{array}{c}
         1\\\textbf{\textit{i}} \\ 0
    \end{array}\!\!\right]\!\!\!\left(\!\!(\!L\hspace{-2.5pt}-\hspace{-2.5pt}l)\overline{\bm{\Omega}}_{l}^{L\hspace{-1pt}-\hspace{-1pt}1,L\text{u}}\!\!-\!z_{1}\overline{\bm{\Omega}}_{l-1}^{L\hspace{-1pt}-\hspace{-1pt}1,L\text{n}}\!\right)\!\!+\!\!\left[\!\!\begin{array}{c}
         0  \\
        0 \\
        1
    \end{array}\!\!\right]\!\!\overline{\bm{\Omega}}_{l}^{LL\text{n}}\label{eq:nabla2up}
\end{eqnarray}
Successive application of these identities for $i=1,2$ yields the relations
\begin{eqnarray}
        \Delta_{i} \overline{\bm{\Omega}}_{l}^{LL\uppi} = \nabla_{i}\!\cdot\!\nabla_{i} \,\overline{\bm{\Omega}}_{l}^{LL\uppi}  &= 0& \label{ldoafi0}\\[5pt]
        \nabla_{1}\!\cdot\!\nabla_{2} \,\overline{\bm{\Omega}}_{l}^{LL\uppi} &= 0& \label{ldoaf30}
\end{eqnarray}
These conditions also hold for the SMBHs $\bm{\Omega}_{l}^{LL\uppi}$ by virtue of Eq.~\eqref{nsbshccnp}. These results are equivalent to Eqs.~(17) and~(18)  of  \textcite{Efros1986}, derived using irreducible tensor algebra. Furthermore, Eq.~(\ref{ldoafi0}) justifies the term ``bispherical harmonics" used for these generators, as these functions are harmonic with respect to the Laplacians of both quasiparticles.

\section{Explicit structure of the RSE\MakeLowercase{s}}\label{method}

\subsection{Expansion of the wave function  \texorpdfstring{$\Psi^{LM\uppi}(\bm{r}_{1},\bm{r}_{2})$}{Psi} in the angular generator bases }\label{ss:psiexp}

We have discussed two distinct bases of angular subspace generators $\mathscr{Y}_{l}^{LM\uppi}(\bm{r}_{1},\bm{r}_{2})$ 
\begin{eqnarray}
\text{Wigner matrices: } && \left\{\mathcal{D}_{L}^{MK}(\alpha,\beta,\gamma): K=-L,\ldots,L\right\} \nonumber\\
\text{SMBHs: }  && \left\{\bm{\Omega}_{l}^{L\hspace{-0.7pt}M\hspace{-0.7pt}\uppi}  \hspace{-0.7pt}(\bm{r}_{1},\bm{r}_{2}): l=d,\ldots,L\right\} \nonumber
\end{eqnarray}
These bases can be used to construct the most general wave function  $\Psi^{LM\uppi}(\bm{r}_{1},\bm{r}_{2})$ with definite values of $L$, $M$, and $\uppi$, previously defined symbolically in Eq.~(\ref{wfa1}). The two distinct expansions associated with these bases can be expressed as follows
\begin{eqnarray}\label{wfa2}
    \hspace{-20pt}\Psi^{LM\uppi}(\bm{r}_{1},\bm{r}_{2}) &\hspace{-4pt}=\hspace{-4pt}& \sum_{l=d}^{L}\psi_{l}^{L\uppi}(r_{1},r_{2},r_{12}) \,\,\bm{\Omega}_{l}^{L\hspace{-0.7pt}M\hspace{-0.7pt}\uppi}  \hspace{-0.7pt}(\bm{r}_{1},\bm{r}_{2}) \\
    \hspace{-20pt}\Psi^{LM\uppi}(\bm{r}_{1},\bm{r}_{2}) &\hspace{-4pt}=\hspace{-1pt}&\hspace{-12pt} \sum_{k=-\frac{L-d}{2}}^{\frac{L-d}{2}}\hspace{-9pt}\varphi_{2k}^{L\uppi}(r_{1},r_{2},r_{12}) \,\mathcal{D}_{L}^{M\hspace{0.2pt}2k}(\alpha,\beta,\gamma) \label{wfa2a}
\end{eqnarray}
where the summation range in the expansion over the Wigner functions follows from the parity properties of $\mathcal{D}_{L}^{M\hspace{0.2pt}2k}(\alpha,\beta,\gamma)$  given by Eq.~(\ref{eq:wignerdpar}). The partial wave components $\psi_{l}^{L\uppi}(r_{1},r_{2},r_{12})$ and $\varphi_{2k}^{L\uppi}(r_{1},r_{2},r_{12})$ do not depend on $M$, owing to the following relations
\begin{eqnarray}
  \hat{L}_\pm\,\, \psi_{l}^{L\uppi}(r_{1},r_{2},r_{12}) & = &0 \label{pmLani}\\
  \hat{L}_\pm\,\, \varphi_{2k}^{L\uppi}(r_{1},r_{2},r_{12}) & = & 0 \label{Lpmani}
\end{eqnarray}
These properties hold because $\hat{L}_\pm$ satisfy\footnote{For details, see Sec.~V of Supplemental Material \cite{supplemental}.} the el\-e\-men\-ta\-ry relations $\hat{L}_\pm\left(r_{1}\right)=\hat{L}_\pm\left(r_{2}\right)=\hat{L}_\pm\left(r_{12}\right)=0$, along with $\hat{L}_\pm\left(\cos \theta\right) = 0$, which can be easily verified using Eqs.~(\ref{eq:costheta}) and~(\ref{ladderlabbsph}).

\subsection{Action of Hamiltonian on \texorpdfstring{$\Psi^{LM\uppi}(\bm{r}_{1},\bm{r}_{2})$}{Psi}  and the separation of the angular generators}\label{actHwf}

To compute the action of the Hamiltonian [cf.\ Eq.~(\ref{secm0})]
\begin{equation}\label{ham}
    \hat{\mathcal{H}}\! =\! \underbrace{-\tfrac{\nabla_{1\vphantom{y}}\hspace{-1pt}\cdot\hspace{-1pt}\nabla_{1}}{2\mu_1}\!-\!\tfrac{\nabla_{2\vphantom{y}}\hspace{-1pt}\cdot\hspace{-1pt}\nabla_{2}}{2\mu_2}\! -\!\tfrac{\nabla_{1\vphantom{y}}\hspace{-1pt}\cdot\hspace{-1pt}\nabla_{2}}{m_{3}}}_{\hat{\mathcal{T}}}  \underbrace{+ \tfrac{q_{1\vphantom{y}}q_{2}}{r_{12}}\!+\!\tfrac{q_{1\vphantom{y}}q_{3}}{r_{1}\vphantom{\mu_2}}\!+\!\tfrac{q_{2\vphantom{y}}q_{3}}{r_{2}}}_{\hat{\mathcal{V}}}
\end{equation}
on $\Psi^{LM\uppi}\equiv\Psi^{LM\uppi}(\bm{r}_{1}\hspace{-1pt},\hspace{-1pt}\bm{r}_{2})$ given by Eq.~(\ref{wfa2}), we first write $\hat{\mathcal{H}}\,\Psi^{LM\uppi}$ in the following form
\begin{equation}\label{twf1}
    \hat{\mathcal{H}}\,\Psi^{LM\uppi} = \tfrac{\overbrace{\hat{L}_{-}\ldots \hat{L}_{-}}^{L-M~\text{times}}}{\prod\limits_{{\scriptscriptstyle \mu=M}}^{{\scriptscriptstyle L-1}}\sqrt{L\left(L+1\right)-\mu\left(\mu+1\right)}}\,\,\hat{\mathcal{H}}\,\Psi^{LL\uppi} 
\end{equation}
which follows from Eqs.~(\ref{pmLani}) and~(\ref{eq:ladderM}), and from the fact that $\hat{L}_{-}$ and $\hat{\mathcal{H}}$ commute [cf.\ the text following Eq.~(\ref{ladderlabbsph})]. The action of $\hat{\mathcal{H}}=\hat{\mathcal{T}}+\hat{\mathcal{V}}$ on $\Psi^{LL\uppi}$ can be computed in a straightforward way as follows. Since the potential $\hat{\mathcal{V}}$ depends only on the interparticle distances, we naturally have
\begin{equation}\label{eq:vonpsi}
   \hat{\mathcal{V}}\,\Psi^{LL\uppi}=\sum_{l=d}^{L}\left[\hat{\mathcal{V}}\,\psi_{l}^{L\uppi}(r_{1},r_{2},r_{12})\right] \,\,\bm{\Omega}_{l}^{L\hspace{-0.7pt}L\hspace{-0.7pt}\uppi}  \hspace{-0.7pt}(\bm{r}_{1},\bm{r}_{2}) 
\end{equation}
The action of the kinetic energy operator $\hat{\mathcal{T}}$ on $\Psi^{LL\uppi}$
\begin{eqnarray}\label{eq:tonpsi}
   \hat{\mathcal{T}}\,\Psi^{LL\uppi}&\hspace{-3pt}=\hspace{-3pt}&\sum_{l=d}^{L}\,\,\hat{\mathcal{T}}\left(\psi_{l}^{L\uppi}(r_{1},r_{2},r_{12}) \,\,\bm{\Omega}_{l}^{LL\uppi}  \hspace{-0.7pt}(\bm{r}_{1},\bm{r}_{2}) \right)\\
   &\hspace{-3pt}=\hspace{-3pt}&\sum_{l=d}^{L}\,N_{l}^{L\uppi}\,\hat{\mathcal{T}}\left( \vphantom{\psi_{l}^{L\uppi}(r_{1},r_{2},r_{12})} \right.\underbrace{\psi_{l}^{L\uppi}(r_{1},r_{2},r_{12})}_{\psi_{l}} \,\,\underbrace{\overline{\bm{\Omega}}_{l}^{LL\uppi}(\bm{r}_{1},\bm{r}_{2})}_{\overline{\bm{\Omega}}_{l}} \left. \vphantom{\psi_{l}^{L\uppi}(r_{1},r_{2},r_{12})} \right) \nonumber
\end{eqnarray}
is most conveniently computed using the USMBHs $\overline{\bm{\Omega}}_{l}\equiv\overline{\bm{\Omega}}_{l}^{LL\uppi}(\bm{r}_{1},\bm{r}_{2})$ implicitly defined in Eq.~(\ref{nsbshccnp}). We have
\begin{equation}\label{eq:tonpsiomega}
    \hat{\mathcal{T}}\left( \psi_{l}\,\overline{\bm{\Omega}}_{l}\right) = -\sum\limits_{i=1}^{2}\tfrac{\nabla_{i\vphantom{y}}\hspace{-1pt}\cdot\hspace{-1pt}\nabla_{i}}{2\mu_i}\left( \psi_{l}\,\overline{\bm{\Omega}}_{l}\right)-\!\tfrac{\nabla_{1\vphantom{y}}\hspace{-1pt}\cdot\hspace{-1pt}\nabla_{2}}{m_{3}} \left( \psi_{l}\,\overline{\bm{\Omega}}_{l}\right)
\end{equation}
The right-hand side of this equation can be computed from the following identity
\begin{eqnarray}
    \hspace{-10pt}\nabla_{\!i}\!\cdot\! \nabla_{\!j}\!\left(\psi_{l}\, \overline{\bm{\Omega}}_{l}\right) &=& \left(\nabla_{\!i}\psi_{l}\right) \!\!\cdot\!\!\left(\nabla_{\!j}\overline{\bm{\Omega}}_{l}\right)+   (\nabla_{\!i}\!\cdot\!\nabla_{\!j}\hspace{1pt}\psi_{l})\,\overline{\bm{\Omega}}_{l}\nonumber\\
    &+& \left(\nabla_{\!j}\psi_{l}\right) \!\!\cdot\!\!\left(\nabla_{\!i}\overline{\bm{\Omega}}_{l}\right)+ \psi_{l}\,(\nabla_{\!i}\!\cdot\!\nabla_{\!j}\hspace{1pt}\overline{\bm{\Omega}}_{l}) \label{eq:fgident}
\end{eqnarray}
For $i,j\in\left\{1,2\right\}$, the $\nabla_{\!i}\!\cdot\!\nabla_{\!j}\hspace{1pt}\overline{\bm{\Omega}}_{l}$ terms vanish owing to Eqs.~(\ref{ldoafi0}) and~(\ref{ldoaf30}), whereas the $\nabla_{\!i}\!\cdot\!\nabla_{\!j}\hspace{1pt}\psi_{l}$ terms can be expressed as
\begin{equation} \label{eq:deldelf}
    \nabla_{\!i}\!\cdot\!\nabla_{\!j}\hspace{1pt}\psi_{l} =\hat{t}_{ij}\,\psi_{l}
\end{equation}
where, for $\left\{i,j\right\}=\left\{1,2\right\}$, the auxiliary operators $\hat{t}_{11}$, $\hat{t}_{22}$,  and $\hat{t}_{12}=\hat{t}_{21}$ are given by
\begin{eqnarray}
    \hspace{-20pt}\hat{t}_{ii} \hspace{-5pt}&=&\hspace{-4pt} \partial_{r_{i}r_{i}} \!\!+\! \tfrac{2}{r_{i}}\partial_{r_i}  \!\! +\! \partial_{r_{12}r_{12}} \!\!+\! \tfrac{2}{r_{12}}\partial_{r_{12}} \!\! +\!\tfrac{r_{i\vphantom{y}}^2-r_{j}^2+r_{12}^2}{r_{i}r_{12}}                       \partial_{r_{i}r_{12}} \label{eq:diff2r1}\\
    \hspace{-20pt}\hat{t}_{ij}\hspace{-5pt}&=&\hspace{-4pt} -\partial_{r_{12}r_{12}}\! - \!\tfrac{2}{r_{12}}\partial_{r_{12}}\!+\!\tfrac{r_{i}^2+r_{j}^2-r_{12}^2}{2r_{i}r_{j}}\partial_{r_{i}r_{j}} \label{eq:diff2r12}\\ 
    \hspace{-20pt}&&   -\tfrac{r_{i}^2-r_{j}^2+r_{12}^2}{2r_{i}r_{12}}~\partial_{r_{i}r_{12}}\!-\! \tfrac{r_{j}^2-r_{i}^2+r_{12}^2}{2r_{j}r_{12}}~\partial_{r_{j}r_{12}}\nonumber
\end{eqnarray}
Equations~\eqref{eq:deldelf}--\eqref{eq:diff2r12} are a consequence of the following first-order differential relation, applicable to any function $\psi_{l}$ that depends only on the internal shape variables $r_1$, $r_2$, and $r_{12}$
\begin{equation}\label{eq:diff1ri}
    \nabla_{i}\,\psi_{l} = \left(\!\tfrac{\bm{r}_{i{\vphantom{y}}}}{r_i} \partial_{r_i}  \!+ \tfrac{\bm{r}_{i{\vphantom{y}}}-\bm{r}_{j}}{r_{12}}\partial_{r_{12}}\!\right)\psi_{l}
\end{equation}
alongside the elementary identities
\begin{eqnarray}\label{eq:eleident}
    \bm{r}_1 \cdot \bm{r}_2 & = & \tfrac{1}{2} \left( r_1^2+r_2^2-r_{12}^2\right) \nonumber\\
    \bm{r}_1 \cdot \left(\bm{r}_1 - \bm{r}_2\right) & = & \tfrac{1}{2} \left( r_1^2-r_2^2+r_{12}^2\right) \nonumber\\
    \bm{r}_2 \cdot \left(\bm{r}_2 - \bm{r}_1\right) & = & \tfrac{1}{2} \left( r_2^2-r_1^2+r_{12}^2\right) \\
    \nabla_{i} \cdot \tfrac{\bm{r}_{j{\vphantom{y}}}}{r_j} & =& \tfrac{2}{r_j} \delta_{ij} \,\,(i,j=1,2) \nonumber\\
    \nabla_{1} \cdot \tfrac{\bm{r}_{1{\vphantom{y}}}-\bm{r}_{2{\vphantom{y}}}}{r_{12}} &=& \nabla_{2} \cdot \tfrac{\bm{r}_{2{\vphantom{y}}}-\bm{r}_{1{\vphantom{y}}}}{r_{12}} = \tfrac{2}{r_{12}}\nonumber
\end{eqnarray}
The remaining terms of the form $\left(\nabla_{\!i}\psi_{l}\right) \!\!\cdot\!\!\left(\nabla_{\!j}\overline{\bm{\Omega}}_{l}\right)$ with $i,j\in \left\{1,2\right\}$ can be computed by combining Eq.~(\ref{eq:diff1ri}) with Eqs.~(\ref{eq:nablanp1})--(\ref{eq:nabla2up}) into the following expression
\begin{eqnarray}
   \left(\nabla_{\!i}\psi_{l}\right) \!\!\cdot\!\!\left(\nabla_{\!j}\overline{\bm{\Omega}}_{l}\right) & = & (l_j \hspace{-3pt}-\hspace{-3pt} |j\hspace{-3pt}-\hspace{-2.5pt}i|\hspace{1pt}d) \left[\left(\!\tfrac{\partial_{r_{i\vphantom{y}}}}{r_i}   \hspace{-3pt}+\hspace{-3pt} \tfrac{\partial_{r_{12\vphantom{y}}}}{r_{12}}\!\right) \psi_{l} \right] \overline{\bm{\Omega}}_{l+j-i} \nonumber \\
   & &  \hspace{-30pt} - (l_j \hspace{-3pt}-\hspace{-3pt} |i\hspace{-3pt}+\hspace{-2.5pt}j\hspace{-3pt}-\hspace{-2.5pt}3|\hspace{1pt}d) \left[\left(\! \tfrac{\partial_{r_{12\vphantom{y}}}}{r_{12}}\!\right) \psi_{l} \right] \overline{\bm{\Omega}}_{l+i+j-3} \label{eq:ninj}
\end{eqnarray}
Here, the partial angular mo\-men\-ta $l_1=l$ and $l_2=L-l+d$ have been used for notational compactness. Note that expressions equivalent to Eq.~(\ref{eq:ninj}) were derived earlier by \textcite{Efros1986} and \textcite{Frolov1996} using irreducible tensor algebra.

The resulting expression for $\hat{\mathcal{T}}\left( \psi_{l}\,\overline{\bm{\Omega}}_{l}\right)$, obtained by combining Eqs.~(\ref{eq:tonpsiomega})--(\ref{eq:deldelf}), (\ref{ldoafi0}), (\ref{ldoaf30}), and~(\ref{eq:ninj}), is given by the following compact formula
\begin{eqnarray}
    \hat{\mathcal{T}}\left( \psi_{l}\,\overline{\bm{\Omega}}_{l}\right)= - \left[\left(\!\tfrac{\hat{t}_{11_{\vphantom{a}}}}{2\mu_1}\hspace{-3pt}+\hspace{-3pt}\tfrac{\hat{t}_{22_{\vphantom{a}}}}{2\mu_2}\hspace{-3pt}+\hspace{-3pt}\tfrac{\hat{t}_{12_{\vphantom{a}}}}{m_3}\!\right) \psi_{l} \right] &&\hspace{-10pt}\overline{\bm{\Omega}}_{l}\hspace{20pt}\label{eq:ninj12} \\
    - \left[\left(\!\tfrac{l_1\vphantom{y}}{\mu_1}\tfrac{\partial_{r_{1\vphantom{y}}}}{r_1}   \hspace{-3pt}+\hspace{-3pt}\tfrac{l_2\vphantom{y}}{\mu_2}\tfrac{\partial_{r_{2\vphantom{y}}}}{r_2}   \hspace{-3pt}+\hspace{-3pt}\left(\tfrac{l_1\vphantom{y}}{m_1}\hspace{-3pt}+\hspace{-3pt}\tfrac{l_2\vphantom{y}}{m_2} \right)\hspace{-3pt}\tfrac{\partial_{r_{12\vphantom{y}}}}{r_{12}}\!\right) \psi_{l} \right] &&\hspace{-10pt}\overline{\bm{\Omega}}_{l} \nonumber\\
    +  (l_2 - d) \left[ \left(\!\tfrac{1}{m_2}\!\tfrac{\partial_{r_{12\vphantom{y}}}}{r_{12}}\hspace{-3pt}-\hspace{-3pt}\tfrac{1}{m_3} \tfrac{\partial_{r_{1\vphantom{y}}}}{r_1}   \!\right) \psi_{l} \right] &&\hspace{-10pt}\overline{\bm{\Omega}}_{l+1} \nonumber \\
    + (l_1 - d)\left[ \left(\! \tfrac{1}{m_1}\!\tfrac{\partial_{r_{12\vphantom{y}}}}{r_{12}}\hspace{-3pt}-\hspace{-3pt}\tfrac{1}{m_3} \tfrac{\partial_{r_{2\vphantom{y}}}}{r_2}   \!\right) \psi_{l} \right] &&\hspace{-10pt}\overline{\bm{\Omega}}_{l-1}  \nonumber
\end{eqnarray}
where the operators $\hat{t}_{11}$, $\hat{t}_{22}$, and $\hat{t}_{12}$ are given by Eqs.~(\ref{eq:diff2r1}) and (\ref{eq:diff2r12}), and $l_1=l$ and $l_2=L-l+d$. The first line of Eq.~(\ref{eq:ninj12}) collects terms of the form $\nabla_{\!i}\!\cdot\!\nabla_{\!j}\hspace{1pt}\psi_{l}$, whereas the remaining three lines collect terms of the form $\left(\nabla_{\!i}\psi_{l}\right) \!\!\cdot\!\!\left(\nabla_{\!j}\overline{\bm{\Omega}}_{l}\right)$. We emphasize that the action of $\hat{\mathcal{T}}$ is closed with respect to the generator set $\left\{ \overline{\bm{\Omega}}_{l}\!: l=d,\ldots,L \right\}$: The value of $l$ is either unchanged or modified by $\pm1$, with the unphysical generators $\overline{\bm{\Omega}}_{d-1}$  and $\overline{\bm{\Omega}}_{L+1}$ never appearing owing to the preceding zero multipliers that annihilate them.

By performing the following operations: (1)~combining the expressions for $\hat{\mathcal{V}}\,\Psi^{LL\uppi}$ in Eq.~(\ref{eq:vonpsi}) and  $\hat{\mathcal{T}}\,\Psi^{LL\uppi}$ in Eq.~(\ref{eq:tonpsi}) with $\hat{\mathcal{T}}\left( \psi_{l}\,\overline{\bm{\Omega}}_{l}\right)$ in Eq.~(\ref{eq:ninj12}), 
(2)~reinstating the SMBHs $\bm{\Omega}_{l}^{LL\uppi}(\bm{r}_{1},\bm{r}_{2})$ using Eq.~(\ref{nsbshccnp}), 
(3)~substituting the explicit forms of the operators $\hat{\mathcal{V}}$ from Eq.~\eqref{ham} and $\hat{t}_{11}$, $\hat{t}_{22}$, and $\hat{t}_{12}$ from Eqs.~(\ref{eq:diff2r1})--(\ref{eq:diff2r12}), and 
(4)~simplifying the results, 
we obtain the following expression for $\hat{\mathcal{H}}\, \Psi^{LL\uppi}$ 
\begin{eqnarray}
    \hspace{-10pt}\hat{\mathcal{H}}\, \Psi^{LL\uppi} \hspace{-2pt}&=&\hspace{-5pt} \sum_{l=d+1}^{L}\hspace{-2.8pt}\left(\hat{h}^{\downarrow}_{l}\hspace{1pt}\psi_{l-1}\right)\bm{\Omega}_{l}^{LL\uppi}\hspace{-0.7pt}(\bm{r}_{1},\bm{r}_{2})\nonumber\\
    &+&\sum_{l=d}^{L}\hspace{6pt}\left(\hat{h}_{l}\hspace{3pt}\psi_{l}\right)\hspace{5pt}\bm{\Omega}_{l}^{LL\uppi}\hspace{-0.7pt}(\bm{r}_{1},\bm{r}_{2})\nonumber\\
    &+&\sum_{l=d}^{L-1}\left(\hat{h}^{\uparrow}_{l}\hspace{1pt}\psi_{l+1}\right)\bm{\Omega}_{l}^{LL\uppi}\hspace{-0.7pt}(\bm{r}_{1},\bm{r}_{2})\label{eq:Hpsi}
\end{eqnarray}
using three auxiliary operators
\begin{eqnarray}
    \hat{h}_{l} &=&-\tfrac{\partial_{r_{1\vphantom{y}}r_{1}}}{2\mu_{1}}
    \hspace{-3pt}-\hspace{-3pt}\tfrac{\partial_{r_{2\vphantom{y}}r_{2}}}{2\mu_{2}}
    \hspace{-3pt}-\hspace{-3pt}\tfrac{\partial_{r_{12\vphantom{y}}r_{12}}}{2\mu_{12}}\hspace{-3pt}+\hspace{-3pt} \tfrac{q_{1\vphantom{y}}q_{2}}{r_{12}}\hspace{-3pt}+\hspace{-3pt}\tfrac{q_{1\vphantom{y}}q_{3}}{r_{1}\vphantom{\mu_2}}\hspace{-3pt}+\hspace{-3pt}\tfrac{q_{2\vphantom{y}}q_{3}}{r_{2}} \label{hkkgnygp2} \\[2pt]
    &\!-\!& \nonumber (1\hspace{-3pt}+\hspace{-1.5pt}l_1)\!\left[\tfrac{\partial_{r_{1\vphantom{y}}}}{\mu_{1}r_{1}}\hspace{-3pt}+\hspace{-3pt}\tfrac{\partial_{r_{12\vphantom{y}}}}{m_{1}r_{12}}\right]
    \hspace{-3pt}-\hspace{-2pt}(1\hspace{-3pt}+\hspace{-1.5pt}l_2)\!\left[\tfrac{\partial_{r_{2\vphantom{y}}}}{\mu_{2}r_{2}}\hspace{-3pt}+\hspace{-3pt}\tfrac{\partial_{r_{12\vphantom{y}}}}{m_{2}r_{12}}\right]\\[2pt]
    &\!-\!\!&
     \tfrac{r_{1\vphantom{y}}^2+r_{12}^2-r_{2}^2}{2m_{1}r_{1}r_{12}}\partial_{r_{1}r_{12}}\hspace{-4pt}-\hspace{-3pt}  \tfrac{r_{2\vphantom{y}}^2+r_{12}^2-r_{1}^2}{2m_{2}r_{2}r_{12}}\partial_{r_{2}r_{12}} \hspace{-4pt}-\hspace{-3pt} \tfrac{r_{1\vphantom{y}}^2+r_{2}^2-r_{12}^2}{2m_{3}r_{1}r_{2}}\partial_{r_{1}r_{2}}\nonumber\\[3pt]
    \hat{h}^{\uparrow}_{l} &=& \sqrt{\tfrac{(l_2-d)(2l_1+3)(l_1-d+1)}{\phantom{(l_2-d)}(2l_2+1)\phantom{(l_1-d+1)}}}\!\left(\!\tfrac{\partial_{r_{12\vphantom{y}}}}{m_{1}\,r_{12}}\hspace{-3pt}-\hspace{-3pt}\tfrac{\partial_{r_{2\vphantom{y}}}}{m_{3}\,r_{2}}\!\right)\label{hkk-1gnygp2}\\
    \hat{h}^{\downarrow}_{l} &=& \sqrt{\tfrac{(l_1-d)(2l_2+3)(l_2-d+1)}{\phantom{(l_1-d)}(2l_1+1)\phantom{(l_2-d+1)}}} \!\left(\!\tfrac{\partial_{r_{12\vphantom{y}}}}{m_{2}\,r_{12}}\hspace{-3pt}-\hspace{-3pt}\tfrac{\partial_{r_{1\vphantom{y}}}}{m_{3}\,r_{1}}\!\right) \label{hkk+1gnygp2}
\end{eqnarray}
Here, $l_1=l$ and $l_2=L-l+d$ (the partial an\-gu\-lar momenta of the quasiparticles)  are used again to highlight the underlying symmetry of the resulting operators. These operators also depend explicitly on the charges and the masses of the particles (with $\mu_{12} = \frac{m_{1\vphantom{y}}m_{2}}{m_{1}+m_{2}}$), as well as on the eigenvalues of the total angular momentum $L$ and parity $\uppi$. The square-root terms in Eqs.~\eqref{hkk-1gnygp2} and~\eqref{hkk+1gnygp2}, corresponding to $(l_1 +1-d) \sfrac{N_{l+1}^{L\uppi}}{N_{l}^{L\uppi}}$ and $(l_2+1-d)\sfrac{N_{l-1}^{L\uppi}}{N_{l}^{L\uppi}}$, respectively, arise from the weighted ratios of the normalization phase factors given by Eq.~(\ref{nlpik}) that emerge when reinstating the SMBHs using Eq.~(\ref{nsbshccnp}).

By substituting Eq.~(\ref{eq:Hpsi}) into Eq.~(\ref{twf1}), employing the fact that the operator $\hat{L}_-$ commutes with an arbitrary term of the form $\hat{h}^{\downarrow}_{l}\hspace{1pt}\psi_{l-1}$, $\hat{h}_{l}\hspace{1pt}\psi_{l}$, or $\hat{h}^{\uparrow}_{l}\hspace{1pt}\psi_{l+1}$,\footnote{Indeed, $\hat{h}^{\downarrow}_{l}\hspace{1pt}\psi_{l-1}$, $\hat{h}_{l}\hspace{1pt}\psi_{l}$, and $\hat{h}^{\uparrow}_{l}\hspace{1pt}\psi_{l+1}$ are functions of the interparticle variables $r_1$, $r_2$, and $r_{12}$, whereas the operator $\hat{L}_-$ depends only on the Euler angles $\alpha$, $\beta$, and $\gamma$. For details, see Sec.~V of Supplemental Material \cite{supplemental}.} and using Eq.~(\ref{eq:ladderM}) $L-M$ times, we arrive at the final formula describing the action of the Hamiltonian $\hat{\mathcal{H}}$ on a general wave function $\Psi^{LM\uppi}$ with definite values of $L$, $M$, and $\uppi$, in the form given by Eq.~(\ref{wfa2})
\begin{eqnarray}
    \hspace{-10pt}\hat{\mathcal{H}}\, \Psi^{LM\uppi} \hspace{-2pt}&=&\hspace{-5pt} \sum_{l=d+1}^{L}\hspace{-2.8pt}\left(\hat{h}^{\downarrow}_{l}\hspace{1pt}\psi_{l-1}\right)\bm{\Omega}_{l}^{LM\uppi}\hspace{-0.7pt}(\bm{r}_{1},\bm{r}_{2})\nonumber\\
    &+&\sum_{l=d}^{L}\hspace{6pt}\left(\hat{h}_{l}\hspace{3pt}\psi_{l}\right)\hspace{5pt}\bm{\Omega}_{l}^{LM\uppi}\hspace{-0.7pt}(\bm{r}_{1},\bm{r}_{2})\nonumber\\
    &+&\sum_{l=d}^{L-1}\left(\hat{h}^{\uparrow}_{l}\hspace{1pt}\psi_{l+1}\right)\bm{\Omega}_{l}^{LM\uppi}\hspace{-0.7pt}(\bm{r}_{1},\bm{r}_{2})\label{eq:Hpsigen}
\end{eqnarray}

\subsection{Matrix representation of the Schr\"{o}dinger equation with definite values of \texorpdfstring{$L$}{L}, \texorpdfstring{$M$}{M}, and \texorpdfstring{$\uppi$}{pi} and comparison with previous works}

Having expressed the action of the Ham\-il\-to\-nian $\hat{\mathcal{H}}$ on the wave function  $\Psi^{LM\uppi}(\bm{r}_{1},\bm{r}_{2})$  in Eq.~(\ref{eq:Hpsigen}) as a closed-form expansion over the generators $\bm{\Omega}_{l}^{LM\uppi}$, and having an analogous expansion of $\Psi^{LM\uppi}(\bm{r}_{1},\bm{r}_{2})$ given by Eq.~(\ref{wfa2}), we can now conveniently rewrite the Schr\"{o}dinger equation in Eq.~(\ref{secm2}) in a structured, transparent tridiagonal matrix operator form 
\begin{widetext}
\begin{eqnarray}\label{hamtme}
    \begin{pmatrix}
        \bm{\Omega}_{d}^{LM\uppi} ~ \bm{\Omega}_{d\hspace{-0.5pt}+\hspace{-1pt}1}^{LM\uppi} \cdots ~\bm{\Omega}_{L}^{LM\uppi}
    \end{pmatrix}
    \left[
    \left(\begin{array}{cclcl}
        \!\!\hat{h}_{d} & \!\!\!\!\hat{h}^{\uparrow}_{d} & \;0\;\;\;\; & \cdots\;\;\;\;\; & \;0\vphantom{\vdots}\\
        \hat{h}^{\downarrow}_{d\hspace{-0.5pt}+\hspace{-1pt}1} & \!\!\hat{h}_{d\hspace{-0.5pt}+\hspace{-1pt}1} & \ddots & \ddots\;\;\;\;\; & \;\vdots\\
        0 &\ddots &\ddots&\ddots\;\;\;\;\;\;& \;0\\ 
        \vdots &\ddots&\ddots & \hat{h}_{L\hspace{-0.6pt}-\hspace{-1pt}1} & \!\!\hat{h}^{\uparrow}_{L\hspace{-0.6pt}-\hspace{-1pt}1}\\ 
        0\vphantom{\vdots} & \cdots\;\; &  \;0& \hat{h}^{\downarrow}_{L} & \hat{h}_{L}
    \end{array} \right)\! -\!E \mathbb{1}\right]
    \left(\!\!\begin{array}{c}
     \psi_{d}\vphantom{\vdots}\\
        \psi_{d\hspace{-0.5pt}+\hspace{-1pt}1}\vphantom{\vdots}\\
        \vdots\\
        \vdots\\
        \psi_{L}\vphantom{\vdots}
    \end{array}\!\! \right) = 0
\end{eqnarray}
\end{widetext}
where the diagonal and off-diagonal operators are defined in Eqs.~\eqref{hkkgnygp2}--\eqref{hkk+1gnygp2}.

For states with $L=0$ and natural parity, the matrix of Hamiltonian operators in Eq.~(\ref{hamtme}) reduces to a single scalar operator $\hat{h}_{0}\equiv \hat{H}_\text{Hyl}$; the off-diagonal operators $\hat{h}^{\uparrow}_{l}$ and $\hat{h}^{\downarrow}_{l}$ do not appear in the formalism. The operator $\hat{H}_\text{Hyl}$, upon the substitutions $l_1=l_2=0$, $q_1=q_2=-1$, $q_3=Z$, $m_1=m_2=1$, and in the limit $m_3\rightarrow\infty$, reduces to the S-state Hamiltonian of the helium-like atom with a clamped nucleus of charge $Z$, first reported by \textcite{Hylleraas1929} and used numerous times since in various forms [see Eqs.~(7) and~(15) of \textcite{Gronwall1932}, Eq.~(1) of \textcite{Bartlett1937}, Eqs.~(1.10), (2.03), and~(2.04) of \textcite{Fock1954,Fock1958}, Eq.~(13) of \textcite{Pluvinage1955}, Eq.~(2.3) of \textcite{Kinoshita1957}, Eq.~(14) of \textcite{Pekeris1958},  Eq.~(8)  of \textcite{Pont1995}, and Eqs.~(5) and~(6) of \textcite{Nakashima2007}]. Consult also \textcite{Hylleraas1960,Morgan1986,Abbott1987,Gottschalk1987a,Gottschalk1987b,He2016,Liverts2010,Liverts2013,Liverts2018,LivertS2022} for a selective review of extensions and applications of this formalism. A generalization of this Hamiltonian to ar\-bi\-trary particle masses was presented as Eq.~(11) of \textcite{frolov1987}. 

For $L > 0$, both the reduced wave function and the RSE  in Eq.~\eqref{hamtme} have a multicomponent structure, except for the $\textnormal{P}^{\textnormal{e}}$ states, for which the matrix Hamiltonian reduces to a single operator. The explicit form of the RSE for the $\textnormal{P}^{\textnormal{e}}$ and $\textnormal{P}^{\textnormal{o}}$ states was first reported by \textcite{Breit1930} as Eq.~(10) and Eqs.~(12), (18), and~(20), respectively, for the helium-like atom with a clamped nucleus. \textcite{Schwartz1961} in Appendix I of his work extended these results to $L = 2$. \textcite{Jackson1954} used the irreducible tensor approach and \textcite{Pont1995} used unnormalized MBHs\footnote{See the paragraph following Eq.~\eqref{Lambdau} for corrections to these results.} to derive the RSE for a two-electron atom with a clamped nucleus. \textcite{Efros1986} in Eqs.~(22)--(24) [corresponding to our Eqs.~(\ref{hkkgnygp2})--(\ref{hkk-1gnygp2})]  presented analogous expressions for the case with a finite nuclear mass [see also Eqs.~(25)--(27) of \textcite{Harris2004} in this context]. \textcite{Bottcher1994} in Eq.~(22) derived an analogous form of the RSE and computed the energy levels for $L=0$ and 1 using the basis-spline collocation method. \textcite{Frolov1996} revisited the work of \textcite{Efros1986} and reported the general structure and explicit form of the Hamiltonian matrix elements. \textcite{Ma1999,Ma2000}  demonstrated how to separate the rotational degrees of freedom for a quantum three-body system using Jacobi coordinates\footnote{The choice of Jacobi coordinates considerably complicates  the analytical form of the potential energy operator [see footnote~\ref{ft:jacc} and Eq.~(9) of \textcite{Chi2007} for details], making it mass-dependent. Note that the evaluation of potential energy integrals over such an operator is numerically challenging.} and reported the RSE [cf.\ Eq.~(12) of \textcite{Ma1999} and Eq.~(17) of \textcite{Ma2000}]; a generalization of this approach to $N$-particle systems was offered by \textcite{Gu2001, Gu2001a}. \textcite{MEREMIANIN2003} developed\footnote{\label{ft:jacc} Note that the form of the potential energy operator in Eq.~(62) of \textcite{MEREMIANIN2003} corresponds to an infinite mass of one of the particles. The actual form of the potential for finite masses of all of the particles is given by
\begin{equation*}
   V\left(\bm{r}_1,\bm{r}_2 \right)= \frac{q_1q_2}{| \bm{r}_1|}+\frac{q_1q_3}{|\frac{m_2}{m_1+m_2} \bm{r}_1+\bm{r}_2|}+\frac{q_2q_3}{|\frac{m_1}{m_1+m_2} \bm{r}_1-\bm{r}_2|} 
\end{equation*}
because $\bm{r}_1$ and $\bm{r}_2$ used by \textcite{MEREMIANIN2003} are the mass-scaled Jacobi vectors and not interparticle coordinates. The evaluation of the corresponding matrix elements involves numerical integration \cite{Chi2007}.} a universal method for separating the angular variables for $N$ particles employing Jacobi vectors and the algebra of irreducible tensors, reporting the explicit forms of the RSE in Eqs.~(67) and~(79) for three and four particles, respectively. 

This brief survey highlights three salient features of the existing literature on the RSE: (i)~most formulations pertain to a
two-electron atom with a clamped nucleus, (ii)~derivations commonly rely on irreducible tensor algebra, in most cases omitting the underlying mathematical details, and (iii)~numerical validation of the resulting expressions---particularly for higher angular momentum states---remains scarce. Extensive numerical studies are available mainly for the $\textnormal{P}^{\textnormal{o}}$ and $\textnormal{P}^{\textnormal{e}}$ states, primarily building on the foundational work of \textcite{Breit1930}. A complete review of such numerical studies is beyond the scope of the current work; representative examples include \textcite{Breit1930c, Traub1959, Pekeris1962a, Perrin1963, Machacek1964, Schiff1965, Scherr1965, Schiff1965a, Rensbergen1972, Drake1979, Drake1988, Ho1990, Pachucki2006, Yerokhin2010, Pachucki2011a, Pachucki2011, Pachucki2015, Patkos2016, Pachucki2017, Zalialiutdinov2017, Zhang2019, Yerokhin2021, Yerokhin2023} for the $\textnormal{P}^{\textnormal{o}}$ states and \textcite{Becker1964, Drake1970, Drake1970a, Hilger1996, dePrunele1992, Banyard1992, Duan2001, Eiglsperger2010, Eiglsperger2010a, Kar2010a, Kar2011} for the $\textnormal{P}^{\textnormal{e}}$ states. For $L > 1$, systematic numerical treatments are considerably fewer, though the studies of \textcite{Warner1978, Conneely1978, Ho1986, Bishop1989, Drake1990, Goodson1991, Duan2002, Kar2009, Kar2008, Kar2009a, Kar2009b, Kar2009c, kar2010,Aznabayev2015, Zalialiutdinov2016, Aznabaev2018,Drake1972, Drake2023, Wienczek2019, Yerokhin2020} provide reliable data for selected higher $L$ states. 

It is important to mention that the discussion above pertains to the separation of angular momentum using  minimal bipolar harmonics; an alternative approach based on Wigner functions $\mathcal{D}$ as the angular basis, leading to an RSE with more intricate differential operators and a generally less clear structure of the resulting multicomponent formalism [see, for example, \textcite{DattaMajumdar1952, BHATIA1964, BHATIA1965, Kalotas1965, Nikitin1985a, Dmitrieva1986, Mukherjee1994, Mukherjee1995}], is not pursued in the current study.

\subsection{Elimination of angular dependence}\label{sec:eleangdep}
We eliminate the angular dependence from Eq.~\eqref{hamtme} by multiplying it from the left by $\begin{pmatrix}\bm{\Omega}_{d}^{LM\uppi} ~ \bm{\Omega}_{d+1}^{LM\uppi} ... ~\bm{\Omega}_{L}^{LM\uppi}\end{pmatrix}^{\dagger}$ and subsequently integrating over the three Euler angles $\alpha$, $\beta$, and $\gamma$. In the next section, we compute the explicit expressions for the elements of the resulting matrix of angular integrals 
\begin{equation}\label{angularint1}
   \mathscr{W}_{ll\hspace{-0.3pt}'\hspace{-0.3pt} }^{L\uppi}  \equiv \mathscr{W}_{ll\hspace{-0.3pt }'\hspace{-0.3pt}}^{L\uppi}(r_{1},r_{2},r_{12}) = \left\langle \left. \bm{\Omega}_{l\hspace{-0.3pt }'\hspace{-0.3pt}}^{LM\uppi}\right| \bm{\Omega}_{l}^{LM\uppi} \right\rangle
\end{equation} 
using the inner product  $\langle \cdot| \cdot \rangle$ defined in Eq.~\eqref{dtau}. The angular integrals $\mathscr{W}_{ll\hspace{-0.3pt}'\hspace{-0.3pt}}^{L\uppi}$ do not depend explicitly on the value of $M$; this follows from Eqs.~(\ref{alpikk1}), (\ref{intlambda}), and~(\ref{eq:Mind}). The procedure delineated above yields the following explicit matrix representation of the RSE, which corresponds formally to a set of coupled PDEs for the partial-wave components $\psi_{l} \equiv \psi_{l}^{L\uppi}(r_{1},r_{2},r_{12})$ with $l=d,\ldots,L$
\begin{widetext}
\begin{eqnarray}\label{rrse}
    \left(\!\begin{array}{cllc}
        \mathscr{W}_{dd\phantom{\hspace{-0.5pt}+\hspace{-1pt}1}}^{L\uppi} & \mathscr{W}_{d\hspace{-0.5pt}+\hspace{-1pt}1\hspace{-0.7pt},d}^{L\uppi} & \!\!\!\!\!\cdots & \mathscr{W}_{Ld}^{L\uppi}\vphantom{\vdots}\\
        \mathscr{W}_{d,d\hspace{-0.5pt}+\hspace{-1pt}1}^{L\uppi} & \mathscr{W}_{d\hspace{-0.5pt}+\hspace{-1pt}1\hspace{-0.7pt},d\hspace{-0.5pt}+\hspace{-1pt}1}^{L\uppi} &\!\!\!\!\!&\vdots\\ 
        \vdots && \!\!\!\!\!\ddots & \vdots\vphantom{\mathscr{W}_{d\hspace{-0.5pt}+\hspace{-1pt}1\hspace{-0.7pt},d}^{L\uppi}}\\ 
        \mathscr{W}_{dL}^{L\uppi} &\hspace{9pt}\cdots &  \!\!\!\!\!\cdots  &\mathscr{W}_{LL}^{L\uppi}\vphantom{\vdots}
    \end{array} \right)\!\! \left[\!
    \left(\begin{array}{cclcl}
        \!\!\hat{h}_{d} & \!\!\!\!\hat{h}^{\uparrow}_{d} & \;0\;\;\;\; & \cdots\;\;\;\;\; & \;0\vphantom{\vdots}\\
        \hat{h}^{\downarrow}_{d\hspace{-0.5pt}+\hspace{-1pt}1} & \!\!\hat{h}_{d\hspace{-0.5pt}+\hspace{-1pt}1} & \ddots & \ddots\;\;\;\;\; & \;\vdots\\
        0 &\ddots &\ddots&\ddots\;\;\;\;\;\;& \;0\\ 
        \vdots &\ddots&\ddots & \hat{h}_{L\hspace{-0.6pt}-\hspace{-1pt}1} & \!\!\hat{h}^{\uparrow}_{L\hspace{-0.6pt}-\hspace{-1pt}1}\\ 
        0\vphantom{\vdots} & \cdots\;\; &  \;0& \hat{h}^{\downarrow}_{L} & \hat{h}_{L}
    \end{array} \right)\! -\!E \mathbb{1}\right]\!\!
    \left(\!\!\begin{array}{c}
     \psi_{d}\vphantom{\vdots}\\
        \psi_{d\hspace{-0.5pt}+\hspace{-1pt}1}\vphantom{\vdots}\\
        \vdots\\
        \vdots\\
        \psi_{L}\vphantom{\vdots}
    \end{array}\!\! \right)\!\! =\!\! \left(\!\begin{array}{c}
     0\vphantom{\vdots}\\
        0\vphantom{\vdots}\\
        \vdots\\
        \vdots\\
        0\vphantom{\vdots}
    \end{array}\! \right)
\end{eqnarray}
\end{widetext}
An alternative method of removing the angular de\-pen\-dence from Eq.~(\ref{hamtme}) was proposed by \textcite{MEREMIANIN2003}, who used the linear independence of MBHs in their Eq.~(64) to obtain the following matrix representation of their RSE in Eq.~(67) 
\begin{equation}\label{rrseMB}
    \scalebox{0.9}{$\hspace{-5pt}\left[\!
    \left(\!\begin{array}{cclcl}
        \!\!\hat{h}_{d} & \!\!\!\!\hat{h}^{\uparrow}_{d} & \;0\;\;\;\; & \cdots\;\;\;\;\; & \;0\vphantom{\vdots}\\
        \hat{h}^{\downarrow}_{d\hspace{-0.5pt}+\hspace{-1pt}1} & \!\!\hat{h}_{d\hspace{-0.5pt}+\hspace{-1pt}1} & \ddots & \ddots\;\;\;\;\; & \;\vdots\\
        0 &\ddots &\ddots&\ddots\;\;\;\;\;\;& \;0\\ 
        \vdots &\ddots&\ddots & \hat{h}_{L\hspace{-0.6pt}-\hspace{-1pt}1} & \!\!\hat{h}^{\uparrow}_{L\hspace{-0.6pt}-\hspace{-1pt}1}\\ 
        0\vphantom{\vdots} & \cdots\;\; &  \;0& \hat{h}^{\downarrow}_{L} & \hat{h}_{L}
    \end{array}\! \right)\hspace{-3pt} -\hspace{-2pt}E \mathbb{1}\right]\!\!
    \left(\!\!\begin{array}{c}
     \psi_{d}\vphantom{\vdots}\\
        \psi_{d\hspace{-0.5pt}+\hspace{-1pt}1}\vphantom{\vdots}\\
        \vdots\\
        \vdots\\
        \psi_{L}\vphantom{\vdots}
    \end{array}\!\! \right)\!\! =\!\! \left(\!\begin{array}{c}
     0\vphantom{\vdots}\\
        0\vphantom{\vdots}\\
        \vdots\\
        \vdots\\
        0\vphantom{\vdots}
    \end{array} \! \right)$}
\end{equation}
This RSE can be formally obtained from the RSE in Eq.~(\ref{rrse}) by multiplying it from the left by the inverse of the matrix $\left[\mathscr{W}_{ll'}^{L\uppi}\right]$.\footnote{The inverse exists because MBHs form a linearly independent (though non-orthogonal) set of tensors.} These two forms of the RSE have distinct advantages. The RSE in Eq.~(\ref{rrse}) supports variational energy calculations,\footnote{The RSE in Eq.~(\ref{rrseMB}) does not permit variational energy calculations. Our attempt to use it for finding the $\textnormal{D}^{\textnormal{e}}$ energy levels yielded energies located \emph{below} the true physical values.} because it is the Euler--Lagrange equation associated with the total energy variation.
The RSE given by Eq.~(\ref{rrseMB}) is a natural departure point for efforts aiming at an exact analytical solution of the SE for states with $L\geq 0$ and arbitrary masses of the particles, following earlier efforts of 
\textcite{Fock1954,Fock1958,Fock1958a,Pluvinage1955,Morgan1986,Abbott1987,Gottschalk1987a,Gottschalk1987b,He2016,Liverts2010,Liverts2013,Liverts2018,LivertS2022,Langner2022} for the two-electron problem with $L=0$ and a clamped nucleus.

\subsection{Determination of angular integrals}\label{sec:angint}

The computation of the angular integrals $\mathscr{W}_{ll\hspace{-0.3pt}'\hspace{-0.3pt}}^{L\uppi}$ in Eq.~(\ref{angularint1}) can be simplified by first separating the $r_1$ and $r_2$ dependence; following Eq.~\eqref{nsbphgen} we have
\begin{equation}\label{wlpikk}
    \mathscr{W}_{ll\hspace{-0.3pt}'\hspace{-0.3pt}}^{L\uppi} = r_{1}^{l+l\hspace{-0.3pt}'\hspace{-0.3pt}}r_{2}^{2L-l-l\hspace{-0.3pt}'\hspace{-0.3pt}\!+2d}\left\langle\left.\Omega_{l\hspace{-0.3pt}'\hspace{-0.3pt}}^{LM\uppi}\right|\Omega_{l}^{LM\uppi}\right\rangle
\end{equation}
By expressing $\Omega_{l'}^{LM\uppi}$ and $\Omega_{l}^{LM\uppi}$ as linear combinations of the Wigner functions $\mathcal{D}_{L}^{MK}$ [see Eq.~\eqref{mbhwdr}] and using the orthonormality of $\mathcal{D}_{L}^{MK}$ [see Eq.~\eqref{oncwd}],  it is straightforward to show that
\begin{equation}\label{alpikk1}
    \left\langle\left.\Omega_{l\hspace{-0.3pt}'\hspace{-0.3pt}}^{LM\uppi}\right|\Omega_{l}^{LM\uppi}\right\rangle = \sum_{K=-L}^{L}\left(\Lambda_{Kl\hspace{-0.3pt}'\hspace{-0.3pt}}^{L\uppi}\right)^\ast\Lambda_{Kl}^{L\uppi}
\end{equation}
with $\Lambda_{Kl}^{L\uppi}$ given by Eqs.~(\ref{eq:Lambdapcs})--(\ref{Lambdau}). We emphasize again that the coefficients $\Lambda_{Kl}^{L\uppi}$ and $\Lambda_{Kl'}^{L\uppi}$ do not depend on $M$, as demonstrated by Eqs.~(\ref{intlambda}) and~(\ref{eq:Mind}). The sum on the right-hand side of Eq.~\eqref{alpikk1}, after a lengthy series of transformations,\footnote{A detailed exposition is provided in Sec.~III of Supplemental Material \cite{supplemental}} can be expressed in terms of Chebyshev polynomials
\begin{equation}\label{cheby}
    T_{n}(x)\hspace{-2pt}=\hspace{-2pt}\sum\limits_{k=0}^{\lfloor \frac{n}{2}\rfloor}\hspace{-2pt}\tfrac{(-1)^{k}}{2}\left[\!\left(\hspace{-4pt}\begin{array}{c}
            \scriptstyle n-k   \\
            \scriptstyle k 
         \end{array}\hspace{-4pt}\right)\hspace{-2pt}+\hspace{-2pt}\left(\hspace{-4pt}\begin{array}{c}
            \scriptstyle n-k-1   \\
            \scriptstyle n-2k 
         \end{array}\hspace{-4pt}\right)\!\right]\left(2x\right)^{n-2k}
\end{equation} 
where the floor symbol $\lfloor\frac{n}{2}\rfloor$ stands for the largest integer that does not exceed $\frac{n}{2}$. Consequently, we have
\begin{equation}\label{alpikk2a}
    \left\langle\left.\Omega_{l\hspace{-0.3pt}'\hspace{-0.3pt}}^{LM\uppi}\right|\Omega_{l}^{LM\uppi}\right\rangle=\sum_{\kappa=d}^{L}\sum_{\texttt{j},\texttt{j}\hspace{-0.3pt}'=0}^{\kappa}\mathcal{C}_{ll\hspace{-0.3pt}'\hspace{-0.3pt}\texttt{j}\texttt{j}\hspace{-0.3pt}'\hspace{-0.3pt}}^{L\kappa\uppi}\hspace{3pt}T_{\texttt{m}}\left(\cos{\theta}\right)
\end{equation}
with $\texttt{m}=\left\vert -l\!+\!l\hspace{-0.3pt}'\hspace{-0.3pt}\!+\!2\texttt{j}\!-\!2\texttt{j}\hspace{-0.3pt}'\hspace{-0.3pt} \right\vert$ and with 
\begin{eqnarray}\label{clullpjjn}
    \mathcal{C}_{ll\hspace{-0.3pt}'\hspace{-0.3pt}\texttt{j}\texttt{j}\hspace{-0.3pt}'\hspace{-0.3pt}}^{L\kappa \text{n}}\hspace{-4pt}&=&\hspace{-4pt}\left[\hspace{-2pt}\tfrac{\left(\!\frac{3}{2}\!\right)_{\!l}\hspace{-2pt}\left(\!\frac{3}{2}\!\right)_{\!L-l}\hspace{-2pt}\left(\!\frac{3}{2}\!\right)_{\!l\hspace{-0.3pt}'}\hspace{-2pt}\left(\!\frac{3}{2}\!\right)_{\!L-l\hspace{-0.3pt}'}}{l!^{\vphantom{l}}\,\left(\!L-l\!\right)!\,l\hspace{-0.3pt}'!\,\left(\!L-l\hspace{-0.3pt}'\!\right)!}\hspace{-2pt}\right]^{\!\frac{1}{2}}\hspace{-4pt}\tfrac{\left(\hspace{-4pt}\begin{array}{c}
         \scriptstyle l \\
         \scriptstyle \texttt{j} 
    \end{array}\hspace{-4pt}\right)\hspace{-2pt}\left(\hspace{-4pt}\begin{array}{c}
         \scriptstyle L\hspace{0pt}-l \\
         \scriptstyle \kappa\hspace{0pt}-\texttt{j} 
    \end{array}\hspace{-4pt}\right)\hspace{-2pt}\left(\hspace{-4pt}\begin{array}{c}
         \scriptstyle l\hspace{-0.3pt}'\hspace{-0.3pt} \\
         \scriptstyle \texttt{j}\hspace{-0.3pt}'\hspace{-0.3pt} 
    \end{array}\hspace{-4pt}\right)\hspace{-2pt}\left(\hspace{-4pt}\begin{array}{c}
         \scriptstyle L\hspace{0pt}-l\hspace{-0.3pt}'\hspace{-0.3pt} \\
         \scriptstyle \kappa\hspace{0pt}-\texttt{j}\hspace{-0.3pt}'\hspace{-0.3pt} 
    \end{array}\hspace{-4pt}\right)}{2\left(2L+1\right)\left(\hspace{-4pt}\begin{array}{c}
         \scriptstyle 2L \\
         \scriptstyle 2\kappa 
    \end{array}\hspace{-4pt}\right)}
\end{eqnarray}
for the natural parity states ($\uppi=\text{n}$) and 
\begin{eqnarray}\label{clullpjju}
    \mathcal{C}_{ll\hspace{-0.3pt}'\hspace{-0.3pt}\texttt{j}\texttt{j}\hspace{-0.3pt}'\hspace{-0.3pt}}^{L\kappa \text{u}}\hspace{-2pt}&=&\hspace{-2pt}\left[\hspace{-2pt}\tfrac{\left(\!\frac{3}{2}\!\right)_{\!l}\hspace{-2pt}\left(\!\frac{3}{2}\!\right)_{\!L-l+1}\hspace{-2pt}\left(\!\frac{3}{2}\!\right)_{\!l\hspace{-0.3pt}'}\hspace{-2pt}\left(\!\frac{3}{2}\!\right)_{\!L-l\hspace{-0.3pt}'+1}}{\left(\!l-1\!\right)!^{\vphantom{l}}\,\left(\!L-l\!\right)!\,\left(\!l\hspace{-0.3pt}'\hspace{-0.3pt}-1\!\right)!\,\left(\!L-l\hspace{-0.3pt}'\!\right)!}\hspace{-2pt}\right]^{\frac{1}{2}}\hspace{-4pt}\tfrac{\left(\!L\texttt{j}\vphantom{\hspace{-0.3pt}'\hspace{-0.3pt}}-\kappa l+\texttt{j}\!\right)\left(\!L\texttt{j}\hspace{-0.3pt}'\hspace{-0.3pt}-\kappa l\hspace{-0.3pt}'\hspace{-0.3pt}+\texttt{j}\hspace{-0.3pt}'\hspace{-0.3pt}\!\right)}{ll\hspace{-0.3pt}'\hspace{-0.3pt}\left(L-l+1\right)^{\vphantom{l}}\left(L-l\hspace{-0.3pt}'\hspace{-0.3pt}+1\right)}\nonumber\\
    && \tfrac{\left(\hspace{-4pt}\begin{array}{c}
         \scriptstyle l \\
         \scriptstyle \texttt{j} 
    \end{array}\hspace{-4pt}\right)\hspace{-2pt}\left(\hspace{-4pt}\begin{array}{c}
         \scriptstyle L-l+1 \\
         \scriptstyle \kappa-\texttt{j} 
    \end{array}\hspace{-4pt}\right)\hspace{-2pt}\left(\hspace{-4pt}\begin{array}{c}
         \scriptstyle l\hspace{-0.3pt}'\hspace{-0.3pt} \\
         \scriptstyle \texttt{j}\hspace{-0.3pt}'\hspace{-0.3pt} 
    \end{array}\hspace{-4pt}\right)\hspace{-2pt}\left(\hspace{-4pt}\begin{array}{c}
         \scriptstyle L-l\hspace{-0.3pt}'\hspace{-0.3pt}+1 \\
         \scriptstyle \kappa-\texttt{j}\hspace{-0.3pt}'\hspace{-0.3pt} 
    \end{array}\hspace{-4pt}\right)}{\left(L+1\right)^{2}\left(\hspace{-4pt}\begin{array}{c}
         \scriptstyle 2L+1 \\
         \scriptstyle 2\kappa-1 
    \end{array}\hspace{-4pt}\right)}
\end{eqnarray}
for the unnatural parity states ($\uppi=\text{u}$).

For practical purposes, we express the angular integral $\mathscr{W}_{ll\hspace{-0.3pt}'\hspace{-0.3pt}}^{L\uppi}$ as a polynomial in $r_1$, $r_2$, and $r_{12}$, which can be useful for implementing the standard Hylleraas scheme for $L\geq 0$. Since $\cos\theta=\frac{r_1^2+r_2^2-r_{12}^2}{2r_1r_2}$, the Chebyshev polynomial can be expressed as
\begin{eqnarray}\label{wlpikkd}
    T_{\texttt{m}}\left(\cos{\theta}\right) &=&\hspace{-2pt} \sum_{\lambda=0}^{\lfloor \frac{\texttt{m}}{2}\rfloor} \sum_{r=0}^{\texttt{m}-2\lambda} \sum_{s=0}^{r}\tfrac{(-1)^{\lambda+s}}{2}\hspace{-2pt}\left[\!\left(\hspace{-4pt}\begin{array}{c}
            \scriptstyle \texttt{m}-\lambda   \\
            \scriptstyle \lambda 
         \end{array}\hspace{-4pt}\right)\hspace{-2pt}+\hspace{-2pt}\left(\hspace{-4pt}\begin{array}{c}
            \scriptstyle \texttt{m}-\lambda-1   \\
            \scriptstyle \texttt{m}-2\lambda 
         \end{array}\hspace{-4pt}\right)\!\right]\hspace{-5pt}\nonumber\\
    &&\hspace{-40pt} \left(\hspace{-4pt}\begin{array}{c}
            \scriptstyle \texttt{m}-2\lambda   \\
            \scriptstyle r 
         \end{array}\hspace{-4pt}\right)\hspace{-3pt}\left(\hspace{-4pt}\begin{array}{c}
            \scriptstyle r   \\
            \scriptstyle s 
         \end{array}\hspace{-4pt}\right)\hspace{1pt} r_{1}^{\texttt{m}-2\lambda-2r}\hspace{2pt}r_{2}^{-\texttt{m}+2\lambda+2r-2s}\hspace{2pt}r_{12}^{2s}
\end{eqnarray}
which yields the following explicit expression for $\mathscr{W}_{ll\hspace{-0.3pt}'\hspace{-0.3pt}}^{L\uppi}$ as a polynomial in $r_1$, $r_2$, and $r_{12}$  
\begin{eqnarray}\label{wlpikkb1}
    \mathscr{W}_{ll\hspace{-0.3pt}'\hspace{-0.3pt}}^{L\uppi}\hspace{-2pt} &=&\frac{1}{2}\hspace{-2pt} \sum_{\kappa=d}^{L}\sum_{\texttt{j},\texttt{j}\hspace{-0.3pt}'=0}^{\kappa} \sum_{\lambda=0}^{\lfloor \frac{\texttt{m}}{2}\rfloor}\hspace{-2pt} \sum_{r=0}^{\texttt{m}-2\lambda}\hspace{-2pt} \sum_{s=0}^{r} (-1)^{\lambda+s}\hspace{2pt}\mathcal{C}_{ll\hspace{-0.3pt}'\hspace{-0.3pt}\texttt{j}\texttt{j}\hspace{-0.3pt}'\hspace{-0.3pt}}^{L\kappa\uppi}\\
    &&\left[\hspace{-1pt}\left(\hspace{-4pt}\begin{array}{c}
            \scriptstyle \texttt{m}-\lambda   \\
            \scriptstyle \lambda 
         \end{array}\hspace{-4pt}\right)\hspace{-2pt}+\hspace{-2pt}\left(\hspace{-4pt}\begin{array}{c}
            \scriptstyle \texttt{m}-\lambda-1   \\
            \scriptstyle \texttt{m}-2\lambda 
         \end{array}\hspace{-4pt}\right)\hspace{-1pt}\right]\hspace{-3pt}\left(\hspace{-4pt}\begin{array}{c}
            \scriptstyle \texttt{m}-2\lambda   \\
            \scriptstyle r 
         \end{array}\hspace{-4pt}\right)\hspace{-3pt}\left(\hspace{-4pt}\begin{array}{c}
            \scriptstyle r   \\
            \scriptstyle s 
         \end{array}\hspace{-4pt}\right)r_{1}^{l+l\hspace{-0.3pt}'\hspace{-0.3pt}+\texttt{m}-2\lambda-2r}\nonumber\\[3pt]
    &&r_{2}^{2L-l-l\hspace{-0.3pt}'\hspace{-0.3pt}\!+2d-\texttt{m}+2\lambda+2r-2s}\hspace{2pt}r_{12}^{2s}\nonumber
\end{eqnarray}

The angular integral $\mathscr{W}_{ll\hspace{-0.3pt}' \hspace{-0.3pt}}^{L\uppi}$ was first introduced by \textcite{Drake1978} in the form given explicitly as Eq.~(15) of \textcite{Frolov1996} 
\begin{eqnarray}\label{angularint20}
    \mathscr{W}_{ll\hspace{-0.3pt}' \hspace{-0.3pt}}^{L\uppi} &=& r_{1}^{l+l\hspace{-0.3pt}' \hspace{-0.3pt}} r_{2}^{2L-l-l\hspace{-0.3pt}'\hspace{-0.3pt}\!+2d} \hspace{2pt}\sum_{\uplambda} (-1)^{L+\uplambda}\left(\uplambda+\tfrac{1}{2}\right)\\
    && \hspace{-30pt} \sqrt{\hspace{-2pt}\left(2L\!+\!2d\!-\!2l\hspace{-0.3pt}'\hspace{-0.3pt}\!+\!1\right)\hspace{-3pt}\left(2L\!+\!2d\!-\!2l\vphantom{'}\!+\!1\right)\hspace{-3pt}\left(2l\hspace{-0.3pt}'\hspace{-0.3pt}\!+\!1\right)\hspace{-3pt}\left(2l\vphantom{'}\!+\!1\right)}\nonumber\\[3pt]
     && \hspace{-30pt} \left(\hspace{-4pt}\begin{array}{ccc}
        \hspace{1pt}\scriptstyle l\hspace{-0.3pt}'\hspace{-0.3pt} & \scriptstyle l & \scriptstyle \uplambda \\[3pt]
        \scriptstyle 0 & \scriptstyle 0 & \scriptstyle 0
    \end{array}\hspace{-4pt}\right)\hspace{-4pt}\left(\hspace{-4pt}\begin{array}{ccc}
        \scriptstyle L+d-l\hspace{-0.3pt}'\hspace{-0.3pt} & \scriptstyle L+d-l & \scriptstyle \uplambda \\[3pt]
        \scriptstyle 0 & \scriptstyle 0 & \scriptstyle 0
    \end{array}\hspace{-4pt}\right)\hspace{-4pt}\left\{\hspace{-5pt}\begin{array}{ccc}
         \scriptstyle l\hspace{-0.3pt}'\hspace{-0.3pt} & \hspace{-4pt}\scriptstyle L+d-l\hspace{-0.3pt}'\hspace{-0.3pt} & \hspace{-4pt}\scriptstyle L \\[3pt]
         \scriptstyle L+d-l & \hspace{-4pt}\scriptstyle l & \hspace{-4pt}\scriptstyle \uplambda
    \end{array}\hspace{-3pt}\right\}P_{\uplambda}(\cos{\theta})\nonumber
\end{eqnarray}
where $(\hspace{-2pt}\begin{array}{ccc}
\scriptscriptstyle a \hspace{-4pt}&\hspace{-4pt} \scriptscriptstyle b \hspace{-4pt}&\hspace{-4pt} \scriptscriptstyle c \\[-6pt]
\scriptscriptstyle\alpha \hspace{-4pt}&\hspace{-4pt} \scriptscriptstyle\beta \hspace{-4pt}&\hspace{-4pt} \scriptscriptstyle\gamma
\end{array}\hspace{-2pt})$ and $\{\hspace{-2pt}\begin{array}{ccc}
\scriptscriptstyle a \hspace{-4pt}&\hspace{-4pt} \scriptscriptstyle b \hspace{-4pt}&\hspace{-4pt} \scriptscriptstyle c \\[-6pt]
\scriptscriptstyle d \hspace{-4pt}&\hspace{-4pt} \scriptscriptstyle e \hspace{-4pt}&\hspace{-4pt} \scriptscriptstyle f
\end{array}\hspace{-2pt}\}$ denote the Wigner $3$-$j$ and $6$-$j$ symbols, respectively.  A detailed, mathematically rigorous demonstration presented in Sec.~IV of Supplemental Material \cite{supplemental} shows that the form of the angular integral obtained here by com\-bin\-ing Eqs.~(\ref{angularint1}), (\ref{wlpikk}), and~(\ref{alpikk2a})
\begin{equation}\label{wlpikkb}
    \mathscr{W}_{ll\hspace{-0.3pt}'\hspace{-0.3pt}}^{L\uppi} = r_{1}^{l+l\hspace{-0.3pt}'\hspace{-0.3pt}}r_{2}^{2L-l-l\hspace{-0.3pt}'\hspace{-0.3pt}\!+2d}\sum_{\kappa=d}^{L}\sum_{\texttt{j},\texttt{j}\hspace{-0.3pt}'=0}^{\kappa}\mathcal{C}_{ll\hspace{-0.3pt}'\hspace{-0.3pt}\texttt{j}\texttt{j}\hspace{-0.3pt}'\hspace{-0.3pt}}^{L\kappa\uppi}\hspace{2pt}T_{\texttt{m}}\left(\cos{\theta}\right)
\end{equation}
with $\texttt{m}=\left\vert -l\!+\!l\hspace{-0.3pt}'\hspace{-0.3pt}\!+\!2\texttt{j}\!-\!2\texttt{j}\hspace{-0.3pt}'\hspace{-0.3pt} \right\vert$ and $\mathcal{C}_{ll\hspace{-0.3pt}'\hspace{-0.3pt}\texttt{j}\texttt{j}\hspace{-0.3pt}'\hspace{-0.3pt}}^{L\kappa\uppi}$ given by Eqs.~(\ref{clullpjjn}) and~(\ref{clullpjju}), is algebraically equivalent to the original definition of \textcite{Drake1978} and \textcite{Frolov1996}.

\subsection{Explicit form of the \text{RSE}s for \texorpdfstring{$\text{S}$, $\text{P}$, $\text{D}$, and $\text{F}$}~ states and the corresponding wave functions with \texorpdfstring{$M=L$}{M=L}}\label{sec:spdf}

The formulas presented in previous sections may appear complicated because they are universal, re\-main\-ing valid for any angular momentum $L$, its projection $M$, and parity $\uppi$. In practice, however, the resulting expressions for small values of $L$ (cf.\ Table~\ref{tab:termsymbols}), which are most commonly used in applications, are quite simple. To illustrate this, we present below several explicit forms of the block Hamiltonians for $L=0$, 1, 2, and 3 and the corresponding wave functions with $M=L$. Analogous wave functions with $M\neq L$ are discussed in the next section.

\begin{table}[tb]
    \centering
    \caption{Term symbols for states with natural ($d=0$) and unnatural ($d=1$) parity for distinct values of $L$.}\label{tab:termsymbols}
    \begin{tabular}{l|@{\quad}l@{\quad}l@{\quad}l@{\quad}l@{\quad}l@{\quad}l@{\quad}l@{\quad}l}
    \hline\hline
    \diagbox{$\vspace{0pt}~d$}{$\vspace{-7pt}~L^{\vphantom{\big|}}$} & 0\footnote{The case of $L=0$ and $d=1$ is unphysical.} & 1 & 2 & 3 & 4 & 5 & 6 & 7 \\
     \hline
      ~0$\vphantom{\Big|}$   & $\textnormal{S}^{\textnormal{e}}$ & $\textnormal{P}^{\textnormal{o}}$ & $\textnormal{D}^{\textnormal{e}}$ & $\textnormal{F}^{\textnormal{o}}$ & $\textnormal{G}^{\textnormal{e}}$ & $\textnormal{H}^{\textnormal{o}}$ & $\textnormal{I}^{\textnormal{e}}$ & $\textnormal{K}^{\textnormal{o}}$\\
      ~1   &  & $\textnormal{P}^{\textnormal{e}}$ & $\textnormal{D}^{\textnormal{o}}$ & $\textnormal{F}^{\textnormal{e}}$ & $\textnormal{G}^{\textnormal{o}}$ & $\textnormal{H}^{\textnormal{e}}$ & $\textnormal{I}^{\textnormal{o}}$ & $\textnormal{K}^{\textnormal{e}}$~ \\
    \hline\hline
    \end{tabular}
\end{table}

\subsubsection{The \textnormal{RSE} for \texorpdfstring{$\textnormal{S}^{\textnormal{e}}$}~ states}
In the simplest case of the $\text{S}^{\textnormal{e}}$ states, the wave function
\begin{equation}
    \Psi^{00\text{n}} = \psi_{0}\hspace{2pt}\bm{\Omega}_{0}^{00\text{n}}
\end{equation}
with $\bm{\Omega}_{0}^{00\text{n}}\!=\! \frac{1}{4\pi}$, can be con\-struct\-ed using only a single reduced component $\psi_{0}\equiv\psi_{0}^{0\text{n}}(r_1,r_2,r_{12})$ considered first by \textcite{Hylleraas1929}. Consequently, the RSE for the $\text{S}^{\textnormal{e}}$ states 
\begin{eqnarray}\label{rrsese}
        \left(\hat{h}_{0}-E\right)\psi_{0} =   0
\end{eqnarray}
with $\hat{h}_{0}\equiv\hat{H}_\text{Hyl}$ obtained from Eq.~\eqref{hkkgnygp2} as
\begin{eqnarray}\label{h00} 
    \hat{H}_\text{Hyl} &\!=\!& 
    -\tfrac{\partial_{r_{1\vphantom{y}}r_{1}}}{2\mu_{1}}
    \hspace{-3pt}-\hspace{-3pt}\tfrac{\partial_{r_{2\vphantom{y}}r_{2}}}{2\mu_{2}}
    \hspace{-3pt}-\hspace{-3pt}\tfrac{\partial_{r_{12\vphantom{y}}r_{12}}}{2\mu_{12}}\hspace{-3pt}+\hspace{-3pt} \tfrac{q_{1\vphantom{y}}q_{2}}{r_{12}}\hspace{-3pt}+\hspace{-3pt}\tfrac{q_{1\vphantom{y}}q_{3}}{r_{1}\vphantom{\mu_2}}\hspace{-3pt}+\hspace{-3pt}\tfrac{q_{2\vphantom{y}}q_{3}}{r_{2}} \nonumber\\[2pt]
    &&  -\tfrac{\partial_{r_{1\vphantom{y}}}}{\mu_{1}r_{1}}\hspace{-3pt}-\hspace{-3pt}\tfrac{\partial_{r_{12\vphantom{y}}}}{m_{1}r_{12}}
    \hspace{-3pt}-\hspace{-2pt}\tfrac{\partial_{r_{2\vphantom{y}}}}{\mu_{2}r_{2}}\hspace{-3pt}-\hspace{-3pt}\tfrac{\partial_{r_{12\vphantom{y}}}}{m_{2}r_{12}}\hspace{-3pt}-\hspace{-3pt} \tfrac{r_{1\vphantom{y}}^2+r_{2}^2-r_{12}^2}{2m_{3}r_{1}r_{2}}\partial_{r_{1}r_{2}}\nonumber\\[2pt]
    && \!-\tfrac{r_{1\vphantom{y}}^2+r_{12}^2-r_{2}^2}{2m_{1}r_{1}r_{12}}\partial_{r_{1}r_{12}}\hspace{-4pt}-\hspace{-3pt}  \tfrac{r_{2\vphantom{y}}^2+r_{12}^2-r_{1}^2}{2m_{2}r_{2}r_{12}}\partial_{r_{2}r_{12}} 
\end{eqnarray}
is the generalization of the original Hylleraas equation for a clamped-nucleus, two-electron system [cf.\ Eq.~(5) of \textcite{Hylleraas1929}, or, for a more recent account, Eq.~(5) of \textcite{Nakashima2007}] to the case of a finite-mass nucleus.\footnote{The original formulation implies that $m_1=m_2=\mu_1=\mu_2=1$, $m_3\rightarrow\infty$, $\mu_{12}=\tfrac{1}{2}$, $q_1=q_2=-1$, and $q_3=Z$.}  The form of the operator $\hat{H}_\text{Hyl}$ in Eq.~\eqref{h00} is consistent with Eq.~(24) of \textcite{Efros1986}. Several variants of Eq.~\eqref{rrsese} were derived in the literature; a list of representative citations has been given above in the paragraph following Eq.~\eqref{hamtme}.

\subsubsection{The \textnormal{RSE} for \texorpdfstring{$\textnormal{P}^{\textnormal{e}}$ and $\textnormal{P}^{\textnormal{o}}$}~ states}
The next value of angular momentum, $L\hspace{-1.5pt}=\hspace{-1.5pt}1$, can be combined with both natural and unnatural parity, which results in the $\text{P}^{\textnormal{o}}$ and $\text{P}^{\textnormal{e}}$ states, respectively. For the $\text{P}^{\textnormal{e}}$ states, the wave function 
\begin{equation}
    \Psi^{11\text{u}} = \psi_{1}\hspace{2pt}\bm{\Omega}_{1}^{11\text{u}}
\end{equation}
with 
\begin{equation}
    \bm{\Omega}_{1}^{11\text{u}} = \tfrac{3}{8\pi}\left[z_{1}(x_{2}+\textit{\textbf{i}}\hspace{1pt}y_{2})-z_{2}(x_{1}+\textit{\textbf{i}}\hspace{1pt}y_{1})\right]\label{Omg111u}
\end{equation}
which follows from Eqs.~(\ref{nsbshccnp})--(\ref{usbshccup}) alongside Eq.~\eqref{nlpik}, can be con\-struct\-ed using only a single reduced component $\psi_{1}\equiv\psi_{1}^{1\text{u}}(r_1,r_2,r_{12})$. The RSE for the $\text{P}^{\textnormal{e}}$ states is
\begin{equation}
   \mathscr{W}_{11}^{1\text{u}}\left(\hat{h}_{1}-E\right)\psi_{1} = 0 
\end{equation}
where
\begin{eqnarray}\label{w111u}
     \mathscr{W}_{11}^{1\text{u}} &=& \tfrac{3}{4}\hspace{2pt}r_{1}^2\hspace{2pt}r_{2}^2\hspace{2pt}\sin^2{\hspace{-1.5pt}\theta} \\[5pt] \nonumber
     &=& \tfrac{3}{16}\left(r_{1}\!+r_{2}\!+r_{12}\right)\left(r_{2}\!-r_{1}\!+r_{12}\right) \\
     && \phantom{-\tfrac{3}{6}} \cdot\!\left(r_{1}\!-r_{2}\!+r_{12}\right)\left(r_{1}\!+r_{2}\!-r_{12}\right) \nonumber
\end{eqnarray}
The explicit form of $\hat{h}_{1}$ follows from Eqs.~\eqref{hkkgnygp2} as
\begin{equation*}
    \hat{h}_{1} = \hat{H}_\text{Hyl}\hspace{-1pt}-\hspace{-1pt}\tfrac{\partial_{r_{1\vphantom{y}}}}{\mu_{1}r_{1}}\hspace{-1pt}-\hspace{-1pt}\tfrac{\partial_{r_{2\vphantom{y}}}}{\mu_{2}r_{2}}\hspace{-1pt}-\hspace{-1pt}\tfrac{\partial_{r_{12\vphantom{y}}}}{\mu_{12}r_{12}}\hspace{-3pt}
\end{equation*}
with $\hat{H}_\text{Hyl}$ defined by Eq.~\eqref{h00}. 

For the $\text{P}^{\textnormal{o}}$ states, the corresponding wave function
\begin{eqnarray}
    \Psi^{11\text{n}} = \psi_{0}\hspace{2pt}\bm{\Omega}_{0}^{11\text{n}}\! + \psi_{1}\hspace{2pt}\bm{\Omega}_{1}^{11\text{n}}
\end{eqnarray}
with  
\begin{eqnarray*}
    \bm{\Omega}_{0}^{11\text{n}} &=& -\tfrac{\sqrt{6}}{8\pi} \left(x_{2}+\textit{\textbf{i}}\, y_{2}\right)\nonumber\\
    \bm{\Omega}_{1}^{11\text{n}} &=& -\tfrac{\sqrt{6}}{8\pi} \left(x_{1}+\textit{\textbf{i}}\, y_{1}\right)
\end{eqnarray*}
which follow from Eqs.~\eqref{nsbshccnp} and~\eqref{usbshccnp} alongside Eq.~\eqref{nlpik}, can be constructed using two reduced components $\psi_{0}\equiv\psi_{0}^{1\text{n}}(r_1,r_2,r_{12})$ and $\psi_{1}\equiv\psi_{1}^{1\text{n}}(r_1,r_2,r_{12})$. The corresponding RSE takes the following two-component form
\begin{equation}\label{rrsepo}
\left(\!\!\begin{array}{cc}
     \mathscr{W}_{00}^{1\text{n}} \hspace{-1pt}& \hspace{-1pt}\mathscr{W}_{10}^{1\text{n}}\\
      \\
     \mathscr{W}_{01}^{1\text{n}} \hspace{-1pt}& \hspace{-1pt} \mathscr{W}_{11}^{1\text{n}}
\end{array}\!\!\right)\hspace{-5pt}\left(\!\begin{array}{lr}
     \hat{h}_{0}\hspace{-2.5pt}-\hspace{-2pt}E \hspace{-9pt}&\hspace{-15pt} \hat{h}^{\uparrow}_{0}\\
        &\\
        \hat{h}^{\downarrow}_{1} \hspace{-9pt}& \hspace{-15pt}\hat{h}_{1}\hspace{-2.5pt}-\hspace{-2pt}E
\end{array}\!\right)\hspace{-5pt}\left(\!\!\begin{array}{c}
     \psi_{0}\\
     \\
    \psi_{1}
\end{array}\!\!\right)\! =\! \left(\!\begin{array}{c}
     0\\
     \\
    0
\end{array}\!\right)
\end{equation}
where
\begin{eqnarray}
    \left(\!\!\begin{array}{cc}
     \mathscr{W}_{00}^{1\text{n}} \hspace{-1pt}& \hspace{-1pt}\mathscr{W}_{10}^{1\text{n}}\\
      \\
     \mathscr{W}_{01}^{1\text{n}} \hspace{-1pt}& \hspace{-1pt} \mathscr{W}_{11}^{1\text{n}}
\end{array}\!\!\right) &=& \tfrac{r_{1}r_{2\vphantom{_{a}}}}{2}\!\left(\!\begin{array}{lr}
     \frac{r_{2\vphantom{_{a}}}}{r_{1}}\hspace{-5pt} & \hspace{-10pt}\cos{\theta}\\
     \\ 
     \cos{\theta} \hspace{-5pt}&\hspace{-10pt} \frac{r_{1\vphantom{_{a}}}}{r_{2}}
\end{array}\!\right)= \left(\!\!\begin{array}{lr}
     \frac{r^{2}_{2\vphantom{_{a}}}}{2} & \hspace*{-32pt}\frac{r_{1\vphantom{_{a}}}^{2}+r_{2}^{2}-r_{12}^{2}}{4}\\
     \\ 
     \frac{r_{1\vphantom{_{a}}}^{2}+r_{2}^{2}-r_{12}^{2}}{4} &\hspace*{-32pt} \frac{r^{2}_{1\vphantom{_{a}}}}{2}
\end{array}\!\!\right) \nonumber
\end{eqnarray}
The explicit form of the operators $\hat{h}_{0}$, $\hat{h}_{1}$, $\hat{h}^{\downarrow}_{1}$, and $\hat{h}^{\uparrow}_{0}$ can be easily found from Eqs.~(\ref{hkkgnygp2})--(\ref{hkk+1gnygp2}) as
\begin{equation*}
\begin{array}{c}
    \hat{h}_{0} = \hat{H}_\text{Hyl}\hspace{-1pt}-\hspace{-1pt}\tfrac{\partial_{r_{2\vphantom{y}}}}{\mu_{2}\,r_{2}}\hspace{-1pt}-\hspace{-1pt}\tfrac{\partial_{r_{12\vphantom{y}}}}{m_{2}\,r_{12}} \hspace{25pt} \hat{h}^{\uparrow}_{0} = \tfrac{\partial_{r_{12\vphantom{y}}}}{m_{1}\,r_{12}}\hspace{-1pt}-\hspace{-1pt}\tfrac{\partial_{r_{2\vphantom{y}}}}{m_{3}\,r_{2}}\\[7pt]
    \hat{h}_{1} = \hat{H}_\text{Hyl}\hspace{-1pt}-\hspace{-1pt}\tfrac{\partial_{r_{1\vphantom{y}}}}{\mu_{1}\,r_{1}}\hspace{-1pt}-\hspace{-1pt}\tfrac{\partial_{r_{12\vphantom{y}}}}{m_{1}\,r_{12}} \hspace{25pt} \hat{h}^{\downarrow}_{1} = \tfrac{\partial_{r_{12\vphantom{y}}}}{m_{2}\,r_{12}}\hspace{-1pt}-\hspace{-1pt}\tfrac{\partial_{r_{1\vphantom{y}}}}{m_{3}\,r_{1}}
\end{array}
\end{equation*}
with $\hat{H}_\text{Hyl}$ defined by Eq.~\eqref{h00}.

\subsubsection{The \textnormal{RSE} for \texorpdfstring{$\textnormal{D}^{\textnormal{e}}$ and $\textnormal{D}^{\textnormal{o}}$}~ states}
The wave function for the $\text{D}^{\text{o}}$ states 
\begin{equation}
    \Psi^{22\text{u}} = \psi_{1}\hspace{2pt}\bm{\Omega}_{1}^{22\text{u}}\! + \psi_{2}\hspace{2pt}\bm{\Omega}_{2}^{22\text{u}} 
\end{equation}
can be constructed using two reduced components $\psi_{1}\equiv\psi_{1}^{2\text{u}}(r_1,r_2,r_{12})$ and $\psi_{2}\equiv\psi_{2}^{2\text{u}}(r_1,r_2,r_{12})$, and two angular generators
\begin{eqnarray*}
    \bm{\Omega}_{1}^{22\text{u}} &=& -\tfrac{\sqrt{15}}{3}\bm{\Omega}_{1}^{11\text{u}}\left(x_{2}+\textit{\textbf{i}}\hspace{1pt}y_{2}\right)\\
    \bm{\Omega}_{2}^{22\text{u}} &=& -\tfrac{\sqrt{15}}{3}\bm{\Omega}_{1}^{11\text{u}}\left(x_{1}+\textit{\textbf{i}}\hspace{1pt}y_{1}\right)
\end{eqnarray*}
with $\bm{\Omega}_{1}^{11\text{u}}$ given by Eq.~\eqref{Omg111u}. The corresponding RSE is
\begin{equation}
\left(\!\begin{array}{cc}
     \mathscr{W}_{11}^{2\text{u}} \hspace{-1pt}& \hspace{-1pt}\mathscr{W}_{21}^{2\text{u}}\\
      \\
     \mathscr{W}_{12}^{2\text{u}} \hspace{-1pt}& \hspace{-1pt} \mathscr{W}_{22}^{2\text{u}}
\end{array}\!\right)\hspace{-5pt}\left(\!\begin{array}{lr}
     \hat{h}_{1}\hspace{-2.5pt}-\hspace{-2pt}E \hspace{-9pt}&\hspace{-15pt} \hat{h}^{\uparrow}_{1}\\
        &\\
        \hat{h}^{\downarrow}_{2} \hspace{-9pt}& \hspace{-15pt}\hat{h}_{2}\hspace{-2.5pt}-\hspace{-2pt}E
\end{array}\!\right)\hspace{-5pt}\left(\!\!\begin{array}{c}
     \psi_{1}\\
     \\
    \psi_{2}
\end{array}\!\!\right)\!\! =\!\! \left(\!\begin{array}{c}
     0\\
     \\
    0
\end{array}\!\right)
\end{equation}
where
\begin{equation*}
    \left(\!\!\begin{array}{cc}
     \mathscr{W}_{11}^{2\text{u}} \hspace{-1pt}& \hspace{-1pt}\mathscr{W}_{21}^{2\text{u}}\\
      \\
     \mathscr{W}_{12}^{2\text{u}} \hspace{-1pt}& \hspace{-1pt} \mathscr{W}_{22}^{2\text{u}}
\end{array}\!\!\right)=r_{1}r_{2}\hspace{2pt} \mathscr{W}_{11}^{1\text{u}}\!\left(\!\begin{array}{cc}
        \frac{r_{2\vphantom{_{a}}}}{r_{1}} \hspace{-2pt}&\hspace{-2pt} \cos{\theta}\\
        &\\
        \cos{\theta} \hspace{-2pt}&\hspace{-2pt} \frac{r_{1\vphantom{_{a}}}}{r_{2}}\\
    \end{array}\!\right) 
\end{equation*}
with $\mathscr{W}_{11}^{1\text{u}}$ defined by Eq.~\eqref{w111u} and the explicit form of the operators $\hat{h}_{1}$, $\hat{h}_{2}$, $\hat{h}^{\uparrow}_{1}$ and $\hat{h}^{\downarrow}_{2}$ given by
\begin{equation*}
\begin{array}{c}
    \hat{h}_{1} =\hspace{1pt} \hat{H}_\text{Hyl}-\tfrac{\partial_{r_{1\vphantom{y}}}}{\mu_{1}\,r_{1}}-\tfrac{\partial_{r_{12\vphantom{y}}}}{m_{1}\,r_{12}}-\tfrac{2\partial_{r_{2\vphantom{y}}}}{\mu_{2}\,r_{2}}-\tfrac{2\partial_{r_{12\vphantom{y}}}}{m_{2}\,r_{12}} \\[4pt]
    \hat{h}_{2} = \hspace{1pt}\hat{H}_\text{Hyl}-\tfrac{2\partial_{r_{1\vphantom{y}}}}{\mu_{1}\,r_{1}}-\tfrac{2\partial_{r_{12\vphantom{y}}}}{m_{1}\,r_{12}}-\hspace{-1pt}\tfrac{\partial_{r_{2\vphantom{y}}}}{\mu_{2}\,r_{2}}-\tfrac{\partial_{r_{12\vphantom{y}}}}{m_{2}\,r_{12}}\\[4pt]
    \hat{h}^{\uparrow}_{1} =\tfrac{\partial_{r_{12\vphantom{y}}}}{m_{1}\,r_{12}}\hspace{-1pt}-\hspace{-1pt}\tfrac{\partial_{r_{2\vphantom{y}}}}{m_{3}\,r_{2}}\hspace{23pt}\hat{h}^{\downarrow}_{2} = \tfrac{\partial_{r_{12\vphantom{y}}}}{m_{2}\,r_{12}}\hspace{-1pt}-\hspace{-1pt}\tfrac{\partial_{r_{1\vphantom{y}}}}{m_{3}\,r_{1}}
\end{array}
\end{equation*}

The wave function for the $\text{D}^{\text{e}}$ states
\begin{eqnarray}
    \Psi^{22\text{n}} = \psi_{0}\hspace{2pt}\bm{\Omega}_{0}^{22\text{n}}\! + \psi_{1}\hspace{2pt}\bm{\Omega}_{1}^{22\text{n}}\! + \psi_{2}\hspace{2pt}\bm{\Omega}_{2}^{22\text{n}}
\end{eqnarray}
can be constructed using three reduced components $\psi_{l}\equiv\psi_{l}^{2\text{n}}(r_1,r_2,r_{12})$ with $l=0$, $1$, and $2$, and three angular generators
\begin{eqnarray*}
    \bm{\Omega}_{0}^{22\text{n}} &=& \tfrac{\sqrt{30}}{16\pi}\left(x_{2}+\textit{\textbf{i}}\,y_{2}\right)^{2}\\
    \bm{\Omega}_{1}^{22\text{n}} &=& \tfrac{3}{8\pi}\left(x_{1}+\textit{\textbf{i}}\,y_{1}\right)\left(x_{2}+\textit{\textbf{i}}\,y_{2}\right)\\
    \bm{\Omega}_{2}^{22\text{n}} &=& \tfrac{\sqrt{30}}{16\pi}\left(x_{1}+\textit{\textbf{i}}\,y_{1}\right)^{2}
\end{eqnarray*}
The corresponding RSE takes the following form
\begin{equation}
    \hspace{-4pt}\left(\!\!\begin{array}{ccc}
        \mathscr{W}_{00}^{2\text{n}} \hspace{-6pt}&\hspace{-6pt} \mathscr{W}_{10}^{2\text{n}} \hspace{-5pt}&\hspace{-5pt} \mathscr{W}_{20}^{2\text{n}}\vspace{-5pt}\\
        &\\
        \mathscr{W}_{01}^{2\text{n}} \hspace{-5pt}&\hspace{-5pt} \mathscr{W}_{11}^{2\text{n}} \hspace{-5pt}&\hspace{-5pt} \mathscr{W}_{21}^{2\text{n}}\vspace{-5pt}\\
        &\\
        \mathscr{W}_{02}^{2\text{n}} \hspace{-5pt}&\hspace{-5pt} \mathscr{W}_{12}^{2\text{n}} \hspace{-5pt}&\hspace{-5pt} \mathscr{W}_{22}^{2\text{n}}
    \end{array}\!\!\right)\hspace{-4.5pt}\left(\!\!\begin{array}{llr}
        \hat{h}_{0}\hspace{-2.5pt}-\hspace{-2pt}E & \hspace{5pt}\hat{h}^{\uparrow}_{0} &  0\hspace*{2pt}\vspace{-5pt}\\
        &\\
        \hat{h}^{\downarrow}_{1} & \hspace{-10pt}\hat{h}_{1}\hspace{-4.5pt}-\hspace{-2pt}E & \hspace{-10pt}\hat{h}^{\uparrow}_{1}\vspace{-5pt}\\
        &\\
        \hspace{2pt}0 & \hspace{-10pt}\hat{h}^{\downarrow}_{2} & \hspace{-10pt}\hat{h}_{2}\hspace{-2.5pt}-\hspace{-2pt}E
    \end{array}\!\!\right)\hspace{-4.5pt}\left(\!\!\begin{array}{c}
    \psi_{0}\vspace{-5pt}\\
    \\
    \psi_{1}\vspace{-5pt}\\
    \\
    \psi_{2}
    \end{array}\!\!\right)\hspace{-4pt} =\hspace{-4pt} \left(\!\begin{array}{c}
    0\vspace{-5pt}\\
    \\
    0\vspace{-5pt}\\
    \\
    0
    \end{array}\!\right)
\end{equation}
where
\begin{equation}
    \left(\!\!\begin{array}{ccc}
        \mathscr{W}_{00}^{2\text{n}} \hspace{-5pt}&\hspace{-5pt} \mathscr{W}_{10}^{2\text{n}} \hspace{-5pt}&\hspace{-5pt} \mathscr{W}_{20}^{2\text{n}}\vspace{-5pt}\\ 
        &\\
        \mathscr{W}_{01}^{2\text{n}} \hspace{-5pt}&\hspace{-5pt} \mathscr{W}_{11}^{2\text{n}} \hspace{-5pt}&\hspace{-5pt} \mathscr{W}_{21}^{2\text{n}}\vspace{-5pt}\\
        &\\
        \mathscr{W}_{02}^{2\text{n}} \hspace{-5pt}&\hspace{-5pt} \mathscr{W}_{12}^{2\text{n}} \hspace{-5pt}&\hspace{-5pt} \mathscr{W}_{22}^{2\text{n}}
    \end{array}\!\!\right)\!\!=\!
     \tfrac{r_{1}^2r_{2\vphantom{_a}}^2}{2}\!\!\left(\!\!\begin{array}{ccc}
        \frac{r_{2\vphantom{_a}}^2}{r_{1}^{2\vphantom{^a}}} \hspace{-5pt}&\hspace{-5pt} \frac{\sqrt{30}\,r_{2\vphantom{_a}}\hspace{-1pt}\cos{\theta}}{5r_{1}} \hspace{-5pt}&\hspace{-5pt} \frac{(3\cos^{2}\!{\theta}-1)}{2}\vspace{-5pt}\\
        &\\
        \frac{\sqrt{30}\,r_{2\vphantom{_a}}\hspace{-1pt}\cos{\theta}}{5r_{1}} \hspace{-5pt}&\hspace{-5pt} \frac{3(\cos^{2}\!{\theta}+3)}{10} \hspace{-5pt}&\hspace{-5pt} \frac{\sqrt{30}\,r_{1\vphantom{_a}}\hspace{-1pt}\cos{\theta}}{5r_{2}}\vspace{-5pt}\\
        &\\
        \frac{(3\cos^{2}\!{\theta}-1)}{2} \hspace{-5pt}&\hspace{-5pt} \frac{\sqrt{30}\,r_{1\vphantom{_a}}\hspace{-1pt}\cos{\theta}}{5r_{2}} \hspace{-5pt}&\hspace{-5pt} \frac{r_{1\vphantom{_a}}^2}{r_{2}^{2\vphantom{^a}}}
    \end{array}\!\!\right) \nonumber
\end{equation}
with $\hat{h}_{0}$, $\hat{h}_{1}$, $\hat{h}_{2}$, $\hat{h}^{\uparrow}_{0}$, $\hat{h}^{\uparrow}_{1}$, $\hat{h}^{\downarrow}_{1}$ and $\hat{h}^{\downarrow}_{2}$ given by
\begin{equation*}
\begin{array}{l}
    \hat{h}_{1} = \hat{H}_\text{Hyl}\hspace{-1pt}-\hspace{-1pt}\tfrac{\partial_{r_{1\vphantom{y}}}}{\mu_{1}\,r_{1}}\hspace{-1pt}-\hspace{-1pt}\tfrac{\partial_{r_{2\vphantom{y}}}}{\mu_{2}\,r_{2}}\hspace{-1pt}-\hspace{-1pt}\tfrac{\partial_{r_{12\vphantom{y}}}}{\mu_{12}\,r_{12}} \\[4pt]
    \hat{h}_{0} = \hat{H}_\text{Hyl}\hspace{-1pt}-\hspace{-1pt}\tfrac{2\partial_{r_{2\vphantom{y}}}}{\mu_{2}\,r_{2}}\hspace{-1pt}-\hspace{-1pt}\tfrac{2\partial_{r_{12\vphantom{y}}}}{m_{2}\,r_{12}} \hspace{20pt} 
    \hat{h}_{2} \hspace{-1pt}=\hspace{-1pt} \hat{H}_\text{Hyl}\hspace{-1pt}-\hspace{-1pt}\tfrac{2\partial_{r_{1\vphantom{y}}}}{\mu_{1}\,r_{1}}\hspace{-1pt}-\hspace{-1pt}\tfrac{2\partial_{r_{12\vphantom{y}}}}{m_{1}\,r_{12}}\\[4pt]
    \hat{h}^{\uparrow}_{0} = \sqrt{\tfrac{6}{5}}\left(\tfrac{\partial_{r_{12\vphantom{y}}}}{m_{1}\,r_{12}}\hspace{-1pt}-\hspace{-1pt}\tfrac{\partial_{r_{2\vphantom{y}}}}{m_{3}\,r_{2}}\right) \hspace{17pt} \hat{h}^{\downarrow}_{1} \hspace{-1pt}=\hspace{-1pt} \sqrt{\tfrac{10}{3}}\hspace{-1pt}\left(\hspace{-1pt}\tfrac{\partial_{r_{12\vphantom{y}}}}{m_{2}\,r_{12}}\hspace{-1pt}-\hspace{-1pt}\tfrac{\partial_{r_{1\vphantom{y}}}}{m_{3}\,r_{1}}\hspace{-1pt}\right) \\[4pt]
    \hat{h}^{\uparrow}_{1}=\sqrt{\tfrac{10}{3}}\left(\tfrac{\partial_{r_{12\vphantom{y}}}}{m_{1}\,r_{12}}\hspace{-1pt}-\hspace{-1pt}\tfrac{\partial_{r_{2\vphantom{y}}}}{m_{3}\,r_{2}}\right) \hspace{13pt} \hat{h}^{\downarrow}_{2} \hspace{-1pt}=\hspace{-1pt} \sqrt{\tfrac{6}{5}}\hspace{-1pt}\left(\hspace{-1pt}\tfrac{\partial_{r_{12\vphantom{y}}}}{m_{2}\,r_{12}}\hspace{-1pt}-\hspace{-1pt}\tfrac{\partial_{r_{1\vphantom{y}}}}{m_{3}\,r_{1}}\hspace{-1pt}\right)
\end{array}
\end{equation*}

\subsubsection{The \textnormal{RSE} for \texorpdfstring{$\textnormal{F}^{\textnormal{e}}$ and $\textnormal{F}^{\textnormal{o}}$}~ states}
The $\text{F}^{\text{e}}$ wave function can be constructed as
\begin{eqnarray}
    \Psi^{33\text{u}} = \psi_{1}\hspace{2pt}\bm{\Omega}_{1}^{33\text{u}}\! + \psi_{2}\hspace{2pt}\bm{\Omega}_{2}^{33\text{u}}\! + \psi_{3}\hspace{2pt}\bm{\Omega}_{3}^{33\text{u}}
\end{eqnarray}
using three reduced components $\psi_{l}\equiv\psi_{l}^{3\text{u}}(r_1,r_2,r_{12})$ with $l=1$, $2$, and $3$, and three angular generators
\begin{eqnarray*}
    \bm{\Omega}_{1}^{33\text{u}} &=& \tfrac{\sqrt{35}}{4}\hspace{1pt}\bm{\Omega}_{1}^{11\text{u}}\left(x_{2}+\textit{\textbf{i}}\hspace{1pt}y_{2}\right)^{2}\\
    \bm{\Omega}_{2}^{33\text{u}} &=& \tfrac{5\sqrt{2}}{4}\hspace{1pt}\bm{\Omega}_{1}^{11\text{u}}\left(x_{1}+\textit{\textbf{i}}\hspace{1pt}y_{1}\right)\left(x_{2}+\textit{\textbf{i}}\hspace{1pt}y_{2}\right)\\
    \bm{\Omega}_{3}^{33\text{u}} &=& \tfrac{\sqrt{35}}{4}\hspace{1pt}\bm{\Omega}_{1}^{11\text{u}}\left(x_{1}+\textit{\textbf{i}}\hspace{1pt}y_{1}\right)^{2}
\end{eqnarray*}
with $\bm{\Omega}_{1}^{11\text{u}}$ given by Eq.~\eqref{Omg111u}. The corresponding RSE has the following form
\begin{equation}
    \mathbb{W}\left(\!\!\begin{array}{llr}
        \hat{h}_{1}\hspace{-2.5pt}-\hspace{-2pt}E & \hspace{5pt}\hat{h}^{\uparrow}_{1} &  0\hspace*{2pt}\vspace{-5pt}\\
        &\\
        \hat{h}^{\downarrow}_{2} & \hspace{-10pt}\hat{h}_{2}\hspace{-2.5pt}-\hspace{-2pt}E & \hspace{-10pt}\hat{h}^{\uparrow}_{2}\vspace{-5pt}\\
        &\\
        \hspace{2pt}0 & \hspace{-10pt}\hat{h}^{\downarrow}_{3} & \hspace{-10pt}\hat{h}_{3}\hspace{-2.5pt}-\hspace{-2pt}E
    \end{array}\!\!\right)\hspace{-4.5pt}\left(\!\!\begin{array}{c}
    \psi_{1}\vspace{-5pt}\\
    \\
    \psi_{2}\vspace{-5pt}\\
    \\
    \psi_{3}
    \end{array}\!\!\right)\hspace{-4pt} =\hspace{-4pt} \left(\!\begin{array}{c}
            0\vspace{-5pt}\\
            \\
            0\vspace{-5pt}\\
            \\
            0
        \end{array}\!\right)
\end{equation}
where $\mathbb{W}=\left[\mathscr{W}_{ij}^{3\text{u}}\right]$ with $i,j=1$, $2$, and $3$ is given by
\begin{equation*}
    \mathbb{W}=r_{1}^2\hspace{1pt}r_{2}^2\hspace{1pt}\mathscr{W}_{11}^{1\text{u}}\left(\!\!\begin{array}{ccc}
        \frac{r_{2\vphantom{_{a}}}^2}{r_{1}^{2\vphantom{^{a}}}} \hspace{-5pt}&\hspace{-5pt} \frac{\sqrt{70}\,r_{2}\hspace{-1pt}\cos{\theta}}{7r_{1}} \hspace{-5pt}&\hspace{-5pt} \frac{(1\hspace{-1pt}-5\cos^{2}\!{\theta})}{4}\vspace{-5pt}\\
        &\\
        \frac{\sqrt{70}\,r_{2\vphantom{_{a}}}\hspace{-1pt}\cos{\theta}}{7r_{1}} \hspace{-5pt}&\hspace{-5pt} \frac{-5(3\cos^2\!{\theta}+5)}{28} \hspace{-5pt}&\hspace{-5pt} \frac{\sqrt{70}\,r_{1\vphantom{_{a}}}\hspace{-1pt}\cos{\theta}}{7r_{2}}\vspace{-5pt}\\
        &\\
        \frac{(1\hspace{-1pt}-5\cos^{2}\!{\theta})}{4} \hspace{-5pt}&\hspace{-5pt} \frac{\sqrt{70}\,r_{1\vphantom{_{a}}}\hspace{-1pt}\cos{\theta}}{7r_{2}} \hspace{-5pt}&\hspace{-5pt} \frac{r_{1\vphantom{_{a}}}^2}{r_{2}^{2\vphantom{^{a}}}}
    \end{array}\!\!\right)
\end{equation*}
with $\mathscr{W}_{11}^{1\text{u}}$ defined by Eq.~\eqref{w111u} and the operators $\hat{h}_{1}$, $\hat{h}_{2}$, $\hat{h}_{3}$, $\hat{h}^{\uparrow}_{1}$, $\hat{h}^{\uparrow}_{2}$, $\hat{h}^{\downarrow}_{2}$ and $\hat{h}^{\downarrow}_{3}$ given by
\begin{equation*}
\begin{array}{l}
    \hat{h}_{1} = \hat{H}_\text{Hyl}\hspace{-1pt}-\hspace{-1pt}\tfrac{\partial_{r_{1\vphantom{y}}}}{\mu_{1}\,r_{1}}\hspace{-1pt}-\hspace{-1pt}\tfrac{\partial_{r_{12\vphantom{y}}}}{m_{1}\,r_{12}}-\hspace{-1pt}\tfrac{3\partial_{r_{2\vphantom{y}}}}{\mu_{2}\,r_{2}}\hspace{-1pt}-\hspace{-1pt}\tfrac{3\partial_{r_{12\vphantom{y}}}}{m_{2}\,r_{12}}\\[4pt]
    \hat{h}_{2} = \hat{H}_\text{Hyl}\hspace{-1pt}-\hspace{-1pt}\tfrac{2\partial_{r_{1\vphantom{y}}}}{\mu_{1}\,r_{1}}\hspace{-1pt}-\hspace{-1pt}\tfrac{2\partial_{r_{12\vphantom{y}}}}{m_{1}\,r_{12}}-\hspace{-1pt}\tfrac{2\partial_{r_{2\vphantom{y}}}}{\mu_{2}\,r_{2}}\hspace{-1pt}-\hspace{-1pt}\tfrac{2\partial_{r_{12\vphantom{y}}}}{m_{2}\,r_{12}}\\[4pt]
    \hat{h}_{3} = \hat{H}_\text{Hyl}\hspace{-1pt}-\hspace{-1pt}\tfrac{3\partial_{r_{1\vphantom{y}}}}{\mu_{1}\,r_{1}}\hspace{-1pt}-\hspace{-1pt}\tfrac{3\partial_{r_{12\vphantom{y}}}}{m_{1}\,r_{12}}-\hspace{-1pt}\tfrac{2\partial_{r_{2\vphantom{y}}}}{\mu_{2}\,r_{2}}\hspace{-1pt}-\hspace{-1pt}\tfrac{2\partial_{r_{12\vphantom{y}}}}{m_{2}\,r_{12}}\\[4pt]
    \hat{h}^{\uparrow}_{1} = \sqrt{\tfrac{10}{7}}\left(\tfrac{\partial_{r_{12\vphantom{y}}}}{m_{1}\,r_{12}}\hspace{-1pt}-\hspace{-1pt}\tfrac{\partial_{r_{2\vphantom{y}}}}{m_{3}\,r_{2}}\right) \hspace{10pt} \hat{h}^{\downarrow}_{2} \hspace{-1pt}=\hspace{-1pt} \sqrt{\tfrac{14}{5}}\left(\tfrac{\partial_{r_{12\vphantom{y}}}}{m_{2}\,r_{12}}\hspace{-1pt}-\hspace{-1pt}\tfrac{\partial_{r_{1\vphantom{y}}}}{m_{3}\,r_{1}}\right)  \\[4pt]
    \hat{h}^{\uparrow}_{2}=\sqrt{\tfrac{14}{5}}\left(\tfrac{\partial_{r_{12\vphantom{y}}}}{m_{1}\,r_{12}}\hspace{-1pt}-\hspace{-1pt}\tfrac{\partial_{r_{2\vphantom{y}}}}{m_{3}\,r_{2}}\right) \hspace{10pt} \hat{h}^{\downarrow}_{3} \hspace{-1pt}=\hspace{-1pt} \sqrt{\tfrac{10}{7}}\left(\tfrac{\partial_{r_{12\vphantom{y}}}}{m_{2}\,r_{12}}\hspace{-1pt}-\hspace{-1pt}\tfrac{\partial_{r_{1\vphantom{y}}}}{m_{3}\,r_{1}}\right)
\end{array}
\end{equation*}

The $\text{F}^{\text{o}}$ wave function can be constructed as
\begin{equation}
    \Psi^{33\text{n}} = \psi_{0}\hspace{1pt}\bm{\Omega}_{0}^{33\text{n}}\! + \psi_{1}\hspace{1pt}\bm{\Omega}_{1}^{33\text{n}}\! + \psi_{2}\hspace{1pt}\bm{\Omega}_{2}^{33\text{n}}\! + \psi_{3}\hspace{1pt}\bm{\Omega}_{3}^{33\text{n}}
\end{equation}
using four reduced components $\psi_{l}\equiv\psi_{l}^{3\text{n}}(r_1,r_2,r_{12})$ with $l=0$, $1$, $2$, and $3$, and four angular generators
\begin{equation}\label{eq:agl3np}
\begin{aligned}
    \bm{\Omega}_{0}^{33\text{n}} \!&=\! -\tfrac{\sqrt{35}}{16\pi}\!\left(x_{2}\!+\!\textit{\textbf{i}}\hspace{1pt}y_{2}\hspace{-1pt}\right)\!^{3}\\
    \bm{\Omega}_{1}^{33\text{n}} \!&=\! -\tfrac{3\sqrt{5}}{16\pi}\!\left(x_{1}\!+\!\textit{\textbf{i}}\hspace{1pt}y_{1}\hspace{-1pt}\right)\!\left(x_{2}\!+\!\textit{\textbf{i}}\hspace{1pt}y_{2}\right)^{2}\\
    \bm{\Omega}_{2}^{33\text{n}} \!&=\! -\tfrac{3\sqrt{5}}{16\pi}\!\left(x_{2}\!+\!\textit{\textbf{i}}\hspace{1pt}y_{2}\hspace{-1pt}\right)\!\left(x_{1}\!+\!\textit{\textbf{i}}\hspace{1pt}y_{1}\hspace{-1pt}\right)\!^{2}\\
    \bm{\Omega}_{3}^{33\text{n}} \!&=\! -\tfrac{\sqrt{35}}{16\pi}\!\left(x_{1}\!+\!\textit{\textbf{i}}\hspace{1pt}y_{1}\hspace{-1pt}\right)\!^{3}
\end{aligned}
\end{equation}
The corresponding RSE takes the following form
\begin{equation}\label{eq:Fostate}
    \mathbb{W}\left(\hspace{-3pt}\begin{array}{cccc}
        \hat{h}_{0}\!-\!E \hspace{-7pt}&\hspace{-7pt} \hat{h}^{\uparrow}_{0} \hspace{-7pt}&\hspace{-7pt} 0 \hspace{-7pt}&\hspace{-7pt} 0\vspace{-5pt}\\
        &\\
        \hat{h}^{\downarrow}_{1} \hspace{-7pt}&\hspace{-7pt} \hat{h}_{1}\!-\!E \hspace{-7pt}&\hspace{-7pt} \hat{h}^{\uparrow}_{1} \hspace{-7pt}&\hspace{-7pt} 0\vspace{-5pt}\\
        &\\
        0 \hspace{-7pt}&\hspace{-7pt} \hat{h}^{\downarrow}_{2} \hspace{-7pt}&\hspace{-7pt} \hat{h}_{2}\!-\!E \hspace{-7pt}&\hspace{-7pt} \hat{h}^{\uparrow}_{2}\vspace{-5pt}\\
        &\\
        0 \hspace{-7pt}&\hspace{-7pt} 0 \hspace{-7pt}&\hspace{-7pt} \hat{h}^{\downarrow}_{3} \hspace{-7pt}&\hspace{-7pt} \hat{h}_{3}\!-\!E
    \end{array}\hspace{-3pt}\right)\hspace{-3.5pt}\left(\hspace{-3pt}\begin{array}{c}
    \psi_{0}\\[8pt]
    
    \psi_{1}\\[8pt]
    
    \psi_{2}\\[8pt]
    
    \psi_{3}
    \end{array}\hspace{-3pt}\right) \hspace{-2pt}=\hspace{-2pt} \left(\hspace{-1pt}\begin{array}{c}
    0\\[8pt]
    
    0\\[8pt]
    
    0\\[8pt]
    
    0
    \end{array}\hspace{-1pt}\right)
\end{equation}
where $\mathbb{W}=\left[\mathscr{W}_{ij}^{3\text{n}}\right]$ with $i,j=0$, $1$, $2$, and $3$ is given by
\begin{equation*}
    \mathbb{W}=\tfrac{3\hspace{0.5pt}r_{1}^3r_{2\vphantom{_{a}}}^3}{2}\hspace{-2pt}\left(\hspace{-5pt}\begin{array}{l}
        \hspace{2pt}\frac{r_{2\vphantom{_{a}}}^3}{3\hspace{0.5pt}r_{1}^{3\vphantom{^{a}}}} \hspace{11pt} \frac{r_{2\vphantom{_{a}}}^2\hspace{-1pt}\cos{\theta}}{\sqrt{7}r_{1}^{2\vphantom{^{a}}}} \hspace{11pt} \frac{r_{2\vphantom{_{a}}}(3\cos^2\!{\theta}-1)}{2\sqrt{7}r_{1}} \hspace{11pt} \frac{\cos{\theta}(5\cos^2\!{\theta}-3)}{6}\\\\
        \frac{r_{2\vphantom{_{a}}}^2\hspace{-1pt}\cos{\theta}}{\sqrt{7}r_{1}^{2\vphantom{^{a}}}} \hspace{4pt} \frac{r_{2\vphantom{_{a}}}(\cos^2\!{\theta}+2)}{7r_{1}} \hspace{4pt} \frac{\cos{\theta}(\cos^2\!{\theta}+5)}{14} \hspace{4pt} \frac{r_{1\vphantom{_{a}}}(3\cos^2\!{\theta}-1)}{2\sqrt{7}r_{2}} \frac{}{}\\\\
        \frac{r_{2\vphantom{_{a}}}(3\cos^2\!{\theta}-1)}{2\sqrt{7}r_{1}} \hspace{4pt} \frac{\cos{\theta}(\cos^2\!{\theta}+5)}{14} \hspace{4pt} \frac{r_{1\vphantom{_{a}}}(\cos^2\!{\theta}+2)}{7r_{2}} \hspace{4pt} \frac{r_{1\vphantom{_{a}}}^2\hspace{-1pt}\cos{\theta}}{\sqrt{7}r_{2}^{2\vphantom{^{a}}}} \\\\
        \hspace{2pt}\frac{\cos{\theta}(5\cos^2\!{\theta}-3)}{6} \hspace{11pt} \frac{r_{1\vphantom{_{a}}}(3\cos^2\!{\theta}-1)}{2\sqrt{7}r_{2}} \hspace{11pt} \frac{r_{1\vphantom{_{a}}}^2\hspace{-1pt}\cos{\theta}}{\sqrt{7}r_{2}^{2\vphantom{^{a}}}} \hspace{11pt} \frac{r_{1\vphantom{_{a}}}^3}{3\hspace{0.5pt}r_{2}^{3\vphantom{^{a}}}}
    \end{array}\hspace{-5.5pt}\right)
\end{equation*}
with the involved operators given by
\begin{equation*}
\begin{array}{l}
    \hat{h}_{1} = \hat{H}_\text{Hyl}\hspace{-1pt}-\hspace{-1pt}\tfrac{\partial_{r_{1\vphantom{y}}}}{\mu_{1}\,r_{1}}\hspace{-1pt}-\hspace{-1pt}\tfrac{\partial_{r_{12\vphantom{y}}}}{m_{1}\,r_{12}}-\hspace{-1pt}\tfrac{2\partial_{r_{2\vphantom{y}}}}{\mu_{2}\,r_{2}}\hspace{-1pt}-\hspace{-1pt}\tfrac{2\partial_{r_{12\vphantom{y}}}}{m_{2}\,r_{12}} \\[4pt]
    \hat{h}_{2} = \hat{H}_\text{Hyl}\hspace{-1pt}-\hspace{-1pt}\tfrac{2\partial_{r_{1\vphantom{y}}}}{\mu_{1}\,r_{1}}\hspace{-1pt}-\hspace{-1pt}\tfrac{2\partial_{r_{12\vphantom{y}}}}{m_{1}\,r_{12}}-\hspace{-1pt}\tfrac{\partial_{r_{2\vphantom{y}}}}{\mu_{2}\,r_{2}}\hspace{-1pt}-\hspace{-1pt}\tfrac{\partial_{r_{12\vphantom{y}}}}{m_{2}\,r_{12}} \\[4pt]
    \hat{h}_{0} = \hat{H}_\text{Hyl}\hspace{-1pt}-\hspace{-1pt}\tfrac{3\partial_{r_{2\vphantom{y}}}}{\mu_{2}\,r_{2}}\hspace{-1pt}-\hspace{-1pt}\tfrac{3\partial_{r_{12\vphantom{y}}}}{m_{2}\,r_{12}} \hspace{14pt} \hat{h}_{3} = \hat{H}_\text{Hyl}\hspace{-1pt}-\hspace{-1pt}\tfrac{3\partial_{r_{1\vphantom{y}}}}{\mu_{1}\,r_{1}}\hspace{-1pt}-\hspace{-1pt}\tfrac{3\partial_{r_{12\vphantom{y}}}}{m_{1}\,r_{12}}\\[4pt]
    \hat{h}^{\uparrow}_{0} = \tfrac{3}{\sqrt{7}}\left(\tfrac{\partial_{r_{12\vphantom{y}}}}{m_{1}\,r_{12}}\hspace{-1pt}-\hspace{-1pt}\tfrac{\partial_{r_{2\vphantom{y}}}}{m_{3}\,r_{2}}\right) \hspace{15pt} \hat{h}^{\downarrow}_{1} = \sqrt{7}\left(\tfrac{\partial_{r_{12\vphantom{y}}}}{m_{2}\,r_{12}}\hspace{-1pt}-\hspace{-1pt}\tfrac{\partial_{r_{1\vphantom{y}}}}{m_{3}\,r_{1}}\right) \\[4pt]
    \hat{h}^{\uparrow}_{1}= 2\left(\tfrac{\partial_{r_{12\vphantom{y}}}}{m_{1}\,r_{12}}\hspace{-1pt}-\hspace{-1pt}\tfrac{\partial_{r_{2\vphantom{y}}}}{m_{3}\,r_{2}}\right) \hspace{23pt} \hat{h}^{\downarrow}_{2} = 2\left(\tfrac{\partial_{r_{12\vphantom{y}}}}{m_{2}\,r_{12}}\hspace{-1pt}-\hspace{-1pt}\tfrac{\partial_{r_{1\vphantom{y}}}}{m_{3}\,r_{1}}\right)\\[4pt]
    \hat{h}^{\uparrow}_{2}=\sqrt{7}\left(\tfrac{\partial_{r_{12\vphantom{y}}}}{m_{1}\,r_{12}}\hspace{-1pt}-\hspace{-1pt}\tfrac{\partial_{r_{2\vphantom{y}}}}{m_{3}\,r_{2}}\right) \hspace{15pt} \hat{h}^{\downarrow}_{3} = \tfrac{3}{\sqrt{7}}\left(\tfrac{\partial_{r_{12\vphantom{y}}}}{m_{2}\,r_{12}}\hspace{-1pt}-\hspace{-1pt}\tfrac{\partial_{r_{1\vphantom{y}}}}{m_{3}\,r_{1}}\right)
\end{array}
\end{equation*} 

\subsection{Explicit wave functions with arbitrary \texorpdfstring{$L$}{L}, \texorpdfstring{$M$}{M}, and \texorpdfstring{$\uppi$}{parity}}

We have explained in the previous section how to construct the wave functions $\Psi^{LL\uppi}$ explicitly for states with an arbitrary parity using the examples of the $\text{S}^{\text{e}}$, $\text{P}^{\text{o}}$, $\text{P}^{\text{e}}$, $\text{D}^{\text{o}}$, $\text{D}^{\text{e}}$, $\text{F}^{\text{o}}$, and $\text{F}^{\text{e}}$   states. These examples are easily extended to higher values of $L$. Furthermore, the presented approach can be easily generalized to an arbitrary value of $M$ using the previously given formula
\begin{equation}\tag{\ref{wfa2}}
    \Psi^{LM\uppi}= \sum_{l=d}^{L}\psi_{l}^{L\uppi}(r_{1},r_{2},r_{12}) \,\,\bm{\Omega}_{l}^{L\hspace{-0.7pt}M\hspace{-0.7pt}\uppi}  \hspace{-0.7pt}(\bm{r}_{1},\bm{r}_{2})
\end{equation}
with $\bm{\Omega}_{l}^{L\hspace{-0.7pt}M\hspace{-0.7pt}\uppi}\equiv \bm{\Omega}_{l}^{L\hspace{-0.7pt}M\hspace{-0.7pt}\uppi}  \hspace{-0.7pt}(\bm{r}_{1},\bm{r}_{2})$ given by Eqs.~(\ref{nsbshccnp})--(\ref{usbshccup}) and $\psi_{l}\equiv \psi_{l}^{L\uppi}(r_{1},r_{2},r_{12})$\footnote{Note again that the partial wave components $\psi_{l}$ do not depend on $M$.} determined as described in Sec.~\ref{sec:spdf}. To illustrate this procedure, an explicit example for $L=3$, $M=1$, and $\uppi=\text{n}$ (i.e., the $M=1$ component of the $\text{F}^{\text{o}}$ state) is provided below.  This component can be constructed as
\begin{equation}
    \Psi^{31\text{n}} = \psi_{0}\hspace{1pt}\bm{\Omega}_{0}^{31\text{n}}\! + \psi_{1}\hspace{1pt}\bm{\Omega}_{1}^{31\text{n}}\! + \psi_{2}\hspace{1pt}\bm{\Omega}_{2}^{31\text{n}}\! + \psi_{3}\hspace{1pt}\bm{\Omega}_{3}^{31\text{n}}
\end{equation}
where the four partial wave components $\psi_{0}$, $\psi_{1}$, $\psi_{2}$, and $\psi_{3}$ are obtained by solving Eq.~(\ref{eq:Fostate}), and the four angular generators are determined from Eq.~(\ref{nsbphgen}) as
\begin{alignat*}{4}
    \bm{\Omega}_{0}^{31\text{n}} & = & & -\hspace{1pt}\tfrac{\sqrt{21}}{16\pi}\hspace{2pt} & (x_{2} & {}+\textit{\textbf{i}}\hspace{1pt}y_{2}\hspace{-1pt}) & (\bm{r_{2}}\! & {}\cdot\bm{r_{2}}-5z_{2}z_{2})\\[1pt]
    \bm{\Omega}_{1}^{31\text{n}} & = & & -\tfrac{\sqrt{3}}{16\pi}\!\Big[\hspace{1.5pt} & (x_{1} & {}+\textit{\textbf{i}}\hspace{1pt}y_{1}\hspace{-1pt})^{\vphantom{l}} & (\bm{r_{2}}\! & {}\cdot\bm{r_{2}}-5z_{2}z_{2}) \\[-1.75pt]
    & & & \hspace{15pt}+\hspace{2pt}2\hspace{2pt} & (x_{2} & {}+\textit{\textbf{i}}\hspace{1pt}y_{2}\hspace{-1pt}) & (\bm{r_{1}}\! & {}\cdot\bm{r_{2}}-5z_{1}z_{2})^{\vphantom{l}}\hspace{1.5pt}\Big]\\[-1.75pt]
    \bm{\Omega}_{2}^{31\text{n}} & = & & -\tfrac{\sqrt{3}}{16\pi}\!\Big[\hspace{1.5pt} & (x_{2} & {}+\textit{\textbf{i}}\hspace{1pt}y_{2}\hspace{-1pt})^{\vphantom{l}} & (\bm{r_{1}}\! & {}\cdot\bm{r_{1}}-5z_{1}z_{1}) \\[-1.75pt]
    & & & \hspace{15pt}+\hspace{2pt}2\hspace{2pt} & (x_{1} & {}+\textit{\textbf{i}}\hspace{1pt}y_{1}\hspace{-1pt}) & (\bm{r_{1}}\! & {}\cdot\bm{r_{2}}-5z_{1}z_{2})^{\vphantom{l}}\hspace{1.5pt}\Big]\\[-1.7pt]
    \bm{\Omega}_{3}^{31\text{n}} & = & & -\tfrac{\sqrt{21}}{16\pi}\hspace{0.8pt} & (x_{1} & {}+\textit{\textbf{i}}\hspace{1pt}y_{1}\hspace{-1pt}) & (\bm{r_{1}}\! & {}\cdot\bm{r_{1}}-5z_{1}z_{1})
\end{alignat*}
Alternatively, one can generate $\bm{\Omega}_{l}^{L\hspace{-0.7pt}M\hspace{-0.7pt}\uppi}$ by acting with the operator $\hat{L}_{-}=\hat{L}_{x}-\textit{\textbf{i}}\hspace{1pt}\hat{L}_{y}$ on $\bm{\Omega}_{l}^{L\hspace{-0.7pt}L\hspace{-0.7pt}\uppi}$ [given by Eq.~(\ref{nsbshccnp})] 
 $L-M$ times in the manner specified by Eq.~(\ref{eq:ladderM}).
 
\section{Numerical validation of the formalism}\label{numerical}

An important aspect of our work is the numerical validation of the resulting RSEs. The correctness of the derived formulas has been verified by implementing them in a computer program, performing extensive calculations for nonrelativistic energy levels of low-lying states of the helium atom with $L\geq 0$ and both natural and unnatural parity, and benchmarking the results against accurate literature values. The details needed to reproduce our results are given below.

\subsection{Technical details}

Upon expanding each of the partial wave components as 
\begin{eqnarray}\label{basis0}
    \psi_{l}^{L\uppi}(r_{1}\hspace{-1pt},\hspace{-1pt}r_{2},\hspace{-1pt}r_{12}) = \sum_{i=1}^{\texttt{N}_{\texttt{a}}}\sum_{j=1}^{\texttt{N}_{\texttt{b}}} \mathscr{C}_{lij}^{L\uppi}\upphi_{ij}^l (r_{1}\hspace{-1pt},\hspace{-1pt}r_{2},\hspace{-1pt}r_{12})
\end{eqnarray}
in a basis consisting of $\texttt{N}_{\texttt{ab}}=\texttt{N}_{\texttt{a}}\texttt{N}_{\texttt{b}}$ suitably chosen functions $\upphi_{ij}^{l}(r_{1}\hspace{-1pt},\hspace{-1pt}r_{2},\hspace{-1pt}r_{12})$ [for details, see Eq.~(\ref{rwf})] and applying the Rayleigh-Ritz variational principle, we can reduce Eq.~\eqref{rrse} to a generalized eigenvalue problem
\begin{eqnarray}\label{gep}
\mathbb{H}~ \mathrm{c} = E~\mathbb{S} ~\mathrm{c}
\end{eqnarray}
where $E$ represents a numerical estimate of each energy eigenvalue and $\mathrm{c}$ is a column vector of dimension $\texttt{N}_{\texttt{ab}}(L\!-\!d\!+\!1)$ containing the expansion coefficients $\mathscr{C}_{lij}^{L\uppi}$ of the corresponding partial wave components $\psi_{l}^{L\uppi}(r_{1},r_{2},r_{12})$. The matrices $\mathbb{S}$ and $\mathbb{H}$ consist of $(L\!-\!d\!+\!1) \times (L\!-\!d\!+\!1)$ blocks $\mathbb{S}^{mn}$ and $\mathbb{H}^{mn}$, respectively, each of size  $\texttt{N}_{\texttt{ab}}\times \texttt{N}_{\texttt{ab}}$, where $m$ and $n$ range from $d$ to $L$. The matrix elements of the blocks are given by
\begin{eqnarray}
    \mathbb{S}^{mn}_{ij} &=& \left\langle b_{i}\left\vert  \mathscr{W}_{nm}^{L\uppi}\right\vert b_{j}\right\rangle\label{smn}\\
    \mathbb{H}^{mn}_{ij} &=& \left\langle b_{i}\left\vert \sum_{k=d}^{L} \mathscr{W}_{mk}^{L\uppi}\hspace{2pt}\hat{h}_{kn}\right\vert b_{j}\right\rangle\label{hmn}
\end{eqnarray}
where the angular integrals $\mathscr{W}_{ij}^{L\uppi}$ are defined in Eq.~\eqref{wlpikkb1} and the operator $\hat{h}_{kn}$ is given by
\begin{equation}
    \hat{h}_{kn} = \left\{\begin{array}{ll}
       \hat{h}_{k}  & \text{~~for $k=n$}\\[2pt]
       \hat{h}^{\uparrow}_{k}  & \text{~~for $k=n-1$}\\[2pt]
       \hat{h}^{\downarrow}_{n}  & \text{~~for $k=n+1$}\\[2pt]
       0  &   \text{~~otherwise}
    \end{array}\right.
\end{equation}
with $\hat{h}_{l}$, $\hat{h}^{\uparrow}_{l}$, and $\hat{h}^{\downarrow}_{l}$ defined in Eqs.~(\ref{hkkgnygp2})--(\ref{hkk+1gnygp2}). The functions 
$b_{k}$ correspond here to the explicitly correlated basis functions $\upphi_{ij}^{l}\equiv \upphi_{ij}^{l}(r_{1}\hspace{-1pt},\hspace{-1pt}r_{2},\hspace{-1pt}r_{12})$ defined as
 \begin{eqnarray}\label{rwf}
    \upphi_{ij}^l = r_{1}^{\texttt{l}_{j}} r_{2}^{\texttt{m}_{j}} r_{12}^{\texttt{n}_{j}}~e^{-\zeta_{i} r_{1}-\eta_{i} r_{2} - \xi_{i} r_{12}}
\end{eqnarray}
where $\texttt{N}_{\texttt{b}} =$ \scalebox{0.7}{${\binom{\texttt{w}+3}{3}}$} distinct triples ($\texttt{l}_{j}$, $\texttt{m}_{j}$, $\texttt{n}_{j}$) of non-negative integers are generated subject to the condition $\texttt{l}_{j}\!+\!\texttt{m}_{j}\!+\!\texttt{n}_{j}\leq \texttt{w}$, and $\texttt{N}_{\texttt{a}}$ distinct triples ($\zeta_{i}$, $\eta_{i}$, $\xi_{i}$) of real exponents satisfy the condition that the sum of any two is always positive. The exponents ($\zeta_{i}$, $\eta_{i}$, $\xi_{i}$) constitute non-linear optimization parameters to be determined together with the linear expansion coefficients $\mathscr{C}_{lij}^{L\uppi}$ during the solution of the variational problem. The numerical solution of the generalized eigenvalue problem in Eq.~(\ref{gep}) is performed using a quadruple-precision code written in the Julia programming language \cite{julia_bezanson_2017}, which is capable of handling arbitrary angular momentum $L$ and parity $\uppi$.  The exponents are optimized using the Nelder-Mead optimization algorithm \cite{Nelder1965, Gao2012}. The integrals arising from the matrix elements in Eqs.~\eqref{smn} and~\eqref{hmn} can be expressed\footnote{See Sec.~III of \textcite{Calais1962} and the text around Eq.~(38) of \textcite{Harris2004} for details.} in the following closed form
\begin{eqnarray}
    &&\int\limits_{0}^{\infty}\hspace{-3pt}dr_{1}\hspace{-3pt}\int\limits_{0}^{\infty}\hspace{-3pt}dr_{2}\hspace{-10pt}\int\limits_{\vert \hspace{-0.5pt}r_{1}\hspace{-1pt}-r_{2}\hspace{-1pt}\vert}^{r_{1}\hspace{-1pt}+r_{2}}\hspace{-10pt}dr_{12} \, r_{1}^{\texttt{k}_{1}}r_{2}^{\texttt{k}_{2}}r_{12}^{\texttt{k}_{3}}\,e^{-\left(\zeta r_{1}+\eta r_{2}+\xi r_{12}\right)}\\[2pt] &=&\frac{\partial^{\texttt{k}_{1}}}{\partial \zeta^{\texttt{k}_{1}}} \frac{\partial^{\texttt{k}_{2}}}{\partial \eta^{\texttt{k}_{2}}} \frac{\partial^{\texttt{k}_{3}}}{\partial \xi^{\texttt{k}_{3}}} \frac{2}{(\zeta+\eta)(\eta+\xi)(\zeta+\xi)} \nonumber\\[3pt] \nonumber
    &=& \sum_{\texttt{j}_{1}=0}^{\texttt{k}_{1}}\sum_{\texttt{j}_{2}=0}^{\texttt{k}_{2}}\sum_{\texttt{j}_{3}=0}^{\texttt{k}_{3}}\!\tfrac{2\,\texttt{k}_{1}\hspace{-1pt}!\hspace{1pt}\texttt{k}_{2}\hspace{-1pt}!\hspace{1pt}\texttt{k}_{3}\hspace{-1pt}!\hspace{1pt}\!\left(\hspace{-4pt}\begin{array}{c}
         \scriptstyle \texttt{k}_{1}\hspace{-1pt}-\hspace{-1pt}\texttt{j}_{1}\hspace{-1pt}+\hspace{-1pt}\texttt{j}_{2}  \\
         \scriptstyle \texttt{j}_{2}
    \end{array}\hspace{-5pt}\right)\!\left(\hspace{-4pt}\begin{array}{c}
         \scriptstyle \texttt{k}_{2}\hspace{-1pt}-\hspace{-1pt}\texttt{j}_{2}\hspace{-1pt}+\hspace{-1pt}\texttt{j}_{3}  \\
         \scriptstyle \texttt{j}_{3}
    \end{array}\hspace{-5pt}\right)\!\left(\hspace{-4pt}\begin{array}{c}
         \scriptstyle \texttt{k}_{3}\hspace{-1pt}-\hspace{-1pt}\texttt{j}_{3}\hspace{-1pt}+\hspace{-1pt}\texttt{j}_{1}  \\
         \scriptstyle \texttt{j}_{1}
    \end{array}\hspace{-5pt}\right)}{(\!\zeta\hspace{-1pt}+\eta\!)^{\texttt{k}_{1}\hspace{-1pt}-\hspace{-1pt}\texttt{j}_{1}\hspace{-1pt}+\hspace{-1pt}\texttt{j}_{2}\hspace{-1pt}+\hspace{-1pt}1^{\vphantom{l}}}(\!\eta+\xi\hspace{-1pt})^{\texttt{k}_{2}\hspace{-1pt}-\hspace{-1pt}\texttt{j}_{2}\hspace{-1pt}+\hspace{-1pt}\texttt{j}_{3}\hspace{-1pt}+\hspace{-1pt}1}(\!\zeta\hspace{-1pt}+\hspace{-1pt}\xi\hspace{-1pt})^{\texttt{k}_{3}\hspace{-1pt}-\hspace{-1pt}\texttt{j}_{3}\hspace{-1pt}+\hspace{-1pt}\texttt{j}_{1}\hspace{-1pt}+\hspace{-1pt}1}}
\end{eqnarray}
which is valid for any arbitrary triple ($\texttt{k}_{1}$, $\texttt{k}_{2}$, $\texttt{k}_{3}$) of non-negative integers, and any arbitrary triple $(\zeta, \eta, \xi)$ of exponents satisfying the condition that the sum of any two is positive.

For systems containing two identical fermions---like the helium atom---the resulting wave functions $\Psi^{LM\uppi}\hspace{-0.7pt}(\bm{r}_{1},\bm{r}_{2})$ are antisymmetric with respect to the exchange of the fermions. This antisymmetry is reflected in the structure of the partial wave components $\psi_{l}^{L\uppi}(r_{1}\hspace{-1pt},\hspace{-1pt}r_{2},\hspace{-1pt}r_{12})$ in a rather non-trivial manner, owing to the antisymmetry properties of the MBHs given by Eq.~(\ref{eq: mbheprop}). Consequently, the analysis of the column vector $\mathrm{c}$ obtained by solving the eigenproblem in Eq.~(\ref{gep}) can be used to classify the corresponding state as a singlet or triplet. A proper spin-adaptation of the formalism---based on the permutational symmetry of the system under particle exchange---can reduce the size of the eigenproblem by about half while simultaneously separating the eigensolutions for the singlet and triplet states. For states with vanishing angular momentum, $L = 0$, the spin-adaptation can be performed at the level of constructing a symmetry-adapted basis set for the singlet and triplet spin states. For higher angular momentum, $L > 0$, the situation is more intricate and requires more detailed considerations going beyond a simple spin-adaptation of the basis functions. We plan to address this issue in detail in future work.

\subsection{Results and discussion}
The computed energy eigenvalues for the singlet ($S=0$) and triplet ($S=1$) states of helium with natural parity and angular momentum $L \leq 7$ are presented in Table~\ref{tab_eng_np}. For each $L$ and $S$, the two lowest energy eigenvalues are computed using single, double, and triple sets of optimized exponents. The alpha-particle-to-electron mass ratio, $\frac{m_{\alpha_{\vphantom{a}}}}{m_{e}}=7294.299\,541\,71$, is  based on CODATA 2022 \cite{codata2022}. We benchmark our results against the accurate reference values of \textcite{Drake1992, Drake1993}, computed using Eq.~(3.3.2) of \textcite{Drake1993} with the same $\frac{m_{\alpha_{\vphantom{a}}}}{m_{e}}$ value as above, and reproduced here as the last column of Table~\ref{tab_eng_np}. Our results are consistent with the reference values to approximately 7 digits of accuracy with the single set of exponents, 11 digits with the double set of exponents, and up to 15 digits with the triple set of exponents, all achieved while keeping the number $\texttt{N}_{\texttt{a}}$ of distinct powers of $r_1$, $r_2$, and $r_{12}$ between 165 and 280.

Table~\ref{tab_eng_up} presents the lowest energy eigenvalues\footnote{Formally, the states of helium with unnatural parity analyzed here are doubly-excited, metastable bound states, usually described in terms of resonance energies and half-lives. However, since the standard Coulomb autoionization decay pathway is symmetry-forbidden for these states \cite{Wu1944,Eiglsperger2010a}, we can estimate their energies in a manner analogous to other bound, singly-excited states.} for the unnatural-parity ${}^{3}\text{P}^{\text{e}}$, ${}^{1}\text{P}^{\text{e}}$, ${}^{1}\text{D}^{\text{o}}$, ${}^{3}\text{D}^{\text{o}}$, ${}^{1}\text{F}^{\text{e}}$, and  ${}^{3}\text{F}^{\text{e}}$ helium states calculated using the double set of optimized exponents and increasing $\texttt{w}$ until convergence to 15 decimal places is achieved. Utilizing the CODATA 2022 \cite{codata2022} value $\frac{m_{\alpha_{\vphantom{a}}}}{m_{e}}=7294.299\,541\,71$, the results in Table~\ref{tab_eng_up} constitute the most accurate estimates of the energy eigenvalues for these states to date. 
Previously reported results \cite{Hilger1996, Hesse2001, Kar2009b} are less accurate, relying either on the clamped-nucleus approximation with $\frac{m_{\alpha_{\vphantom{a}}}}{m_{e}}=\infty$ or on an imprecise alpha-particle-to-electron mass ratio $\frac{m_{\alpha_{\vphantom{a}}}}{m_{e}}=7294.261\,824\,1$ used by \textcite{Hesse2001}. Table~\ref{tab_eng_up} demonstrates that our method can also successfully reproduce these quantities when the appropriate values of $\frac{m_{\alpha_{\vphantom{a}}}}{m_{e}}$ are used.

\definecolor{mygray}{gray}{0.65} % 0 is black, 1 is white. 
\begin{table*}
\begin{scriptsize}
\caption{Nonrelativistic energy eigenvalues (in a.u.) for the low-lying states of helium atom with $0 \leq L\leq7$ and natural parity computed using single, double, and triple set of exponents. The alpha-particle-to-electron mass ratio, $\frac{m_{\alpha}}{m_{e}}=7294.299\,541\,71$, is  taken from CODATA 2022 \cite{codata2022}.
}\label{tab_eng_np}
\begin{tabular}{c@{\quad}l@{\quad}c@{\quad}c@{\quad}c@{\quad}c@{\quad}c@{\quad}c@{\quad}c@{\quad}c@{\quad}c@{\quad}}
\hline\hline
\vspace{0.02cm}\\
 &  & &   & Energy eigenvalues &   & Energy eigenvalues &  & Energy eigenvalues &  & \\
$L$ & States\footnote{The state labels follow the notation used by \textcite{Drake1993}.} & $\texttt{w}$ & $\texttt{N}_{\texttt{ab}}$  & for single exponent & $\texttt{N}_{\texttt{ab}}$  & for double exponent & $\texttt{N}_{\texttt{ab}}$ & for triple exponent &  & Reference values\footnote{\textcite{Drake1992, Drake1993}: Energy eigenvalues evaluated using Eq.~(3.3.2) of \textcite{Drake1993} with double exponent basis considering the same alpha particle-to-electron mass ratio as in the present work.}\\
    &      &    &      & basis ($\texttt{N}_{\texttt{a}}=1$) &      & basis ($\texttt{N}_{\texttt{a}}=2$) &      & basis ($\texttt{N}_{\texttt{a}}=3$) &     & \\
\vspace{0.05cm}\\
\hline
\vspace{0.1cm}\\
%0 & $1\,{}^{1}\text{S}^{\text{e}}$ & 4 & 35 & $-2.903\, 2\textcolor{mygray}{97\, 408\, 054}$ & 70 & $-2.903\, 304\, 55\textcolor{mygray}{2\, 017\, 192}$ & 105 & $-2.903\, 304\, 557\, \textcolor{mygray}{583\, 364}$ &  & $-2.903\, 304\, 557\, 729\, 880$\\[3pt]
%0 & $1\,{}^{1}\text{S}^{\text{e}}$ & 7 &    &                                                &    &                                                      & 360 & $-2.903\, 304\, 557\, 729\, 8\textcolor{mygray}{35}$ &  & $-2.903\, 304\, 557\, 729\, 880$\\[3pt]
0 & $1\,{}^{1}\text{S}^{\text{e}}$ & 9 & 220 & $-2.903\, 304\, 54\textcolor{mygray}{8\, 546\, 840}$ & 440 & $-2.903\, 304\, 557\, 729\, \textcolor{mygray}{306}$ & 660 & $-2.903\, 304\, 557\, 729\, 879$                      &  & $-2.903\, 304\, 557\, 729\, 880$\\[3pt]
%0 & $2\,{}^{3}\text{S}^{\text{e}}$ & 20 & $-$\texttt{2.174 92}8 642 722 & 40 & $-$\texttt{2.174 930 1}85 745 714 & 60 & $-$\texttt{2.174 930 190} 161 045 &  & $-$\texttt{2.174 930 190 712 523} \\[3pt]
%0 & $2\,{}^{3}\text{S}^{\text{e}}$ & 4 & 35  & $-2.174\, 92\textcolor{mygray}{9\, 982\, 189}$ & 70  & $-2.174\, 930\, 190\, \textcolor{mygray}{527\, 338}$ & 105 & $-2.174\, 930\, 190\, 7\textcolor{mygray}{04\, 091}$ &  & $-2.174\, 930\, 190\, 712\, 523$ \\[3pt]
%0 & $2\,{}^{3}\text{S}^{\text{e}}$ & 7 &    &                                                 &     &                                                      & 360 & $-2.174\, 930\, 190\, 712\, 522$                     &  & $-2.174\, 930\, 190\, 712\, 523$ \\[3pt]
0 & $2\,{}^{3}\text{S}^{\text{e}}$ & 8 & 165 & $-2.174\, 930\, 18\textcolor{mygray}{9\, 809\, 936}$ & 330 & $-2.174\, 930\, 190\, 712\, 5\textcolor{mygray}{02}$ & 495 & $-2.174\, 930\, 190\, 712\, 523$                     &  & $-2.174\, 930\, 190\, 712\, 523$ \\[3pt]
%0 & $2\,{}^{1}\text{S}^{\text{e}}$ & 20 & $-$\texttt{2.145} 461 736 338 & 40 & $-$\texttt{2.145 678} 295 209 647 & 60 & $-$\texttt{2.145 678 57}8 472 952 &  & $-$\texttt{2.145 678 587 580 789} \\[3pt]
%0 & $2\,{}^{1}\text{S}^{\text{e}}$ & 4 & 35  & $-2.145\, 6\textcolor{mygray}{52\, 007\, 879}$ & 70  & $-2.145\, 678\, 5\textcolor{mygray}{16\, 207\, 170}$ & 105 & $-2.145\, 678\, 587\, \textcolor{mygray}{203\, 916}$ &  & $-2.145\, 678\, 587\, 580\, 789$ \\[3pt]
%0 & $2\,{}^{1}\text{S}^{\text{e}}$ & 7 &     &                                                &     &                                                      & 360 & $-2.145\, 678\, 587\, 580\, \textcolor{mygray}{592}$ &  & $-2.145\, 678\, 587\, 580\, 789$ \\[3pt]
%0 & $2\,{}^{1}\text{S}^{\text{e}}$ & 8 &     &                                                &     &                                                      & 495 & $-2.145\, 678\, 587\, 580\, 7\textcolor{mygray}{64}$ &  & $-2.145\, 678\, 587\, 580\, 789$ \\[3pt]
0 & $2\,{}^{1}\text{S}^{\text{e}}$ & 9 & 220 & $-2.145\, 678\, \textcolor{mygray}{360\, 871\, 400}$ & 440 & $-2.145\, 678\, 587\, 5\textcolor{mygray}{75\, 083}$ & 660 & $-2.145\, 678\, 587\, 580\, 78\textcolor{mygray}{4}$ &  & $-2.145\, 678\, 587\, 580\, 789$ \\[3pt]
%1 & $2\,{}^{3}\text{P}^{\text{o}}$ & 4 & 70 & $-2.132\, 87\textcolor{mygray}{8\, 137\, 229}$ & 140 & $-2.132\, 880\, 64\textcolor{mygray}{0\, 369\, 481}$ & 210 & $-2.132\, 880\, 642\, 09\textcolor{mygray}{4\, 853}$ &  & $-2.132\, 880\, 642\, 103\, 275$ \\[3pt]
%1 & $2\,{}^{3}\text{P}^{\text{o}}$ & 5 &    &                                                &     &                                                      & 336 & $-2.132\, 880\, 642\, 103\, \textcolor{mygray}{045}$ &  & $-2.132\, 880\, 642\, 103\, 275$ \\[3pt]
%1 & $2\,{}^{3}\text{P}^{\text{o}}$ & 6 &    &                                                &     &                                                      & 504 & $-2.132\, 880\, 642\, 103\, 27\textcolor{mygray}{1}$ &  & $-2.132\, 880\, 642\, 103\, 275$ \\[3pt]
1 & $2\,{}^{3}\text{P}^{\text{o}}$ & 7 & 240 & $-2.132\, 880\, 6\textcolor{mygray}{21\, 599\, 233}$ & 480 & $-2.132\, 880\, 642\, 10\textcolor{mygray}{1\, 701}$ & 720 & $-2.132\, 880\, 642\, 103\, 284$                     &  & $-2.132\, 880\, 642\, 103\, 275$ \\[3pt]
%1 & $2\,{}^{1}\text{P}^{\text{o}}$ & 4 & 70 & $-2.123\, 54\textcolor{mygray}{3\, 253\, 962}$ & 140 & $-2.123\, 545\, 65\textcolor{mygray}{2\, 626\, 601}$ & 210 & $-2.123\, 545\, 654\, 11\textcolor{mygray}{8\, 246}$ &  & $-2.123\, 545\, 654\, 127\, 433$ \\[3pt]
%1 & $2\,{}^{1}\text{P}^{\text{o}}$ & 5 &    &                                                &     &                                                      & 336 & $-2.123\, 545\, 654\, 127\, \textcolor{mygray}{084}$ &  & $-2.123\, 545\, 654\, 127\, 433$ \\[3pt]
%1 & $2\,{}^{1}\text{P}^{\text{o}}$ & 6 &    &                                                &     &                                                      & 504 & $-2.123\, 545\, 654\, 127\, 4\textcolor{mygray}{24}$ &  & $-2.123\, 545\, 654\, 127\, 433$ \\[3pt]
1 & $2\,{}^{1}\text{P}^{\text{o}}$ & 7 & 240 & $-2.123\, 545\, 6\textcolor{mygray}{27\, 061\, 478}$ & 480 & $-2.123\, 545\, 654\, 12\textcolor{mygray}{5\, 234}$ & 720 & $-2.123\, 545\, 654\, 127\, 442$                    &  & $-2.123\, 545\, 654\, 127\, 433$ \\[3pt]
%1 & $3\,{}^{3}\text{P}^{\text{o}}$ & 40 & $-$\texttt{2.057 7}67 820 558 & 80 & $-$\texttt{2.057 801 4}68 389 607 & 120 & $-$\texttt{2.057 801 492} 233 858 &  & $-$\texttt{2.057 801 492 606 401} \\[3pt]
%1 & $3\,{}^{3}\text{P}^{\text{o}}$ & 4 & 70  & $-2.057\, 7\textcolor{mygray}{91\, 370\, 514}$ & 140 & $-2.057\, 801\, 4\textcolor{mygray}{84\, 562\, 486}$ & 210 & $-2.057\, 801\, 492\, \textcolor{mygray}{593\, 018}$ &  & $-2.057\, 801\, 492\, 606\, 401$ \\[3pt] 
%1 & $3\,{}^{3}\text{P}^{\text{o}}$ & 5 &     &                                                &     &                                                      & 336 & $-2.057\, 801\, 492\, 605\, \textcolor{mygray}{868}$ &  & $-2.057\, 801\, 492\, 606\, 401$ \\[3pt] 
%1 & $3\,{}^{3}\text{P}^{\text{o}}$ & 6 &     &                                                &     &                                                      & 504 & $-2.057\, 801\, 492\, 606\, \textcolor{mygray}{355}$ &  & $-2.057\, 801\, 492\, 606\, 401$ \\[3pt]
1 & $3\,{}^{3}\text{P}^{\text{o}}$ & 7 & 240 & $-2.057\, 801\, \textcolor{mygray}{122\, 645\, 887}$ & 480 & $-2.057\, 801\, 492\, 5\textcolor{mygray}{98\, 137}$ & 720 & $-2.057\, 801\, 492\, 606\, 39\textcolor{mygray}{3}$ &  & $-2.057\, 801\, 492\, 606\, 401$ \\[3pt]
%1 & $3\,{}^{1}\text{P}^{\text{o}}$ & 40 & $-$\texttt{2.054 8}15 165 956 & 80 & $-$\texttt{2.054 862 6}26 932 422 & 120 & $-$\texttt{2.054 862 660} 735 560 &  & $-$\texttt{2.054 862 661 148 483} \\[3pt]
%1 & $3\,{}^{1}\text{P}^{\text{o}}$ & 4 & 70  & $-2.054\, 8\textcolor{mygray}{53\, 231\, 633}$ & 140 & $-2.054\, 862\, 6\textcolor{mygray}{53\, 598\, 027}$ & 210 & $-2.054\, 862\, 661\, 1\textcolor{mygray}{29\, 732}$ &  & $-2.054\, 862\, 661\, 148\, 483$ \\[3pt]
%1 & $3\,{}^{1}\text{P}^{\text{o}}$ & 5 &     &                                                &     &                                                      & 336 & $-2.054\, 862\, 661\, 147\, \textcolor{mygray}{721}$ &  & $-2.054\, 862\, 661\, 148\, 483$ \\[3pt]
%1 & $3\,{}^{1}\text{P}^{\text{o}}$ & 6 &     &                                                &     &                                                      & 504 & $-2.054\, 862\, 661\, 148\, 4\textcolor{mygray}{33}$ &  & $-2.054\, 862\, 661\, 148\, 483$ \\[3pt]
1 & $3\,{}^{1}\text{P}^{\text{o}}$ & 7 & 240 & $-2.054\, 862\, \textcolor{mygray}{281\, 047\, 834}$ & 480 & $-2.054\, 862\, 661\, 13\textcolor{mygray}{7\, 923}$ & 720 & $-2.054\, 862\, 661\, 148\, 483$ &  & $-2.054\, 862\, 661\, 148\, 483$ \\[3pt]
%2 & $3\,{}^{3}\text{D}^{\text{e}}$ & 4 & 105 & $-2.055\, 354\, \textcolor{mygray}{255\, 608}$ & 210 & $-2.055\, 354\, 531\, 4\textcolor{mygray}{83\, 362}$ & 315 & $-2.055\, 354\, 531\, 560\, \textcolor{mygray}{707}$ &  & $-2.055\, 354\, 531\, 561\, 215$ \\[3pt]
%2 & $3\,{}^{3}\text{D}^{\text{e}}$ & 5 &     &                                                &     &                                                      & 504 & $-2.055\, 354\, 531\, 561\, 2\textcolor{mygray}{05}$ &  & $-2.055\, 354\, 531\, 561\, 215$ \\[3pt]
2 & $3\,{}^{3}\text{D}^{\text{e}}$ & 6 & 252 & $-2.055\, 354\, 5\textcolor{mygray}{13\, 861\, 960}$ & 504 & $-2.055\, 354\, 531\, 560\, \textcolor{mygray}{538}$ & 756 & $-2.055\, 354\, 531\, 561\, 215$                     &  & $-2.055\, 354\, 531\, 561\, 215$ \\[3pt]
%2 & $3\,{}^{1}\text{D}^{\text{e}}$ & 4 & 105 & $-2.055\, 338\, \textcolor{mygray}{678\, 948}$ & 210 & $-2.055\, 338\, 994\, 7\textcolor{mygray}{22\, 575}$ & 315 & $-2.055\, 338\, 994\, 793\, \textcolor{mygray}{401}$ &  & $-2.055\, 338\, 994\, 793\, 989$ \\[3pt]
%2 & $3\,{}^{1}\text{D}^{\text{e}}$ & 5 &     &                                                &     &                                                      & 504 & $-2.055\, 338\, 994\, 793\, 9\textcolor{mygray}{74}$ &  & $-2.055\, 338\, 994\, 793\, 989$ \\[3pt]
2 & $3\,{}^{1}\text{D}^{\text{e}}$ & 6 & 252 & $-2.055\, 338\, 9\textcolor{mygray}{75\, 311\, 650}$ & 504 & $-2.055\, 338\, 994\, 793\, \textcolor{mygray}{343}$ & 756 & $-2.055\, 338\, 994\, 793\, 989$                     &  & $-2.055\, 338\, 994\, 793\, 989$ \\[3pt]
%2 & $4\,{}^{3}\text{D}^{\text{e}}$ & 60 & $-$\texttt{2.031 00}6 949 976 & 120 & $-$\texttt{2.031 010 40}2 349 470 & 180 & $-$\texttt{2.031 010 405} 940 523 &  & $-$\texttt{2.031 010 406 012 130} \\[3pt]
%2 & $4\,{}^{3}\text{D}^{\text{e}}$ & 4 & 105 & $-2.031\, 00\textcolor{mygray}{9\, 289\, 770}$ & 210 & $-2.031\, 010\, 405\, \textcolor{mygray}{590\, 770}$ & 315 & $-2.031\, 010\, 406\, 01\textcolor{mygray}{0\, 637}$ &  & $-2.031\, 010\, 406\, 012\, 130$ \\[3pt]
%2 & $4\,{}^{3}\text{D}^{\text{e}}$ & 5 &     &                                                &     &                                                      & 504 & $-2.031\, 010\, 406\, 012\, 0\textcolor{mygray}{94}$ &  & $-2.031\, 010\, 406\, 012\, 130$ \\[3pt]
2 & $4\,{}^{3}\text{D}^{\text{e}}$ & 6 & 252 & $-2.031\, 010\, \textcolor{mygray}{312\, 245\, 074}$ & 504 & $-2.031\, 010\, 406\, 00\textcolor{mygray}{7\, 487}$ & 756 & $-2.031\, 010\, 406\, 012\, 130$                     &  & $-2.031\, 010\, 406\, 012\, 130$ \\[3pt]
%2 & $4\,{}^{1}\text{D}^{\text{e}}$ & 60 & $-$\texttt{2.030 99}7 395 479 & 120 & $-$\texttt{2.031 001 42}3 682 911 & 180 & $-$\texttt{2.031 001 427 6}10 633&  & $-$\texttt{2.031 001 427 688 724} \\[3pt]
%2 & $4\,{}^{1}\text{D}^{\text{e}}$ & 4 & 105 & $-2.031\, 00\textcolor{mygray}{0\, 126\, 984}$ & 210 & $-2.031\, 001\, 427\, \textcolor{mygray}{284\, 442}$ & 315 & $-2.031\, 001\, 427\, 68\textcolor{mygray}{7\, 169}$ &  & $-2.031\, 001\, 427\, 688\, 724$ \\[3pt]
%2 & $4\,{}^{1}\text{D}^{\text{e}}$ & 5 &     &                                                &     &                                                      & 504 & $-2.031\, 001\, 427\, 688\, 6\textcolor{mygray}{72}$ &  & $-2.031\, 001\, 427\, 688\, 724$ \\[3pt]
2 & $4\,{}^{1}\text{D}^{\text{e}}$ & 6 & 252 & $-2.031\, 001\, \textcolor{mygray}{309\, 153\, 051}$ & 504 & $-2.031\, 001\, 427\, 68\textcolor{mygray}{4\, 188}$ & 756 & $-2.031\, 001\, 427\, 688\, 724$                     &  & $-2.031\, 001\, 427\, 688\, 724$ \\[3pt]
%3 & $4\,{}^{3}\text{F}^{\text{o}}$ & 4 & 140 & $-2.030\, 976\, 7\textcolor{mygray}{09\, 625}$ & 280 & $-2.030\, 976\, 736\, 8\textcolor{mygray}{96\, 226}$ & 420 & $-2.030\, 976\, 736\, 900\, 6\textcolor{mygray}{22}$ &  & $-2.030\, 976\, 736\, 900\, 653$ \\[3pt]
3 & $4\,{}^{3}\text{F}^{\text{o}}$ & 5 & 224 & $-2.030\, 976\, 73\textcolor{mygray}{1\, 112\, 664}$ & 448 & $-2.030\, 976\, 736\, 900\, \textcolor{mygray}{278}$ & 672 & $-2.030\, 976\, 736\, 900\, 65\textcolor{mygray}{2}$ &  & $-2.030\, 976\, 736\, 900\, 653$ \\[3pt]
%3 & $4\,{}^{1}\text{F}^{\text{o}}$ & 4 & 140 & $-2.030\, 976\, 6\textcolor{mygray}{85\, 816}$ & 280 & $-2.030\, 976\, 712\, 92\textcolor{mygray}{6\, 618}$ & 420 & $-2.030\, 976\, 712\, 930\, 9\textcolor{mygray}{50}$ &  & $-2.030\, 976\, 712\, 930\, 983$ \\[3pt]
3 & $4\,{}^{1}\text{F}^{\text{o}}$ & 5 & 224 & $-2.030\, 976\, 70\textcolor{mygray}{7\, 280\, 918}$ & 448 & $-2.030\, 976\, 712\, 930\, \textcolor{mygray}{600}$ & 672 & $-2.030\, 976\, 712\, 930\, 983$                     &  & $-2.030\, 976\, 712\, 930\, 983$ \\[3pt]
%3 & $5\,{}^{3}\text{F}^{\text{o}}$ & 80 & $-$\texttt{2.019 725} 717 764 & 160 & $-$\texttt{2.019 726 066} 970 066 & 240 & $-$\texttt{2.019 726 067 4}62 175 &  & $-$\texttt{2.019 726 067 470 835}\\[3pt]
%3 & $5\,{}^{3}\text{F}^{\text{o}}$ & 4 & 140 & $-2.019\, 725\, \textcolor{mygray}{940\, 080}$ & 280 & $-2.019\, 726\, 067\, 4\textcolor{mygray}{24\, 319}$ & 420 & $-2.019\, 726\, 067\, 470\, \textcolor{mygray}{118}$ &  & $-2.019\, 726\, 067\, 470\, 835$\\[3pt]
3 & $5\,{}^{3}\text{F}^{\text{o}}$ & 5 & 224 & $-2.019\, 726\, 0\textcolor{mygray}{25\, 373\, 027}$ & 448 & $-2.019\, 726\, 067\, 46\textcolor{mygray}{8\, 560}$ & 672 & $-2.019\, 726\, 067\, 470\, 83\textcolor{mygray}{2}$ &  & $-2.019\, 726\, 067\, 470\, 835$\\[3pt]
%3 & $5\,{}^{1}\text{F}^{\text{o}}$ & 80 & $-$\texttt{2.019 72}5 697 649 & 160 & $-$\texttt{2.019 726 046} 760 566 & 240 & $-$\texttt{2.019 726 047 2}85 432 &  & $-$\texttt{2.019 726 047 295 738} \\[3pt]
%3 & $5\,{}^{1}\text{F}^{\text{o}}$ & 4 & 140 & $-2.019\, 725\, \textcolor{mygray}{920\, 904}$ & 280 & $-2.019\, 726\, 047\, 2\textcolor{mygray}{48\, 011}$ & 420 & $-2.019\, 726\, 047\, 295\, \textcolor{mygray}{638}$ &  & $-2.019\, 726\, 047\, 295\, 738$ \\[3pt]
3 & $5\,{}^{1}\text{F}^{\text{o}}$ & 5 & 224 & $-2.019\, 726\, 0\textcolor{mygray}{05\, 694\, 717}$ & 448 & $-2.019\, 726\, 047\, 29\textcolor{mygray}{3\, 805}$ & 672 & $-2.019\, 726\, 047\, 295\, 73\textcolor{mygray}{5}$ &  & $-2.019\, 726\, 047\, 295\, 738$ \\[3pt]
%4 & $5\,{}^{3}\text{G}^{\text{e}}$ & 4 & 175 & $-2.019\, 723\, 81\textcolor{mygray}{7\, 782}$ & 350 & $-2.019\, 723\, 820\, 777\, \textcolor{mygray}{339}$ & 525 & $-2.019\, 723\, 820\, 777\, 92\textcolor{mygray}{9}$ &  & $-2.019\, 723\, 820\, 777\, 932$ \\[3pt]
4 & $5\,{}^{3}\text{G}^{\text{e}}$ & 5 & 280 & $-2.019\, 723\, 81\textcolor{mygray}{9\, 989\, 996}$ & 560 & $-2.019\, 723\, 820\, 777\, \textcolor{mygray}{898}$ & 840 & $-2.019\, 723\, 820\, 777\, 932$                     &  & $-2.019\, 723\, 820\, 777\, 932$ \\[3pt]
%4 & $5\,{}^{1}\text{G}^{\text{e}}$ & 4 & 175 & $-2.019\, 723\, 81\textcolor{mygray}{7\, 755}$ & 350 & $-2.019\, 723\, 820\, 750\, \textcolor{mygray}{680}$ & 525 & $-2.019\, 723\, 820\, 751\, 23\textcolor{mygray}{0}$ &  & $-2.019\, 723\, 820\, 751\, 233$ \\[3pt]
4 & $5\,{}^{1}\text{G}^{\text{e}}$ & 5 & 280 & $-2.019\, 723\, 81\textcolor{mygray}{9\, 962\, 222}$ & 560 & $-2.019\, 723\, 820\, 751\, \textcolor{mygray}{204}$ & 840 & $-2.019\, 723\, 820\, 751\, 233$                     &  & $-2.019\, 723\, 820\, 751\, 233$ \\[3pt]
%4 & $6\,{}^{3}\text{G}^{\text{e}}$ & 100 & $-$\texttt{2.013 613 2}48 745 & 200 & $-$\texttt{2.013 613 292 7}09 037 & 300 & $-$\texttt{2.013 613 292 79}6 013 &  & $-$\texttt{2.013 613 292 798 135} \\[3pt]
%4 & $6\,{}^{3}\text{G}^{\text{e}}$ & 4 & 175 & $-2.013\, 613\, 2\textcolor{mygray}{76\, 805}$ & 350 & $-2.013\, 613\, 292\, 79\textcolor{mygray}{3\, 127}$ & 525 & $-2.013\, 613\, 292\, 798\, 1\textcolor{mygray}{29}$ &  & $-2.013\, 613\, 292\, 798\, 135$ \\[3pt]
4 & $6\,{}^{3}\text{G}^{\text{e}}$ & 5 & 280 & $-2.013\, 613\, 28\textcolor{mygray}{7\, 465\, 260}$ & 560 & $-2.013\, 613\, 292\, 797\, \textcolor{mygray}{894}$ & 840 & $-2.013\, 613\, 292\, 798\, 134$                     &  & $-2.013\, 613\, 292\, 798\, 135$ \\[3pt]
%4 & $6\,{}^{1}\text{G}^{\text{e}}$ & 100 & $-$\texttt{2.013 613 2}48 698 & 200 & $-$\texttt{2.013 613 292 6}76 268 & 300 & $-$\texttt{2.013 613 292 76}6 647 &  & $-$\texttt{2.013 613 292 768 561} \\[3pt]
%4 & $6\,{}^{1}\text{G}^{\text{e}}$ & 4 & 175 & $-2.013\, 613\, 2\textcolor{mygray}{76\, 768}$ & 350 & $-2.013\, 613\, 292\, 76\textcolor{mygray}{3\, 498}$ & 525 & $-2.013\, 613\, 292\, 767\, \textcolor{mygray}{955}$ &  & $-2.013\, 613\, 292\, 768\, 561$ \\[3pt]
4 & $6\,{}^{1}\text{G}^{\text{e}}$ & 5 & 280 & $-2.013\, 613\, 28\textcolor{mygray}{7\, 436\, 964}$ & 560 &  $-2.013\, 613\, 292\, 768\, \textcolor{mygray}{327}$ & 840 & $-2.013\, 613\, 292\, 768\, 561$                     &  & $-2.013\, 613\, 292\, 768\, 561$ \\[3pt]
5 & $6\,{}^{3}\text{H}^{\text{o}}$ & 4 & 210 & $-2.013\, 612\, 981\, \textcolor{mygray}{697\, 309}$ & 420 & $-2.013\, 612\, 982\, 094\, \textcolor{mygray}{643}$ & 630 & $-2.013\, 612\, 982\, 094\, 738$ &  & $-2.013\, 612\, 982\, 094\, 738$ \\[3pt]
5 & $6\,{}^{1}\text{H}^{\text{o}}$ & 4 & 210 & $-2.013\, 612\, 981\, \textcolor{mygray}{697\, 286}$ & 420 & $-2.013\, 612\, 982\, 094\, \textcolor{mygray}{617}$ & 630 & $-2.013\, 612\, 982\, 094\, 717$ &  & $-2.013\, 612\, 982\, 094\, 717$ \\[3pt]
5 & $7\,{}^{3}\text{H}^{\text{o}}$ & 4 & 210 & $-2.009\, 928\, 63\textcolor{mygray}{2\, 669\, 666}$ & 420 & $-2.009\, 928\, 635\, 16\textcolor{mygray}{3\, 627}$  & 630 & $-2.009\, 928\, 635\, 164\, 958$ &  & $-2.009\, 928\, 635\, 164\, 959$ \\[3pt]
%5 & $7\,{}^{3}\text{H}^{\text{o}}$ & 120 & $-$\texttt{2.009 928 62}7 676 & 240 & $-$\texttt{2.009 928 635 1}47 460 & 360 & $-$\texttt{2.009 928 635 164} 629 &  & $-$\texttt{2.009 928 635 164 959} \\[3pt]
%5 & $7\,{}^{1}\text{H}^{\text{o}}$ & 120 & $-$\texttt{2.009 928 62}7 676 & 240 & $-$\texttt{2.009 928 635 1}47 418 & 360 & $-$\texttt{2.009 928 635 164} 629 &  & $-$\texttt{2.009 928 635 164 929} \\[3pt]
5 & $7\,{}^{1}\text{H}^{\text{o}}$ & 4 & 210 & $-2.009\, 928\, 63\textcolor{mygray}{2\, 669\, 680}$ & 420 & $-2.009\, 928\, 635\, 164\, \textcolor{mygray}{069}$ & 630 & $-2.009\, 928\, 635\, 164\, 92\textcolor{mygray}{5}$ &  & $-2.009\, 928\, 635\, 164\, 929$ \\[3pt]
6 & $7\,{}^{3}\text{I}^{\text{e}}$ & 4 & 245 & $-2.009\, 928\, 572\, \textcolor{mygray}{890\, 879}$ & 490 & $-2.009\, 928\, 572\, 956\, 2\textcolor{mygray}{28}$ & 735 & $-2.009\, 928\, 572\, 956\, 256$ &  & $-2.009\, 928\, 572\, 956\, 256$ \\[3pt]
6 & $7\,{}^{1}\text{I}^{\text{e}}$ & 4 & 245 & $-2.009\, 928\, 572\, \textcolor{mygray}{890\, 879}$ & 490 & $-2.009\, 928\, 572\, 956\, 2\textcolor{mygray}{26}$ & 735 & $-2.009\, 928\, 572\, 956\, 256$ &  & $-2.009\, 928\, 572\, 956\, 256$ \\[3pt]
%6 & $8\,{}^{3}\text{I}^{\text{e}}$ & 140 & $-$\texttt{2.007 537 30}7 023 & 280 & $-$\texttt{2.007 537 308 67}4 221 & 420 & $-$\texttt{2.007 537 308 67}7 338 &  & $-$\texttt{2.007 537 308 678 299} \\[3pt]
6 & $8\,{}^{3}\text{I}^{\text{e}}$ & 4 & 245 & $-2.007\, 537\, 308\, \textcolor{mygray}{181\, 872}$ & 490 & $-2.007\, 537\, 308\, 677\, \textcolor{mygray}{983}$ & 735 & $-2.007\, 537\, 308\, 678\, 299$ &  & $-2.007\, 537\, 308\, 678\, 299$ \\[3pt]
%6 & $8\,{}^{1}\text{I}^{\text{e}}$ & 140 & $-$\texttt{2.007 537 30}7 023 & 280 & $-$\texttt{2.007 537 308 67}4 142 & 420 & $-$\texttt{2.007 537 308 67}7 318 &  & $-$\texttt{2.007 537 308 678 299} \\[3pt]
6 & $8\,{}^{1}\text{I}^{\text{e}}$ & 4 & 245 & $-2.007\, 537\, 308\, \textcolor{mygray}{181\, 872}$ & 490 & $-2.007\, 537\, 308\, 678\, \textcolor{mygray}{140}$ & 735 & $-2.007\, 537\, 308\, 678\, 299$ &  & $-2.007\, 537\, 308\, 678\, 299$ \\[3pt]
7 & $8\,{}^{3}\text{K}^{\text{o}}$ & 4 & 280 & $-2.007\, 537\, 292\, 6\textcolor{mygray}{83\, 336}$ & 560 & $-2.007\, 537\, 292\, 696\, 7\textcolor{mygray}{60}$ & 840 & $-2.007\, 537\, 292\, 696\, 771$ &  & $-2.007\, 537\, 292\, 696\, 771$ \\[3pt]
7 & $8\,{}^{1}\text{K}^{\text{o}}$ & 4 & 280 & $-2.007\, 537\, 292\, 6\textcolor{mygray}{83\, 336}$ & 560 & $-2.007\, 537\, 292\, 696\, 7\textcolor{mygray}{61}$ & 840 & $-2.007\, 537\, 292\, 696\, 771$ &  & $-2.007\, 537\, 292\, 696\, 771$ \\[3pt]
%7 & $9\,{}^{3}\text{K}^{\text{o}}$ & 160 & $-$\texttt{2.005 897 853} 491 & 320 & $-$\texttt{2.005 897 853 94}6 127 & 480 & $-$\texttt{2.005 897 853 947 3}68 &  & $-$\texttt{2.005 897 853 947 405} \\[3pt]
7 & $9\,{}^{3}\text{K}^{\text{o}}$ & 4 & 280 & $-2.005\, 897\, 853\, \textcolor{mygray}{825\, 104}$ & 560 & $-2.005\, 897\, 853\, 947\, \textcolor{mygray}{347}$ & 840 & $-2.005\, 897\, 853\, 947\, 405$ &  & $-2.005\, 897\, 853\, 947\, 405$ \\[3pt]
%7 & $9\,{}^{1}\text{K}^{\text{o}}$ & 160 & $-$\texttt{2.005 897 853} 491 & 320 & $-$\texttt{2.005 897 853 94}6 127 & 480 & $-$\texttt{2.005 897 853 947 3}68 &  & $-$\texttt{2.005 897 853 947 405} \\[3pt]
7 & $9\,{}^{1}\text{K}^{\text{o}}$ & 4 & 280 & $-2.005\, 897\, 853\, \textcolor{mygray}{825\, 104}$ & 560 & $-2.005\, 897\, 853\, 947\, \textcolor{mygray}{ 355}$ & 840 &  $-2.005\, 897\, 853\, 947\, 3\textcolor{mygray}{78}$ &  & $-2.005\, 897\, 853\, 947\, 405$ \\[3pt]
&&&&&\\
\hline\hline
\end{tabular}
\end{scriptsize}
\end{table*}

\begin{table*}
\begin{scriptsize}
\caption{Convergence study of the nonrelativistic energy eigenvalues (in a.u.) for few selective low-lying unnatural parity states of the helium atom with angular momentum quantum number $L$ ranging from 1 to 3. Calculations are performed for a finite nuclear mass of helium ($\frac{m_{\alpha_{\vphantom{a}}}}{m_{e}}=7294.299\,541\,71$) \cite{codata2022} and for clamped nucleus using double exponent basis ($\texttt{N}_{\texttt{a}}=2$). Additional results for the ${}^{3}\text{P}^{\text{e}}$ and ${}^{1}\text{P}^{\text{e}}$ states, with less accurate alpha particle mass ($\frac{m_{\alpha_{\vphantom{a}}}}{m_{e}}=7294.261\,824\,1$) used earlier by \textcite{Hesse2001}, are also included for comparison.}\label{tab_eng_up}
\begin{tabular}{c@{\quad\quad}c@{\quad\quad}c@{\quad\quad}r@{\quad\quad}l@{\quad\quad}l@{\quad\quad}l@{\quad\quad}}
\hline\hline
\vspace{-0.1cm}\\
     &       &            &            & \multicolumn{3}{c}{Energy eigenvalues for}\vspace{-0.1cm}\\
     $L$ & State & $\texttt{w}$ & $\texttt{N}_{\texttt{ab}}$ & \\
     \vspace{-0.25cm}\\
     &       &            &            & $\frac{m_{\alpha_{\vphantom{a}}}}{m_{e}}=7294.299\,541\,71$ & \multicolumn{1}{c}{$\frac{m_{\alpha_{\vphantom{a}}}}{m_{e}}\rightarrow\infty$\footnote{We model the clamped-nucleus limit by setting $m_{3}=10^{21}$ a.u., whereas cited works achieve this by omitting the mass-polarization term and equating the quasiparticle reduced mass to the electron mass.}} & $\hspace{5pt}\frac{m_{\alpha_{\vphantom{a}}}}{m_{e}}=7294.261\,824\,1$\\
     &       &            &            & \cite{codata2022}         &                                        & \cite{Hesse2001}    \\[3pt]
\hline\\
1  &  $2\,{}^{3}\text{P}^{\text{e}}$ & 1 & 8   & $-0.710\, 3\textcolor{mygray}{87\, 858\, 051\, 759}$ & $-0.710\, 4\textcolor{mygray}{91\, 540\, 385\, 704}$ & $-0.710\, 3\textcolor{mygray}{87\, 857\, 515\, 723}$  \\
   &                                 & 2 & 20  & $-0.710\, 396\, 3\textcolor{mygray}{90\, 526\, 887}$ & $-0.710\, 500\, 0\textcolor{mygray}{88\, 655\, 640}$ & $-0.710\, 396\, \textcolor{mygray}{389\, 990\, 769}$  \\
   &                                 & 3 & 40  & $-0.710\, 396\, 45\textcolor{mygray}{4\, 899\, 369}$ & $-0.710\, 500\, 15\textcolor{mygray}{3\, 021\, 917}$ & $-0.710\, 396\, 45\textcolor{mygray}{4\, 363\, 251}$  \\
   &                                 & 4 & 70  & $-0.710\, 396\, 457\, \textcolor{mygray}{434\, 880}$ & $-0.710\, 500\, 155\, \textcolor{mygray}{556\, 159}$ & $-0.710\, 396\, 456\, \textcolor{mygray}{898\, 762}$  \\
   &                                 & 5 & 112 & $-0.710\, 396\, 457\, 55\textcolor{mygray}{0\, 609}$ & $-0.710\, 500\, 155\, 67\textcolor{mygray}{1\, 073}$ & $-0.710\, 396\, 457\, \textcolor{mygray}{014\, 491}$  \\
   &                                 & 6 & 168 & $-0.710\, 396\, 457\, 557\, \textcolor{mygray}{322}$ & $-0.710\, 500\, 155\, 677\, \textcolor{mygray}{740}$ & $-0.710\, 396\, 457\, 021\, \textcolor{mygray}{204}$  \\
   &                                 & 7 & 240 & $-0.710\, 396\, 457\, 557\, \textcolor{mygray}{841}$ & $-0.710\, 500\, 155\, 678\, \textcolor{mygray}{256}$ & $-0.710\, 396\, 457\, 021\, 7\textcolor{mygray}{22}$  \\
   &                                 & 8 & 330 & $-0.710\, 396\, 457\, 557\, 9\textcolor{mygray}{04}$ & $-0.710\, 500\, 155\, 678\, 3\textcolor{mygray}{23}$ & $-0.710\, 396\, 457\, 021\, 7\textcolor{mygray}{86}$  \\
   &                                 & 9 & 440 & $-0.710\, 396\, 457\, 557\, 91\textcolor{mygray}{0}$ & $-0.710\, 500\, 155\, 678\, 3\textcolor{mygray}{29}$ & $-0.710\, 396\, 457\, 021\, 79\textcolor{mygray}{2}$ \\
   &                                & 10 & 572 & $-0.710\, 396\, 457\, 557\, 91\textcolor{mygray}{2}$ & $-0.710\, 500\, 155\, 678\, 33\textcolor{mygray}{1}$ & $-0.710\, 396\, 457\, 021\, 79\textcolor{mygray}{4}$ \\
   %&                                & 11 & 728 & $-$\texttt{0.710 396 457 557 912} & $-$\texttt{0.710 500 155 678 331} & $-$\texttt{0.710 396 457 021 794} \\
   %&                                & 12 & 910 & \textbf{$-$0.710 396 457 557 912} & $-$\texttt{0.710 500 155 678 331} & $-$\texttt{0.710 396 457 021 794} \\
   %&                                 &   &     &                        &                                       &                                     \\
   %&                                 & 8 & 330 & $-$0.710 396 457 557 796 & $-$\texttt{0.710 500 155 678} 214       & $-$\texttt{0.710 396 457 021} 678     \\
   &                                 &   &     &                        & $-0.710\, 500\, 155\, 678\, 33\,$\footnote{\textcite{Hilger1996}: Hylleraas CI calculation using 540 term basis set.\label{fn:b}}  & $-0.710\, 396\, 457\, 021\, 8\textcolor{mygray}{1}\,$\footnote{\textcite{Hesse2001}: Lagrange-mesh method using $12\,600$ terms for the ${}^{3}\text{P}^{\text{e}}$ state and $11\,900$ terms for the ${}^{1}\text{P}^{\text{e}}$ state.\label{fn:c}} \\[3pt]
1  &  $3\,{}^{1}\text{P}^{\text{e}}$ & 1 & 8   & $-0.580\, 1\textcolor{mygray}{20\, 139\, 169\, 499}$ & $-0.580\, 2\textcolor{mygray}{00\, 853\, 605\, 484}$ & $-0.580\, 1\textcolor{mygray}{20\, 138\, 752\, 229}$ \\
   &                                 & 2 & 20  & $-0.580\, 165\, 7\textcolor{mygray}{51\, 382\, 360}$ & $-0.580\, 246\, 4\textcolor{mygray}{55\, 612\, 226}$ & $-0.580\, 165\, 7\textcolor{mygray}{50\, 965\, 113}$ \\
   &                                 & 3 & 40  & $-0.580\, 165\, 768\, \textcolor{mygray}{096\, 868}$ & $-0.580\, 246\, 47\textcolor{mygray}{1\, 964\, 013}$ & $-0.580\, 165\, 76\textcolor{mygray}{7\, 679\, 621}$ \\
   &                                 & 4 & 70  & $-0.580\, 165\, 768\, 7\textcolor{mygray}{02\, 010}$ & $-0.580\, 246\, 472\, 5\textcolor{mygray}{71\, 402}$ & $-0.580\, 165\, 768\, \textcolor{mygray}{284\, 763}$ \\
   &                                 & 5 & 112 & $-0.580\, 165\, 768\, 72\textcolor{mygray}{3\, 826}$ & $-0.580\, 246\, 472\, 59\textcolor{mygray}{2\, 397}$ & $-0.580\, 165\, 768\, 30\textcolor{mygray}{6\, 579}$ \\
   &                                 & 6 & 168 & $-0.580\, 165\, 768\, 725\, \textcolor{mygray}{535}$ & $-0.580\, 246\, 472\, 594\, \textcolor{mygray}{263}$ & $-0.580\, 165\, 768\, 308\, \textcolor{mygray}{288}$ \\
   &                                 & 7 & 240 & $-0.580\, 165\, 768\, 725\, 6\textcolor{mygray}{37}$ & $-0.580\, 246\, 472\, 594\, 3\textcolor{mygray}{74}$ & $-0.580\, 165\, 768\, 308\, 3\textcolor{mygray}{90}$ \\
   &                                 & 8 & 330 & $-0.580\, 165\, 768\, 725\, 6\textcolor{mygray}{47}$ & $-0.580\, 246\, 472\, 594\, 38\textcolor{mygray}{4}$ & $-0.580\, 165\, 768\, 308\, 40\textcolor{mygray}{0}$ \\
   &                                 & 9 & 440 & $-0.580\, 165\, 768\, 725\, 64\textcolor{mygray}{8}$ & $-0.580\, 246\, 472\, 594\, 38\textcolor{mygray}{5}$ & $-0.580\, 165\, 768\, 308\, 40\textcolor{mygray}{1}$ \\
   &                                & 10 & 572 & $-0.580\, 165\, 768\, 725\, 648$ & $-0.580\, 246\, 472\, 594\, 385$ & $-0.580\, 165\, 768\, 308\, 401$ \\
   %&                                & 11 & 728 & $-$\texttt{0.580 165 768 725 648} & $-$\texttt{0.580 246 472 594 385} & $-$\texttt{0.580 165 768 308 401} \\
   %&                                & 12 & 910 & $-$\texttt{0.580 165 768 725 648} & $-$\texttt{0.580 246 472 594 385} & $-$\texttt{0.580 165 768 308 401} \\
%   &                                 & 8 & 330 & $-$0.580 165 768 725 556 & $-$\texttt{0.580 246 472 594 3}02       & $-$\texttt{0.580 165 768 308} 316       \\
   &                                 &   &     &                        & $-0.580\, 246\, 472\, 594\, 38\,$\footref{fn:b} & $-0.580\, 165\, 768\, 308\, 3\textcolor{mygray}{9}\,$\footref{fn:c}  \\[3pt]
  % &                                 &   &     &                        &                                       &                                       \\
2  &  $3\,{}^{1}\text{D}^{\text{o}}$ & 1 & 16  & $-0.563\, \textcolor{mygray}{667\, 098\, 051\, 029}$ & $-0.563\, 7\textcolor{mygray}{96\, 435\, 839\, 906}$ &                               \\
   &                                 & 2 & 40  & $-0.563\, 7\textcolor{mygray}{14\, 742\, 964\, 218}$ & $-0.563\, 800\, \textcolor{mygray}{355\, 646\, 005}$ &                                       \\
   &                                 & 3 & 80  & $-0.563\, 725\, \textcolor{mygray}{021\, 876\, 851}$ & $-0.563\, 800\, 4\textcolor{mygray}{17\, 719\, 023}$ &                                       \\
   &                                 & 4 & 140 & $-0.563\, 725\, 592\, \textcolor{mygray}{506\, 259}$ & $-0.563\, 800\, 420\, \textcolor{mygray}{356\, 275}$ &                                       \\
   &                                 & 5 & 224 & $-0.563\, 725\, 592\, 6\textcolor{mygray}{02\, 907}$ & $-0.563\, 800\, 420\, 4\textcolor{mygray}{57\, 190}$ &                                       \\
   &                                 & 6 & 336 & $-0.563\, 725\, 592\, 61\textcolor{mygray}{1\, 733}$ & $-0.563\, 800\, 420\, 462\, \textcolor{mygray}{004}$ &                                       \\
   &                                 & 7 & 480 & $-0.563\, 725\, 592\, 612\, \textcolor{mygray}{396}$ & $-0.563\, 800\, 420\, 462\, \textcolor{mygray}{340}$ &                                       \\
   &                                 & 8 & 660 & $-0.563\, 725\, 592\, 612\, 4\textcolor{mygray}{58}$ & $-0.563\, 800\, 420\, 462\, 36\textcolor{mygray}{4}$ &                                       \\
   &                                 & 9 & 880 & $-0.563\, 725\, 592\, 612\, 46\textcolor{mygray}{7}$                     & $-0.563\, 800\, 420\, 462\, 367$                     &                                       \\
  &                                & 10 & 1144 & $-0.563\, 725\, 592\, 612\, 468$ & $-0.563\, 800\, 420\, 462\, 367$ &                                       \\
  % &                                 & 4 & 140 & $-$0.563 725 592 254 888 & $-$\texttt{0.563 800 420} 081 822       &                               \\
   %&                                 &   &     &                        & $-$\texttt{0.563 800 420} 26$^{f}$      &                               \\
   &                                 &   &     &                        & $-0.563\, 800\, 420\, 462\,$\footnote{\textcite{Kar2009b}: Variational calculation with correlated exponential wave functions with 1\,100 terms for ${}^{1}\text{D}^{\text{o}}$ and ${}^{3}\text{D}^{\text{o}}$ states and 2\,200 terms for ${}^{1}\text{F}^{\text{e}}$ and ${}^{3}\text{F}^{\text{e}}$ states.\label{fn:d}}      &                               \\[3pt]
2  &  $3\,{}^{3}\text{D}^{\text{o}}$ & 1 & 16  & $-0.559\, 2\textcolor{mygray}{43\, 591\, 256\, 649}$ & $-0.559\, 32\textcolor{mygray}{4\, 252\, 504\, 229}$ &                               \\
   &                                 & 2 & 40  & $-0.559\, 248\, \textcolor{mygray}{243\, 961\, 630}$ & $-0.559\, 328\, \textcolor{mygray}{196\, 908\, 495}$ &                               \\
   &                                 & 3 & 80  & $-0.559\, 248\, 3\textcolor{mygray}{07\, 156\, 082}$ & $-0.559\, 328\, 26\textcolor{mygray}{0\, 154\, 298}$ &                               \\
   &                                 & 4 & 140 & $-0.559\, 248\, 3\textcolor{mygray}{09\, 993\, 418}$ & $-0.559\, 328\, 26\textcolor{mygray}{2\, 959\, 022}$ &                               \\
   &                                 & 5 & 224 & $-0.559\, 248\, 310\, 0\textcolor{mygray}{86\, 529}$ & $-0.559\, 328\, 263\, 0\textcolor{mygray}{88\, 623}$ &                               \\
   &                                 & 6 & 336 & $-0.559\, 248\, 310\, 09\textcolor{mygray}{3\, 681}$ & $-0.559\, 328\, 263\, 09\textcolor{mygray}{6\, 574}$ &                               \\
   &                                 & 7 & 480 & $-0.559\, 248\, 310\, 094\, 0\textcolor{mygray}{66}$ & $-0.559\, 328\, 263\, 097\, 2\textcolor{mygray}{00}$ &                               \\
   &                                 & 8 & 660 & $-0.559\, 248\, 310\, 094\, 09\textcolor{mygray}{9}$ & $-0.559\, 328\, 263\, 097\, 24\textcolor{mygray}{2}$ &                               \\
   &                                 & 9 & 880 & $-0.559\, 248\, 310\, 094\, 102$                     & $-0.559\, 328\, 263\, 097\, 247$                     &                                       \\
   &                               & 10 & 1144 & $-0.559\, 248\, 310\, 094\, 102$ & $-0.559\, 328\, 263\, 097\, 247$ &                                     \\
   %&                                 & 4 & 140 & $-$0.559 248 307 926 455 & $-$\texttt{0.559 328 26}1 398 964       &                               \\
   %&                                 &   &     &                        & $-$\texttt{0.559 328 262 819}$^{f}$     &                               \\ 
   &                                 &   &     &                        & $-0.559\, 328\, 263\, 096\,$\footref{fn:d}    &                               \\[3pt]
3  & $4\,{}^{1}\text{F}^{\text{e}}$  & 1 & 24  & $-0.531\, \textcolor{mygray}{894\, 604\, 658\, 559}$ & $-0.531\, 9\textcolor{mygray}{67\, 673\, 049\, 904}$ &                               \\
   &                                 & 2 & 60  & $-0.531\, 922\, 3\textcolor{mygray}{47\, 053\, 019}$ & $-0.531\, 995\, 4\textcolor{mygray}{22\, 503\, 136}$ &                               \\
   &                                 & 3 & 120 & $-0.531\, 922\, 367\, \textcolor{mygray}{449\, 804}$ & $-0.531\, 995\, 436\, \textcolor{mygray}{341\, 146}$ &                               \\
   &                                 & 4 & 210 & $-0.531\, 922\, 367\, 9\textcolor{mygray}{50\, 899}$ & $-0.531\, 995\, 436\, 9\textcolor{mygray}{35\, 757}$ &                               \\
   &                                 & 5 & 336 & $-0.531\, 922\, 367\, 96\textcolor{mygray}{6\, 984}$ & $-0.531\, 995\, 436\, 95\textcolor{mygray}{1\, 294}$ &                               \\
   &                                 & 6 & 504 & $-0.531\, 922\, 367\, 968\, \textcolor{mygray}{035}$ & $-0.531\, 995\, 436\, 952\, \textcolor{mygray}{629}$ &                               \\
   &                                 & 7 & 720 & $-0.531\, 922\, 367\, 968\, 11\textcolor{mygray}{6}$ & $-0.531\, 995\, 436\, 952\, 71\textcolor{mygray}{5}$ &                               \\
   &                                 & 8 & 990 & $-0.531\, 922\, 367\, 968\, 124$                     & $-0.531\, 995\, 436\, 952\, 724$                     &                               \\
   %&                                 & 4 & 210 & $-$0.531 922 303 453 196 & $-$\texttt{0.531 995} 378 174 180       &                               \\
   &                                 &   &     &                        & $-0.531\, 995\, 436\, 950\, 9$\footref{fn:d} &                               \\[3pt]
%   &                                 &   &     &                        & $-0.531\, 995\, 5\,$\footnote{\textcite{Kar2008}: Using the stabilization method with CI-type basis functions, the ${}^{3}\text{F}^{\text{e}}$ (${}^{1}\text{F}^{\text{e}}$) states were calculated with WFs of 1233 (1160) terms.\label{fn:e}}           &                               \\[3pt]
3  &  $4\,{}^{3}\text{F}^{\text{e}}$ & 1 & 24  & $-0.531\, 9\textcolor{mygray}{04\, 704\, 203\, 036}$ & $-0.531\, 99\textcolor{mygray}{0\, 468\, 695\, 483}$ &                               \\
   &                                 & 2 & 60  & $-0.531\, 918\, \textcolor{mygray}{294\, 710\, 703}$ & $-0.531\, 991\, 3\textcolor{mygray}{13\, 413\, 311}$ &                               \\
   &                                 & 3 & 120 & $-0.531\, 918\, 31\textcolor{mygray}{1\, 914\, 017}$ & $-0.531\, 991\, 32\textcolor{mygray}{5\, 813\, 023}$ &                               \\
   &                                 & 4 & 210 & $-0.531\, 918\, 312\, 4\textcolor{mygray}{45\, 571}$ & $-0.531\, 991\, 326\, 3\textcolor{mygray}{35\, 037}$ &                               \\
   &                                 & 5 & 336 & $-0.531\, 918\, 312\, 46\textcolor{mygray}{0\, 673}$ & $-0.531\, 991\, 326\, 34\textcolor{mygray}{7\, 176}$ &                               \\
   &                                 & 6 & 504 & $-0.531\, 918\, 312\, 461\, \textcolor{mygray}{857}$ & $-0.531\, 991\, 326\, 348\, \textcolor{mygray}{272}$ &                               \\
   &                                 & 7 & 720 & $-0.531\, 918\, 312\, 461\, 92\textcolor{mygray}{9}$ & $-0.531\, 991\, 326\, 348\, 32\textcolor{mygray}{7}$ &                               \\
   &                                 & 8 & 990 & $-0.531\, 918\, 312\, 461\, 937$                     & $-0.531\, 991\, 326\, 348\, 335$                     &                               \\
   %&                                 & 4 & 210 & $-$0.531 918 308 689 176 & $-$\texttt{0.531 991} 325 554 257       &                               \\
   &                                 &   &     &                        & $-0.531\, 991\, 326\, 346\, 5$\footref{fn:d} &                               \\[3pt]
%   &                                 &   &     &                        & $-0.531\, 991\, 5\,$\footref{fn:e}           &                               \\[3pt] 
\hline\hline
\end{tabular}
\end{scriptsize}
\end{table*}

\section{Conclusion}\label{conc}
We have established a comprehensive, mathematically rigorous, and straightforward procedure for eliminating the rotational degrees of freedom from the nonrelativistic Schr\"{o}dinger equation for a three-body Coulomb system using a basis of solid minimal bipolar harmonics (MBH). The resulting reduced Schr\"{o}dinger equation (RSE), which admits arbitrary masses, charges, angular momentum quantum numbers $L$ and $M$, as well as parity $\uppi$, takes a matrix-operator form, that provides a natural foundation for the numerical computation of the corresponding energy levels and wave functions. The algebraically transparent form of the RSE also provides a suitable departure point for an exact analytical treatment of the quantum three-body problem. 

The effectiveness of our approach hinges largely on expressing the MBHs as linear combinations of Wigner functions  $\mathcal{D}$, which simplifies the evaluation of the angular integral compared to conventional techniques. To validate the derived ex\-pres\-sions, we computed the nonrelativistic energies for several singlet and triplet helium states with natural and unnatural parities using an explicitly correlated multi-exponent variational Hylleraas-type framework. The re\-ported energy eigenvalues for natural-parity states agree well with previous benchmarks. For unnatural-parity states, our results currently establish the most accurate estimates of energy eigenvalues for this system.

This exposition provides a critical summary of literature results spanning the past century on the subject. Our work establishes a complete and self-contained reference for the derivation of the RSE in the MBH basis. It also lays the groundwork for constructing analogous theories for systems with more than three quantum particles.

\begin{acknowledgments}
%The authors are grateful to Prof. G. W. F. Drake for critical reading of the manuscript and for valuable suggestions. 
AS acknowledges the financial support from National Science and Technology Council (NSTC), Taiwan under grant numbers NSTC 111-2811-M-A49-558, NSTC 112-2811-M-A49-538, NSTC 113-2811-M-A49-531, and NSTC 114-2811-M-A49-529. HW acknowledges the financial support from Ministry of Science and Technology (MOST), Taiwan under grant number MOST 111-2113-M-A49-017 and National Science and Technology Council (NSTC), Taiwan under grant numbers NSTC 112-2113-M-A49-033, NSTC 113-2113-M-A49-001, and NSTC 114-2113-M-A49-018. The symbolic computations in this paper are performed using Maple™.  
\end{acknowledgments}

\bibliography{biblio_rmp}
%\bibliography{biblio_pra}
\end{document}

\preprint{APS/123-QED}

\title{Supplementary Material for `Elimination of angular dependency in quantum three-body problem made easy'}

\author{Anjan Sadhukhan}
\affiliation{Department of Applied Chemistry, National Yang Ming Chiao Tung University, 
1001 University Road, Hsinchu, 300, Taiwan}
\author{Grzegorz Pestka}
\affiliation{Faculty of Humanities, Social and Information Technology Sciences, The Mazovian University in Płock, 
Pl. Dabrowskiego 2, 09-402 Płock, Poland}
\author{Rafał Podeszwa}
\affiliation{Institute of Chemistry, University of Silesia, ul. Szkolna 9, 40-006 Katowice, Poland}
\author{Henryk A. Witek}
\email{hwitek@nycu.edu.tw}
\affiliation{Department of Applied Chemistry, National Yang Ming Chiao Tung University, 1001 University Road, Hsinchu, 300, Taiwan}
\date{\today}
\begin{abstract}

\end{abstract}

\maketitle

The main body of the current study reports several mathematical results, whose derivations are too lengthy or too technical to be included directly in the main text (abbreviated as \textbf{1} in the following). The details of these derivations, important for the completeness of our work, are provided here, where---not being bound by the space limitations---we can present appropriate demonstrations of these derivations.

\section{Expansion of minimal bipolar harmonics in terms of the Wigner functions \texorpdfstring{$\mathcal{D}$}{D} and evaluation of the corresponding linear expansion coefficients in closed form }

The minimal bipolar harmonics (MBHs) $\Omega_{l}^{LM\uppi}(\theta_{1},\phi_{1},\theta_{2},\phi_{2})$ [explicitly defined by Eq.~(66) of \textbf{1}] can be expressed as linear combinations of Wigner functions $\mathcal{D}_{L}^{MK}(\alpha,\beta,\gamma)$ [explicitly defined by Eq.~(35)  of \textbf{1}] as 
\begin{equation}\label{mbhwdr}
    \Omega_{l}^{LM\uppi}(\theta_{1},\phi_{1},\theta_{2},\phi_{2})\hspace{-2pt} =\hspace{-6pt}\sum_{K=-L}^{L}\hspace{-6pt}\Lambda_{Kl}^{L\uppi}\hspace{-2pt}\left(\theta\right)\mathcal{D}_{L}^{MK}\left(\alpha,\beta,\gamma\right)
\end{equation}
where the expansion coefficients $\Lambda_{Kl}^{L\uppi}\hspace{-2pt}\left(\theta\right)$ are functions of the interparticle angle $\theta$ [defined by Eqs.~(24) and~(25) of \textbf{1}], as demonstrated below. The coefficients $\Lambda_{Kl}^{L\uppi}\hspace{-2pt}\left(\theta\right)$, by virtue of Eq.~(77) of \textbf{1}, have been shown to be $M$-independent and equal to
\begin{equation}\label{intlambda}
    \Lambda_{Kl}^{L\uppi}\left(\theta\right) = \left\langle\left.\mathcal{D}_{L}^{LK}\right|\Omega_{l}^{LL\uppi}\right\rangle
\end{equation}
with the inner product understood as 
\begin{equation}\label{dtau}
    \left\langle f\vphantom{g} \right|\left.\!g\vphantom{f} \right\rangle =\!\!\int\limits_{0}^{2\pi}\hspace{-4pt} d\alpha\! \int\limits_{0}^{\pi}\hspace{-3pt}\sin{\beta}\,d\beta \hspace{-3pt}\int\limits_{0}^{2\pi}\hspace{-3pt} d\gamma \left(\,\,\hspace{-3pt}f^\ast\hspace{-3pt} \,\, g\right)
\end{equation}

Substituting $M=L$ in Eq.~(35)  of \textbf{1} produces a particularly simple expression for the Wigner function
\begin{equation}\label{wignerdmm}
    \mathcal{D}^{LK}_{L}\!(\!\alpha,\beta,\gamma\!)\!=\!\mathscr{N}^{LK}_{L} \!\! \left(\!\cos{\tfrac{\beta}{2}}\!\right)^{\!L\hspace{-0.5pt}+\hspace{-0.5pt}K}\!\!\left(\!\sin{\tfrac{\beta}{2}}\!\right)^{\!L\hspace{-0.5pt}-\hspace{-0.5pt}K}\hspace{-2pt}e^{\textbf{\textit{i}} L \alpha} e^{\textbf{\textit{i}} K \gamma} 
\end{equation}
with the normalization factor 
\begin{equation}\label{eq:nllk}
    \mathscr{N}^{LK}_{L} = \tfrac{(-1)^{K}}{4\pi}\left[\tfrac{2\left(2L+1\right)!}{\left(L+K\right)!\left(L-K\right)!}\right]^{\frac{1}{2}}
\end{equation}
following from Eq.~(38)  of \textbf{1}. 

The corresponding MBHs with $M=L$ can be expressed in Euler angles by combining Eqs.~(90) and (91)  of \textbf{1} with Eq.~(89) of \textbf{1}
\begin{eqnarray}
    \Omega_{l}^{LL\uppi} &=&N_{l}^{L\uppi}\sum_{\kappa=0}^{d}(-1)^{\kappa}\left(\hat{x}_{1}+\textbf{\textit{i}}\hat{y}_{1}\right)^{l-\kappa}\left(\hat{z}_{1}\right)^{\kappa}\nonumber\\ 
    && \left(\hat{x}_{2}+\textbf{\textit{i}}\hat{y}_{2}\right)^{L-l+\kappa}\left(\hat{z}_{2}\right)^{d-\kappa}
\end{eqnarray}
and substituting into them the expressions given implicitly by Eq.~(16) of \textbf{1}
\begin{equation}\label{ctetrans}
    \left[\begin{array}{c}
         \hat{x}_{i} \\
         \hat{y}_{i} \\
         \hat{z}_{i} 
    \end{array}\right]\!\! = \!\!\left[\begin{array}{l}
         +\cos\gamma_{i}\cos\beta\cos\alpha-\sin\gamma_{i}\sin\alpha  \\
         +\cos\gamma_{i}\cos\beta\sin\alpha+\sin\gamma_{i}\cos\alpha \\
         -\cos\gamma_{i}\sin\beta
    \end{array}\right]
\end{equation}
with $\gamma_{i}=\left(\gamma+(-1)^{i}\,\frac{\theta}{2}\right)$,  $i=1,2$, where the normalization factor 
\begin{equation}\label{nlpik}
    N_{l}^{L\uppi} = \frac{(-1)^L}{4\pi}\left[\!\left(\!\tfrac{2}{L+\hspace{-1pt}1}\!\right)^{\!d}\!\tfrac{\left(\frac{3}{2}\right)_{\hspace{-1pt}l}\left(\frac{3}{2}\right)_{\hspace{-1pt}L-l_{\vphantom{a}}+d}}{(l\hspace{-1pt}-d)!(L\hspace{-1pt}-l)!}\!\right]^{\!\frac{1}{2}}
\end{equation}
follows directly from Eq.~(84) of \textbf{1}. Finally, one obtains
\begin{eqnarray}\label{mbhsLLpi}
    \Omega_{l}^{LL\uppi} \! &=&\! N^{L\uppi}_{l}e^{\textit{\textbf{i}}L\alpha}\hspace{-2pt}\sum_{\kappa=0}^{d}\hspace{-2pt}\left(\!-1\!\right)^{\kappa}\hspace{-3pt}\left(\hspace{-4pt}\left(\!\cos{\tfrac{\beta}{2}}\!\right)^{\!2}\hspace{-4pt}e^{\textit{\textbf{i}}\gamma_{1}}\hspace{-3pt}-\hspace{-3pt}\left(\sin{\tfrac{\beta}{2}}\right)^{\!2}\hspace{-4pt}e^{-\textit{\textbf{i}}\gamma_{1}}\hspace{-4pt}\right)^{\!l-\kappa}\nonumber\\
    && \left(\!\hspace{-2pt}\left(\!\cos{\tfrac{\beta}{2}}\!\right)^{2}\hspace{-4pt}e^{\textit{\textbf{i}}\gamma_{2}}\hspace{-3pt}-\hspace{-3pt}\left(\sin{\tfrac{\beta}{2}}\right)^{2}\hspace{-4pt}e^{-\textit{\textbf{i}}\gamma_{2}}\!\right)^{\!L-l+\kappa}\nonumber\\
    && \left(-\sin{\beta}\cos{\gamma_{1}}\right)^{\kappa}\left(-\sin{\beta}\cos{\gamma_{2}}\right)^{d-\kappa}
\end{eqnarray}
By expanding the binomial factors, rewriting $\cos{\gamma_{i}}$ in exponential form, and carrying out a sequence of straightforward algebraic simplifications, one arrives at the following representation of $\Omega_{l}^{LL\uppi}$
\begin{eqnarray}\label{mbhsLLpi1}
    &&\Omega_{l}^{LL\uppi} \! =\! N^{L\uppi}_{l}e^{\textit{\textbf{i}}L\alpha}\hspace{-2pt}\sum_{\kappa=0}^{d}\hspace{-2pt}\sum_{\texttt{j}_{1}=0}^{l-\kappa}\hspace{-1pt}\sum_{\texttt{j}_{2}=0}^{L-l+\kappa}\hspace{-8pt}\left(\!-1\!\right)^{d\hspace{-0.5pt}+\hspace{-0.5pt}\kappa\hspace{-0.5pt}+\hspace{-1pt}\texttt{j}_{\hspace{-0.5pt}1}\hspace{-1pt}+\hspace{-1pt}\texttt{j}_{\hspace{-0.5pt}2}}\hspace{-4pt}\left(\hspace{-5pt}\begin{array}{c}
         \scriptstyle l-\kappa \\
         \scriptstyle \texttt{j}_{1} 
    \end{array}\hspace{-5pt}\right)\!\!\!\left(\hspace{-5pt}\begin{array}{c}
         \scriptstyle L-l+\kappa \\
         \scriptstyle \texttt{j}_{2} 
    \end{array}\hspace{-5pt}\right)\nonumber\\[3pt]
    &&\hspace{15pt} \left(\hspace{-2pt}\cos{\tfrac{\beta}{2}}\hspace{-2pt}\right)^{2L+d-2\texttt{j}_{\hspace{-0.5pt}1}\hspace{-1pt}-\hspace{-0.5pt}2\texttt{j}_{\hspace{-0.5pt}2}}\left(\hspace{-2pt}\sin{\tfrac{\beta}{2}}\hspace{-2pt}\right)^{d+2\texttt{j}_{\hspace{-0.5pt}1}\hspace{-1pt}+2\texttt{j}_{\hspace{-0.5pt}2}\hspace{-0.5pt}}\\
    &&\hspace{15pt}\hspace{-4pt}\sum_{\eta=0}^{d}e^{\textit{\textbf{i}}\left(L+d-\hspace{-0.5pt}2\texttt{j}_{\hspace{-0.5pt}1}\hspace{-1pt}-\hspace{-1pt}2\texttt{j}_{\hspace{-0.5pt}2}\hspace{-1pt}-\hspace{-1pt}2\eta\right)\gamma} \, e^{\textit{\textbf{i}}\left(\hspace{-1pt}-\hspace{-0.5pt}2l+L+d+\hspace{-0.5pt}2\texttt{j}_{\hspace{-0.5pt}1}\hspace{-1pt}-\hspace{-0.5pt}2\texttt{j}_{\hspace{-0.5pt}2}\hspace{-0.5pt}-\hspace{-1pt}(\!-1\!)^{\kappa}2\eta\right)\frac{\theta}{2}}\nonumber
\end{eqnarray}
with the summation over $\eta$ originating from the expansion of $\left(\cos{\gamma_{1}}\right)^{\kappa}\left(\cos{\gamma_{2}}\right)^{d-\kappa}$.
This representation of $\Omega_{l}^{LL\uppi}$ can  be simplified further by introducing the following sum\-ma\-tion identity
\begin{equation*}
    \sum_{\texttt{j}_{1}=0}^{l\!-\kappa}\hspace{-1pt}\sum_{\texttt{j}_{2}=0}^{L\!-l+\kappa}\hspace{-3pt}\left(\hspace{-5pt}\begin{array}{c}
         \scriptstyle l-\kappa \\
         \scriptstyle \texttt{j}_{1} 
    \end{array}\hspace{-5pt}\right)\!\!\!\left(\hspace{-5pt}\begin{array}{c}
         \scriptstyle L-l+\kappa \\
         \scriptstyle \texttt{j}_{2} 
    \end{array}\hspace{-5pt}\right)  \hspace{-2pt} A_{\texttt{j}_{1}}^{\texttt{j}_{2}} = \hspace{-2pt}\sum_{\texttt{J}=0}^{L}\hspace{-1pt}\sum_{\texttt{j}=0}^{\texttt{J}}\hspace{-3pt}\left(\hspace{-5pt}\begin{array}{c}
         \scriptstyle l-\kappa \\
         \scriptstyle \texttt{j} 
    \end{array}\hspace{-5pt}\right)\!\!\!\left(\hspace{-5pt}\begin{array}{c}
         \scriptstyle L-l+\kappa \\
         \scriptstyle \texttt{J}-\texttt{j} 
    \end{array}\hspace{-5pt}\right)\hspace{-2pt} A_{\texttt{j}}^{\texttt{J}-\texttt{j}}
\end{equation*}
valid for all relevant values of $L$, $l$, $d$, and $\kappa$, owing to the presence of the binomial factors.
With this transformation, Eq.~\eqref{mbhsLLpi1} reduces to
\begin{eqnarray}\label{mbhsLLpi2}
    \Omega_{l}^{LL\uppi} \! &=&\! N^{L\uppi}_{l}\,e^{\textit{\textbf{i}}L\alpha}\hspace{-1pt}\sum_{\eta=0}^{d}\hspace{-1pt}\sum_{\texttt{J}=0}^{L}\hspace{-1pt}\left(\hspace{-2pt}\cos{\tfrac{\beta}{2}}\hspace{-2pt}\right)^{2L+d-2\texttt{J}}\left(\hspace{-2pt}\sin{\tfrac{\beta}{2}}\hspace{-2pt}\right)^{2\texttt{J}+d}\nonumber\\
    &&  e^{\textit{\textbf{i}}\left(L+d-2\texttt{J}-2\eta\right)\gamma}\,\texttt{T}^{\texttt{J}\uppi}_{l\eta}\hspace{-2pt}\left(\theta\right)
\end{eqnarray}
where the factor 
\begin{eqnarray}\label{tfactor}
    \texttt{T}^{\texttt{J}\uppi}_{l\eta}\hspace{-2pt}\left(\theta\right) &=& \sum_{\zeta=0}^{d}\sum_{\texttt{j}=0}^{\texttt{J}}\left(-1\right)^{\texttt{J}+d+\zeta}\left(\hspace{-5pt}\begin{array}{c}
         \scriptstyle l-\zeta \\
         \scriptstyle \texttt{j}
    \end{array}\hspace{-5pt}\right)\!\!\!\left(\hspace{-5pt}\begin{array}{c}
         \scriptstyle L-l+\zeta \\
         \scriptstyle \texttt{J}-\texttt{j} 
    \end{array}\hspace{-5pt}\right)\nonumber\\
    &&  e^{\textit{\textbf{i}}\left(-2l+L+d+4\texttt{j}-2\texttt{J}-(\hspace{-1pt}-1\hspace{-1pt})^{\zeta}2\eta\right)\frac{\theta}{2}}
\end{eqnarray}
does not depend on the Euler angles.

Substituting Eqs.~\eqref{wignerdmm} and~\eqref{mbhsLLpi2} into Eq.~\eqref{intlambda}, and explicitly performing the integrations over the Euler angles allows us to express $\Lambda_{Kl}^{L\uppi}\!\left(\theta\right)$ in the following form
\begin{equation}\label{ECpi0}
    \Lambda_{Kl}^{L\uppi}\!\left(\theta\right)=\mathscr{N}^{LK}_{L}N_{l}^{L\uppi}\sum_{\eta=0}^{d}\hspace{-1pt}\sum_{\texttt{J}=0}^{L}\,\texttt{T}^{\texttt{J}\uppi}_{l\eta}\hspace{-2pt}\left(\theta\right)\mathbb{I}_{\alpha}\,\mathbb{I}_{\beta}\,\mathbb{I}_{\gamma}
\end{equation}
where the Euler angles integrals are given by
\begin{eqnarray}
    \hspace{-18pt}\mathbb{I}_{\alpha}\hspace{-2pt}  &=& \hspace{-3pt}\int\limits_{0}^{2\pi}\hspace{-2pt}d\alpha\, e^{\textit{\textbf{i}}L\alpha}\,e^{-\textit{\textbf{i}}L\alpha} = 2\pi\label{intalpha}\\
    \hspace{-18pt}\mathbb{I}_{\gamma}  \hspace{-2pt}  &=& \hspace{-3pt} \int\limits_{0}^{2\pi}\hspace{-2pt}d\gamma \,e^{\textit{\textbf{i}}\left(L\hspace{-0.5pt}-\hspace{-0.5pt}K\hspace{-1pt}+d\hspace{-0.5pt}-\hspace{-0.5pt}2\texttt{J}\hspace{-0.5pt}-\hspace{-0.5pt}2\eta\right)\gamma}=2\pi\,\delta_{\texttt{J},\frac{L-K+d}{2}-\eta}\label{intgamma}\\
    \hspace{-18pt}\mathbb{I}_{\beta}  \hspace{-2pt}  &=& \hspace{-3pt} \int\limits_{0}^{\pi}\hspace{-2pt}\sin{\beta}\,d\beta\,\left(\!\cos{\tfrac{\beta}{2}}\!\right)^{\!3L+\hspace{-0.5pt}K\hspace{-1pt}+d-2\texttt{J}}\hspace{-2pt}\left(\!\sin{\tfrac{\beta}{2}}\!\right)^{\!L-\hspace{-1pt}K+d+2\texttt{J}}\label{intbeta}
\end{eqnarray}
The Kronecker delta function $\delta$ in the integral $\mathbb{I}_{\gamma}$ vanishes for every odd value of $L\!-\!K\!+\!d$, as both $\texttt{J}$ and $\eta$ can assume only integer values. This gives
\begin{equation}\label{ECpi1}
    \Lambda_{Kl}^{L\uppi}\hspace{-2pt}\left(\!\theta\hspace{-1pt}\right) = \left\{\begin{array}{lc}
      \overline{\Lambda}_{Kl}^{L\uppi}\hspace{-2pt}\left(\!\theta\hspace{-1pt}\right)   & \text{for $L-K+d$ even}  \\[5pt]
       0  & \text{for $L-K+d$ odd} 
    \end{array}\right.
\end{equation}
with 
\begin{equation}\label{ECpi}
    \overline{\Lambda}_{Kl}^{L\uppi}\!\left(\theta\right)=4\pi^2\mathscr{N}^{LK}_{L}N_{l}^{L\uppi}\sum_{\eta=0}^{d}\hspace{-1pt}\,\texttt{T}^{\texttt{J}\uppi}_{l\eta}\hspace{-2pt}\left(\theta\right)\overline{{}_{\hspace{2pt}}\mathbb{I}_{\beta}\hspace{-3pt}}
\end{equation}
where $\texttt{J}=\tfrac{L-K+d}{2}-\eta$ [as stipulated by Eq.~(\ref{intgamma})] and where $\overline{{}_{\hspace{2pt}}\mathbb{I}_{\beta}\hspace{-3pt}}\hspace{3pt} \equiv \left. \mathbb{I}_{\beta}\right|_{2\texttt{J}=L-K+d-2\eta}$ is given by
\begin{eqnarray}
    \overline{{}_{\hspace{2pt}}\mathbb{I}_{\beta}\hspace{-3pt}}\hspace{3pt} &=& \int\limits_{0}^{\pi}\hspace{-2pt}\sin{\beta}\,d\beta\,\left(\!\cos{\tfrac{\beta}{2}}\!\right)^{\!2(L+K+\eta)}\hspace{-2pt}\left(\!\sin{\tfrac{\beta}{2}}\!\right)^{\!2(L-K+d-\eta)}\nonumber\\
    &=&  2 \left(\hspace{-2pt}(2L+1+d)\hspace{-2pt}\left(\hspace{-5pt}\begin{array}{c}
         \scriptstyle 2L+d \\
         \scriptstyle L+K+\eta
    \end{array}\hspace{-5pt}\right)_{\vphantom{l_{l}}}\hspace{-4pt}\right)^{-1}\label{intbeta1}
\end{eqnarray}
with the integration result following from the formulae~3.621(5) and~8.384(1) of \textcite{Gradshteyn2015}. Substituting this expression into Eq.~\eqref{ECpi} gives us  the final expression for $\overline{\Lambda}_{Kl}^{L\uppi}\hspace{-2pt}\left(\!\theta\hspace{-1pt}\right)$ as
\begin{equation}
     \label{ECpi1a} \overline{\Lambda}_{Kl}^{L\uppi}\hspace{-2pt}\left(\!\theta\hspace{-1pt}\right) = \frac{8\pi^2\mathscr{N}^{LK}_{L}N_{l_{\vphantom{l}}}^{L\uppi}}{(2L+1+d) }\sum_{\eta=0}^{d} \frac{\texttt{T}^{\texttt{J}\uppi}_{l_{\vphantom{l}}\eta}\hspace{-2pt}\left(\theta\right)}{\left(\hspace{-5pt}\begin{array}{c}
         \scriptstyle 2L+d \\
         \scriptstyle L+K+\eta
    \end{array}\hspace{-5pt}\right)}
\end{equation}
with $\texttt{J}=\tfrac{L-K+d}{2}-\eta$. The explicit form of $\overline{\Lambda}_{Kl}^{L\uppi}\hspace{-2pt}\left(\!\theta\hspace{-1pt}\right)$ for natural and unnatural  parity can be obtained by substituting the normalization factors from Eqs.~\eqref{eq:nllk} and~\eqref{nlpik} and $\texttt{T}^{\texttt{J}\uppi}_{l\eta}\hspace{-2pt}\left(\theta\right)$ from Eq.~\eqref{tfactor} into Eq.~\eqref{ECpi1a}. For natural parity (decoded by $\uppi=\text{n}$ and $d=0$), after a sequence of simplifications, we obtain
\begin{eqnarray}\label{ECn2}
    \overline{\Lambda}_{Kl}^{L\text{n}}\hspace{-2pt}\left(\!\theta\hspace{-1pt}\right)\hspace{-2pt} &=& \hspace{-2pt}\left(\!-1\!\right)^{\frac{L-K}{2}}\hspace{-4pt} \left[\hspace{-2pt}\tfrac{\left(\!L+\!K\!\right)!\left(\!L-\!K\!\right)!}{2\left(\!2L+\!1\!\right)!}\tfrac{\left(\!\frac{3}{2}\!\right)_{\!l}\hspace{-2pt}\left(\!\frac{3}{2}\!\right)_{\!L-l}}{l!\left(L-l\right)!}\hspace{-2pt}\right]^{\frac{1}{2}}\hspace{-2pt} \\
    &&\hspace{0pt} \hspace{-2pt}\sum\limits_{\texttt{j}=0}^{\frac{L-K}{2}}\hspace{-2pt}\left(\hspace{-4pt}\begin{array}{c}
        \scriptstyle l \\
        \scriptstyle \texttt{j} 
    \end{array}\hspace{-4pt}\right)\hspace{-4pt}\left(\hspace{-4pt}\begin{array}{c}
         \scriptstyle L-l \\
         \scriptstyle \frac{L-K}{2}\!-\!\texttt{j} 
    \end{array}\hspace{-4pt}\right) e^{\textit{\textbf{i}}\left(\!4\texttt{j}-2l+K\!\right)\frac{\theta}{2}}\nonumber
\end{eqnarray}
while for unnatural parity ($\uppi=\text{u}$ and $d=1$), we obtain
\begin{eqnarray}\label{ECu2}
    \overline{\Lambda}_{Kl}^{L\text{u}}\hspace{-2pt}\left(\!\theta\hspace{-1pt}\right)\hspace{-2pt} &=& \hspace{-2pt}\left(\!-1\!\right)^{\frac{L-K+1}{2}}\hspace{-4pt} \left[\hspace{-2pt}\tfrac{\left(\!L+\!K\!\right)!\left(\!L-\!K\!\right)!}{\left(\!L+\!1 \!\right)\left(\!2L+\!1\!\right)!}\tfrac{\left(\!\frac{3}{2}\!\right)_{\!l}\hspace{-2pt}\left(\!\frac{3}{2}\!\right)_{\!L-l+1}}{\left(l-1\right)!\left(L-l\right)!}\hspace{-2pt}\right]^{\frac{1}{2}}\hspace{-2pt} \\
    &&\hspace{-10pt} \hspace{-6pt}\sum\limits_{\texttt{j}=0}^{\frac{L-K+1}{2}}\hspace{-6pt}\tfrac{2\texttt{j}\left(L+1\right)-l\left(L-K+1\right)}{2l\left(L-l+1\right)}\left(\hspace{-4pt}\begin{array}{c}
        \scriptstyle l \\
        \scriptstyle \texttt{j} 
    \end{array}\hspace{-4pt}\right)\hspace{-4pt}\left(\hspace{-4pt}\begin{array}{c}
         \scriptstyle L-l+1 \\
         \scriptstyle \frac{L-K+1}{2}\!-\!\texttt{j} 
    \end{array}\hspace{-4pt}\right) e^{\textit{\textbf{i}}\left(\!4\texttt{j}-2l\hspace{-1pt}+\hspace{-1pt}K\!\right)\frac{\theta}{2}}\nonumber
\end{eqnarray}

\section{Verification of the Results Reported by Pont and Shakeshaft}
%\section{Verification of the Results Reported by Pont and Shakeshaft \texorpdfstring{\cite{Pont1995}}{}} 
The expansion of minimal bipolar harmonics in terms of Wigner functions and evaluation of the corresponding linear expansion coefficients in closed form was previously attempted by
\textcite{Pont1995}, who proposed a similar relationship to that given in Eq.~\eqref{mbhwdr} [see Eq.~(A22) and Eq.~(A23) of \textcite{Pont1995}  for natural and unnatural parity, respectively]. They reported a closed-form expression, Eq.~(A13) of \textcite{Pont1995}, for their expansion coefficients $C^{K}_{l_{1}\hspace{-1pt},\,l_{2}}(\theta)$. However, we have identified several issues associated with this expression, as explained in detail in the following paragraphs. Note that the discussion below uses the original notation of \textcite{Pont1995}.

The expression for $C^{K}_{l_{1}\hspace{-1pt},\,l_{2}}(\theta)$ reported by \textcite{Pont1995} is given by the following formula
\begin{eqnarray}\label{Cl1l2K}
    C^{K}_{l_{1}\hspace{-1pt},\,l_{2}}(\theta) &=& \tfrac{(-1)^{\frac{l_{1_{\vphantom{a}}}+l_{2}-K}{2^{\vphantom{l}}}}}{N_{L}^{K\vphantom{^a}}}\tfrac{l_{1_{\vphantom{l}}}!}{\left[\frac{l_{1_{\vphantom{l}}}-l_{2}-K\vphantom{^a}}{2^{\vphantom{l}}}\right]^{\vphantom{l}}!\left[\frac{l_{1_{\vphantom{l}}}+l_{2}+K\vphantom{^a}}{2^{\vphantom{l}}}\right]!}\nonumber\\
    &&\hspace{-30pt}  {}_2F_1\left[\!\begin{array}{c}
         {\scriptstyle -\frac{l_{1_{\vphantom{l}}}+l_{2}+K}{2^{\vphantom{l}}},\, -l_{2}} \\[2pt]
         {\scriptstyle{\frac{l_{1_{\vphantom{l}}}-l_{2}-K}{2^{\vphantom{l}}}+1}}
    \end{array};e^{2\textbf{\textit{i}}\theta}\right]e^{\textit{\textbf{i}}\left(l_{2}+\frac{K}{2}\right)\theta}
\end{eqnarray}
with
\begin{equation}
    N_{L}^{K}=(-1)^{L}\left[\tfrac{(2L+1)}{8\pi^2}\tfrac{(2L)!}{(L-K)!(L+K)!}\right]
\end{equation}
and with $L=l_{1}+l_{2}$ for natural parity and $L=l_{1}+l_{2}+1$ for unnatural parity. Note that the second of these expressions is erroneous: by Clebsch-Gordan coupling of two angular momenta $l_1$ and $l_2$ it is not possible to obtain the composite angular momentum $L=l_{1}+l_{2}+1$. The correct formula, first derived by Schwartz in Appendix~I of \textcite{Schwartz1961}, should read $L=l_{1}+l_{2}-1$. This correction needs to be introduced in many places of the paper by Pont and Shakeshaft.

The analytic form of Eq.~\eqref{Cl1l2K} is not always numerically stable in the usual sense. For example, considering a valid set of  parameters ($l_{1}=0$, $l_{2}=1$, $K=1$,  $L=1$) Eq.~\eqref{Cl1l2K} reduces to 
\begin{eqnarray}\label{Cl1l2K1}
    C^{1}_{0,1}(\theta)=-\frac{8\pi^{2}}{3}e^{-\textit{\textbf{i}}\frac{3}{2}\theta}\frac{{}_2F_1\left[\!\begin{array}{c}
          {\scriptstyle -1,\, -1} \\
          {\scriptstyle{0}}
     \end{array};e^{2\textbf{\textit{i}}\theta}\right]}{\Gamma(0)}
\end{eqnarray}
which is not a well-defined expression with a singular factor $\Gamma(0)$ and zero as the lower parameter of the hypergeometric function ${}_2F_1$. However, grouping these two terms together, it is possible to interpret Eq.~\eqref{Cl1l2K1} as a limit
\begin{equation*}\label{Cl1l2K2}
    \frac{{}_2F_1\left[\!\begin{array}{c}
          {\scriptstyle -1,\, -1} \\
          {\scriptstyle{0}}
     \end{array};e^{2\textbf{\textit{i}}\theta}\right]}{\Gamma(0)}=\lim_{\epsilon \to 0}\frac{{}_2F_1\left[\!\begin{array}{c}
          {\scriptstyle -1,\, -1} \\
          {\scriptstyle{\epsilon}}
     \end{array};e^{2\textbf{\textit{i}}\theta}\right]}{\Gamma(\epsilon)} = e^{2\textbf{\textit{i}}\theta}
\end{equation*}
Similar ambiguities can also arise for other parameter sets. To circumvent this technical  difficulty, it is possible to  use one of the hypergeometric identities \cite{wolfram}
\begin{equation*}
    {}_2F_1\!\left[\!\begin{array}{c}
         {\scriptstyle -n,\, a} \\
         {\scriptstyle a-b+1}
    \end{array}\!\!;\frac{1}{1-z}\right]\! =\! \frac{\left(1-b\right)_{n}\vphantom{^{l}}}{\left(a-b+1\right)_{n\vphantom{_{l}}}}\left(1\!-\!z\right)^{a} {}_2F_1\left[\!\begin{array}{c}
         {\scriptstyle a,\, b} \\
         {\scriptstyle b-n}
    \end{array}\!\!;z\right]
\end{equation*}
with $ a=-\sfrac{\left(l_{1}\!+\!l_{2}\!+\!K\right)}{2}$, $b=-l_{1}$, $n=l_{2}$ and $z=1-e^{\textit{-\textbf{i}}2\theta}$. Applying this identity to Eq.~\eqref{Cl1l2K}, we can obtain a numerically stable expression
\begin{eqnarray}\label{Cl1l2Kstab}
    C^{K}_{l_{1}\hspace{-1pt},\,l_{2}}(\theta)\! &=&\! e^{\textit{\textbf{i}}\left(l_{1}+\frac{K}{2}\right)\theta}{}_2F_1\left[\!\begin{array}{c}
         {\scriptstyle -l_{1},\, -\frac{l_{1_{\vphantom{l}}}+l_{2}+K}{2^{\vphantom{l}}}} \nonumber\\
         {\scriptstyle{-l_{1}-l_{2}}}
    \end{array};1\!-\!e^{-2\textit{\textbf{i}}\theta}\right]\\
    &&\hspace{-50pt}\tfrac{4\pi^{\frac{3}{2}}(-1)^{\frac{l_{1_{\vphantom{a}}}}{2}+\frac{l_{2_{\vphantom{a}}}}{2}+\frac{K}{2}}\left(\frac{l_{1}}{2}+\frac{l_{2}}{2}+\frac{K}{2}-\frac{1}{2}\right)!\left(\frac{l_{1}}{2}+\frac{l_{2}}{2}-\frac{K}{2}-\frac{1}{2}\right)!}{\left(l_{1}+l_{2}+\frac{1}{2}\right)!}
\end{eqnarray}
replacing the original formula Eq.~(A13) of \textcite{Pont1995}. Unfortunately, this expression does not produce correct expansion of minimal bipolar harmonics in terms of Wigner functions. We show it  below on several examples with $L=l_{1}+l_{2}$.

The explicit values of the coefficients $C^{K}_{l_{1}\hspace{-1pt},\,l_{2}}(\theta)$ can be computed directly from Eq.~(A22) of \textcite{Pont1995}
\begin{equation}\label{eqa22}
    \zeta_{1}^{l_{1}}\zeta_{2}^{l_{2}}=\hspace{-15pt}\sum_{\substack{K=-L,\\(-1)^{K}=(-1)^{L}}}^{L}\hspace{-15pt}\mathscr{D}_{L}^{LK}(\alpha,\beta,\gamma)\,\,C^{K}_{l_{1}\hspace{-1pt},\,l_{2}}(\theta)
\end{equation}
with $\zeta_{1}$ and $\zeta_{2}$ given by Eqs. (A20) and (A21), respectively, of \textcite{Pont1995} as
\begin{equation}\label{zeta1}
    \zeta_{1}\!=\!e^{\textit{\textbf{i}}\alpha}\!\!\left[\!\left(\cos{\tfrac{\beta}{2}}\right)^{2}e^{\textit{\textbf{i}}\left(\gamma-\tfrac{\theta}{2}\right)}\!-\!\left(\sin{\tfrac{\beta}{2}}\right)^{2}e^{-\textit{\textbf{i}}\left(\gamma-\tfrac{\theta}{2}\right)}\!\right]
\end{equation}
\begin{equation}\label{zeta2}
    \zeta_{2}\!=\!e^{\textit{\textbf{i}}\alpha}\!\!\left[\!\left(\cos{\tfrac{\beta}{2}}\right)^{2}e^{\textit{\textbf{i}}\left(\gamma+\tfrac{\theta}{2}\right)}\!-\!\left(\sin{\tfrac{\beta}{2}}\right)^{2}e^{-\textit{\textbf{i}}\left(\gamma+\tfrac{\theta}{2}\right)}\!\right]
\end{equation}
and with the Wigner functions $\mathscr{D}_{L}^{LK}(\alpha,\beta,\gamma)$ defined explicitly by Eq.~(A19) of \textcite{Pont1995} as
\begin{eqnarray}\label{wignerdpsml}
    \mathscr{D}_{L}^{LK}(\alpha,\beta,\gamma) &=& (-1)^{L}\sqrt{\tfrac{(2L+1)!}{8\pi^2(L+K)!(L-K)!}}e^{\textit{\textbf{i}}L\alpha}e^{\textit{\textbf{i}}K\gamma}\nonumber\\
    && \left(\cos{\tfrac{\beta}{2}}\right)^{L+K}\left(\sin{\tfrac{\beta}{2}}\right)^{L-K}
\end{eqnarray}
Note that $\mathscr{D}_{L}^{LK}(\alpha,\beta,\gamma)$ is identical to our Wigner function $\mathcal{D}_{L}^{LK}(\alpha,\beta,\gamma)$ given by Eq.~(\ref{wignerdmm}) except for the phase factor
\begin{equation*}\label{wignerdps}
    \mathscr{D}_{L}^{LK}\equiv\mathscr{D}_{L}^{LK}(\alpha,\beta,\gamma)=(-1)^{L+K}\,\mathcal{D}_{L}^{LK}(\alpha,\beta,\gamma)
\end{equation*}

\begin{table}[t]
    \centering
    \caption{The expansion coefficients $C^{K}_{l_{1}\hspace{-1pt},\,l_{2}}(\theta)$ computed directly from  Eq.~\eqref{eqa22a} for several valid sets of parameters do not agree with the analogous results computed from the  closed-form expression given by Pont and Shakeshaft (see Eq.~\eqref{Cl1l2Kstab} and the discussion therein). } \label{tab:tab1}\vspace{2pt}
    \begin{tabular}{c@{\quad\quad}c@{\quad\quad}c@{\quad\quad}c@{\quad\quad\quad}c@{\quad\quad\quad}c@{\quad}}
    \hline\hline\\
      &  &  &  & $C^{K}_{l_{1}\hspace{-1pt},\,l_{2}}(\theta)$ & $C^{K}_{l_{1}\hspace{-1pt},\,l_{2}}(\theta)$\\
      L & $l_{1}$ & $l_{2}$ & $K$ & from & from\\
       & &  &  & \eqref{eqa22a} & \eqref{Cl1l2Kstab}\\
      \\
    \hline\\
     0 & 0 & 0 & 0 & 2$\sqrt{2}\pi$ & 8$\pi^{2}$ \\
     1 & 1 & 0 & $\pm 1$ & $\mp\tfrac{2\sqrt{2}\pi}{\sqrt{3}} e^{\mp\textit{\textbf{i}}\tfrac{\theta}{2}}$ & $\mp\tfrac{8\pi^{2}}{3} e^{\mp\textit{\textbf{i}}\tfrac{\theta}{2}}$ \\
     1 & 0 & 1 & $\pm 1$ & $\mp\tfrac{2\sqrt{2}\pi}{\sqrt{3}} e^{\pm\textit{\textbf{i}}\tfrac{\theta}{2}}$ & $\mp\tfrac{8\pi^{2}}{3} e^{\pm\textit{\textbf{i}}\tfrac{\theta}{2}}$ \\
     2 & 1 & 1 & 0 & $-\tfrac{4\pi}{\sqrt{15}}\cos{\theta}$ & $-\tfrac{8\pi^{2}}{15}\cos{\theta}$ \\
     2 & 1 & 1 & $\pm 2$ & $\tfrac{2\sqrt{2}\pi}{\sqrt{5}}$ & $\tfrac{8\pi^{2}}{5}$ \\
     2 & 2 & 0 & 0 & $-\tfrac{4\pi}{\sqrt{15}}$ & $-\tfrac{8\pi^{2}}{15}$ \\
     2 & 0 & 2 & 0 & $-\tfrac{4\pi}{\sqrt{15}}$ & $-\tfrac{8\pi^{2}}{15}$ \\
     2 & 2 & 0 & $\pm 2$ & $\tfrac{2\sqrt{2}\pi}{\sqrt{5}} e^{\mp\textit{\textbf{i}}\theta}$ & $\tfrac{8\pi^{2}}{5} e^{\mp\textit{\textbf{i}}\theta}$ \\
     2 & 0 & 2 & $\pm 2$ & $\tfrac{2\sqrt{2}\pi}{\sqrt{5}} e^{\pm\textit{\textbf{i}}\theta}$ & $\tfrac{8\pi^{2}}{5} e^{\pm\textit{\textbf{i}}\theta}$ \\
     \\
    \hline\hline
    \end{tabular}
\end{table}

Due to the orthogonality of the Wigner functions $\mathscr{D}_{L}^{LK}$ (see Eq.~(41) of~\textbf{1}), Eq.~\eqref{eqa22} can be rewritten as
\begin{equation}\label{eqa22a}
    C^{K}_{l_{1}\hspace{-1pt},\,l_{2}}(\theta) = \int\limits_{\alpha=0}^{2\pi}\hspace{-5pt}d\alpha\hspace{-5pt}\int\limits_{\beta=0}^{\pi}\hspace{-3pt} d\beta\sin{\beta} \hspace{-5pt}\int\limits_{\gamma=0}^{2\pi}\hspace{-5pt} d\gamma~\left(\mathscr{D}_{L}^{LK}\right)^{\ast}\zeta_{1}^{l_{1}}\zeta_{2}^{l_{2}}
\end{equation}
We have evaluated the values of $C^{K}_{l_{1}\hspace{-1pt},\,l_{2}}(\theta)$ for several valid sets of $l_{1}$, $l_{2}$, and $K$ corresponding to natural parity using Eqs.~\eqref{Cl1l2Kstab} and \eqref{eqa22a}. The results are listed in  Table~\ref{tab:tab1}; it is evident that these  values  are mutually inconsistent, showing that the formula in Eq.~(A13) of \textcite{Pont1995} is incorrect. We have analyzed the possible sources of errors in the formulation given by Pont and Shakeshaft and found that the main discrepancy between the results given in Table~\ref{tab:tab1} originates from the missing\footnote{For a proper placement of the missing symbol, see Eq.~(\ref{eq:nllk})} square root sign in Eq.~(A14) of \textcite{Pont1995}.  Correcting for this error brings the results given in both columns of Table~\ref{tab:tab1} almost to an agreement, but in several cases a spurious factor of $\sqrt{2}$ remains for some of the parameter sets; the origin of this factor remains unclear. A similar validation for unnatural parity has not been attempted here. 

\section{Explicit form of the angular integral \texorpdfstring{$\mathscr{W}_{ll\hspace{-0.3pt}'\hspace{-0.3pt}}^{L\uppi}(r_{1},r_{2},\theta)$}~}\label{sec:angint}

The angular integrals $\mathscr{W}_{ll\hspace{-0.3pt}'\hspace{-0.3pt}}^{L\uppi}\equiv \mathscr{W}_{ll\hspace{-0.3pt}'\hspace{-0.3pt}}^{L\uppi}(r_{1},r_{2},\theta)$ are implicitly defined by Eq.~(123) of \textbf{1} as 
\begin{equation}\label{wlpikk}
    \mathscr{W}_{ll\hspace{-0.3pt}'\hspace{-0.3pt}}^{L\uppi} = r_{1}^{l+l\hspace{-0.3pt}'\hspace{-0.3pt}}r_{2}^{2L-l-l\hspace{-0.3pt}'\hspace{-0.3pt}\!+2d}\mathcal{F}_{ll\hspace{-0.3pt}'\hspace{-0.3pt}}^{L\uppi}(\theta)
\end{equation}
where
\begin{equation}\label{flpillp}
    \mathcal{F}_{ll\hspace{-0.3pt}'\hspace{-0.3pt}}^{L\uppi}(\theta)=\left\langle\left.\Omega_{l\hspace{-0.3pt}'\hspace{-0.3pt}}^{LL\uppi}\right|\Omega_{l}^{LL\uppi}\right\rangle
\end{equation}
is evaluated by expanding $\Omega_{\lambda}^{LL\uppi}\equiv\Omega_{\lambda}^{LL\uppi}(\theta_{1},\phi_{1},\theta_{2},\phi_{2})$ with $\lambda=l,l'$ in terms of the Wigner functions $\mathcal{D}_{L}^{LK}(\alpha,\beta,\gamma)$ following Eq.~(74) of \textbf{1}. 
Using the orthonormality of the Wigner functions $\mathcal{D}_{L}^{MK}(\alpha,\beta,\gamma)$ (see Eq.~(39) of \textbf{1}) it follows that
\begin{equation}\label{flpillp1}
    \mathcal{F}_{ll\hspace{-0.3pt}'\hspace{-0.3pt}}^{L\uppi}(\theta)=\sum_{K=-L}^{L}\left(\Lambda_{Kl\hspace{-0.3pt}'\hspace{-0.3pt}}^{L\uppi}\right)^{\!\ast}\Lambda_{Kl}^{L\uppi}
\end{equation}

Substituting the explicit form of $\Lambda_{Kl}^{L\uppi}$, given by Eq.~\eqref{ECpi1}, into Eq.~\eqref{flpillp1} and introducing a new summation index $\kappa=\tfrac{L-K+d}{2}$ (which accounts for the fact that $\Lambda_{Kl}^{L\uppi}=0$ for odd values of $L-K+d$), the expression for $\mathcal{F}_{ll\hspace{-0.3pt}'\hspace{-0.3pt}}^{L\uppi}(\theta)$ in Eq.~\eqref{flpillp1} can be rewritten as
\begin{equation}\label{flpillp2}
    \mathcal{F}_{ll\hspace{-0.3pt}'\hspace{-0.3pt}}^{L\uppi}(\theta)=\sum_{\kappa=d}^{L}\left(\overline{\Lambda}_{L-2\kappa+d,l\hspace{-0.3pt}'\hspace{-0.3pt}}^{L\uppi}\right)^{\hspace{-2pt}\ast}\,\overline{\Lambda}_{L-2\kappa+d,l}^{L\uppi}
\end{equation}
where the explicit form of $\overline{\Lambda}_{L-2\kappa+d,\lambda}^{L\uppi}$ with $\lambda=l,l'$ follows from Eq.~(79) of \textbf{1}. Substituting it into Eq.~\eqref{flpillp2} and using the explicit form of $\mathscr{N}^{L,L-2\kappa+d}_{L}$ from Eq.~\eqref{eq:nllk}, the following representation of $\mathcal{F}_{ll\hspace{-0.3pt}'\hspace{-0.3pt}}^{L\uppi}(\theta)$ can be derived
\begin{equation}\label{flpillp3}
    \mathcal{F}_{ll\hspace{-0.3pt}'\hspace{-0.3pt}}^{L\uppi}(\theta)\hspace{-2pt}=\hspace{-2pt}\tfrac{8\pi^2(2L+1)}{(2l+1+d)^2}N_{l\hspace{-0.3pt}'}^{L\uppi}N_{l}^{L\uppi}\hspace{-2pt}\sum_{\kappa=d}^{L}\sum_{\substack{\eta\hspace{-0.3pt}'\hspace{-0.3pt},\zeta\hspace{-0.3pt}'\hspace{-2.3pt},\\\eta,\zeta=0}}^{d}\hspace{-2pt}\mathcal{T}_{ll\hspace{-0.3pt}'\hspace{-0.3pt}\eta\eta\hspace{-0.3pt}'\hspace{-0.3pt}\zeta\zeta\hspace{-0.3pt}'\hspace{-0.3pt}}^{L\kappa \uppi}(\theta)
\end{equation}
with
\begin{eqnarray}\nonumber
  \mathcal{T}_{ll\hspace{-0.3pt}'\hspace{-0.3pt}\eta\eta\hspace{-0.3pt}'\hspace{-0.3pt}\zeta\zeta\hspace{-0.3pt}'\hspace{-0.3pt}}^{L\kappa \uppi}(\theta) &\!=\!& (-1)^{\zeta\hspace{-0.3pt}'\hspace{-0.3pt}+\zeta-\eta\hspace{-0.3pt}'\hspace{-0.3pt}-\eta} \,e^{\textit{\textbf{i}}\left(l\hspace{-0.3pt}'\hspace{-0.3pt}-l+\eta-\eta\hspace{-0.3pt}'\hspace{-0.3pt}+(-1)^\zeta\hspace{-0.3pt}'\hspace{-0.3pt}\eta\hspace{-0.3pt}'\hspace{-0.3pt}-(-1)^\zeta\eta\right)\theta}\\[5pt]
    &&\hspace{-45pt}\tfrac{\left(\hspace{-4pt}\begin{array}{c}
         \scriptstyle 2L \\
         \scriptstyle 2\kappa-d 
    \end{array}\hspace{-4pt}\right)}{\left(\hspace{-4pt}\begin{array}{c}
         \scriptstyle 2L+d \\
         \scriptstyle 2\kappa-\eta\hspace{-0.3pt}' 
    \end{array}\hspace{-4pt}\right)\left(\hspace{-4pt}\begin{array}{c}
         \scriptstyle 2L+d \\
         \scriptstyle 2\kappa -\eta
    \end{array}\hspace{-4pt}\right)}  \,\,\texttt{G}^{L,l\hspace{-0.3pt}'\hspace{-0.3pt}-\zeta\hspace{-0.3pt}'\hspace{-0.3pt}}_{\kappa-\eta\hspace{-0.3pt}'\hspace{-0.3pt}}\hspace{-2pt}\left(e^{-\textit{\textbf{i}}2\theta}\right)\,\texttt{G}^{L,l-\zeta}_{\kappa-\eta}\hspace{-2pt}\left(e^{+\textit{\textbf{i}}2\theta}\right)\label{coeft}
\end{eqnarray}
where the function $\texttt{G}_{\alpha}^{L\lambda}\left(z\right)$ is defined as
\begin{equation}\label{tg}
    \texttt{G}_{\alpha}^{L\lambda}\left(z\right)=\sum_{\texttt{j}=0}^{\alpha}\left(\hspace{-4pt}\begin{array}{c}
         \scriptstyle \lambda  \\
         \scriptstyle \texttt{j} 
    \end{array}\hspace{-4pt}\right)\hspace{-4pt}\left(\hspace{-4pt}\begin{array}{c}
         \scriptstyle L-\lambda  \\
         \scriptstyle \alpha-\texttt{j} 
    \end{array}\hspace{-4pt}\right)z^{\texttt{j}}
\end{equation}

We are going to show now that the function $\texttt{G}_{\alpha}^{L\lambda}\left(z\right)$ obeys the following symmetry relation
\begin{equation}\label{symrel1}
    \texttt{G}_{\alpha}^{L\lambda}\left(z\right)=z^{\lambda}\,\,\texttt{G}_{L-\alpha}^{L\lambda}\left(z^{-1}\right)
\end{equation}
a property\footnote{A careful reader, interested in the origin of such a property, might be willing to learn that $\texttt{G}_{\alpha}^{L\lambda}\left(z\right)$ can also be interpreted as the coefficient of $t^{\alpha}$ in the polynomial
\begin{equation*}
    \text{F}^{L}_{\lambda}\left(t;z\right)=\left(1+t\right)^{L-\lambda}\left(1+zt\right)^{\lambda}
\end{equation*}
of order $L$, which can be expressed using the standard notation as
\begin{equation*}
    \texttt{G}_{\alpha}^{L\lambda}\left(z\right)\equiv\left[t^{\alpha}\right]\text{F}^{L}_{\lambda}\left(t;z\right)
\end{equation*}
According to the reverse polynomial identity [consult Section 7.2 of \textcite{Brualdi2010}] the following identity holds
\begin{equation*}
    \left[t^{\alpha}\right]\text{F}^{L}_{\lambda}\left(t;z\right)=\left[t^{L-\alpha}\right]t^{L}\,\text{F}^{L}_{\lambda}\left(\tfrac{1}{t};z\right)
\end{equation*}
which upon expansion and simplification leads to Eq.~(\ref{symrel1}).} which is substantial for further simplification of Eq.~(\ref{coeft}).
We start the demonstration of the symmetry relation by showing that not every value of $\texttt{j}$ contributes to the sum in Eq.~\eqref{tg}. Since $d\leq \kappa,l,l\hspace{-0.3pt}'\hspace{-0.3pt}\leq L$ and $0\leq \eta,\eta\hspace{-0.3pt}'\hspace{-1.3pt},\zeta,\zeta\hspace{-0.3pt}'\hspace{-0.3pt}\leq d$, all the permissible values of $\alpha\in\{\kappa-\eta\hspace{-0.3pt}'\hspace{-0.3pt},\kappa-\eta\}$ and $\lambda\in \{l\hspace{-0.3pt}'\hspace{-0.3pt}-\zeta\hspace{-0.3pt}'\hspace{-0.3pt},l-\zeta\}$ satisfy
\begin{equation*}
    0\leq \lambda,\alpha\leq L
\end{equation*}
which signifies that the upper parameters in the binomial coefficients in Eq.~\eqref{tg} are always non-negative integers. Consequently, the binomial coefficients appearing in Eq.~\eqref{tg} do not vanish only within the following ranges of $\texttt{j}$
\begin{equation}
   \label{jrange}
   0 \leq \texttt{j} \leq \lambda \hspace{10pt} \text{ and }  \hspace{10pt} \lambda+\alpha-L \leq \texttt{j} \leq \alpha
\end{equation}
By combining these two constraints, the summation range in Eq.~\eqref{tg} can be modified to encompass only non-zero contributions, giving 
\begin{equation}\label{tg1}
    \texttt{G}_{\alpha}^{L\lambda}\left(z\right)=\sum_{\texttt{j}=\text{max}(0,\lambda+\alpha-L)}^{\text{min}(\lambda,\alpha)}\left(\hspace{-4pt}\begin{array}{c}
         \scriptstyle \lambda  \\
         \scriptstyle \texttt{j} 
    \end{array}\hspace{-4pt}\right)\hspace{-4pt}\left(\hspace{-4pt}\begin{array}{c}
         \scriptstyle L-\lambda  \\
         \scriptstyle \alpha-\texttt{j} 
    \end{array}\hspace{-4pt}\right)z^{\texttt{j}}
\end{equation}

Let us now consider the expression $z^{\lambda}\,\,\texttt{G}_{L-\alpha}^{L\lambda}\left(z^{-1}\right)$ appearing in the RHS of Eq.~(\ref{symrel1}); by virtue of Eq.~(\ref{tg}) we have
\begin{equation}\label{tg2a}
    z^{\lambda}\,\,\texttt{G}_{L-\alpha}^{L\lambda}\left(z^{-1}\right) = \sum_{\texttt{j}'=0}^{L-\alpha}\left(\hspace{-4pt}\begin{array}{c}
         \scriptstyle \lambda  \\
         \scriptstyle \texttt{j}' 
    \end{array}\hspace{-4pt}\right)\hspace{-4pt}\left(\hspace{-4pt}\begin{array}{c}
         \scriptstyle L-\lambda  \\
         \scriptstyle L-\alpha-\texttt{j}' 
    \end{array}\hspace{-4pt}\right)z^{\lambda-\texttt{j}'}
\end{equation}
Introducing a new summation index $\texttt{j}= \lambda-\texttt{j}'$ and using a binomial identity
\begin{equation}\label{binomiden}
    \left(\hspace{-4pt}\begin{array}{c}
         \scriptstyle n  \\
         \scriptstyle m 
    \end{array}\hspace{-4pt}\right)=\left(\hspace{-4pt}\begin{array}{c}
         \scriptstyle n  \\
         \scriptstyle n-m 
    \end{array}\hspace{-4pt}\right)
\end{equation}
for both binomial coefficients bring us to the following formula
\begin{equation}\label{tg4a}
    z^{\lambda}\,\,\texttt{G}_{L-\alpha}^{L\lambda}\left(z^{-1}\right) = \sum_{\texttt{j}=\alpha+\lambda-L}^{\lambda}\left(\hspace{-4pt}\begin{array}{c}
         \scriptstyle \lambda  \\
         \scriptstyle \texttt{j} 
    \end{array}\hspace{-4pt}\right)\hspace{-4pt}\left(\hspace{-4pt}\begin{array}{c}
         \scriptstyle L-\lambda  \\
         \scriptstyle \alpha-\texttt{j} 
    \end{array}\hspace{-4pt}\right)z^{\texttt{j}}
\end{equation}
The binomial coefficients appearing in Eq.~\eqref{tg4a} do not vanish only within the range of $\texttt{j}$ implied by the inequalities specified by Eq.~(\ref{jrange}), giving
\begin{equation}\label{tg1a}
    z^{\lambda}\,\,\texttt{G}_{L-\alpha}^{L\lambda}\left(z^{-1}\right) =\sum_{\texttt{j}=\text{max}(0,\lambda+\alpha-L)}^{\text{min}(\lambda,\alpha)}\left(\hspace{-4pt}\begin{array}{c}
         \scriptstyle \lambda  \\
         \scriptstyle \texttt{j} 
    \end{array}\hspace{-4pt}\right)\hspace{-4pt}\left(\hspace{-4pt}\begin{array}{c}
         \scriptstyle L-\lambda  \\
         \scriptstyle \alpha-\texttt{j} 
    \end{array}\hspace{-4pt}\right)z^{\texttt{j}}
\end{equation}
The RHSs of Eqs.~(\ref{tg1}) and~(\ref{tg1a}) are identical, which proves the symmetry relation in Eq.~(\ref{symrel1}).

Using the symmetry relation given in Eq.~\eqref{symrel1} and the binomial identity in Eq.~\eqref{binomiden} one finds that
\begin{equation}\label{coeft1}
    \mathcal{T}_{ll\hspace{-0.3pt}'\hspace{-0.3pt}\eta\eta\hspace{-0.3pt}'\hspace{-0.3pt}\zeta\zeta\hspace{-0.3pt}'\hspace{-0.3pt}}^{L\kappa \uppi}(\theta) = e^{\textit{\textbf{i}}\upepsilon\theta}\left(\mathcal{T}_{ll\hspace{-0.3pt}'\hspace{-0.3pt},d-\eta,d-\eta\hspace{-0.3pt}'\hspace{-0.3pt},\zeta\zeta\hspace{-0.3pt}'\hspace{-0.3pt}}^{L,L-\kappa+d, \uppi}(\theta)\right)^\ast
\end{equation}
with
\begin{equation}\label{kvalue}
    \upepsilon = 2(\zeta\hspace{-0.3pt}'\hspace{-0.3pt}-\zeta)-(-1)^{\zeta}d+(-1)^{\zeta\hspace{-0.3pt}'\hspace{-0.3pt}}d
\end{equation}
However, for all the values of the parameters $d$, $\zeta$, and $\zeta\hspace{-0.3pt}'\hspace{-0.3pt}$ associated with Eq.~\eqref{flpillp3}, we have
\begin{equation}
   \upepsilon = 0 \hspace{10pt} \text{ for }  \hspace{10pt} d\in\{0,1\} \text{ and } \zeta,\zeta\hspace{-0.3pt}'\hspace{-0.3pt} \in \{0,d\}
\end{equation}
which implies $e^{\textit{\textbf{i}}\upepsilon\theta}=1$ and consequently 
\begin{equation}\label{coeft2}
    \mathcal{T}_{ll\hspace{-0.3pt}'\hspace{-0.3pt}\eta\eta\hspace{-0.3pt}'\hspace{-0.3pt}\zeta\zeta\hspace{-0.3pt}'\hspace{-0.3pt}}^{L\kappa \uppi}(\theta)=\left(\mathcal{T}_{ll\hspace{-0.3pt}'\hspace{-1.3pt},d-\eta,d-\eta\hspace{-0.3pt}'\hspace{-0.3pt},\zeta\zeta\hspace{-0.3pt}'\hspace{-0.3pt}}^{L,L-\kappa+d, \uppi}(\theta)\right)^\ast
\end{equation}
This relation can be rewritten as
\begin{eqnarray}
    \hspace{-8pt}\mathfrak{Re}\left\{\mathcal{T}_{ll\hspace{-0.3pt}'\hspace{-0.3pt}\eta\eta\hspace{-0.3pt}'\hspace{-0.3pt}\zeta\zeta\hspace{-0.3pt}'\hspace{-0.3pt}}^{L\kappa \uppi}(\theta)\right\}&=&+\mathfrak{Re}\left\{\mathcal{T}_{ll\hspace{-0.3pt}'\hspace{-1.3pt},d-\eta,d-\eta\hspace{-0.3pt}'\hspace{-0.3pt},\zeta\zeta\hspace{-0.3pt}'\hspace{-0.3pt}}^{L,L-\kappa+d, \uppi}(\theta)\right\}\\
    \hspace{-8pt}\mathfrak{Im}\left\{\mathcal{T}_{ll\hspace{-0.3pt}'\hspace{-1.3pt}\eta\eta\hspace{-0.3pt}'\hspace{-0.3pt}\zeta\zeta\hspace{-0.3pt}'\hspace{-0.3pt}}^{L\kappa \uppi}(\theta)\right\}&=&-\mathfrak{Im}\left\{\mathcal{T}_{ll\hspace{-0.3pt}'\hspace{-1.3pt},d-\eta,d-\eta\hspace{-0.3pt}'\hspace{-0.3pt},\zeta\zeta\hspace{-0.3pt}'\hspace{-0.3pt}}^{L,L-\kappa+d, \uppi}(\theta)\right\}
\end{eqnarray}
Since the terms $\mathcal{T}_{ll\hspace{-0.3pt}'\hspace{-0.3pt}\eta\eta\hspace{-0.3pt}'\hspace{-0.3pt}\zeta\zeta\hspace{-0.3pt}'\hspace{-0.3pt}}^{L\kappa \uppi}(\theta)$ and $\mathcal{T}_{ll\hspace{-0.3pt}'\hspace{-1.3pt},d-\eta,d-\eta\hspace{-0.3pt}'\hspace{-0.3pt},\zeta\zeta\hspace{-0.3pt}'\hspace{-0.3pt}}^{L,L-\kappa+d, \uppi}(\theta)$ always appear in pairs in the 
quintuple summation in Eq.~\eqref{flpillp3} and since 
\begin{equation} \label{eq:cancelim} 
    \mathfrak{Im}\left\{\mathcal{T}_{ll\hspace{-0.3pt}'\hspace{-1.3pt}\eta\eta\hspace{-0.3pt}'\hspace{-0.3pt}\zeta\zeta\hspace{-0.3pt}'\hspace{-0.3pt}}^{L\kappa \uppi}(\theta) + \mathcal{T}_{ll\hspace{-0.3pt}'\hspace{-1.3pt},d-\eta,d-\eta\hspace{-0.3pt}'\hspace{-0.3pt},\zeta\zeta\hspace{-0.3pt}'\hspace{-0.3pt}}^{L,L-\kappa+d, \uppi}(\theta)\right\}=0
\end{equation}
the imaginary part of $\mathcal{F}_{ll\hspace{-0.3pt}'\hspace{-0.3pt}}^{L\uppi}(\theta)$ (stemming from the sum of the imaginary parts of all $\mathcal{T}_{ll\hspace{-0.3pt}'\hspace{-0.3pt}\eta\eta\hspace{-0.3pt}'\hspace{-0.3pt}\zeta\zeta\hspace{-0.3pt}'\hspace{-0.3pt}}^{L\kappa \uppi}(\theta)$) vanishes identically
\begin{equation}\label{rlnllpim}
     \mathfrak{Im}\left\{\mathcal{F}_{ll\hspace{-0.3pt}'\hspace{-0.3pt}}^{L\uppi}(\theta)\right\}=0
\end{equation}

Having established that  $\mathcal{F}_{ll\hspace{-0.3pt}'\hspace{-0.3pt}}^{L\uppi}(\theta)$  is real, we now rewrite Eq.~\eqref{flpillp3} in a more transparent form
\begin{eqnarray}\label{flpillp5}
    &&\mathcal{F}_{ll\hspace{-0.3pt}'\hspace{-0.3pt}}^{L\uppi}(\theta)\hspace{-2pt}=\sum_{\kappa=d}^{L}\sum_{\eta\hspace{-0.3pt}'\hspace{-0.3pt},\zeta\hspace{-0.3pt}'\hspace{-0.3pt}=0}^{d}\sum_{\eta,\zeta=0}^{d}\sum_{\texttt{j}\hspace{-0.3pt}'=0}^{\kappa-\eta\hspace{-0.3pt}'}\sum_{\texttt{j}=0}^{\kappa-\eta}\mathscr{B}_{ll\hspace{-0.3pt}'\hspace{-0.3pt}\texttt{j}\texttt{j}\hspace{-0.3pt}'\hspace{-0.3pt}}^{L\kappa \uppi}\, e^{\textit{\textbf{i}}\,\upmu\theta}\nonumber\\
    &&=\underbrace{\sum_{\kappa=d}^{L}\sum_{\eta\hspace{-0.3pt}'\hspace{-0.3pt},\zeta\hspace{-0.3pt}'\hspace{-0.3pt}=0}^{d}\sum_{\eta,\zeta=0}^{d}\sum_{\texttt{j}\hspace{-0.3pt}'=0}^{\kappa-\eta\hspace{-0.3pt}'}\sum_{\texttt{j}=0}^{\kappa-\eta}\mathscr{B}_{ll\hspace{-0.3pt}'\hspace{-0.3pt}\texttt{j}\texttt{j}\hspace{-0.3pt}'\hspace{-0.3pt}}^{L\kappa \uppi}\, \cos(\upmu\,\theta)\nonumber}_{\mathfrak{Re}\left\{\mathcal{F}_{ll\hspace{-0.3pt}'\hspace{-0.3pt}}^{L\uppi}(\theta)\right\}}\nonumber\\
    &&+\textit{\textbf{i}}\underbrace{\sum_{\kappa=d}^{L}\sum_{\eta\hspace{-0.3pt}'\hspace{-0.3pt},\zeta\hspace{-0.3pt}'\hspace{-0.3pt}=0}^{d}\sum_{\eta,\zeta=0}^{d}\sum_{\texttt{j}\hspace{-0.3pt}'=0}^{\kappa-\eta\hspace{-0.3pt}'}\sum_{\texttt{j}=0}^{\kappa-\eta}\mathscr{B}_{ll\hspace{-0.3pt}'\hspace{-0.3pt}\texttt{j}\texttt{j}\hspace{-0.3pt}'\hspace{-0.3pt}}^{L\kappa \uppi}\, \sin(\upmu\,\theta)}_{\mathfrak{Im}\left\{\mathcal{F}_{ll\hspace{-0.3pt}'\hspace{-0.3pt}}^{L\uppi}(\theta)\right\}}
\end{eqnarray}
in which the real and imaginary components have been separated, and  where
\begin{eqnarray}\label{blpillpjjp}
    \mathscr{B}_{ll\hspace{-0.3pt}'\hspace{-0.3pt}\texttt{j}\texttt{j}\hspace{-0.3pt}'}^{L\kappa \uppi}\hspace{-2pt}&=&(-1)^{\zeta\hspace{-0.3pt}'\hspace{-0.3pt}+\zeta-\eta\hspace{-0.3pt}'\hspace{-0.3pt}-\eta}\hspace{2pt}\frac{8\pi^2(2L+1)}{(2l+1+d)^2}\hspace{2pt}\frac{N_{l\hspace{-0.3pt}'}^{L\uppi}N_{l}^{L\uppi}\left(\hspace{-4pt}\begin{array}{c}
         \scriptstyle 2L \\
         \scriptstyle 2\kappa-d 
    \end{array}\hspace{-4pt}\right)}{\left(\hspace{-4pt}\begin{array}{c}
         \scriptstyle 2L+d \\
         \scriptstyle 2\kappa-\eta\hspace{-0.3pt}' 
    \end{array}\hspace{-4pt}\right)\left(\hspace{-4pt}\begin{array}{c}
         \scriptstyle 2L+d \\
         \scriptstyle 2\kappa -\eta
    \end{array}\hspace{-4pt}\right)}\nonumber\\
    &&\left(\hspace{-4pt}\begin{array}{c}
         \scriptstyle l\hspace{-0.3pt}'-\zeta\hspace{-0.3pt}' \\
         \scriptstyle \texttt{j}\hspace{-0.3pt}'
    \end{array}\hspace{-4pt}\right)\hspace{-2pt}\left(\hspace{-4pt}\begin{array}{c}
         \scriptstyle L-l\hspace{-0.3pt}'+\zeta\hspace{-0.3pt}' \\
         \scriptstyle \kappa-\eta\hspace{-0.3pt}'-\texttt{j}\hspace{-0.3pt}' 
    \end{array}\hspace{-4pt}\right)\hspace{-2pt}\left(\hspace{-4pt}\begin{array}{c}
         \scriptstyle l-\zeta \\
         \scriptstyle \texttt{j} 
    \end{array}\hspace{-4pt}\right)\hspace{-2pt}\left(\hspace{-4pt}\begin{array}{c}
         \scriptstyle L-l+\zeta \\
         \scriptstyle \kappa-\eta-\texttt{j} 
    \end{array}\hspace{-4pt}\right)
\end{eqnarray}
and
\begin{equation} \label{defM}
    \upmu=  l\hspace{-0.3pt}'\hspace{-0.3pt}\!-\eta\hspace{-0.3pt}'\!+(-1)^{\zeta\hspace{-0.3pt}'}\eta\hspace{-0.3pt}'\!-l\!+\eta\!-(-1)^{\zeta}\eta\!+\!2\texttt{j}\!-\!2\texttt{j}\hspace{-0.3pt}'\hspace{-0.3pt}
\end{equation}
As a consequence of Eq.~\eqref{rlnllpim}, it follows that Eq.~\eqref{flpillp5} can be rewritten as
\begin{equation}\label{flpillp6}
    \mathcal{F}_{ll\hspace{-0.3pt}'\hspace{-0.3pt}}^{L\uppi}(\theta)\hspace{-2pt}=\hspace{-2pt}\sum_{\kappa=d}^{L}\sum_{\substack{\eta\hspace{-0.3pt}'\hspace{-0.3pt},\zeta\hspace{-0.3pt}'\hspace{-2.3pt},\\\eta,\zeta=0}}^{d}\sum_{\texttt{j}\hspace{-0.3pt}'=0}^{\kappa-\eta\hspace{-0.3pt}'}\sum_{\texttt{j}=0}^{\kappa-\eta}\mathcal{B}_{ll\hspace{-0.3pt}'\hspace{-0.3pt}\texttt{j}\texttt{j}\hspace{-0.3pt}'\hspace{-0.3pt}}^{L\kappa \uppi}\, \cos(\upmu\,\theta)
\end{equation}
Introducing the explicit form of $N_{\lambda}^{L\uppi}$ (for $\lambda=l,l\hspace{-0.3pt}'$) given by Eq.~(84) of \textbf{1}, and performing the sums over $\eta$, $\zeta$, $\eta\hspace{-0.3pt}'\hspace{-0.3pt}$, and $\zeta\hspace{-0.3pt}'\hspace{-0.3pt}$, Eq.~\eqref{flpillp6} can be expressed in a more compact form 
\begin{equation}\label{flpillp7}
    \mathcal{F}_{ll\hspace{-0.3pt}'\hspace{-0.3pt}}^{L\uppi}(\theta)=\sum_{\kappa=d}^{L}\,\sum_{\texttt{j},\texttt{j}\hspace{-0.3pt}'=0}^{\kappa}\mathcal{C}_{ll\hspace{-0.3pt}'\hspace{-0.3pt}\texttt{j}\texttt{j}\hspace{-0.3pt}'\hspace{-0.3pt}}^{L\kappa \uppi}\, \cos(\texttt{m}\,\theta)
\end{equation}
with the new notation
\begin{equation} \label{defm}
    \texttt{m}=  \left|l\hspace{-0.3pt}'\hspace{-0.3pt}\!-l\!+\!2\texttt{j}\!-\!2\texttt{j}\hspace{-0.3pt}'\hspace{-0.3pt}\right|
\end{equation}
Note that the absolute value is introduced for convenience of further developments (see Eq.~(\ref{eq:tm})); this can be done because $\cos$ is an even function. The numerical coefficients $\mathcal{C}_{ll\hspace{-0.3pt}'\hspace{-0.3pt}\texttt{j}\texttt{j}\hspace{-0.3pt}'\hspace{-0.3pt}}^{L\kappa \uppi}$ in Eq.~(\ref{flpillp7}) are given by
\begin{equation}\label{clnllpjjp}
    \mathcal{C}_{ll\hspace{-0.3pt}'\hspace{-0.3pt}\texttt{j}\texttt{j}\hspace{-0.3pt}'}^{L\kappa \text{n}}\hspace{-2pt}=\hspace{-2pt}\left[\hspace{-2pt}\tfrac{\left(\!\frac{3}{2}\!\right)_{\!l}\hspace{-2pt}\left(\!\frac{3}{2}\!\right)_{\!L-l}\hspace{-2pt}\left(\!\frac{3}{2}\!\right)_{\!l\hspace{-0.3pt}'}\hspace{-2pt}\left(\!\frac{3}{2}\!\right)_{\!L-l\hspace{-0.3pt}'}}{l!\,\left(\!L-l\!\right)!\,l\hspace{-0.3pt}'!\,\left(\!L-l\hspace{-0.3pt}'\!\right)!}\hspace{-2pt}\right]^{\!\frac{1}{2}}\hspace{-4pt} \tfrac{\left(\hspace{-4pt}\begin{array}{c}
         \scriptstyle l \\
         \scriptstyle \texttt{j} 
    \end{array}\hspace{-4pt}\right)\hspace{-2pt}\left(\hspace{-4pt}\begin{array}{c}
         \scriptstyle L\hspace{0pt}-l \\
         \scriptstyle \kappa\hspace{0pt}-\texttt{j} 
    \end{array}\hspace{-4pt}\right)\hspace{-2pt}\left(\hspace{-4pt}\begin{array}{c}
         \scriptstyle l\hspace{-0.3pt}'\hspace{-0.3pt} \\
         \scriptstyle \texttt{j}\hspace{-0.3pt}'\hspace{-0.3pt} 
    \end{array}\hspace{-4pt}\right)\hspace{-2pt}\left(\hspace{-4pt}\begin{array}{c}
         \scriptstyle L\hspace{0pt}-l\hspace{-0.3pt}'\hspace{-0.3pt} \\
         \scriptstyle \kappa\hspace{0pt}-\texttt{j}\hspace{-0.3pt}'\hspace{-0.3pt} 
    \end{array}\hspace{-4pt}\right)}{2\left(\!2L+\!1\!\right)\left(\hspace{-4pt}\begin{array}{c}
         \scriptstyle 2L \\
         \scriptstyle 2\kappa 
    \end{array}\hspace{-4pt}\right)}
\end{equation}
for natural parity (i.e., $d=0$ and $\uppi=\text{n}$) and by
\begin{eqnarray}\label{clullpjjp}
    \mathcal{C}_{ll\hspace{-0.3pt}'\hspace{-0.3pt}\texttt{j}\texttt{j}\hspace{-0.3pt}'\hspace{-0.3pt}}^{L\kappa \text{u}}\hspace{-2pt}&=&\hspace{-2pt}\left[\hspace{-2pt}\tfrac{\left(\!\frac{3}{2}\!\right)_{\!l}\hspace{-2pt}\left(\!\frac{3}{2}\!\right)_{\!L-l+1}\hspace{-2pt}\left(\!\frac{3}{2}\!\right)_{\!l\hspace{-0.3pt}'}\hspace{-2pt}\left(\!\frac{3}{2}\!\right)_{\!L-l\hspace{-0.3pt}'+1}}{\left(\!l-1\!\right)!\,\left(\!L-l\!\right)!\,\left(\!l\hspace{-0.3pt}'\hspace{-0.3pt}-1\!\right)!\,\left(\!L-l\hspace{-0.3pt}'\!\right)!}\hspace{-2pt}\right]^{\!\frac{1}{2}}\hspace{-2pt}\tfrac{\left(\hspace{-4pt}\begin{array}{c}
         \scriptstyle l \\
         \scriptstyle \texttt{j} 
    \end{array}\hspace{-4pt}\right)\hspace{-2pt}\left(\hspace{-4pt}\begin{array}{c}
         \scriptstyle L\hspace{0pt}-l+1 \\
         \scriptstyle \kappa\hspace{0pt}-\texttt{j} 
    \end{array}\hspace{-4pt}\right)\hspace{-2pt}\left(\hspace{-4pt}\begin{array}{c}
         \scriptstyle l\hspace{-0.3pt}'\hspace{-0.3pt} \\
         \scriptstyle \texttt{j}\hspace{-0.3pt}'\hspace{-0.3pt} 
    \end{array}\hspace{-4pt}\right)\hspace{-2pt}\left(\hspace{-4pt}\begin{array}{c}
         \scriptstyle L\hspace{0pt}-l\hspace{-0.3pt}'\hspace{-0.3pt}+1 \\
         \scriptstyle \kappa\hspace{0pt}-\texttt{j}\hspace{-0.3pt}'\hspace{-0.3pt} 
    \end{array}\hspace{-4pt}\right)}{\left(\!L+\!1\!\right)^2\left(\hspace{-4pt}\begin{array}{c}
         \scriptstyle 2L+1 \\
         \scriptstyle 2\kappa-1 
    \end{array}\hspace{-4pt}\right)}\nonumber\\
    && \tfrac{\left(\!L\texttt{j}\vphantom{\hspace{-0.3pt}'\hspace{-0.3pt}}-\kappa l+\texttt{j}\!\right)\left(\!L\texttt{j}\hspace{-0.3pt}'\hspace{-0.3pt}-\kappa l\hspace{-0.3pt}'\hspace{-0.3pt}+\texttt{j}\hspace{-0.3pt}'\hspace{-0.3pt}\!\right)}{ll\hspace{-0.3pt}'\hspace{-0.3pt}\left(L-l+1\right)\left(L-l\hspace{-0.3pt}'\hspace{-0.3pt}+1\right)}
\end{eqnarray}
for unnatural parity (i.e., $d=1$ and $\uppi=\text{u}$). 
Eq.~\eqref{flpillp7} can be rewritten further in terms of Chebyshev polynomials of the first kind 
\begin{equation} \label{eq:tm}
    T_{\texttt{m}}(\cos{\theta})=\cos(\texttt{m}\,\theta)
\end{equation} as
\begin{equation}\label{flpillp8}
    \mathcal{F}_{ll\hspace{-0.3pt}'\hspace{-0.3pt}}^{L\uppi}(\theta)\hspace{-2pt}=\sum_{\kappa=d}^{L}\sum_{\texttt{j},\texttt{j}\hspace{-0.3pt}'=0}^{\kappa}\mathcal{C}_{ll\hspace{-0.3pt}'\hspace{-0.3pt}\texttt{j}\texttt{j}\hspace{-0.3pt}'\hspace{-0.3pt}}^{L\kappa \uppi}\hspace{2pt}T_{\texttt{m}}(\cos{\theta})
\end{equation}

Substituting Eq.~\eqref{flpillp8} into Eq.~\eqref{wlpikk} the explicit form of the angular integrals $\mathscr{W}_{ll\hspace{-0.3pt}'\hspace{-0.3pt}}^{L\uppi}$ can be expressed as
\begin{equation}\label{wlpikk1}
    \mathscr{W}_{ll\hspace{-0.3pt}'\hspace{-0.3pt}}^{L\uppi} = r_{1}^{l+l\hspace{-0.3pt}'\hspace{-0.3pt}}r_{2}^{2L-l-l\hspace{-0.3pt}'\hspace{-0.3pt}\!+2d}\sum_{\kappa=d}^{L}\sum_{\texttt{j},\texttt{j}\hspace{-0.3pt}'=0}^{\kappa}\mathcal{C}_{ll\hspace{-0.3pt}'\hspace{-0.3pt}\texttt{j}\texttt{j}\hspace{-0.3pt}'\hspace{-0.3pt}}^{L\uppi}\hspace{2pt}T_{\texttt{m}}(\cos{\theta})
\end{equation}
where the numerical coefficients $\mathcal{C}_{ll\hspace{-0.3pt}'\hspace{-0.3pt}\texttt{j}\texttt{j}\hspace{-0.3pt}'\hspace{-0.3pt}}^{L\uppi}$ are defined in Eqs.~\eqref{clnllpjjp} and~\eqref{clullpjjp} for natural and unnatural parity, respectively and $\texttt{m}$ is defined in Eq.~\eqref{defm}. 

\section{Correspondence between the present approach and the classical results of Drake and Calais-L\"{o}wdin}

It is important to establish the correspondence between the computational framework used in our paper for numerical calculations and the scheme developed by \textcite{Drake1978} building upon the earlier work of \textcite{Calais1962} (hereafter referred to as the \texttt{DCL} scheme), which is frequently adopted in the literature for practical computations \cite{Drake1979, Drake1988a, Drake1988, Drake1990, Yan1996, Goldman1997, Yan1997, Liu1997, Lee2003, Harris2005, Harris2005a, Drake2005, Kar2009, Kar2009a, Kar2009b, Kar2009c, Kar2010, Kar2008, Laughlin2009, Kar2012, Kar2013, Zhong2013, Kruppa2014, Jiao2014, Frolov2015, Jiao2015, Kar2015, Jiang2015, Kar2015a, Aznabayev2015, Zalialiutdinov2016, HU2016, Zalialiutdinov2017, Aznabaev2018, Jiao2019, Zhang2019, Korobov2020, Sims2020, Korobov2022, Jiao2022, Drake2023, Qi2024}. The \texttt{DCL} scheme  relies on expanding the bra, the ket, and all the relevant operators as symmetry-adapted products of spherical harmonics and radial components. The application of standard angular momentum coupling techniques involving $3n$-$j$ symbols reduces the general integrals to radial integrals, evaluated either using recurrence relations [see Sec. IV of \textcite{Drake1978}] or the Calais and L\"{o}wdin integration scheme [cf.\ Eq.~(28) of \textcite{Calais1962}]. Our approach eliminates the angular degrees of freedom from the theory, leaving their remnants in the form of additional angular factors $\mathscr{W}_{ll\hspace{-0.3pt}'\hspace{-0.3pt}}^{L\uppi}$ appearing in the reduced radial Schr\"{o}dinger equation.  Therefore, it should be possible to express our angular factors $\mathscr{W}_{ll\hspace{-0.3pt}'\hspace{-0.3pt}}^{L\uppi}$ via the $3n$-$j$ symbols in the spirit of the \texttt{DCL} scheme. We attempt to find this connection in this section.

To achieve this goal, it is sufficient to consider a simple overlap integral
\begin{equation}\label{ovint}
    \mathbb{I} = \hspace{-3pt} \int \hspace{-3pt}d\mathbf{r}_{1}\hspace{-3pt}\int \hspace{-3pt} d\mathbf{r}_{2} \left(\mathscr{R}'\hspace{3pt} \Omega_{l\hspace{-0.3pt}'\hspace{-0.3pt}}^{LL\uppi}\right)^{\ast}\hspace{1pt}\left(\mathscr{R}\hspace{3pt}\Omega_{l}^{LL\uppi}\right)
\end{equation}
and evaluate it twice: once using the conventional \texttt{DCL} scheme \cite{Calais1962,Drake1978} and once in the spirit of the current work. A comparison between these results should yield a formula that expresses the angular factors $\mathscr{W}_{ll\hspace{-0.3pt}'\hspace{-0.3pt}}^{L\uppi}$ via the $3n$-$j$ symbols. In Eq.~(\ref{ovint}), $\mathscr{R}\equiv \mathscr{R}(r_{1},r_{2},\theta)$ and $\mathscr{R'}\equiv \mathscr{R'}(r_{1},r_{2},\theta)$ represent radial components and 
\begin{eqnarray}\label{eq:minbh3j}
   \Omega_{\lambda}^{LL\uppi} &\equiv& \Omega_{\lambda}^{LL\uppi}(\theta_{1},\phi_{1},\theta_{2},\phi_{2})\\[5pt] \nonumber
   & = & (-1)^{d}\hspace{1pt}\sqrt{2L+1} \\\nonumber &&\sum_{\mu=\lambda-d}^{\lambda}\hspace{-3pt}\left(\hspace{-3pt}\begin{array}{ccc}
        \scriptstyle \lambda & \hspace{-3pt}\scriptstyle L+d-\lambda & \hspace{-3pt}\scriptstyle L \\[3pt]
        \scriptstyle \mu & \hspace{-3pt}\scriptstyle L-\mu & \hspace{-3pt}\scriptstyle -L
    \end{array}\hspace{-3pt}\right) Y_{\mu}^{\lambda}\hspace{-1pt}(\theta_{1}\hspace{-0.5pt},\hspace{-1pt}\phi_{1}\hspace{-0.7pt}) \,
Y_{L\!-\!\mu}^{L\hspace{-0.6pt}+\hspace{-0.6pt}d\hspace{-0.8pt}-\hspace{-0.8pt}\lambda}\hspace{-1pt}(\theta_{2},\!\phi_{2}\hspace{-0.5pt})
\end{eqnarray}
with $\lambda=l,l'$ represents a  minimal bipolar harmonic. Eq.~(\ref{eq:minbh3j}) follows directly from Eq.~(66) of \textbf{1} and the relation between the 3-$j$
coefficients and the Clebsch-Gordan coefficients given by Eq.~8.1.2(12) of \textcite{Varshalovich1988}. 

\subsection{The conventional \texttt{DCL} scheme}\label{dscheme}
In the conventional \texttt{DCL} scheme \cite{Drake1978,Calais1962},  the integral $\mathbb{I}$ can be expressed in bispherical coordinates $(r_{1},r_{2},\theta_{1},\theta_{2},\phi_{1},\phi_{2})$ as
\begin{eqnarray}\nonumber
    \mathbb{I} &=& \int\limits_{0}^{\infty}\hspace{-3pt}r_1^2\hspace{1pt}dr_{1}\hspace{-2pt}\int\limits_{0}^{\infty}\hspace{-3pt}r_2^2\hspace{1pt}dr_{2}\hspace{-2pt}\int\limits_{0}^{\pi}\hspace{-3pt}\sin{\theta_{1}}\hspace{0.5pt} d\theta_{1}\hspace{-2pt}\int\limits_{0}^{\pi}\hspace{-3pt}\sin{\theta_{2}}\hspace{0.5pt} d\theta_{2}\hspace{-2pt}\int\limits_{0}^{2\pi}\hspace{-3pt}d\phi_{1}\hspace{-2pt}\int\limits_{0}^{2\pi} \hspace{-3pt}d\phi_{2} \\[3pt]
    \hspace{-20pt}&&\hspace{-20pt}\mathscr{R}'^{\ast}\hspace{1pt}\mathscr{R}\hspace{-4pt}\sum_{\mu\hspace{-0.3pt}'\hspace{-0.3pt}=l\hspace{-0.3pt}'\hspace{-0.3pt}-d}^{l\hspace{-0.3pt}'\hspace{-0.3pt}}\sum_{\mu=l-d}^{l}\hspace{-2pt} \left(2L+1\right) \hspace{-2pt}\left(\hspace{-5pt}\begin{array}{ccc}
        \scriptstyle l\hspace{-0.3pt}'\hspace{-0.3pt} & \hspace{-3pt}\scriptstyle L+d-l\hspace{-0.3pt}'\hspace{-0.3pt} & \hspace{-3pt}\scriptstyle L \\[3pt]
        \scriptstyle \mu\hspace{-0.3pt}'\hspace{-0.3pt} & \hspace{-3pt}\scriptstyle L-\mu\hspace{-0.3pt}'\hspace{-0.3pt} & \hspace{-3pt}\scriptstyle -L
    \end{array}\hspace{-5pt}\right)\hspace{-3pt}\left(\hspace{-5pt}\begin{array}{ccc}
        \scriptstyle l & \hspace{-3pt}\scriptstyle L+d-l & \hspace{-3pt}\scriptstyle L \\[3pt]
        \scriptstyle \mu & \hspace{-3pt}\scriptstyle L-\mu & \hspace{-3pt}\scriptstyle -L
    \end{array}\hspace{-5pt}\right) \nonumber\\[3pt]
    &&\hspace{-20pt}\left(Y_{\mu\hspace{-0.3pt}'\hspace{-0.3pt}}^{l\hspace{-0.3pt}'\hspace{-0.3pt}}\hspace{-1pt}(\hat{r}_{1}\hspace{-0.7pt})\hspace{2pt}Y_{L-\mu\hspace{-0.3pt}'\hspace{-0.3pt}}^{L+d-l\hspace{-0.3pt}'\hspace{-0.3pt}}\hspace{-1pt}(\hat{r}_{2}\hspace{-0.5pt})\right)^{\ast} Y_{\mu}^{l}\hspace{-1pt}(\hat{r}_{1}\hspace{-0.7pt})\hspace{2pt}Y_{L-\mu}^{L+d-l}\hspace{-1pt}(\hat{r}_{2}\hspace{-0.5pt})\label{ovint1}
\end{eqnarray}
By using the fact that $\left(Y_{m}^{l}\right)^{\!\ast}=(-1)^{m}\,Y_{\!-m}^{l}$ [cf. Eq.~5.1(11) of \textcite{Varshalovich1988}] and by linearizing the product of two spherical harmonics  [cf. Eq.~5.6.2(9) of \textcite{Varshalovich1988}]
\begin{eqnarray}\label{pdsp0a}
    Y_{m_{1}}^{l_{1}}\hspace{-1pt}(\hat{r})\hspace{2pt}Y_{m_{2}}^{l_{2}}\hspace{-1pt}(\hat{r})\hspace{3pt}&=& \sqrt{\tfrac{(2l_{1}+1)(2l_{2}+1)}{4\pi}}\hspace{-4pt}\sum\limits_{l_3=\left|l_1-l_2\right|}^{l_1+l_2}\hspace{2pt}\sum\limits_{m_3=-l_3}^{l_3}(-1)^{m_3}\nonumber\\[5pt]
    &&\hspace{-60pt}\sqrt{2l_3+1}\left(\!\!\begin{array}{ccc}
        \scriptstyle l_{1} & \scriptstyle l_{2} & \scriptstyle l_3\\[3pt]
        \scriptstyle 0 & \scriptstyle 0 & \scriptstyle 0 
    \end{array}\!\!\right)\hspace{-4pt}\left(\!\!\begin{array}{ccc}
        \scriptstyle l_{1} & \scriptstyle l_{2} & \scriptstyle l_3\\[3pt]
        \scriptstyle m_{1} & \scriptstyle m_{2} & \scriptstyle -m_3 
    \end{array}\!\!\right)\hspace{3pt}Y_{m_3}^{l_3}\hspace{-1pt}(\hat{r})
\end{eqnarray}
we can express  $\mathbb{I}$ as
\begin{equation}\label{ovint2}
    \mathbb{I} = \hspace{-6pt}\sum_{\mu\hspace{-0.3pt}'\hspace{-0.3pt}=l\hspace{-0.3pt}'\hspace{-0.3pt}-d}^{l\hspace{-0.3pt}'\hspace{-0.3pt}}\hspace{1pt}\sum_{\mu=l-d}^{l}\hspace{3pt}  \sum_{L_{1}=\left|l-l\hspace{-0.3pt}'\hspace{-0.3pt}\right|}^{l+l\hspace{-0.3pt}'\hspace{-0.3pt}}\sum_{L_{2}=\left|l-l\hspace{-0.3pt}'\hspace{-0.3pt}\right|}^{2L+2d-l-l\hspace{-0.3pt}'\hspace{-0.3pt}}\sum_{M_{1}=-L_1}^{L_1}\sum_{M_{2}=-L_2}^{L_2}\hspace{-3pt}\mathscr{M}\hspace{3pt}\mathbb{J}
\end{equation}
with
\begin{eqnarray}
    \mathscr{M} &=& \tfrac{(-1)^{L+M_{1}+M_{2}}}{4\pi}\left(2L+1\right)\hspace{-2pt}\left[\left(\!2l\hspace{-0.3pt}'\hspace{-0.3pt}\!+\!1\!\right)\!\left(\!2l\vphantom{'}\!+\!1\!\right)\!\left(\!2L_{1}\vphantom{'}\!\!+\!1\!\right)\right.\nonumber\\[2pt]
    &&\hspace{-20pt}\left.\left(\!2L_{2}\vphantom{'}\!\!+\!1\!\right)\!\left(\!2L\!+\!2d\!-\!2l\hspace{-0.3pt}'\hspace{-0.3pt}\!+\!1\!\right)\!\left(\!2L\!+\!2d\!-\!2l\vphantom{'}\!+\!1\!\right)\right]^{\frac{1}{2}}\nonumber\\[2pt]
    &&\hspace{-20pt}\hspace{-2pt}\left(\hspace{-5pt}\begin{array}{ccc}
        \scriptstyle l & \hspace{-3pt}\scriptstyle L+d-l & \hspace{-3pt}\scriptstyle L \\[3pt]
        \scriptstyle \mu & \hspace{-3pt} \scriptstyle L-\mu & \hspace{-3pt}\scriptstyle -L
    \end{array}\hspace{-5pt}\right)\hspace{-4pt}\left(\hspace{-5pt}\begin{array}{ccc}
        \scriptstyle l\hspace{-0.3pt}'\hspace{-0.3pt} & \hspace{-3pt}\scriptstyle L+d-l\hspace{-0.3pt}'\hspace{-0.3pt} & \hspace{-3pt}\scriptstyle L \\[3pt]
        \scriptstyle \mu\hspace{-0.3pt}'\hspace{-0.3pt} & \hspace{-3pt}\scriptstyle L-\mu\hspace{-0.3pt}'\hspace{-0.3pt} & \hspace{-3pt}\scriptstyle -L
    \end{array}\hspace{-5pt}\right)\hspace{-4pt}\left(\hspace{-4pt}\begin{array}{ccc}
        \scriptstyle l\hspace{-0.3pt}'\hspace{-0.3pt} & \scriptstyle l & \scriptstyle L_{1} \\[3pt]
        \scriptstyle 0 & \scriptstyle 0 & \scriptstyle 0
    \end{array}\hspace{-6pt}\right)\hspace{-4pt}\left(\hspace{-4pt}\begin{array}{ccc}
        \scriptstyle l\hspace{-0.3pt}'\hspace{-0.3pt} & \scriptstyle l & \scriptstyle L_{1} \\[3pt]
        \scriptstyle -\mu\hspace{-0.3pt}'\hspace{-0.3pt} & \scriptstyle \mu & \scriptstyle -M_{1}
    \end{array}\hspace{-4pt}\right)\nonumber\\[2pt]
    &&\hspace{-20pt}\hspace{-2pt}\left(\hspace{-4pt}\begin{array}{ccc}
        \scriptstyle L+d-l\hspace{-0.3pt}'\hspace{-0.3pt} & \scriptstyle L+d-l & \scriptstyle L_{2} \\[3pt]
        \scriptstyle 0 & \scriptstyle 0 & \scriptstyle 0
    \end{array}\hspace{-6pt}\right)\hspace{-4pt}\left(\hspace{-4pt}\begin{array}{ccc}
        \scriptstyle L+d-l\hspace{-0.3pt}'\hspace{-0.3pt} & \scriptstyle L+d-l & \scriptstyle L_{2} \\[3pt]
        \scriptstyle \mu\hspace{-0.3pt}'\hspace{-0.3pt}-L & \scriptstyle L-\mu & \scriptstyle -M_{2}
    \end{array}\hspace{-4pt}\right)
\end{eqnarray}
and 
\begin{eqnarray}\label{ovint3}
    \mathbb{J} &=& \int\limits_{0}^{\infty}\hspace{-3pt}r_1^2\hspace{1pt}dr_{1}\hspace{-2pt}\int\limits_{0}^{\infty}\hspace{-3pt}r_2^2\hspace{1pt}dr_{2}\hspace{-2pt}\int\limits_{0}^{\pi}\hspace{-3pt}\sin{\theta_{1}}\hspace{0.5pt} d\theta_{1}\hspace{-2pt}\int\limits_{0}^{\pi}\hspace{-3pt}\sin{\theta_{2}}\hspace{0.5pt} d\theta_{2}\hspace{-2pt}\int\limits_{0}^{2\pi}\hspace{-3pt}d\phi_{1}\hspace{-2pt}\int\limits_{0}^{2\pi} \hspace{-3pt}d\phi_{2}\nonumber\\[3pt]
    && \mathscr{R}'^{\ast}\hspace{3pt}\mathscr{R}\hspace{4pt}Y_{M_{1}}^{L_{1}}\hspace{-1pt}(\theta_{1},\phi_{1}\hspace{-0.7pt})\,Y_{M_{2}}^{L_{2}}\hspace{-1pt}(\theta_{2},\phi_{2}\hspace{-0.5pt})
\end{eqnarray}

To evaluate the integral $\mathbb{J}$, let us consider for a moment a  Cartesian integral $\mathbb{K}$ over a general integrand $K(\bm{r}_{1},\bm{r}_{2})$, given by the expression
\begin{equation}
    \mathbb{K} = \int\hspace{-3pt}d\bm{r}_{1}\hspace{-2pt}\int\hspace{-3pt}d\bm{r}_{2}\hspace{3pt} K(\bm{r}_{1},\bm{r}_{2})
\end{equation}
with $\bm{r}_{i}=\left[ x_i,y_i,z_i\right]^\mathsf{T}$.
We transform this integral to two alternative coordinate systems. First we consider a transformation to the bispherical coordinate system $(r_1,\theta_1,\phi_1,r_2,\theta_2,\phi_2)$ given by the following mapping
\begin{equation*}
    \left[ x_i,y_i,z_i\right]^\mathsf{T} \rightarrow \left[ r_i \sin\theta_i \cos\phi_i, r_i \sin\theta_i \sin\phi_i, r_i \cos\theta_i\right]^\mathsf{T}    
\end{equation*}
with $i=1,2$, which results in the following representation of $\mathbb{K} $
\begin{eqnarray}\label{ovintK1}
    \mathbb{K} &=& \int\limits_{0}^{\infty}\hspace{-3pt}r_1^2\hspace{1pt}dr_{1}\hspace{-2pt}\int\limits_{0}^{\infty}\hspace{-3pt}r_2^2\hspace{1pt}dr_{2}\hspace{-2pt}\int\limits_{0}^{\pi}\hspace{-3pt}\sin{\theta_{1}}\hspace{0.5pt} d\theta_{1}\hspace{-2pt}\int\limits_{0}^{\pi}\hspace{-3pt}\sin{\theta_{2}}\hspace{0.5pt} d\theta_{2}\hspace{-2pt}\int\limits_{0}^{2\pi}\hspace{-3pt}d\phi_{1}\hspace{-2pt}\int\limits_{0}^{2\pi} \hspace{-3pt}d\phi_{2}\nonumber\\[3pt]
    && K(r_{1},\theta_1,\phi_1,r_{2},\theta_2,\phi_2)
\end{eqnarray}
Now we consider a transformation\footnote{A similar transformation with $\mathbf{R}_{i}=\left[ r_i\cos{\frac{\theta}{2}},(-1)^i r_i\sin{\frac{\theta}{2}},0\right]^\mathsf{T}$ has been used in Eqs.~(12--17) of \textbf{1} to derive the angular momentum operator $\hat{\bm{L}}_{xyz}$ in the $(r_{1},r_{2},\theta,\alpha,\beta,\gamma)$ coordinates.} to the second (internal + Euler angles) coordinate system $(r_{1},r_{2},\theta,\alpha,\beta,\gamma)$ given by the mapping
$\mathbf{r}_{i} \rightarrow  \mathbb{R}_{\alpha\beta\gamma}\hspace{2pt} \mathbf{R}_{i}$ with $i=1,2$, where 
\begin{equation}\label{eq:defRi}
   \mathbf{R}_{1} =  \left[ 0,0,r_1\right]^\mathsf{T} \text{ and } \hspace{3pt}\mathbf{R}_{2} =  \left[ r_2\sin{\theta},0,r_2\cos{\theta}\right]^\mathsf{T}
\end{equation}
and where 
\begin{equation} \label{eq:rotmat} 
  \mathbb{R}_{\alpha\beta\gamma} \hspace{-3pt}= \hspace{-4pt}\left[\!\!\begin{array}{ccc}
       \cos{\alpha}  & \!\!-\!\sin{\alpha} & \!0\\
       \sin{\alpha} & \!\!+\!\cos{\alpha} & \!0\\
       0 & 0 & \!1
    \end{array}\!\!\right]\hspace{-5pt}\left[\!\!\begin{array}{ccc}
       \!+\!\cos{\beta}  & \hspace{-3pt}0 & \hspace{-3pt}\sin{\beta}\\
       0 & \hspace{-3pt}1 & \hspace{-3pt}0\\
       \!-\!\sin{\beta}  & \hspace{-3pt}0 & \hspace{-3pt}\cos{\beta}
    \end{array}\!\!\right]\hspace{-5pt}\left[\!\!\begin{array}{ccc}
       \cos{\gamma}  & \!\!-\!\sin{\gamma} & \!0\\
       \sin{\gamma} & \!\!+\!\cos{\gamma} & \!0\\
       0 & 0 & \!1
    \end{array}\!\!\right]
\end{equation}
which results in the following representation of $\mathbb{K} $
\begin{eqnarray}\label{ovintK2}
    \mathbb{K} &=& \int\limits_{0}^{\infty}\hspace{-3pt}r_1^2\hspace{1pt}dr_{1}\hspace{-2pt}\int\limits_{0}^{\infty}\hspace{-3pt}r_2^2\hspace{1pt}dr_{2}\hspace{-2pt}\int\limits_{0}^{\pi}\hspace{-3pt}\sin{\theta}\hspace{0.5pt} d\theta\hspace{-2pt}\int\limits_{0}^{2\pi}\hspace{-3pt}d\alpha\hspace{-2pt}\int\limits_{0}^{\pi}\hspace{-3pt}\sin{\beta}\hspace{0.5pt} d\beta\hspace{-2pt}\int\limits_{0}^{2\pi} \hspace{-3pt}d\gamma\nonumber\\[3pt]
    && K(r_{1},r_{2},\theta,\alpha,\beta,\gamma)
\end{eqnarray}
The fact that both integrals in Eqs.~(\ref{ovintK1}) and~(\ref{ovintK2}) are equal to $\mathbb{K}$, allows us to rewrite the integral $\mathbb{J}$ in Eq.~(\ref{ovint3}) in the (internal + Euler angles) coordinate system $(r_{1},r_{2},\theta,\alpha,\beta,\gamma)$ as 
\begin{eqnarray}\label{ovintJ2}
    \mathbb{J} &=& \int\limits_{0}^{\infty}\hspace{-3pt}r_1^2\hspace{1pt}dr_{1}\hspace{-2pt}\int\limits_{0}^{\infty}\hspace{-3pt}r_2^2\hspace{1pt}dr_{2}\hspace{-2pt}\int\limits_{0}^{\pi}\hspace{-3pt}\sin{\theta}\hspace{0.5pt} d\theta\hspace{-2pt}\hspace{-2pt}\int\limits_{0}^{2\pi}\hspace{-3pt}d\alpha\int\limits_{0}^{\pi}\hspace{-3pt}\sin{\beta}\hspace{0.5pt} d\beta\hspace{-2pt}\int\limits_{0}^{2\pi} \hspace{-3pt}d\gamma\nonumber\\[3pt]
    && J(r_{1},r_{2},\theta,\alpha,\beta,\gamma)
\end{eqnarray}
where $J(r_{1},r_{2},\theta,\alpha,\beta,\gamma)$ is the representation of the integrand
\begin{equation*}
    J(r_1,\theta_1,\phi_1,r_2,\theta_2,\phi_2)
= \mathscr{R}'^{\ast}\hspace{3pt}\mathscr{R}\hspace{4pt}Y_{M_{1}}^{L_{1}}\hspace{-1pt}(\theta_{1},\phi_{1}\hspace{-0.7pt})\,Y_{M_{2}}^{L_{2}}\hspace{-1pt}(\theta_{2},\phi_{2}\hspace{-0.5pt})
\end{equation*}
in the (internal + Euler angles) coordinate system. 

The function $\mathscr{R}$ (as well as $\mathscr{R}'$) has been defined as an Euler-angle-independent expansion coefficient of the total wave function $\Psi(\bm{r}_{1},\bm{r}_{2})$ in the basis of angular generator functions (cf.\ Eq.~(1) of \textbf{1}). Consequently, $\mathscr{R}$ and $\mathscr{R}'$ depend solely on the internal variables $r_{1}$, $r_{2}$ and $\theta$. The explicit form of this dependence is irrelevant for the forthcoming analysis. 

The mapping $\mathbf{r}_{i} \rightarrow  \mathbb{R}_{\alpha\beta\gamma}\hspace{2pt} \mathbf{R}_{i}$ induces the following transformation of the spherical harmonics $Y_{M_{1}}^{L_{1}}\hspace{-1pt}(\theta_{1},\phi_{1}\hspace{-0.7pt})$ and $Y_{M_{2}}^{L_{2}}\hspace{-1pt}(\theta_{2},\phi_{2}\hspace{-0.5pt})$ to the new coordinates $(r_{1},r_{2},\theta,\alpha,\beta,\gamma)$ 
\begin{eqnarray}
    \hspace{-10pt}Y_{M_{1}}^{L_{1}}\hspace{-1pt}(\theta_{1},\phi_{1}\hspace{-0.7pt}) &=& \hspace{-8pt}\sum_{M_{1}'=-L_{1}}^{L_{1}}\hspace{-8pt}\left(\hspace{-2pt}\mathcal{D}_{L_{1}}^{M_{1}M_{1}'}\hspace{-1pt}\left(\alpha,\beta,\gamma\right)\hspace{-2pt}\right)^{\!\ast}Y_{\!M_{1}'}^{L_{1}}\hspace{-1pt}(0,0)\label{sphrf1}\\
    \hspace{-10pt}Y_{M_{2}}^{L_{2}}\hspace{-1pt}(\theta_{2},\phi_{2}\hspace{-0.5pt}) &=& \hspace{-8pt}\sum_{M_{2}'=-L_{2}}^{L_{2}}\hspace{-8pt}\left(\hspace{-2pt}\mathcal{D}_{L_{2}}^{M_{2}M_{2}'}\hspace{-1pt}\left(\alpha,\beta,\gamma\right)\hspace{-2pt}\right)^{\!\ast}Y_{\!M_{2}'}^{L_{2}}\hspace{-1pt}(\theta,0)\label{sphrf2}
\end{eqnarray}
These formulas are a consequence of the following two facts: (1)~The mapping $\mathbf{r}_{i} \rightarrow  \mathbb{R}_{\alpha\beta\gamma}\hspace{2pt} \mathbf{R}_{i}$ implicitly defines two reference frames: the laboratory-fixed frame and the body-fixed frame, in which the direction vectors of the quasiparticles are given by
\[ \hat{\bm{r}}_{i}=\left[ \begin{array}{l}
     \sin\theta_i \cos\phi_i  \\
     \sin\theta_i \sin\phi_i  \\ \cos\theta_i\end{array}\right] \text{ and } 
     \hat{\bm{R}}_{i}=\left[ \begin{array}{l}
     \sin\Theta_i \cos\Phi_i  \\
     \sin\Theta_i \sin\Phi_i  \\ \cos\Theta_i\end{array}\right]\]
respectively. 
The corresponding angular coordinates $(\theta_{i}, \phi_i)$  and $(\Theta_{i}, \Phi_i)$ serve as  arguments for the spherical harmonics in Eqs.~(\ref{sphrf1}) and~(\ref{sphrf2}). The specific values of $(\Theta_{i}, \Phi_i)$
follow from the definition of $\hat{\bm{R}}_i=\bm{R}_i/r_i$ with $\bm{R}_{i}$ given by Eq.~(\ref{eq:defRi}); we have
\begin{equation*}
   \hat{\bm{R}}_1 \hspace{-2pt}= \hspace{-2pt} \left[ \hspace{-2pt}\begin{array}{c}
    0 \\ 0 \\ 1 
   \end{array}\hspace{-2pt}\right]\hspace{-2pt}=\hspace{-2pt}\underbrace{\left[ \hspace{-2pt}\begin{array}{l}
     \sin0 \cos0  \\
     \sin0 \sin0  \\ \cos0\end{array}\hspace{-2pt}\right]}_{(\Theta_{1}=0, \Phi_1=0)} \text{, }
     \hat{\bm{R}}_2 \hspace{-2pt}=\hspace{-2pt} \left[ \hspace{-2pt}\begin{array}{c}
    \sin{\theta} \\ 0 \\ \cos{\theta} 
   \end{array}\hspace{-2pt}\right]\hspace{-2pt}=\hspace{-2pt}\underbrace{\left[\hspace{-2pt} \begin{array}{l}
     \sin{\theta} \cos0  \\
     \sin{\theta} \sin0  \\ \cos{\theta}\end{array}\hspace{-2pt}\right]}_{(\Theta_{2}=\theta, \Phi_2=0)}
\end{equation*}
(2)~The transformation of a spherical harmonic from one Cartesian reference frame to another reference frame differing only by an Euler rotation $\mathbb{R}_{\alpha\beta\gamma}$ is given by [cf.\ Eq.~4.1(5) of \textcite{Varshalovich1988} and the text therein]
\begin{equation}\label{sphrf0}
    Y_M^L\hspace{-1pt}(\theta,\phi) = \sum_{m=-L}^{L}\left(\mathcal{D}_{L}^{Mm}\hspace{-1pt}\left(\alpha,\beta,\gamma\right)\right)^{\!\ast}\hspace{2pt}Y_{m}^L\hspace{-1pt}(\Theta,\Phi)
\end{equation}
where the angular coordinates $(\theta,\phi)$ and $(\Theta,\Phi)$ correspond to the particle's direction vectors in the initial and rotated frames, respectively. Performing this transformation for both of our quasiparticles with the specific values of $(\Theta_{i}, \Phi_i)$, reproduces Eqs.~(\ref{sphrf1}) and~(\ref{sphrf2}). 

Taking into account these considerations, the integrand in Eq.~\eqref{ovintJ2} can be rewritten as
\begin{eqnarray}
    \hspace{-5pt}J(r_{1},r_{2},\theta,\alpha,\beta,\gamma) &=& \mathscr{R}'^{\ast}\hspace{3pt}\mathscr{R}\hspace{-1pt}\sum_{M_{2}',M_{1}'}\hspace{-4pt}\left(\hspace{-2pt}\mathcal{D}_{L_{1}}^{M_{1}M_{1}'}\hspace{-1pt}\left(\alpha,\beta,\gamma\right)\hspace{-2pt}\right)^{\!\ast}\nonumber\\
&&\hspace{-60pt}\left(\hspace{-2pt}\mathcal{D}_{L_{2}}^{M_{2}M_{2}'}\hspace{-1pt}\left(\alpha,\beta,\gamma\right)\hspace{-2pt}\right)^{\!\ast}Y_{\!M_{1}'}^{L_{1}}\hspace{-1pt}(0,0)\hspace{4pt}Y_{\!M_{2}'}^{L_{2}}\hspace{-1pt}(\theta,0)
\end{eqnarray}
Since [cf.\ Eq.~5.14(1) of \textcite{Varshalovich1988}]
\begin{equation}\label{sph00}
    Y_{\!M_{1}'}^{L_{1}}\hspace{-1pt}(0,0)=\sqrt{\frac{2L_{1}+1}{4\pi}}\,\delta_{M_{1\!}',0}
\end{equation}
and [cf. Eq.~4.4(2) of \textcite{Varshalovich1988}]
\begin{equation}\label{wdsymp}
    \left(\hspace{-2pt}\mathcal{D}_{L_{2}}^{M_{2}M_{2}'}\hspace{-1pt}(\alpha,\beta,\gamma)\hspace{-2pt}\right)^{\!\ast}\hspace{-2pt} = \frac{(-1)^{M_{2}'}}{(-1)^{M_{2}}}\hspace{2pt}\mathcal{D}_{L_{2}}^{-M_{2},-M_{2}'}\hspace{-1pt}(\alpha,\beta,\gamma)
\end{equation}
the integral $\mathbb{J}$ in Eq.~\eqref{ovintJ2} can be rewritten as
\begin{eqnarray}\label{ovintJ3}
    \mathbb{J} &\hspace{4pt}=\hspace{4pt}& \sqrt{\frac{2L_{1}+1}{4\pi}}\sum_{M_{2}'=-L_{2}}^{L_{2}}   (-1)^{M_{2}'-M_2}\\[3pt]
    &&\hspace{-17pt} \int\limits_{0}^{\infty}\hspace{-2pt}r_1^2\hspace{1pt}dr_{1}\hspace{-2pt}\int\limits_{0}^{\infty}\hspace{-2pt}r_2^2\hspace{1pt}dr_{2}\hspace{-2pt}\int\limits_{0}^{\pi}\hspace{-2pt}\sin{\theta}\hspace{0.5pt} d\theta \hspace{4pt} \mathscr{R}'^{\ast}\hspace{2pt}\mathscr{R} \hspace{5pt} Y_{\!M_{2}'}^{L_{2}}\hspace{-1pt}(\theta,0)\nonumber\\
    &&\hspace{-17pt}\int\limits_{0}^{2\pi} \hspace{-3pt}d\alpha\hspace{-2pt}\int\limits_{0}^{\pi}\hspace{-3pt}\sin{\beta}\hspace{0.5pt} d\beta\hspace{-2pt}\int\limits_{0}^{2\pi}\hspace{-3pt}d\gamma\left(\hspace{-2pt}\mathcal{D}_{L_{1}}^{M_{1}0}\hspace{-1pt}\left(\alpha,\beta,\gamma\right)\hspace{-2pt}\right)^{\!\ast}\mathcal{D}_{L_{2}}^{-M_{2},-M_{2}'}\hspace{-1pt}(\alpha,\beta,\gamma)\nonumber
\end{eqnarray}
where the last line simplifies further owing to the orthonormality\footnote{Note that, the Wigner functions $\mathcal{D}$ follow the normalization conventions different from those employed in \textbf{1}. Specifically, we follow the conventions outlined in \cite{Varshalovich1988}.} of Wigner functions $\mathcal{D}$  [cf.\ Eq.~4.10(5) of \textcite{Varshalovich1988}]
\begin{eqnarray}
    \int\limits_{0}^{2\pi} \hspace{-3pt}d\alpha\hspace{-2pt}\int\limits_{0}^{\pi}\hspace{-3pt}\sin{\beta}\hspace{0.5pt} d\beta\hspace{-2pt}\int\limits_{0}^{2\pi}\hspace{-3pt}d\gamma && \hspace{-10pt}\left(\hspace{-2pt}\mathcal{D}_{L_{1}}^{M_{1}0}\hspace{-1pt}\left(\alpha,\beta,\gamma\right)\hspace{-2pt}\right)^{\!\ast}\mathcal{D}_{L_{2}}^{-M_{2},-M_{2}'}\hspace{-1pt}(\alpha,\beta,\gamma)\nonumber\\
    &&\hspace{-25pt}= \delta_{L_{1},L_{2}}\hspace{3pt}\delta_{M_{1},-M_{2}}\hspace{3pt}\delta_{M_{2}',0}\hspace{3pt}\frac{8\pi^{2}}{2L_{1}+1}
\end{eqnarray}
Consequently, the integral $\mathbb{J}$ in Eq.~\eqref{ovintJ3} reduces to
\begin{eqnarray}\label{ovint5}
    \mathbb{J} &\hspace{2pt}=\hspace{2pt}& (-1)^{M_{1}}\hspace{3pt}\delta_{L_{1},L_{2}}\hspace{3pt}\delta_{M_{1},-M_{2}}\hspace{3pt}\sqrt{\frac{16\pi^3}{2L_{1}+1}}\\
    && \int\limits_{0}^{\infty}\hspace{-2pt}r_1^2\hspace{1pt}dr_{1}\hspace{-2pt}\int\limits_{0}^{\infty}\hspace{-2pt}r_2^2\hspace{1pt}dr_{2}\hspace{-2pt}\int\limits_{0}^{\pi}\hspace{-2pt}\sin{\theta}\hspace{0.5pt} d\theta \hspace{4pt} \mathscr{R}'^{\ast}\hspace{2pt}\mathscr{R} \hspace{5pt}Y_{0}^{L_{1}}\hspace{-1pt}(\theta,0)\nonumber
\end{eqnarray}
Expressing the spherical harmonic $Y_{0}^{L_{1}}\hspace{-1pt}(\theta,0)$ in terms of the Legendre polynomial $P_{L_{1}}\hspace{-2pt}\left(\cos{\theta}\right)$ [cf.\ Eq.~5.13.2(7) of \textcite{Varshalovich1988}]
\begin{equation}\label{sphl20}
    Y_{0}^{L_{1}}\hspace{-1pt}(\theta,0) = \sqrt{\frac{2L_{1}+1}{4\pi}} P_{L_{1}}\hspace{-2pt}\left(\cos{\theta}\right)
\end{equation}
the integral $\mathbb{J}$ in Eq.~\eqref{ovint5} reduces to
\begin{eqnarray}\label{ovint6}
    \mathbb{J} &\hspace{2pt}=\hspace{2pt}& 2\pi\hspace{3pt}(-1)^{M_{1}}\hspace{3pt}\delta_{L_{1},L_{2}}\hspace{3pt}\delta_{M_{1},-M_{2}}\\[3pt]
    && \int\limits_{0}^{\infty}\hspace{-2pt}r_1^2\hspace{1pt}dr_{1}\hspace{-2pt}\int\limits_{0}^{\infty}\hspace{-2pt}r_2^2\hspace{1pt}dr_{2}\hspace{-2pt}\int\limits_{0}^{\pi}\hspace{-2pt}\sin{\theta}\hspace{0.5pt} d\theta \hspace{4pt} \mathscr{R}'^{\ast}\hspace{2pt}\mathscr{R} \hspace{5pt}P_{L_{1}}\hspace{-2pt}\left(\cos{\theta}\right) \nonumber
\end{eqnarray}
Substituting $\mathbb{J}$ into Eq.~\eqref{ovint2}, the integral $\mathbb{I}$ can be expressed as
\begin{equation}\label{ovint8}
    \mathbb{I} =\hspace{-2pt}  \int\limits_{0}^{\infty}\hspace{-3pt}r_1^2\hspace{1pt}dr_{1}\hspace{-2pt}\int\limits_{0}^{\infty}\hspace{-3pt}r_2^2\hspace{1pt}dr_{2}\hspace{-3pt}\int\limits_{0}^{\pi}\hspace{-3pt}\sin{\theta}\hspace{0.5pt} d\theta \hspace{2pt}\mathscr{R}'^{\ast}\hspace{2pt}\mathscr{R}\hspace{2pt}F_{\hspace{-1pt}ll\hspace{-0.3pt}'\hspace{-0.3pt}}^{L\uppi}(\theta)
\end{equation}
where
\begin{eqnarray}\label{fdef}
    F_{\hspace{-1pt}ll\hspace{-0.3pt}'\hspace{-0.3pt}}^{L\uppi}(\theta) \hspace{2pt} &=& \hspace{-4pt}\sum_{\uplambda}\sum_{\mu\hspace{-0.3pt}'\hspace{-0.3pt}=l\hspace{-0.3pt}'\hspace{-1.3pt}-d}^{l\hspace{-0.3pt}'\hspace{-0.3pt}}\hspace{2pt}\sum_{\mu=l-d}^{l}\hspace{2pt}\sum_{\nu=-\uplambda}^{\uplambda} \hspace{-3pt}\tfrac{(-1)^{L+\nu}\left(2L+1\right)\left(2\uplambda+1\right)}{2}\hspace{-2pt}\nonumber\\[3pt]
    && \hspace{-45pt} \sqrt{\left(2L\!+\!2d\!-\!2l\hspace{-0.3pt}'\hspace{-0.3pt}\!+\!1\right)\left(2L\!+\!2d\!-\!2l\vphantom{'}\!+\!1\right)\hspace{-2pt}\left(2l\hspace{-0.3pt}'\hspace{-0.3pt}\!+\!1\right)\hspace{-2pt}\left(2l\vphantom{'}\!+\!1\right)}\hspace{-2pt}\nonumber\\[3pt]
    &&\hspace{-45pt} \left(\hspace{-4pt}\begin{array}{ccr}
        \scriptstyle l & \hspace{-3pt}\scriptstyle L+d-l & \hspace{-3pt}\scriptstyle L\hspace{1pt} \\[3pt]
        \scriptstyle \mu & \hspace{-3pt} \scriptstyle L-\mu & \hspace{-3pt}\scriptstyle -L
    \end{array}\hspace{-4pt}\right)\hspace{-4pt}\left(\hspace{-4pt}\begin{array}{ccr}
        \scriptstyle l\hspace{-0.3pt}'\hspace{-0.3pt} & \hspace{-3pt}\scriptstyle L+d-l\hspace{-0.3pt}'\hspace{-0.3pt} & \hspace{-3pt}\scriptstyle L\hspace{1pt} \\[3pt]
        \scriptstyle \mu\hspace{-0.3pt}'\hspace{-0.3pt} & \hspace{-3pt}\scriptstyle L-\mu\hspace{-0.3pt}'\hspace{-0.3pt} & \hspace{-3pt}\scriptstyle -L
    \end{array}\hspace{-5pt}\right)\hspace{-4pt}\left(\hspace{-4pt}\begin{array}{ccc}
        \hspace{1pt}\scriptstyle l\hspace{-0.3pt}'\hspace{-0.3pt} & \scriptstyle l & \scriptstyle \uplambda \\[3pt]
        \scriptstyle 0 & \scriptstyle 0 & \scriptstyle 0
    \end{array}\hspace{-4pt}\right)\hspace{-4pt}\left(\hspace{-4pt}\begin{array}{ccr}
        \scriptstyle l\hspace{-0.3pt}'\hspace{-0.3pt} & \scriptstyle l & \scriptstyle \uplambda \\[3pt]
        \scriptstyle -\mu\hspace{-0.3pt}'\hspace{-0.3pt} & \scriptstyle \mu & \scriptstyle \hspace{-2pt}-\nu
    \end{array}\hspace{-4pt}\right)\nonumber\\[3pt]
    &&\hspace{-45pt}\left(\hspace{-4pt}\begin{array}{ccc}
        \scriptstyle L+d-l\hspace{-0.3pt}'\hspace{-0.3pt} & \scriptstyle L+d-l & \scriptstyle \uplambda \\[3pt]
        \scriptstyle 0 & \scriptstyle 0 & \scriptstyle 0
    \end{array}\hspace{-4pt}\right)\hspace{-4pt}\left(\hspace{-4pt}\begin{array}{ccc}
        \scriptstyle L+d-l\hspace{-0.3pt}'\hspace{-0.3pt} & \scriptstyle L+d-l & \scriptstyle \uplambda \\[3pt]
        \scriptstyle \mu\hspace{-0.3pt}'\hspace{-0.3pt}-L & \scriptstyle L-\mu & \scriptstyle \nu
    \end{array}\hspace{-4pt}\right) P_{\uplambda}\hspace{-2pt}\left(\cos{\theta}\right)
\end{eqnarray}
with %\[\uplambda_{\text{min}}=|l-l\hspace{-0.3pt}'\hspace{-0.3pt}| \text{ and }\uplambda_{\text{max}}=\text{min}\left(l+l\hspace{-0.3pt}'\hspace{-0.3pt},2L+2d-l-l\hspace{-0.3pt}'\hspace{-0.3pt}\right)\]
\begin{equation}\label{lrange}
    |l-l\hspace{-0.3pt}'\hspace{-0.3pt}| \leq \uplambda \leq \text{min}\left(l+l\hspace{-0.3pt}'\hspace{-0.3pt},2L+2d-l-l\hspace{-0.3pt}'\hspace{-0.3pt}\right)
\end{equation}
The form of the integral $\mathbb{I}$---given by Eq.~\eqref{ovint8} with $F_{\hspace{-1pt}ll\hspace{-0.3pt}'\hspace{-0.3pt}}^{L\uppi}(\theta)$ and explicitly expressed in terms of the Wigner $3$-$j$ symbols---is equivalent to Eq.~(21) of \textcite{Calais1962}. It should be noted that in their work, they did not consider the minimal bipolar harmonics as the angular generator functions; rather they considered the product of two spherical harmonics instead. However, the integral $\mathbb{J}$, given by Eq.~\eqref{ovint6}, is similar to the integral in Eq.~(17) of \textcite{Calais1962}. 

Considering the definition of the Wigner $6$-$j$ symbol [cf.\ Eq~12.1.4(8) of \textcite{Varshalovich1988}], 
\begin{eqnarray}
    \left\{\hspace{-5pt}\begin{array}{ccc}
         \scriptstyle l\hspace{-0.3pt}'\hspace{-0.3pt} & \hspace{-4pt}\scriptstyle L+d-l\hspace{-0.3pt}'\hspace{-0.3pt} & \hspace{-4pt}\scriptstyle L \\[3pt]
         \scriptstyle L+d-l & \hspace{-4pt}\scriptstyle l & \hspace{-4pt}\scriptstyle \uplambda
    \end{array}\hspace{-5pt}\right\} &=& \left(2L+1\right)\hspace{-7pt}\sum_{\mu\hspace{-0.3pt}'\hspace{-0.3pt}=l\hspace{-0.3pt}'\hspace{-0.3pt}-d}^{l\hspace{-0.3pt}'\hspace{-0.3pt}}\sum_{\mu=l-d}^{l}  \sum_{\nu=-\uplambda}^{\uplambda}(-1)^{\nu-\uplambda}\nonumber\\[3pt]
    &&\hspace{-80pt}\hspace{-3pt} \left(\hspace{-5pt}\begin{array}{ccc}
        \scriptstyle l & \hspace{-4pt}\scriptstyle L+d-l & \hspace{-4pt}\scriptstyle L \\[3pt]
        \scriptstyle \mu & \hspace{-4pt} \scriptstyle L-\mu & \hspace{-4pt}\scriptstyle -L
    \end{array}\hspace{-5pt}\right)\hspace{-5pt}\left(\hspace{-5pt}\begin{array}{ccc}
        \scriptstyle l\hspace{-0.3pt}'\hspace{-0.3pt} & \hspace{-4pt}\scriptstyle L+d-l\hspace{-0.3pt}'\hspace{-0.3pt} & \hspace{-4pt}\scriptstyle L \\[3pt]
        \scriptstyle \mu\hspace{-0.3pt}'\hspace{-0.3pt} & \hspace{-4pt}\scriptstyle L-\mu\hspace{-0.3pt}'\hspace{-0.3pt} & \hspace{-4pt}\scriptstyle -L
    \end{array}\hspace{-5pt}\right)\hspace{-5pt} \left(\hspace{-4pt}\begin{array}{ccc}
        \scriptstyle l\hspace{-0.3pt}'\hspace{-0.3pt} & \hspace{-4pt}\scriptstyle l & \hspace{-4pt}\scriptstyle \uplambda \\[3pt]
        \scriptstyle -\mu\hspace{-0.3pt}'\hspace{-0.3pt} & \hspace{-4pt}\scriptstyle \mu & \hspace{-4pt}\scriptstyle -\nu
    \end{array}\hspace{-4pt}\right)\hspace{-5pt}\left(\hspace{-4pt}\begin{array}{ccc}
        \scriptstyle L+d-l\hspace{-0.3pt}'\hspace{-0.3pt} & \hspace{-4pt}\scriptstyle L+d-l & \hspace{-4pt}\scriptstyle \uplambda \\[3pt]
        \scriptstyle \mu\hspace{-0.3pt}'\hspace{-0.3pt}-L & \hspace{-4pt}\scriptstyle L-\mu & \hspace{-4pt}\scriptstyle \nu
    \end{array}\hspace{-4pt}\right)\nonumber
\end{eqnarray}
$F_{\hspace{-1pt}ll\hspace{-0.3pt}'\hspace{-0.3pt}}^{L\uppi}(\theta)$, given by \eqref{fdef}, can also be expressed in a more compact form as
\begin{equation}\label{fdef1}
   F_{\hspace{-1pt}ll\hspace{-0.3pt}'\hspace{-0.3pt}}^{L\uppi}(\theta) \hspace{2pt} = \sum_{\uplambda} \mathcal{B}^{L\uppi}_{ll\hspace{-0.3pt}'\uplambda} \hspace{2pt}P_{\uplambda}\left(\cos{\theta}\right)
\end{equation}
with
\begin{eqnarray}\label{defB1}
    \mathcal{B}^{L\uppi}_{ll\hspace{-0.3pt}'\hspace{-1.0pt}\uplambda}  &=& \sqrt{\hspace{-2pt}\left(2L\!+\!2d\!-\!2l\hspace{-0.3pt}'\hspace{-0.3pt}\!+\!1\right)\hspace{-3pt}\left(2L\!+\!2d\!-\!2l\vphantom{'}\!+\!1\right)\hspace{-3pt}\left(2l\hspace{-0.3pt}'\hspace{-0.3pt}\!+\!1\right)\hspace{-3pt}\left(2l\vphantom{'}\!+\!1\right)}\nonumber\\[3pt]
    && \hspace{-2pt} \left(\hspace{-4pt}\begin{array}{ccc}
        \hspace{1pt}\scriptstyle l\hspace{-0.3pt}'\hspace{-0.3pt} & \scriptstyle l & \scriptstyle \uplambda \\[3pt]
        \scriptstyle 0 & \scriptstyle 0 & \scriptstyle 0
    \end{array}\hspace{-4pt}\right)\hspace{-4pt}\left(\hspace{-4pt}\begin{array}{ccc}
        \scriptstyle L+d-l\hspace{-0.3pt}'\hspace{-0.3pt} & \scriptstyle L+d-l & \scriptstyle \uplambda \\[3pt]
        \scriptstyle 0 & \scriptstyle 0 & \scriptstyle 0
    \end{array}\hspace{-4pt}\right)\hspace{-4pt}\left\{\hspace{-5pt}\begin{array}{ccc}
         \scriptstyle l\hspace{-0.3pt}'\hspace{-0.3pt} & \hspace{-4pt}\scriptstyle L+d-l\hspace{-0.3pt}'\hspace{-0.3pt} & \hspace{-4pt}\scriptstyle L \\[3pt]
         \scriptstyle L+d-l & \hspace{-4pt}\scriptstyle l & \hspace{-4pt}\scriptstyle \uplambda
    \end{array}\hspace{-5pt}\right\}\nonumber\\[3pt]
    &&\frac{1}{2}(-1)^{L+\uplambda}\left(2\uplambda+1\right)
\end{eqnarray}
The integral $\mathbb{I}$, given by Eq.~\eqref{ovint8}, combined with this form of $F_{\hspace{-1pt}ll\hspace{-0.3pt}'\hspace{-0.3pt}}^{L\uppi}(\theta)$, explicitly expressed in terms of the Wigner $3$-$j$ and $6$-$j$ symbols, is identical to the overlap integral in Eq.~(21) together with Eq.~(20) of \textcite{Drake1978}. Specifically, the parameters in  Eqs.~(20-21) of \textcite{Drake1978} are mapped to our notation as $l_{1}\rightarrow l$, $l_{2}\rightarrow L+d-l$, $l\hspace{-0.3pt}'\hspace{-2.3pt}_{1}\rightarrow l\hspace{-0.3pt}'\hspace{-0.3pt}$, $l\hspace{-0.3pt}'\hspace{-2.3pt}_{2}\rightarrow L+d-l\hspace{-0.3pt}'\hspace{-0.3pt}$ and $\Lambda\rightarrow\uplambda$.

\subsection{Present approach}\label{oscheme}
In the spirit of the present work the integral 
\begin{equation*}\tag{\ref{ovint}}
    \mathbb{I} = \hspace{-3pt} \int \hspace{-3pt}d\mathbf{r}_{1}\hspace{-3pt}\int \hspace{-3pt} d\mathbf{r}_{2} \left(\mathscr{R}'\hspace{3pt} \Omega_{l\hspace{-0.3pt}'\hspace{-0.3pt}}^{LL\uppi}\right)^{\ast}\hspace{1pt}\left(\mathscr{R}\hspace{3pt}\Omega_{l}^{LL\uppi}\right)
\end{equation*}
is evaluated by a  transformation to the $(r_{1},r_{2},\theta,\alpha,\beta,\gamma)$ coordinates, given by the mapping $\mathbf{r}_{i} \rightarrow  \mathbb{R}_{\alpha\beta\gamma}\hspace{2pt} \mathbf{R}_{i}$ with $i=1,2$, where $\mathbb{R}_{\alpha\beta\gamma}$ is given by Eq.~(\ref{eq:rotmat}) and where (cf.\ Eq.~(12) of \textbf{1})
\begin{equation} \label{maplb}
    \mathbf{R}_{i} =  \left[  r_{i} \cos\tfrac{\theta}{2} , (-1)^i r_{i} \sin\tfrac{\theta}{2}  ,0\right]^\mathsf{T}    
\end{equation}
Then, the minimal bipolar harmonics in Eq.~(\ref{ovint}) can be expressed as linear combinations of the Wigner functions $\mathcal{D}$ (cf.\ Eq.~(\ref{mbhwdr}) and Eq.~(74) of \textbf{1} with $M=L$)
\begin{equation*}\tag{\ref{mbhwdr}a}\label{mbhwdra}
    \Omega_{l}^{LL\uppi}(\theta_{1},\phi_{1},\theta_{2},\phi_{2})\hspace{-2pt} =\hspace{-6pt}\sum_{K=-L}^{L}\hspace{-6pt}\Lambda_{Kl}^{L\uppi}\hspace{-2pt}\left(\theta\right)\mathcal{D}_{L}^{LK}\left(\alpha,\beta,\gamma\right)
\end{equation*}

These transformations bring  $\mathbb{I}$ into the following form
\begin{eqnarray}\label{ovintI}
    \mathbb{I} \hspace{-1pt} &=& \hspace{-5pt}\int\limits_{0}^{\infty}\hspace{-3pt}r_1^2\hspace{1pt}dr_{1}\hspace{-2pt}\int\limits_{0}^{\infty}\hspace{-3pt}r_2^2\hspace{1pt}dr_{2}\hspace{-2pt}\int\limits_{0}^{\pi}\hspace{-3pt}\sin{\theta}\hspace{0.5pt} d\theta\hspace{4pt}\mathscr{R}'^{\ast}\hspace{2pt}\mathscr{R}\sum_{K,K\hspace{-0.3pt}'\hspace{-0.3pt}}\left(\hspace{-1pt}\Lambda_{K\hspace{-0.3pt}'\hspace{-0.3pt}l\hspace{-0.3pt}'\hspace{-0.3pt}}^{L\uppi}\hspace{-2pt}\left(\theta\right)\hspace{-1pt}\right)^{\!\ast}\Lambda_{Kl}^{L\uppi}\hspace{-2pt}\left(\theta\right)\nonumber\\
    && \hspace{-13pt}\hspace{-2pt}\int\limits_{0}^{2\pi}\hspace{-3pt}d\alpha\hspace{-2pt}\int\limits_{0}^{\pi}\hspace{-3pt}\sin{\beta}\hspace{0.5pt} d\beta\hspace{-2pt}\int\limits_{0}^{2\pi} \hspace{-3pt}d\gamma\left(\hspace{-1pt}\mathcal{D}_{L}^{LK\hspace{-0.3pt}'\hspace{-0.3pt}}\left(\alpha,\beta,\gamma\right)\hspace{-1pt}\right)^{\!\ast}\mathcal{D}_{L}^{LK}\left(\alpha,\beta,\gamma\right)
\end{eqnarray}
which simplifies further due to the orthonormality of Wigner functions $\mathcal{D}$  (cf.\ Eq.~(39) of \textbf{1})
\begin{equation*}
    \int\limits_{0}^{2\pi} \hspace{-3pt}d\alpha\hspace{-2pt}\int\limits_{0}^{\pi}\hspace{-3pt}\sin{\beta}\hspace{0.5pt} d\beta\hspace{-2pt}\int\limits_{0}^{2\pi}\hspace{-3pt}d\gamma \hspace{2pt}\left(\hspace{-2pt}\mathcal{D}_{L}^{LK\hspace{-0.3pt}'\hspace{-0.3pt}}\hspace{-1pt}\left(\alpha,\beta,\gamma\right)\hspace{-2pt}\right)^{\!\ast}\mathcal{D}_{L}^{LK}\hspace{-1pt}(\alpha,\beta,\gamma) \hspace{-1pt}=\hspace{-1pt} \delta_{K,K\hspace{-0.3pt}'\hspace{-0.3pt}}
\end{equation*}
as
\begin{equation}\label{ovintI1}
    \mathbb{I} = \int\limits_{0}^{\infty}\hspace{-3pt}r_1^2\hspace{1pt}dr_{1}\hspace{-2pt}\int\limits_{0}^{\infty}\hspace{-3pt}r_2^2\hspace{1pt}dr_{2}\hspace{-2pt}\int\limits_{0}^{\pi}\hspace{-3pt}\sin{\theta}\hspace{0.5pt} d\theta\hspace{4pt}\mathscr{R}'^{\ast}\hspace{2pt}\mathscr{R}\hspace{2pt}\mathcal{F}_{ll\hspace{-0.3pt}'\hspace{-0.3pt}}^{L\uppi}(\theta)
\end{equation}
with
\begin{equation}\tag{\ref{flpillp1}}
    \mathcal{F}_{ll\hspace{-0.3pt}'\hspace{-0.3pt}}^{L\uppi}(\theta)=\sum_{K=-L}^{L}\left(\hspace{-1pt}\Lambda_{Kl\hspace{-0.3pt}'\hspace{-0.3pt}}^{L\uppi}\hspace{-2pt}\left(\theta\right)\hspace{-1pt}\right)^{\!\ast}\Lambda_{Kl}^{L\uppi}\hspace{-2pt}\left(\theta\right)
\end{equation}
The explicit form of $\mathcal{F}_{ll\hspace{-0.3pt}'\hspace{-0.3pt}}^{L\uppi}(\theta)$ has been derived in Section \ref{sec:angint} in terms of Chebyshev polynomials (cf.\ Eq.~\eqref{flpillp8}).

Since Eqs.~\eqref{ovintI1} and~\eqref{ovint8} represent the same integral $\mathbb{I}$, the integrant $\mathcal{F}_{ll\hspace{-0.3pt}'\hspace{-0.3pt}}^{L\uppi}(\theta)$ given by Eq.~\eqref{flpillp1} must equal the integrant $F_{\hspace{-1pt}ll\hspace{-0.3pt}'\hspace{-0.3pt}}^{L\uppi}(\theta)$, given by Eq.~\eqref{fdef1}. Thus, $\mathcal{F}_{ll\hspace{-0.3pt}'\hspace{-0.3pt}}^{L\uppi}\equiv \mathcal{F}_{ll\hspace{-0.3pt}'\hspace{-0.3pt}}^{L\uppi}(\theta)=F_{\hspace{-1pt}ll\hspace{-0.3pt}'\hspace{-0.3pt}}^{L\uppi}(\theta)$ can be first expressed in terms of Wigner $3n$-$j$ symbols and Legendre polynomials $P_{\uplambda}\left(\cos{\theta}\right)$, and then equated to the expression (cf.\ Eq.~\eqref{wlpikk1}) in terms of Chebyshev polynomials $T_{\texttt{m}}(\cos{\theta})$---as derived in our formalism---as
\begin{eqnarray}
    \mathcal{F}_{ll\hspace{-0.3pt}'\hspace{-0.3pt}}^{L\uppi} \hspace{-2pt} &=&\hspace{-2pt}  \sqrt{\hspace{-2pt}\left(2L\!+\!2d\!-\!2l\hspace{-0.3pt}'\hspace{-0.3pt}\!+\!1\right)\hspace{-3pt}\left(2L\!+\!2d\!-\!2l\vphantom{'}\!+\!1\right)\hspace{-3pt}\left(2l\hspace{-0.3pt}'\hspace{-0.3pt}\!+\!1\right)\hspace{-3pt}\left(2l\vphantom{'}\!+\!1\right)}\nonumber\\[3pt]
    && \hspace{-2pt} \sum_{\uplambda}\left(\hspace{-4pt}\begin{array}{ccc}
        \hspace{1pt}\scriptstyle l\hspace{-0.3pt}'\hspace{-0.3pt} & \scriptstyle l & \scriptstyle \uplambda \\[3pt]
        \scriptstyle 0 & \scriptstyle 0 & \scriptstyle 0
    \end{array}\hspace{-4pt}\right)\hspace{-4pt}\left(\hspace{-4pt}\begin{array}{ccc}
        \scriptstyle L+d-l\hspace{-0.3pt}'\hspace{-0.3pt} & \scriptstyle L+d-l & \scriptstyle \uplambda \\[3pt]
        \scriptstyle 0 & \scriptstyle 0 & \scriptstyle 0
    \end{array}\hspace{-4pt}\right)\hspace{-4pt}\left\{\hspace{-5pt}\begin{array}{ccc}
         \scriptstyle l\hspace{-0.3pt}'\hspace{-0.3pt} & \hspace{-4pt}\scriptstyle L+d-l\hspace{-0.3pt}'\hspace{-0.3pt} & \hspace{-4pt}\scriptstyle L \\[3pt]
         \scriptstyle L+d-l & \hspace{-4pt}\scriptstyle l & \hspace{-4pt}\scriptstyle \uplambda
    \end{array}\hspace{-5pt}\right\}\nonumber\\[3pt]
    &&\frac{1}{2}(-1)^{L+\uplambda}\left(2\uplambda+1\right)\hspace{2pt}P_{\uplambda}(\cos{\theta})\label{flpillp9}\\[3pt]
    &=&\sum_{\kappa=d}^{L}\sum_{\texttt{j},\texttt{j}\hspace{-0.3pt}'=0}^{\kappa}\mathcal{C}_{ll\hspace{-0.3pt}'\hspace{-0.3pt}\texttt{j}\texttt{j}\hspace{-0.3pt}'\hspace{-0.3pt}}^{L\kappa \uppi}\hspace{2pt}T_{\texttt{m}}(\cos{\theta})\label{wlpikk2}
\end{eqnarray}
Here, the range of the summation over $\uplambda$ in Eq.~\eqref{flpillp9} is determined by Eq.~\eqref{lrange}. The numerical coefficients $\mathcal{C}_{ll\hspace{-0.3pt}'\hspace{-0.3pt}\texttt{j}\texttt{j}\hspace{-0.3pt}'\hspace{-0.3pt}}^{L\kappa \uppi}$ in Eq.~(\ref{wlpikk2}) are given by Eqs.~\eqref{clnllpjjp} and~\eqref{clullpjjp} for natural and unnatural parity, respectively, and the parameter $\texttt{m}$ is given by Eq.~\eqref{defm}.

Consequently, the angular factors
\begin{equation}\tag{\ref{wlpikk1}}
     \mathscr{W}_{ll\hspace{-0.3pt}'\hspace{-0.3pt}}^{L\uppi} \hspace{2pt} = \hspace{2pt} r_{1}^{l+l\hspace{-0.3pt}'\hspace{-0.3pt}}r_{2}^{2L-l-l\hspace{-0.3pt}'\hspace{-0.3pt}\!+2d}\sum_{\kappa=d}^{L}\sum_{\texttt{j},\texttt{j}\hspace{-0.3pt}'=0}^{\kappa}\mathcal{C}_{ll\hspace{-0.3pt}'\hspace{-0.3pt}\texttt{j}\texttt{j}\hspace{-0.3pt}'\hspace{-0.3pt}}^{L\uppi}\hspace{2pt}T_{\texttt{m}}(\cos{\theta})
\end{equation}
expressed previously in terms of Chebyshev polynomials, can also be represented in terms of Legendre polynomials as (cf.\ Eq.~\eqref{wlpikk})
\begin{equation}
    \mathscr{W}_{ll\hspace{-0.3pt}'\hspace{-0.3pt}}^{L\uppi} \hspace{2pt} = \hspace{2pt} r_{1}^{l+l\hspace{-0.3pt}'\hspace{-0.3pt}}r_{2}^{2L-l-l\hspace{-0.3pt}'\hspace{-0.3pt}\!+2d}\hspace{2pt}\sum_{\uplambda} \mathcal{B}^{L\uppi}_{ll\hspace{-0.3pt}'\uplambda} \hspace{2pt}P_{\uplambda}(\cos{\theta})\label{angularfact}
\end{equation}
where, $\mathcal{B}^{L\uppi}_{ll\hspace{-0.3pt}'\uplambda}$ is explicitly given by Eq.~\eqref{defB1} in terms of Wigner $3$-$j$ and $6$-$j$ symbols. It should be noted that, Eq.~\eqref{angularfact} is consistent with the explicit expression of the angular integral given in Eq.~(15) of \textcite{Frolov1996}.

\section{Angular momentum operators in  \texorpdfstring{$(r_{1},r_{2}, r_{12},\alpha,\beta,\gamma)$}~ coordinates}

\begin{figure}[t]
    \centering
    \begin{minipage}[b]{0.40\columnwidth}
        \centering
        
        \scalebox{0.43}{\input{lab_frame3}} 
        
        %\vspace{0.0cm}
        (a)
    \end{minipage}\hfill 
    \begin{minipage}[b]{0.58\columnwidth}
        \centering
        
        \scalebox{0.55}{\input{body_frame1}} 
        
        %\vspace{0.0cm}
        (b)
    \end{minipage}
    \caption{(a) The relative orientation of the body-fixed \textcolor{yellow!10!green}{$(\hspace{-0.8pt}XY\hspace{-1pt}Z)$} and the laboratory-fixed \textcolor{cyan}{$(x\,y\,z)$} reference frames is described by the three Euler angles  $\left(\alpha,\beta,\gamma\right)$ introduced in Eq.~(13) of \textbf{1}. (b) The vectors $\bm{r}_{1}$ and $\bm{r}_{2}$ lie in the $XY$ plane, with the $X$ axis bisecting the angle $\theta$ between $\bm{r}_{1}$ and $\bm{r}_{2}$, and the $Z$ axis aligned with the vector $\bm{r}_{1}\times\bm{r}_{2}$. The interparticle distances are defined as $r_1=\left| \bm{r}_{1} \right|$, $r_2=\left| \bm{r}_{2} \right|$, and $r_{12}=\left| \bm{r}_{1} - \bm{r}_{2} \right|$. }
    \label{fig:comparison1}
\end{figure}

Certain commutation relations between the operators appearing in \textbf{1} are most transparent in $(r_{1},r_{2},r_{12},\alpha,\beta,\gamma)$ coordinates introduced in the caption of Fig.~\ref{fig:comparison1}. Elementary geometrical considerations show that 
\begin{eqnarray}
    \hspace{-10pt}\sin\left(\tfrac{\theta}{2}\right) & = & \sqrt{\tfrac{(r_{1}-r_{2}+r_{12})(r_{2}-r_{1}+r_{12})}{4\hspace{1pt}r_{1}r_{2}}} \\
    \hspace{-10pt}\cos\left(\tfrac{\theta}{2}\right) & = & \sqrt{\tfrac{(r_{1}+r_{2}+r_{12})(r_{1}+r_{2}-r_{12})}{4\hspace{1pt}r_{1}r_{2}}}
\end{eqnarray}
Consequently, the positions of the quasiparticles $i=1,2$  in the  $(\hspace{-0.8pt}XY\hspace{-1pt}Z)$ frame are 
\begin{equation} \label{bodyfixed}
    \left[ \begin{array}{c} X_{i} \\ Y_{i} \\ Z_{i} \end{array}\right] = \left[ \begin{array}{c} \phantom{(-1)^i~ }r_{i} \hspace{3pt}\sqrt{\frac{(r_{1}+r_{2}+r_{12})(r_{1}+r_{2}-r_{12})}{4\hspace{1pt}r_{1}r_{2}}} \\[2pt] (-1)^i~ r_{i} \hspace{3pt}\sqrt{\frac{(r_{12}-r_{1}+r_{2})(r_{12}+r_{1}-r_{2})}{4\hspace{1pt}r_{1}r_{2}}}  \\[2pt] \phantom{(-1)^i~ }0 ~\phantom{\sin\left(\frac{\theta}{2}\right)}\end{array}\right]
\end{equation}
and the corresponding positions in the  $(xyz)$ frame are
\begin{equation}\label{eulerrotation1}
    \left[\begin{array}{c} x_i \\ y_i \\ z_i \end{array}\right] = \bm{R}_{\texttt{z}}^{\alpha}\,\bm{R}_{\texttt{y}}^{\beta}\,\bm{R}_{\texttt{z}}^{\gamma}\left[\begin{array}{c} X_i \\ Y_i \\ Z_i \end{array}\right]
\end{equation}
where the rotation matrices $\bm{R}_{\texttt{z}}^{\alpha}$, $\bm{R}_{\texttt{y}}^{\beta}$, and $\bm{R}_{\texttt{z}}^{\gamma}$ have been defined in Eq.~(15) of \textbf{1}. Using the transformation relations between the $\left(x_{1},y_{1},z_{1},x_{2},y_{2},z_{2}\right)$ and $(r_{1},r_{2},r_{12},\alpha,\beta,\gamma)$ coordinate systems stemming from Eq.~(\ref{eulerrotation1}), it is possible to show that the operators $\hat{\bm{L}}_{xyz}$, $\hat{L}_{\pm}$, and $\hat{L}^2$ have exactly the same form as in coordinates $(r_{1},r_{2},\theta,\alpha,\beta,\gamma)$, and are given by Eqs.~(17), (21), and~(18), respectively, of \textbf{1}. Straightforward reasoning shows that for $\hat{g}\in \{r_1,r_2, r_{12},\partial_{r_1},\partial_{r_2},\partial_{r_{12}}\}$
\begin{align}
    \left[\hat{L}_x,\hat{g}\right]  = 0 & \hspace{10pt} & \left[\hat{L}_y,\hat{g}\right]  = 0 & \hspace{10pt} & \left[\hat{L}_z,\hat{g}\right]  = 0 \\
    \left[\hat{L}_+,\hat{g}\right]  = 0 & \hspace{10pt} & \left[\hat{L}_-,\hat{g}\right]  = 0 & \hspace{10pt} & \left[\hat{L}^2,\hat{g}\right]  = 0
\end{align}
since both sets of operators, $\{r_1,r_2, r_{12},\partial_{r_1},\partial_{r_2},\partial_{r_{12}}\}$ and $\{\hat{L}_x,\hat{L}_y,\hat{L}_z,\hat{L}_+,\hat{L}_-,\hat{L}^2\}$, involve different subsets of coordinates. It is exactly for this reason the operator $\hat{L}_-$ commutes with the operators $ \hat{h}_{l}$, $ \hat{h}^{\uparrow}_{l}$, and $ \hat{h}^{\downarrow}_{l}$ defined in Eqs.~(118)--(120), respectively.

\bibliography{biblio_rmp}

%% file: lab_frame1.tex
\begin{tikzpicture}[scale=2, line cap=round]

% Define styles
\tikzset{
    %axis/.style={->, very thick, blue},
    axis1/.style={->, ultra thick, blue, >=stealth},
    vector/.style={->, very thick},
    particle/.style={fill=red, draw=black, thick},
    CMM/.style={fill=blue, draw=black, thick},
    labelstyle/.style={font=\Large} 
}

% Adjust arrowhead size
\tikzset{every arrow/.append style={scale=2}}

% Draw the axes
\draw[axis1] (0,0) -- (2.5,0) node[labelstyle, anchor=north east, xshift=5pt, yshift=-5pt] {$\bar{x}$};
\draw[axis1] (0,0) -- (0,2.5) node[labelstyle, anchor=south, xshift=-5pt, yshift=5pt] {$\bar{y}$};
\draw[axis1] (0,0) -- (-1,-1.0) node[labelstyle, anchor=north, yshift=-7pt] {$\bar{z}$};

% Particle radius
\def\particleradius{0.06}

% Coordinates for particles
\coordinate (m1) at (-1.00,1.50);
\coordinate (m2) at (1.5,-0.75);
\coordinate (m3) at (2.0,1.0);
\coordinate (cmm) at (0.33,1.08);

% Compute scaling factors for each vector
\pgfmathsetmacro{\scaleFactorMOne}{1 - \particleradius / sqrt((1.5)^2 + (2)^2)}
\pgfmathsetmacro{\scaleFactorMTwo}{1 - \particleradius / sqrt((1.5)^2 + (0.75)^2)}
\pgfmathsetmacro{\scaleFactorMThree}{1 - \particleradius / sqrt((1.0)^2 + (2)^2)}

% Adjusted vector endpoints
\path (0,0) -- (m1) coordinate[pos=\scaleFactorMOne] (t1);
\path (0,0) -- (m2) coordinate[pos=\scaleFactorMTwo] (t2);
\path (0,0) -- (m3) coordinate[pos=\scaleFactorMThree] (t3);

% Draw the particles as red circles with black borders
\draw[particle] (m1) circle (\particleradius) node[labelstyle, anchor=east, xshift=-5pt] {$\textcolor{red}{\mathbf{1}}$};
\draw[particle] (m2) circle (\particleradius) node[labelstyle, anchor=west, xshift=5pt] {$\textcolor{red}{\mathbf{2}}$};
\draw[particle] (m3) circle (\particleradius) node[labelstyle, anchor=west, xshift=5pt] {$\textcolor{red}{\mathbf{3}}$};

% Draw vectors from the origin to touch each circle
\draw[vector] (0,0) -- (t1) node[labelstyle, midway, anchor=south west, xshift=-5pt, yshift=5pt] {$\bar{\bm{r}}_1$};
\draw[vector] (0,0) -- (t2) node[labelstyle, midway, anchor=north, xshift=5pt, yshift=-5pt] {$\bar{\bm{r}}_2$};
\draw[vector] (0,0) -- (t3) node[labelstyle, midway, anchor=south, xshift=5pt, yshift=5pt] {$\bar{\bm{r}}_3$};

% Draw the CM as a gray solid circle with black border
\draw[CMM] (cmm) circle (\particleradius) node[labelstyle, anchor=west, xshift=5pt] {$CM$};

% Draw vector from origin to CM
\draw[vector] (0,0) -- (0.31,1.01) node[midway, labelstyle, anchor=north west, xshift=-5pt, yshift=5pt] {$\bm{R}$};

\end{tikzpicture}

%% file: lab_frame2.tex
\begin{tikzpicture}[scale=2, line cap=round]

% Define styles
\tikzset{
    axis1/.style={->, ultra thick, blue, >=stealth},
    axis2/.style={->, ultra thick, cyan, >=stealth},
    vector/.style={->, very thick},
    particle/.style={fill=red, draw=black, thick},
    labelstyle/.style={font=\Large} 
}

% Adjust arrowhead size
\tikzset{every arrow/.append style={scale=2}}

% Draw the axes1
\draw[axis1] (0,0) -- (2.5,0) node[labelstyle, anchor=north east, xshift=5pt, yshift=-5pt] {$\bar{x}$};
\draw[axis1] (0,0) -- (0,2.5) node[labelstyle, anchor=south, xshift=-5pt, yshift=5pt] {$\bar{y}$};
\draw[axis1] (0,0) -- (-1,-1.0) node[labelstyle, anchor=north, yshift=-7pt] {$\bar{z}$};

% Draw the axes2
\draw[axis2] (2.0,1.0) -- (3.0,1.0) node[labelstyle, anchor=north east, xshift=5pt, yshift=-5pt] {$x$};
\draw[axis2] (2.0,1.0) -- (2.0,2.1) node[labelstyle, anchor=south, xshift=-5pt, yshift=4pt] {$y$};
\draw[axis2] (2.0,1.0) -- (1.4,0.4) node[labelstyle, anchor=north, yshift=-1pt] {$z$};

% Particle radius
\def\particleradius{0.06}

% Numeric coordinates for particles
\def\mOneX{-1.0}
\def\mOneY{1.5}
\def\mTwoX{1.5}
\def\mTwoY{-0.75}
\def\mThreeX{2.0}
\def\mThreeY{1.0}

% Compute scaling factors to adjust arrow tips to touch the circumference of m1 and m2
\pgfmathsetmacro{\scaleFactorMOne}{1 - \particleradius / sqrt((\mOneX-\mThreeX)^2 + (\mOneY-\mThreeY)^2)}
\pgfmathsetmacro{\scaleFactorMTwo}{1 - \particleradius / sqrt((\mTwoX-\mThreeX)^2 + (\mTwoY-\mThreeY)^2)}

% Adjusted vector endpoints
\path (\mThreeX,\mThreeY) -- (\mOneX,\mOneY) coordinate[pos=\scaleFactorMOne] (t1);
\path (\mThreeX,\mThreeY) -- (\mTwoX,\mTwoY) coordinate[pos=\scaleFactorMTwo] (t2);

% Draw vectors from m3 to m1 and m2 with adjusted endpoints
\draw[vector] (\mThreeX,\mThreeY) -- (t1) node[labelstyle, midway, anchor=south west, xshift=-5pt, yshift=5pt] {$\bm{r}_1$};
\draw[vector] (\mThreeX,\mThreeY) -- (t2) node[labelstyle, midway, anchor=south west, xshift=5pt, yshift=-5pt] {$\bm{r}_2$};

% Draw the particles as red circles with black borders
\draw[particle] (\mOneX,\mOneY) circle (\particleradius) node[labelstyle, anchor=east, xshift=-5pt] {$\textcolor{red}{\mathbf{1}}$};
\draw[particle] (\mTwoX,\mTwoY) circle (\particleradius) node[labelstyle, anchor=west, xshift=5pt] {$\textcolor{red}{\mathbf{2}}$};

% Draw the circle for m3 AFTER the vectors to cover the inner portions
\draw[particle] (\mThreeX,\mThreeY) circle (\particleradius) node[labelstyle, anchor=north west, xshift=5pt] {$\textcolor{red}{\mathbf{3}}$};

\end{tikzpicture}

%% file: body_frame.tex
\begin{tikzpicture}[scale=2, line cap=round]

% Define styles
\tikzset{
    %axis/.style={->, very thick, blue},
    axis1/.style={->, ultra thick, yellow!10!green, >=stealth},
    vector/.style={->, very thick},
    particle/.style={fill=red, draw=black, thick},
    CMM/.style={fill=blue, draw=black, thick},
    labelstyle/.style={font=\Large} 
}

% Adjust arrowhead size
\tikzset{every arrow/.append style={scale=2}}

% Draw the axes
\draw[axis1] (0,0) -- (2.0,0) node[labelstyle, anchor=north east, xshift=5pt, yshift=-5pt] {$X$};
\draw[axis1] (0,0) -- (0,1.5) node[labelstyle, anchor=south, xshift=-5pt, yshift=5pt] {$Y$};
\draw[axis1] (0,0) -- (-0.8,-0.3) node[labelstyle, anchor=north, yshift=-7pt] {$Z$};

% Particle radius
\def\particleradius{0.06}

% Coordinates for particles
\coordinate (m1) at (1.0,-1.0);
\coordinate (m2) at (1.0,1.0);
\coordinate (m3) at (0.0,0.0);

% Compute scaling factors for each vector
\pgfmathsetmacro{\scaleFactorMOne}{1 - \particleradius / sqrt((1.5)^2 + (2)^2)}
\pgfmathsetmacro{\scaleFactorMTwo}{1 - \particleradius / sqrt((1.5)^2 + (0.75)^2)}
\pgfmathsetmacro{\scaleFactorMThree}{1 - \particleradius / sqrt((1.0)^2 + (2)^2)}

% Adjusted vector endpoints
\path (0,0) -- (m1) coordinate[pos=\scaleFactorMOne] (t1);
\path (0,0) -- (m2) coordinate[pos=\scaleFactorMTwo] (t2);

% Draw the particles as red circles with black borders
\draw[particle] (m1) circle (\particleradius) node[labelstyle, anchor=west, xshift=5pt] {$\textcolor{red}{\mathbf{1}}$};
\draw[particle] (m2) circle (\particleradius) node[labelstyle, anchor=west, xshift=5pt] {$\textcolor{red}{\mathbf{2}}$};
\draw[particle] (m3) circle (\particleradius) node[labelstyle, anchor=south east, xshift=-5pt] {$\textcolor{red}{\mathbf{3}}$};

% Draw vectors from the origin to touch each circle
\draw[vector] (0,0) -- (0.95,-0.95) node[labelstyle, midway, anchor=north, xshift=-5pt, yshift=-5pt] {$\bm{r}_1$};
\draw[vector] (0,0) -- (0.95,+0.95) node[labelstyle, midway, anchor=south, xshift=-5pt, yshift=5pt] {$\bm{r}_2$};

% Draw curved lines to indicate angles
\draw[thick, dashed, <->] (0.5,0) arc[start angle=0, end angle=-45, radius=0.5];
\draw[thick, dashed, <->] (0.5,0) arc[start angle=0, end angle=45, radius=0.5];

% Label the angles
\node[labelstyle] at (0.6,-0.27) {$\frac{\theta}{2}$};
\node[labelstyle] at (0.6,0.27) {$\frac{\theta}{2}$};

% Draw the circle for m3 AFTER the vectors to cover the inner portions
\draw[particle] (0,0) circle (\particleradius);

\end{tikzpicture}

%% file: lab_frame3.tex
\begin{tikzpicture}[x={(-0.6cm,-0.4cm)}, y={(1cm,0cm)}, z={(0cm,1cm)}, 
                    line join=round, line cap=round]

    % 1. Draw the base x-y plane (z=0) in Cyan
    %\fill[cyan!40, opacity=0.2] 
    %    (-2.5,-2.5,0) -- (2.5,-2.5,0) -- (2.5,2.5,0) -- (-2.5,2.5,0) -- cycle;
    
    % Optional: Subtle border for the base plane
   % \draw[cyan!40, thin] 
   %     (-2.5,-2.5,0) -- (2.5,-2.5,0) -- (2.5,2.5,0) -- (-2.5,2.5,0) -- cycle;

    % 2. Draw the Inclined Plane in yellow!10!green
    \fill[yellow!10!green, opacity=0.1] 
        %(-1.5,-3.5,-2.0) -- (3.5,1.5,-2.0) -- (1.5,3.5,2.0) -- (-3.5,-1.5,2.0) -- cycle;
        %(-4.375, -1.875, 2.5) -- (1.875, 4.375, 2.5) -- (3.247, 3.003, -0.244) -- (-3.003, -3.247, -0.244) -- cycle;
        (-5.688, -2.438, 3.250) -- (2.438, 5.688, 3.250) -- (4.221, 3.904, -0.317) -- (-3.904, -4.221, -0.317) -- cycle;
        
    % Optional: Subtle border for the inclined plane
    %\draw[yellow!10!green, thin] 
        %(-1.5,-3.5,-2.0) -- (3.5,1.5,-2.0) -- (1.5,3.5,2.0) -- (-3.5,-1.5,2.0) -- cycle;
        %(-4.375, -1.875, 2.5) -- (1.875, 4.375, 2.5) -- ($(3.125, 3.125, 0)!10pt!(4.375, 1.875, -2.5)$) -- ($(-3.125, -3.125, 0)!10pt!(-1.875, -4.375, -2.5)$) -- cycle;

    % 3. Draw the 3D Axes 
    \draw[-latex, ultra thick, cyan] (0,0,0) -- (5.5,0,0) node[anchor=north east] {\Large $x$};
    \draw[-latex, ultra thick, cyan] (0,0,0) -- (0,5.5,0) node[anchor=west] {\Large $y$};
    \draw[-latex, ultra thick, cyan] (0,0,0) -- (0,0,5.5) node[anchor=south] {\Large $z$};

    % NEW: Draw the Z-axis (Perpendicular to X and Y)
    % Using the cross product of X (-1.5, 1.5, 3.0) and Y (-4.0, -4.0, 0)
    % Direction vector is (1, -1, 1). Scaled to (2.5, -2.5, 2.5)
    \draw[-latex, ultra thick, yellow!10!green] (0,0,0) -- (2.5,-2.5,2.5) node[anchor=south west] {\Large $Z$};

    % 4. Draw the X-axis through the midpoint of particles 1 and 2
    \draw[-latex, ultra thick, yellow!10!green] (0,0,0) -- (-1.625, 1.625, 3.250) node[anchor=south west, shift={(-0.25cm, 0.0cm)}] {\Large $X$};
    %\draw[-latex, ultra thick, yellow!10!green] (0,0,0) -- (-1.5,1.5,3.0) node[anchor=south] {\Large $X$};
    
    % 5. Draw the Y-axis along the intersection of the two planes
    \draw[-latex, ultra thick, yellow!10!green] (0,0,0) -- (-4.0,-4.0,0) node[anchor=south] {\Large $Y$};

    % 6. Draw the vectors (Drawn before the particles so arrowheads don't cover them)
    
    % Vector connecting Particle 1 to Particle 2 (r12)
    \draw[-latex, very thick, darkgray] (1.5,3.5,2.0) -- (-3.5,-1.5,2.0) node[midway, shift={(-0.1cm,0.6cm)}] {\large $\bm{r}_{12}$};
    
    % Vector to Particle 1 (r1)
    \draw[-latex, very thick, black] (0,0,0) -- (1.5,3.5,2.0) node[midway, shift={(0.0cm,-0.4cm)}] {\large $\bm{r}_{1}$};
    
    % Vector to Particle 2 (r2)
    \draw[-latex, very thick, black] (0,0,0) -- (-3.5,-1.5,2.0) node[midway, shift={(0.5cm,0.4cm)}] {\large $\bm{r}_{2}$};

    % 7. Draw the solid red circles and enlarged bold red labels
    
    % Origin (Particle 3)
    \filldraw[fill=red, draw=black, thin] (0,0,0) circle (4pt) node[anchor=north west, text=red, shift={(0.0cm,-0.2cm)}] {\Large $\bm{3}$};
    
    % Top corners (Particle 1 and 2)
    \filldraw[fill=red, draw=black, thin] (1.5,3.5,2.0) circle (4pt) node[anchor=south west, text=red, shift={(0.2cm,-0.5cm)}] {\Large $\bm{1}$};
    \filldraw[fill=red, draw=black, thin] (-3.5,-1.5,2.0) circle (4pt) node[anchor=south east, text=red, shift={(0.0cm,0.2cm)}] {\Large $\bm{2}$};

\end{tikzpicture}

%% file: body_frame1.tex
\begin{tikzpicture}[scale=2, line cap=round]

% Define styles
\tikzset{
    line/.style={-, thick},
    axis1/.style={->, ultra thick, yellow!10!green, >=stealth},
    vector/.style={->, very thick},
    particle/.style={fill=red, draw=black, thick},
    CMM/.style={fill=blue, draw=black, thick},
    labelstyle/.style={font=\Large} 
}

% Adjust arrowhead size
\tikzset{every arrow/.append style={scale=2}}

% Draw the axes
\draw[axis1] (0,0) -- (2.0,0) node[labelstyle, anchor=north east, xshift=5pt, yshift=-5pt] {$X$};
\draw[axis1] (0,0) -- (0,1.5) node[labelstyle, anchor=south, xshift=-5pt, yshift=5pt] {$Y$};
\draw[axis1] (0,0) -- (-0.8,-0.8) node[labelstyle, anchor=north, yshift=-7pt] {$Z$};

% Particle radius
\def\particleradius{0.06}

% Coordinates for particles
\coordinate (m1) at (1.0,-1.0);
\coordinate (m2) at (1.0,1.0);
\coordinate (m3) at (0.0,0.0);

% Compute scaling factors for each vector
\pgfmathsetmacro{\scaleFactorMOne}{1 - \particleradius / sqrt((1.5)^2 + (2)^2)}
\pgfmathsetmacro{\scaleFactorMTwo}{1 - \particleradius / sqrt((1.5)^2 + (0.75)^2)}
\pgfmathsetmacro{\scaleFactorMThree}{1 - \particleradius / sqrt((1.0)^2 + (2)^2)}

% Adjusted vector endpoints
\path (0,0) -- (m1) coordinate[pos=\scaleFactorMOne] (t1);
\path (0,0) -- (m2) coordinate[pos=\scaleFactorMTwo] (t2);

% Draw the particles as red circles with black borders
\draw[particle] (m1) circle (\particleradius) node[labelstyle, anchor=west, xshift=5pt] {$\textcolor{red}{\mathbf{1}}$};
\draw[particle] (m2) circle (\particleradius) node[labelstyle, anchor=west, xshift=5pt] {$\textcolor{red}{\mathbf{2}}$};
\draw[particle] (m3) circle (\particleradius) node[labelstyle, anchor=south east, xshift=-5pt] {$\textcolor{red}{\mathbf{3}}$};

% Draw vectors from the origin to touch each circle
\draw[line] (0,0) -- (0.95,-0.95) node[labelstyle, midway, anchor=north, xshift=-5pt, yshift=-5pt] {$r_1$};
\draw[line] (0,0) -- (0.95,+0.95) node[labelstyle, midway, anchor=south, xshift=-5pt, yshift=5pt] {$r_2$};

\draw[line] (0.99,+0.95) -- (0.99,-0.95) node[labelstyle, midway, anchor=north, xshift=15pt, yshift=20pt] {$r_{12}$};

% Draw curved lines to indicate angles
\draw[thick, dashed, <->] (0.5,0) arc[start angle=0, end angle=-45, radius=0.5];
\draw[thick, dashed, <->] (0.5,0) arc[start angle=0, end angle=45, radius=0.5];

% Label the angles
\node[labelstyle] at (0.6,-0.27) {$\frac{\theta}{2}$};
\node[labelstyle] at (0.6,0.27) {$\frac{\theta}{2}$};

% Draw the circle for m3 AFTER the vectors to cover the inner portions
\draw[particle] (0,0) circle (\particleradius);

\end{tikzpicture}

%% file: biblio_rmp.bib
@article{Hylleraas1929,
author={Hylleraas, E A},
title={Neue Berechnung der Energie des Heliums im Grundzustande, sowie des tiefsten Terms von Ortho-Helium},
journal={Z. Phys.},
year={1929},
month={May},
day={01},
volume={54},
number={5},
pages={347-366},
issn={0044-3328},
doi={10.1007/BF01375457},
url={https://doi.org/10.1007/BF01375457}
}

@article{Breit1930,
  title = {Separation of Angles in the Two-Electron Problem},
  author = {Breit, G},
  journal = {Phys. Rev.},
  volume = {35},
  issue = {6},
  pages = {569--578},
  numpages = {0},
  year = {1930},
  month = {Mar},
  publisher = {American Physical Society},
  doi = {10.1103/PhysRev.35.569},
  url = {https://link.aps.org/doi/10.1103/PhysRev.35.569}
}

@misc{codata2022,
  author = {Peter, J M and Eite, T and David, B N and Barry, N T},
  title = {Codata Internationally Recommended 2022 Values of the Fundamental Physical Constants},
  year = {2024},
  month = {2024-05-08 04:05:00},
  publisher = {Codata Internationally Recommended 2022 Values of the Fundamental Physical Constants},
  url = {https://tsapps.nist.gov/publication/get_pdf.cfm?pub_id=958002}
}

@article{hirschfelder1935,
  title={Separation of Rotational Co{\"o}rdinates from the {Schr{\"o}dinger} Equation for ${N}$ Particles},
  author={Hirschfelder, J O and Wigner, E},
  journal={Proc. Natl. Acad. Sci.},
  volume={21},
  number={2},
  pages={113--119},
  year={1935},
  publisher={National Acad Sciences},
  doi = {10.1073/pnas.21.2.113},
  url = {https://doi.org/10.1073/pnas.21.2.113}
}

@article{BHATIA1964,
  title = {Symmetric {Euler}-Angle Decomposition of the Two-Electron Fixed-Nucleus Problem},
  author = {Bhatia, A K and Temkin, A},
  journal = {Rev. Mod. Phys.},
  volume = {36},
  issue = {4},
  pages = {1050--1064},
  numpages = {0},
  year = {1964},
  month = {Oct},
  publisher = {American Physical Society},
  doi = {10.1103/RevModPhys.36.1050},
  url = {https://link.aps.org/doi/10.1103/RevModPhys.36.1050}
}

@article{BHATIA1965,
  title = {Decomposition of the {Schr\"odinger} Equation for Two Identical Particles and a Third Particle of Finite Mass},
  author = {Bhatia, A K and Temkin, A},
  journal = {Phys. Rev.},
  volume = {137},
  issue = {5A},
  pages = {A1335--A1343},
  numpages = {0},
  year = {1965},
  month = {Mar},
  publisher = {American Physical Society},
  doi = {10.1103/PhysRev.137.A1335},
  url = {https://link.aps.org/doi/10.1103/PhysRev.137.A1335}
}

@article{Pont1995,
  title = {Decomposition of the two-electron-atom eigenvalue problem},
  author = {Pont, M and Shakeshaft, R},
  journal = {Phys. Rev. A},
  volume = {51},
  issue = {1},
  pages = {257--265},
  numpages = {0},
  year = {1995},
  month = {Jan},
  publisher = {American Physical Society},
  doi = {10.1103/PhysRevA.51.257},
  url = {https://link.aps.org/doi/10.1103/PhysRevA.51.257}
}

@book{wigner2013group,
  title={Group Theory: And Its Application to the Quantum Mechanics of Atomic Spectra},
  author={Wigner, E P},
  isbn={9781483275765},
  url={https://books.google.com.tw/books?id=UITNCgAAQBAJ},
  year={2013},
  publisher={Elsevier Science}
}

@article{Schwartz1961,
  title = {Lamb Shift in the Helium Atom},
  author = {Schwartz, C},
  journal = {Phys. Rev.},
  volume = {123},
  issue = {5},
  pages = {1700--1705},
  numpages = {0},
  year = {1961},
  month = {Sep},
  publisher = {American Physical Society},
  doi = {10.1103/PhysRev.123.1700},
  url = {https://link.aps.org/doi/10.1103/PhysRev.123.1700}
}

@book{Varshalovich1988,
author = {Varshalovich, D A and Moskalev, A N and Khersonskii, V K},
title = {Quantum Theory of Angular Momentum},
publisher = {World Scientific},
year = {1988},
doi = {10.1142/0270},
address = {},
edition   = {},
URL = {https://www.worldscientific.com/doi/abs/10.1142/0270}}

@article{Drake1990,
  title = {Variational eigenvalues for the {Rydberg} states of helium: {Comparison} with experiment and with asymptotic expansions},
  author = {Drake, G W F},
  journal = {Phys. Rev. Lett.},
  volume = {65},
  issue = {22},
  pages = {2769--2772},
  numpages = {0},
  year = {1990},
  month = {Nov},
  publisher = {American Physical Society},
  doi = {10.1103/PhysRevLett.65.2769},
  url = {https://link.aps.org/doi/10.1103/PhysRevLett.65.2769}
}

@article{Frolov1996,
  title = {Bound states with arbitrary angular momenta in nonrelativistic three-body systems},
  author = {Frolov, A M and Smith, Jr., V H},
  journal = {Phys. Rev. A},
  volume = {53},
  issue = {6},
  pages = {3853--3864},
  numpages = {0},
  year = {1996},
  month = {Jun},
  publisher = {American Physical Society},
  doi = {10.1103/PhysRevA.53.3853},
  url = {https://link.aps.org/doi/10.1103/PhysRevA.53.3853}
}

@article{Frolov1987,
author = {Frolov, A M},
year = {1987},
month = {06},
pages = {1100-1110},
title = {Variational expansions for three-body {Coulomb} problem},
volume = {65},
journal = {Sov. Phys. JETP},
url = {}
}

@article{Efros1986,
author = {\'{E}fros, V},
year = {1986},
month = {01},
pages = {5},
title = {The three-body problem: {Generalized} exponential expansion; {Arbitrary} states in a correlated basis, and the binding energy of muonic molecules},
volume = {63},
journal = {Sov. Phys. JETP},
url = {http://www.jetp.ras.ru/cgi-bin/e/index/e/63/1/p5?a=list}
}

@article{MEREMIANIN2003,
title = {The irreducible tensor approach in the separation of collective angles in the quantum ${N}$-body problem},
journal = {Phys. Rep.},
volume = {384},
number = {4},
pages = {121-195},
year = {2003},
issn = {0370-1573},
doi = {https://doi.org/10.1016/S0370-1573(03)00262-X},
url = {https://www.sciencedirect.com/science/article/pii/S037015730300262X},
author = {Meremianin, A V and Briggs, J S}
}

@article{King1967,
    author = {King, H F},
    title = {Some Theorems Concerning Symmetry, Angular Momentum, and Completeness of Atomic Geminals with Explicit {$r_{12}$} Dependence},
    journal = {J. Chem. Phys.},
    volume = {46},
    number = {2},
    pages = {705-713},
    year = {1967},
    month = {01},
    issn = {0021-9606},
    doi = {10.1063/1.1840730},
    url = {https://doi.org/10.1063/1.1840730}
}

@article{Manakov1998,
  title = {Invariant representations of finite rotation matrices and some applications},
  author = {Manakov, N L and Meremianin, A V and Starace, A F},
  journal = {Phys. Rev. A},
  volume = {57},
  issue = {5},
  pages = {3233--3244},
  numpages = {0},
  year = {1998},
  month = {May},
  publisher = {American Physical Society},
  doi = {10.1103/PhysRevA.57.3233},
  url = {https://link.aps.org/doi/10.1103/PhysRevA.57.3233}
}

@article{Drake1987,
  title = {New variational techniques for the $1snd$ states of helium},
  author = {Drake, G W F},
  journal = {Phys. Rev. Lett.},
  volume = {59},
  issue = {14},
  pages = {1549--1552},
  numpages = {0},
  year = {1987},
  month = {Oct},
  publisher = {American Physical Society},
  doi = {10.1103/PhysRevLett.59.1549},
  url = {https://link.aps.org/doi/10.1103/PhysRevLett.59.1549}
}

@article{Calais1962,
title = {A simple method of treating atomic integrals containing functions of $r_{12}$},
journal = {J. Mol. Spectr.},
volume = {8},
number = {1},
pages = {203-211},
year = {1962},
issn = {0022-2852},
doi = {https://doi.org/10.1016/0022-2852(62)90021-8},
url = {https://www.sciencedirect.com/science/article/pii/0022285262900218},
author = {Calais, J -L and Löwdin, P -O}
}

@article{Drake1978,
  title = {Angular integrals and radial recurrence relations for two-electron matrix elements in {Hylleraas} coordinates},
  author = {Drake, G W F},
  journal = {Phys. Rev. A},
  volume = {18},
  issue = {3},
  pages = {820--826},
  numpages = {0},
  year = {1978},
  month = {Sep},
  publisher = {American Physical Society},
  doi = {10.1103/PhysRevA.18.820},
  url = {https://link.aps.org/doi/10.1103/PhysRevA.18.820}
}

@article{Shimpuku1963,
title = {On the expressions of {Clebsch}-{Gordan} coefficients},
journal = {J. Math. Anal. Appl.},
volume = {7},
number = {3},
pages = {397-419},
year = {1963},
issn = {0022-247X},
doi = {https://doi.org/10.1016/0022-247X(63)90060-X},
url = {https://www.sciencedirect.com/science/article/pii/0022247X6390060X},
author = {Shimpuku, T}
}

@article{Mukherjee1994,
  title = {Variational equation of states of arbitrary angular momentum for two-particle systems},
  author = {Mukherjee, T K and Mukherjee, P K},
  journal = {Phys. Rev. A},
  volume = {50},
  issue = {1},
  pages = {850--853},
  numpages = {0},
  year = {1994},
  month = {Jul},
  publisher = {American Physical Society},
  doi = {10.1103/PhysRevA.50.850},
  url = {https://link.aps.org/doi/10.1103/PhysRevA.50.850}
}

@article{Pestka2008,
doi = {10.1088/1751-8113/41/23/235202},
url = {https://dx.doi.org/10.1088/1751-8113/41/23/235202},
year = {2008},
month = {may},
publisher = {},
volume = {41},
number = {23},
pages = {235202},
author = {Pestka, G},
title = {The {Schrödinger} equation for a spherical two-particle system in $r_1$, $r_2$, $r_{12}$ variables},
journal = {J. Phys. A: Math. Theor.}
}

@article{Mukherjee1995,
  title = {Variational equation of states of arbitrary angular momenta for three-particle systems},
  author = {Mukherjee, T K and Mukherjee, P K},
  journal = {Phys. Rev. A},
  volume = {51},
  issue = {5},
  pages = {4276--4278},
  numpages = {0},
  year = {1995},
  month = {May},
  publisher = {American Physical Society},
  doi = {10.1103/PhysRevA.51.4276},
  url = {https://link.aps.org/doi/10.1103/PhysRevA.51.4276}
}

@article{Yerokhin2021,
  title = {Atomic Structure Calculations of Helium with Correlated Exponential Functions},
  author = {Yerokhin, V A and Patkóš, V and Pachucki, K},
  journal = {Symmetry},
  volume = {13},
  issue = {7},
  pages = {1246},
  numpages = {},
  year = {2021},
  month = {Jul},
  publisher = {},
  doi = {10.3390/sym13071246},
  url = {https://doi.org/10.3390/sym13071246}
}

@article{Aznabaev2018,
  title = {Nonrelativistic energy levels of helium atoms},
  author = {Aznabaev, D T and Bekbaev, A K and Korobov, V I},
  journal = {Phys. Rev. A},
  volume = {98},
  issue = {1},
  pages = {012510},
  numpages = {7},
  year = {2018},
  month = {Jul},
  publisher = {American Physical Society},
  doi = {10.1103/PhysRevA.98.012510},
  url = {https://link.aps.org/doi/10.1103/PhysRevA.98.012510}
}

@article{HU2016,
title = {Variational calculation of energy levels for metastable states of antiprotonic helium},
journal = {Chem. Phys. Lett.},
volume = {654},
pages = {114-118},
year = {2016},
issn = {0009-2614},
doi = {https://doi.org/10.1016/j.cplett.2016.05.017},
url = {https://www.sciencedirect.com/science/article/pii/S0009261416303074},
author = {Hu, M -H and Yao, S -M and Wang, Y and Li, W and Gu, Y -Y and Zhong, Z -X}
}

@article{Kar2009,
author={Kar, S and Ho, Y K},
title={D-Wave Resonances in Three-Body System {Ps$^-$} with Pure {Coulomb} and Screened {Coulomb} ({Yukawa}) Potentials},
journal={Few-Body Syst.},
year={2009},
month={Apr},
day={01},
volume={45},
number={1},
pages={43-49},
doi={10.1007/s00601-008-0007-2},
url={https://doi.org/10.1007/s00601-008-0007-2}
}

@Article{Kalotas1965,
author={Kalotas, T M},
title={A Symmetric Decomposition of the Two-Electron Wave Equation for Finite Nuclear Mass},
journal={Aust. J. Phys.},
year={1965},
volume={18},
number={6},
pages={511-520},
doi={10.1071/PH650511},
url={https://doi.org/10.1071/PH650511}
}

@article{Bottcher1994,
  title = {Correlated two-electron wave functions of any symmetry},
  author = {Bottcher, C and Schultz, D R and Madison, D H},
  journal = {Phys. Rev. A},
  volume = {49},
  issue = {3},
  pages = {1714--1723},
  numpages = {0},
  year = {1994},
  month = {Mar},
  publisher = {American Physical Society},
  doi = {10.1103/PhysRevA.49.1714},
  url = {https://link.aps.org/doi/10.1103/PhysRevA.49.1714}
}

@article{Nelder1965,
    author = {Nelder, J A and Mead, R},
    title = {A Simplex Method for Function Minimization},
    journal = {Comput. J.},
    volume = {7},
    number = {4},
    pages = {308-313},
    year = {1965},
    month = {01},
    issn = {0010-4620},
    doi = {10.1093/comjnl/7.4.308},
    url = {https://doi.org/10.1093/comjnl/7.4.308}
}

@article{Drake1992,
  title = {Energies and relativistic corrections for the {Rydberg} states of helium: Variational results and asymptotic analysis},
  author = {Drake, G W F and Yan, Z -C},
  journal = {Phys. Rev. A},
  volume = {46},
  issue = {5},
  pages = {2378--2409},
  numpages = {0},
  year = {1992},
  month = {Sep},
  publisher = {American Physical Society},
  doi = {10.1103/PhysRevA.46.2378},
  url = {https://link.aps.org/doi/10.1103/PhysRevA.46.2378}
}

@article{Harris2004,
title = {Current Methods for {Coulomb} Few-Body Problems},
journal =  {Adv. Quantum Chem.},
publisher = {Academic Press},
volume = {47},
pages = {129-155},
year = {2004},
doi = {https://doi.org/10.1016/S0065-3276(04)47008-7},
url = {https://www.sciencedirect.com/science/article/pii/S0065327604470087},
author = {Harris, F E}
}

@article{Hsiang1997,
author = {Hsiang, W -Y},
title = {Kinematic geometry of mass-triangles and reduction of {Schr\"{o}dinger’s} equation of three-body systems to partial differential equations solely defined on triangular parameters},
journal = {Proc. Natl. Acad. Sci.},
volume = {94},
number = {17},
pages = {8936-8938},
year = {1997},
doi = {10.1073/pnas.94.17.8936},
URL = {https://www.pnas.org/doi/abs/10.1073/pnas.94.17.8936}}

@article{Gu2001,
author = {Gu, X -Y and Duan, B and Ma, Z -Q},
title = {SEPARATION OF ROTATIONAL DEGREES OF FREEDOM IN A QUANTUM THREE-BODY PROBLEM},
journal = {Int. J. Mod. Phys. E},
volume = {10},
number = {01},
pages = {69-82},
year = {2001},
doi = {10.1142/S0218301301000411},
URL = {https://doi.org/10.1142/S0218301301000411}
}

@article{Ma2000,
author={Ma, Z -Q},
title={Quantum three-body problems},
journal={Sci. China Ser. A: Math},
year={2000},
month={Oct},
day={01},
volume={43},
number={10},
pages={1093-1107},
issn={1862-2763},
doi={10.1007/BF02898245},
url={https://doi.org/10.1007/BF02898245}
}

@article{Hsiang2007,
author = {Hsiang, W -Y and Straume, E},
journal = {Lobachevskii J. Math.},
pages = {9-130},
publisher = {Maik Nauka Interperiodica (Pleiades Publishing); Springer, New York},
title = {Kinematic geometry of triangles and the study of the three-body problem.},
url = {http://eudml.org/doc/228460},
volume = {25},
year = {2007},
}

@article{Chi2007,
title = {Kinetic energy operator approach to the quantum three-body problem with {Coulomb} interactions},
journal = {Solid State Commun.},
volume = {141},
number = {4},
pages = {173-177},
year = {2007},
issn = {0038-1098},
doi = {https://doi.org/10.1016/j.ssc.2006.10.031},
url = {https://www.sciencedirect.com/science/article/pii/S0038109806009641},
author = {Chi, X and Fang, A and Hsiang, W and Sheng, P}
}

@article{Hughes1930,
  title = {The Effect of the Motion of the Nucleus on the Spectra of {Li I} and {Li II}},
  author = {Hughes, D S and Eckart, C},
  journal = {Phys. Rev.},
  volume = {36},
  issue = {4},
  pages = {694--698},
  numpages = {0},
  year = {1930},
  month = {Aug},
  publisher = {American Physical Society},
  doi = {10.1103/PhysRev.36.694},
  url = {https://link.aps.org/doi/10.1103/PhysRev.36.694}
}

@article{Hylleraas1928,
author={Hylleraas, E A},
title={\"{U}ber den Grundzustand des Heliumatoms},
journal={Z. Phys.},
year={1928},
month={Jul},
day={01},
volume={48},
number={7},
pages={469-494},
issn={0044-3328},
doi={10.1007/BF01340013},
url={https://doi.org/10.1007/BF01340013}
}

@article{Ma1999,
author={Ma, Z -Q},
title={Exact Solution to the {Schr{\"o}dinger} Equation for the Quantum Rigid Body},
journal={Found. Phys. Lett.},
year={1999},
month={Dec},
day={01},
volume={12},
number={6},
pages={561-570},
issn={1572-9524},
doi={10.1023/A:1021647209315},
url={https://doi.org/10.1023/A:1021647209315}
}

@article{Jackson1954, 
title={Separation of angle variables for helium}, 
volume={50}, 
DOI={10.1017/S0305004100029364}, 
number={2}, 
journal={Math. Proc. Camb. Philos. Soc.}, 
author={Jackson, T A S}, 
year={1954}, 
pages={298–304}}

@article{Gao2012,
author={Gao, F and Han, L},
title={Implementing the {Nelder-Mead} simplex algorithm with adaptive parameters},
journal={Comput. Optim. Appl},
year={2012},
month={Jan},
day={01},
volume={51},
number={1},
pages={259-277},
issn={1573-2894},
doi={10.1007/s10589-010-9329-3},
url={https://doi.org/10.1007/s10589-010-9329-3}
}

@article{Fock1954,
author={Fock, V A},
title={The {Sch\"{o}dinger} equation for helium atom},
journal={Izv. Akad. SSSR, Ser. Fiz.},
year={1954},
volume={18},
pages={161}
}

@phdthesis{Langner2022,
  author       = {J Langner},
  title        = {Towards an exact solution to the {Schr\"{o}dinger} equation of helium atom},
  school       = {National Yang Ming Chiao Tung University},
  year         = {2022},
  address      = {Hsinchu, Taiwan},
  type         = {{Ph.D.} thesis}}

@article{Bartlett1955,
  title = {Helium Wave Equation},
  author = {Bartlett, J H},
  journal = {Phys. Rev.},
  volume = {98},
  issue = {4},
  pages = {1067--1070},
  numpages = {0},
  year = {1955},
  month = {May},
  publisher = {American Physical Society},
  doi = {10.1103/PhysRev.98.1067},
  url = {https://link.aps.org/doi/10.1103/PhysRev.98.1067}
}

@article{Frost1964a,
    author = {Frost, A A},
    title = {Approximate Series Solutions of Nonseparable {Schr\"{o}dinger} Equations. {I}. {Mathematical} Formulation},
    journal = {J. Chem. Phys.},
    volume = {41},
    number = {2},
    pages = {478-482},
    year = {1964},
    month = {07},
    issn = {0021-9606},
    doi = {10.1063/1.1725893},
    url = {https://doi.org/10.1063/1.1725893}
}

@article{Frost1964b,
    author = {Frost, A A and Inokuti, M and Lowe, J P},
    title = {Approximate Series Solutions of Nonseparable {Schr\"{o}dinger} Equations. {II}. {General} Three-particle System with {Coulomb} Interaction},
    journal = {J. Chem. Phys.},
    volume = {41},
    number = {2},
    pages = {482-489},
    year = {1964},
    month = {07},
    issn = {0021-9606},
    doi = {10.1063/1.1725894},
    url = {https://doi.org/10.1063/1.1725894}
}

@article{Frost1964c,
    author = {Frost, A A and Harriss, D K and Scargle, J D},
    title = {Approximate Series Solutions of Nonseparable {Schr\"{o}dinger} Equations. {III}. {B} Matrix Method},
    journal = {J. Chem. Phys.},
    volume = {41},
    number = {2},
    pages = {489-494},
    year = {1964},
    month = {07},
    issn = {0021-9606},
    doi = {10.1063/1.1725895},
    url = {https://doi.org/10.1063/1.1725895}
}

@article{White1970,
    author = {White, R J and Stillinger Jr, F H},
    title = {Analytic Approach to Electron Correlation in Atoms},
    journal = {J. Chem. Phys.},
    volume = {52},
    number = {11},
    pages = {5800-5814},
    year = {1970},
    month = {06},
    issn = {0021-9606},
    doi = {10.1063/1.1672862},
    url = {https://doi.org/10.1063/1.1672862}
}

@article{Haftel1983,
title = {Exact solution of coupled equations and the hyperspherical formalism: Calculation of expectation values and wavefunctions of three {Coulomb-bound} particles},
journal = {Ann. Phys.},
volume = {150},
number = {1},
pages = {48-91},
year = {1983},
issn = {0003-4916},
doi = {https://doi.org/10.1016/0003-4916(83)90004-0},
url = {https://www.sciencedirect.com/science/article/pii/0003491683900040},
author = {Haftel, M I and Mandelzweig, V B}
}

@article{Liverts2010,
title = {The two-electron atomic systems. {S}-states},
journal = {Comput. Phys. Commun.},
volume = {181},
number = {1},
pages = {206-212},
year = {2010},
issn = {0010-4655},
doi = {https://doi.org/10.1016/j.cpc.2009.09.012},
url = {https://www.sciencedirect.com/science/article/pii/S0010465509003026},
author = {Liverts, E Z and Barnea, N}
}

@article{Liverts2013,
title = {Three-body systems with {Coulomb} interaction. {Bound} and quasi-bound {S-states}},
journal = {Comput. Phys. Commun.},
volume = {184},
number = {11},
pages = {2596-2603},
year = {2013},
issn = {0010-4655},
doi = {https://doi.org/10.1016/j.cpc.2013.06.013},
url = {https://www.sciencedirect.com/science/article/pii/S0010465513002075},
author = {Liverts, E Z and Barnea, N}
}

@Article{DattaMajumdar1952,
author={Datta Majumdar, S},
title={The problem of three bodies in quantum mechanics},
journal={Z. Phys.},
year={1952},
month={Dec},
day={01},
volume={131},
number={4},
pages={528-537},
issn={0044-3328},
doi={10.1007/BF01333405},
url={https://doi.org/10.1007/BF01333405}
}

@Inbook{Drake1993,
author="Drake, G W F",
editor="Levin, F S
and Micha, D A",
chapter="High-Precision Calculations for the {Rydberg} States of Helium",
bookTitle="Long-Range Casimir Forces: Theory and Recent Experiments on Atomic Systems",
year="1993",
publisher="Springer US",
address="Boston, MA",
pages="107--217",
isbn="978-1-4899-1228-2",
doi="10.1007/978-1-4899-1228-2_3",
url="https://doi.org/10.1007/978-1-4899-1228-2_3"
}

@article{julia_bezanson_2017,
  title={Julia: A fresh approach to numerical computing},
  author={Bezanson, J and Edelman, A and Karpinski, S and Shah, V B},
  journal={SIAM review},
  volume={59},
  number={1},
  pages={65--98},
  year={2017},
  publisher={SIAM},
  url={https://doi.org/10.1137/141000671}
}

@Inbook{Landau1977ch3,
chapter = {{III} {Schr\"{o}dinger's} equation},
author = {Landau, L D and Lifshitz, E M},
booktitle = {Quantum Mechanics},
publisher = {Pergamon},
edition = {{Third}},
pages = {50-81},
year = {1977},
isbn = {978-0-08-020940-1},
doi = {https://doi.org/10.1016/B978-0-08-020940-1.50010-4},
url = {https://www.sciencedirect.com/science/article/pii/B9780080209401500104}
}

@article{Iwai1987,
    author = {Iwai, T},
    title = {A geometric setting for internal motions of the quantum three-body system},
    journal = {J. Math. Phys.},
    volume = {28},
    number = {6},
    pages = {1315-1326},
    year = {1987},
    month = {06},
    issn = {0022-2488},
    doi = {10.1063/1.527534},
    url = {https://doi.org/10.1063/1.527534}
}

@article{Kemeny1964,
    author = {Kemeny, G and Walsh, P J},
    title = {Some Mathematical Generalizations of the {Schr\"{o}dinger} Equation for Two-electron Atoms},
    journal = {J. Chem. Phys.},
    volume = {41},
    number = {1},
    pages = {81-84},
    year = {1964},
    month = {07},
    issn = {0021-9606},
    doi = {10.1063/1.1725654},
    url = {https://doi.org/10.1063/1.1725654}
}

@article{Majumdar1964,
author={Datta Majumdar, S},
title={Wave equations in momentum space},
journal={Acta Phys. Hung. Tom. XVIII Fasc. I.},
year={1964},
month={Oct},
day={01},
volume={18},
number={1},
pages={19-26},
issn={0001-6705},
doi={10.1007/BF03158615},
url={https://doi.org/10.1007/BF03158615}
}

@article{Gu2001a,
  title = {Independent eigenstates of angular momentum in a quantum {$N$}-body system},
  author = {Gu, X -Y and Duan, B and Ma, Z -Q},
  journal = {Phys. Rev. A},
  volume = {64},
  issue = {4},
  pages = {042108},
  numpages = {14},
  year = {2001},
  month = {Sep},
  publisher = {American Physical Society},
  doi = {10.1103/PhysRevA.64.042108},
  url = {https://link.aps.org/doi/10.1103/PhysRevA.64.042108}
}

@article{Proriol1967,
title = {Euler angles for three and four identical particle systems: {(II)}. {Relations} between two sets of angular momentum functions},
journal = {Nucl. Phys. A},
volume = {98},
number = {1},
pages = {196-205},
year = {1967},
issn = {0375-9474},
doi = {https://doi.org/10.1016/0375-9474(67)90910-4},
url = {https://www.sciencedirect.com/science/article/pii/0375947467909104},
author = {Proriol, J}
}

@book{slater1966,
  title={Generalized Hypergeometric Functions},
  author={Slater, L J},
  isbn={9780521090612},
  lccn={66010050},
  url={https://books.google.com.tw/books?id=Hq0NAQAAIAAJ},
  year={1966},
  publisher={Cambridge University Press}
}

@book{Biedenharn1984, 
place={Cambridge}, 
series={Encyclopedia of Mathematics and its Applications}, 
title={Angular Momentum in Quantum Physics: Theory and Application}, publisher={Cambridge University Press}, 
author={Biedenharn, L C and Louck, J D}, 
year={1984}, 
volume={8},
url={https://doi.org/10.1017/CBO9780511759888},
collection={Encyclopedia of Mathematics and its Applications}
}

@book{Edmonds1974, 
place={Princeton}, 
title={Angular Momentum in Quantum Mechanics}, 
publisher={Princeton University Press}, 
author={Edmonds, A R}, 
year={1974}, 
url={}
}

@article{Manakov1996,
doi = {10.1088/0953-4075/29/13/010},
url = {https://iopscience.iop.org/article/10.1088/0953-4075/29/13/010},
year = {1996},
volume = {29},
pages = {2711-2737},
author = {Manakov, N L and Marmo, S I and Meremianin, A V},
title = {A new technique in the theory of angular distributions in atomic processes: {The} angular distribution of photoelectrons in single and double photoionization},
journal = {J. Phys. B: At. Mol. Opt. Phys.}
}

@article{Nakashima2007,
    author = {Nakashima, H and Nakatsuji, H},
    title = {Solving the {Schr\"{o}dinger} equation for helium atom and its isoelectronic ions with the free iterative complement interaction ({ICI}) method},
    journal = {J. Chem. Phys.},
    volume = {127},
    number = {22},
    pages = {224104},
    year = {2007},
    month = {12},
    doi = {10.1063/1.2801981},
    url = {https://doi.org/10.1063/1.2801981}
}

@article{Gronwall1932,
 URL = {http://www.jstor.org/stable/1968330},
 author = {Gronwall, T H},
 journal = {Ann. Math.},
 number = {2},
 pages = {279--293},
 publisher = {[Annals of Mathematics, Trustees of Princeton University on Behalf of the Annals of Mathematics, Mathematics Department, Princeton University]},
 title = {A Special Conformally {Euclidean} Space of Three Dimensions Occurring in Wave Mechanics},
 urldate = {2025-03-26},
 volume = {33},
 year = {1932}
}

@Article{Morgan1986,
author={Morgan, J D},
title={Convergence properties of {Fock's} expansion for {S}-state eigenfunctions of the helium atom},
journal={Theor. Chim. Acta.},
year={1986},
month={Apr},
day={01},
volume={69},
number={3},
pages={181-223},
doi={10.1007/BF00526420},
url={https://doi.org/10.1007/BF00526420}
}

@article{Fock1958,
  title={On the {Schr\"{o}dinger} equation of helium atom {I}},
  author={Fock, V A},
  journal={D. Kngl. Norske Videnskab. Selsk. Forh.},
  volume={31},
  number={},
  pages={138},
  year={1958}
}

@article{Fock1958a,
  title={On the {Schr\"{o}dinger} equation of helium atom {II}},
  author={Fock, V A},
  journal={D. Kngl. Norske Videnskab. Selsk. Forh.},
  volume={31},
  number={},
  pages={145},
  year={1958}
}

@article{Bartlett1937,
  title = {The Helium Wave Equation},
  author = {Bartlett, J H},
  journal = {Phys. Rev.},
  volume = {51},
  issue = {8},
  pages = {661--669},
  numpages = {0},
  year = {1937},
  month = {Apr},
  publisher = {American Physical Society},
  doi = {10.1103/PhysRev.51.661},
  url = {https://link.aps.org/doi/10.1103/PhysRev.51.661}
}

@article{Pluvinage1955,
  title = {Approximations systématiques dans la résolution de l'équation de {Schrödinger} des atomes à deux électrons. {I}. {Principe} de la méthode. {Etats} {S} symétriques},
  author = {Pluvinage, P},
  journal = {J. Physique Rad.},
  volume = {16},
  issue = {8-9},
  pages = {675-680},
  year = {1955},
  month = {September},
  doi = {10.1051/jphysrad:01955001608-9067500},
  url = {https://doi.org/10.1051/jphysrad:01955001608-9067500}
}

@article{Kinoshita1957,
  title = {Ground State of the Helium Atom},
  author = {Kinoshita, T},
  journal = {Phys. Rev.},
  volume = {105},
  issue = {5},
  pages = {1490--1502},
  numpages = {0},
  year = {1957},
  month = {Mar},
  publisher = {American Physical Society},
  doi = {10.1103/PhysRev.105.1490},
  url = {https://link.aps.org/doi/10.1103/PhysRev.105.1490}
}

@article{Pekeris1958,
  title = {Ground State of Two-Electron Atoms},
  author = {Pekeris, C L},
  journal = {Phys. Rev.},
  volume = {112},
  issue = {5},
  pages = {1649--1658},
  numpages = {0},
  year = {1958},
  month = {Dec},
  publisher = {American Physical Society},
  doi = {10.1103/PhysRev.112.1649},
  url = {https://link.aps.org/doi/10.1103/PhysRev.112.1649}
}

@article{Abbott1987,
doi = {10.1088/0305-4470/20/8/023},
url = {https://dx.doi.org/10.1088/0305-4470/20/8/023},
year = {1987},
month = {jun},
volume = {20},
number = {8},
pages = {2043},
author = {Abbott, P C and Maslen, E N},
title = {Coordinate systems and analytic expansions for three-body atomic wavefunctions. {I}. {Partial} summation for the {Fock} expansion in hyperspherical coordinates},
journal = {J. Phys. A: Math. Gen.}
}

@article{Gottschalk1987a,
doi = {10.1088/0305-4470/20/8/024},
url = {https://dx.doi.org/10.1088/0305-4470/20/8/024},
year = {1987},
month = {jun},
publisher = {},
volume = {20},
number = {8},
pages = {2077},
author = {Gottschalk, J E and Abbott, P C and Maslen, E N},
title = {Coordinate systems and analytic expansions for three-body atomic wavefunctions. {II}. {Closed} form wavefunction to second order in $r$},
journal = {J. Phys. A: Math. Gen.}
}

@article{Gottschalk1987b,
doi = {10.1088/0305-4470/20/10/022},
url = {https://dx.doi.org/10.1088/0305-4470/20/10/022},
year = {1987},
month = {jul},
publisher = {},
volume = {20},
number = {10},
pages = {2781},
author = {Gottschalk, J E and Maslen, E N},
title = {Coordinate systems and analytic expansions for three-body atomic wavefunctions. {III}. {Derivative} continuity via solutions to {Laplace's} equation},
journal = {J. Phys. A: Math. Gen.}
}

@article{He2016,
author = {He, B-H and Witek, H A},
title = {Toward Exact Analytical Wave Function of Helium Atom: Two Techniques for Constructing Homogeneous Functions of Kinetic Energy Operator},
journal = {J. Chin. Chem. Soc.},
volume = {63},
number = {1},
pages = {69-82},
doi = {https://doi.org/10.1002/jccs.201500086},
url = {https://onlinelibrary.wiley.com/doi/abs/10.1002/jccs.201500086},
year = {2016}
}

@article{Liverts2018,
doi = {10.1088/1751-8121/aaa2ce},
url = {https://dx.doi.org/10.1088/1751-8121/aaa2ce},
year = {2018},
month = {jan},
publisher = {IOP Publishing},
volume = {51},
number = {8},
pages = {085204},
author = {Liverts, E Z and Barnea, N},
title = {The {Green’s} function approach to the {Fock} expansion calculations of two-electron atoms},
journal = {J. Phys. A: Math. Theor.}
}

@article{Liverts2022,
doi = {10.3390/atoms10010004},
url = {https://doi.org/10.3390/atoms10010004},
year = {2022},
month = {},
publisher = {},
volume = {10},
number = {4},
pages = {1-11},
author = {Liverts, E Z and Barnea, N},
title = {Accurate Exponential Representations for the Ground State Wave Functions of the Collinear Two-Electron Atomic Systems},
journal = {Atoms}
}

@article{Hylleraas1960,
author={Hylleraas, E A},
title={On the formal solution of the two-electron wave equation},
journal={Phys. Math. Univ. Osloensis Inst. Rep.},
year={1960},
volume={6},
number={},
pages={272},
doi={},
url={}
}

@article{Dmitrieva1986,
author = {Dmitrieva, I K and Plindov, G I},
title = {{Schrödinger} equation for two-electron atomic states with conserved angular momentum and parity},
DOI= {10.1051/jphys:019860047090149300},
url= {https://doi.org/10.1051/jphys:019860047090149300},
journal = {J. Phys. France},
year = {1986},
volume = {47},
number = {9},
pages = {1493-1501},
}

@misc{supplemental,
  note = {See Supplemental Material at [URL will be inserted by publisher] for additional derivations.},
  year = {2025},
  author = {Sadhukhan, A and Pestka, G and Podeszwa, R and Witek, H A},
  url   = {}
}

@misc{wolfram,
  author = {{Wolfram Research}},
  title  = {Home page of functions.wolfram.com, Hypergeometric Functions, formula identification number: 07.23.17.0048.01},
  year   = {2001},
  url    = {http://functions.wolfram.com/07.23.17.0048.01}
}

@article{Traub1959,
  title = {Variational Calculations of Energy and Fine Structure for the $2^{3}${P} State of Helium},
  author = {Traub, J and Foley, H M},
  journal = {Phys. Rev.},
  volume = {116},
  issue = {4},
  pages = {914--919},
  numpages = {0},
  year = {1959},
  month = {Nov},
  publisher = {American Physical Society},
  doi = {10.1103/PhysRev.116.914},
  url = {https://link.aps.org/doi/10.1103/PhysRev.116.914}
}

@article{Pekeris1962a,
  title = {Fine Structure of the $2^{3}${P} and $3^{3}${P} States of Helium},
  author = {Pekeris, C L and Schiff, B and Lifson, H},
  journal = {Phys. Rev.},
  volume = {126},
  issue = {3},
  pages = {1057--1058},
  numpages = {0},
  year = {1962},
  month = {May},
  publisher = {American Physical Society},
  doi = {10.1103/PhysRev.126.1057},
  url = {https://link.aps.org/doi/10.1103/PhysRev.126.1057}
}

@article{Breit1930c,
  title = {The Fine Structure of {He} as a Test of the Spin Interactions of Two Electrons},
  author = {Breit, G},
  journal = {Phys. Rev.},
  volume = {36},
  issue = {3},
  pages = {383--397},
  numpages = {0},
  year = {1930},
  month = {Aug},
  publisher = {American Physical Society},
  doi = {10.1103/PhysRev.36.383},
  url = {https://link.aps.org/doi/10.1103/PhysRev.36.383}
}

@article{Perrin1963,
doi = {10.1088/0370-1328/81/1/307},
url = {https://dx.doi.org/10.1088/0370-1328/81/1/307},
year = {1963},
month = {jan},
publisher = {},
volume = {81},
number = {1},
pages = {28},
author = {Perrin, R and Stewart, A L},
title = {The Correlation Energies of Excited States of the Helium Sequence},
journal = {Proc. Phys. Soc.}
}

@article{Machacek1964,
  title = {Finite-Mass Helium Atoms. {I}. {The} {$2^{1}$P} State},
  author = {Machacek, M and Sanders, F C and Scherr, C W},
  journal = {Phys. Rev.},
  volume = {136},
  issue = {3A},
  pages = {A680--A683},
  numpages = {0},
  year = {1964},
  month = {Nov},
  publisher = {American Physical Society},
  doi = {10.1103/PhysRev.136.A680},
  url = {https://link.aps.org/doi/10.1103/PhysRev.136.A680}
}

@article{Schiff1965,
  title = {Fine Structure of the $2^{3}${P} and $3^{3}${P} States of Helium},
  author = {Schiff, B and Pekeris, C L and Lifson, H},
  journal = {Phys. Rev.},
  volume = {137},
  issue = {6A},
  pages = {A1672--A1675},
  numpages = {0},
  year = {1965},
  month = {Mar},
  publisher = {American Physical Society},
  doi = {10.1103/PhysRev.137.A1672},
  url = {https://link.aps.org/doi/10.1103/PhysRev.137.A1672}
}

@article{Scherr1965,
  title = {P States of Systems of Three {Coulombic} Particles},
  author = {Scherr, C W and Machacek, M},
  journal = {Phys. Rev.},
  volume = {138},
  issue = {2A},
  pages = {A371--A381},
  numpages = {0},
  year = {1965},
  month = {Apr},
  publisher = {American Physical Society},
  doi = {10.1103/PhysRev.138.A371},
  url = {https://link.aps.org/doi/10.1103/PhysRev.138.A371}
}

@article{Schiff1965a,
  title = {$2^{1,3}${P}, $3^{1,3}${P}, and $4^{1,3}${P} States of {He} and the $2^{1}${P} State of {Li$^{+}$}},
  author = {Schiff, B and Lifson, H and Pekeris, C L and Rabinowitz, P},
  journal = {Phys. Rev.},
  volume = {140},
  issue = {4A},
  pages = {A1104--A1121},
  numpages = {0},
  year = {1965},
  month = {Nov},
  publisher = {American Physical Society},
  doi = {10.1103/PhysRev.140.A1104},
  url = {https://link.aps.org/doi/10.1103/PhysRev.140.A1104}
}

@article{Rensbergen1972,
title = {A low-lying resonance in the spectrum of {H}$^-$ {II}. {The} $2^{1}${P} state},
journal = {J. Quant. Spectrosc. Radiat. Transf.},
volume = {12},
number = {7},
pages = {1105-1113},
year = {1972},
issn = {0022-4073},
doi = {https://doi.org/10.1016/0022-4073(72)90013-1},
url = {https://www.sciencedirect.com/science/article/pii/0022407372900131},
author = {Rensbergen, W V}
}

@article{Ho1990,
doi = {10.1088/0953-4075/23/15/007},
url = {https://dx.doi.org/10.1088/0953-4075/23/15/007},
year = {1990},
month = {aug},
publisher = {},
volume = {23},
number = {15},
pages = {L419},
author = {Ho, Y K},
title = {P-wave resonances in positron-hydrogen scattering},
journal = {J. Phys. B: At. Mol. Opt. Phys.}
}

@article{Warner1978,
title = {Relativistic corrections in a series of helium excited states},
journal = {Chem. Phys. Lett.},
volume = {56},
number = {1},
pages = {164-166},
year = {1978},
issn = {0009-2614},
doi = {https://doi.org/10.1016/0009-2614(78)80211-5},
url = {https://www.sciencedirect.com/science/article/pii/0009261478802115},
author = {Warner, J W and Blinder, S M}
}

@article{Drake1970,
  title = {$2{p}^{2}$ {${}^{3}$P} and $2p3p$ {${}^{1}$P} States of the Helium Isoelectronic Sequence},
  author = {Drake, G W F and Dalgarno, A},
  journal = {Phys. Rev. A},
  volume = {1},
  issue = {5},
  pages = {1325--1329},
  numpages = {0},
  year = {1970},
  month = {May},
  publisher = {American Physical Society},
  doi = {10.1103/PhysRevA.1.1325},
  url = {https://link.aps.org/doi/10.1103/PhysRevA.1.1325}
}

@article{Drake1988,
author = {Drake, G W F and Makowski, A J},
journal = {J. Opt. Soc. Am. B},
number = {10},
pages = {2207--2214},
publisher = {Optica Publishing Group},
title = {High-precision eigenvalues for the $1s2p$ $^1${P} and $^3${P} states of helium},
volume = {5},
month = {Oct},
year = {1988},
url = {https://opg.optica.org/josab/abstract.cfm?URI=josab-5-10-2207},
doi = {10.1364/JOSAB.5.002207}
}

@article{Kar2009a,
doi = {10.1088/0953-4075/42/5/055001},
url = {https://dx.doi.org/10.1088/0953-4075/42/5/055001},
year = {2009},
month = {feb},
volume = {42},
number = {5},
pages = {055001},
author = {Kar, S and Ho, Y K},
title = {Isotope shift for the {${}^{1}\text{D}^{\text{e}}$} autodetaching resonance in {H$^-$} and {D$^-$}},
journal = {J. Phys. B: At. Mol. Opt. Phys.}
}

@article{Kar2010,
author = {Kar, S and Ho, Y K},
title = {Calculations of {D}-wave bound states and resonance states of the screened helium atom using correlated exponential wave functions},
journal = {Int. J. Quant. Chem.},
volume = {110},
number = {5},
pages = {993-1002},
doi = {https://doi.org/10.1002/qua.22074},
url = {https://onlinelibrary.wiley.com/doi/abs/10.1002/qua.22074},
year = {2009}
}

@article{Kar2009b,
  title = {Doubly excited nonautoionizing {P}, {D}, and {F} states of helium with {Coulomb} and screened {Coulomb} potentials},
  author = {Kar, S and Ho, Y K},
  journal = {Phys. Rev. A},
  volume = {79},
  issue = {6},
  pages = {062508},
  numpages = {7},
  year = {2009},
  month = {Jun},
  publisher = {American Physical Society},
  doi = {10.1103/PhysRevA.79.062508},
  url = {https://link.aps.org/doi/10.1103/PhysRevA.79.062508}
}

@article{Kar2009c,
doi = {10.1088/0953-4075/42/18/185005},
url = {https://dx.doi.org/10.1088/0953-4075/42/18/185005},
year = {2009},
month = {sep},
volume = {42},
number = {18},
pages = {185005},
author = {Kar, S and Ho, Y K},
title = {Doubly excited {P}, {D} and {F} unnatural parity states of hydrogen negative ion using correlated wavefunctions},
journal = {J. Phys. B: At. Mol. Opt. Phys.}
}

@article{Aznabayev2015,
author={Aznabayev, D T
and Bekbaev, A K
and Ishmukhamedov, I S
and Korobov, V I},
title={Energy levels of a helium atom},
journal={Phys. Part. Nucl. Lett.},
year={2015},
month={Sep},
day={01},
volume={12},
number={5},
pages={689-694},
doi={10.1134/S1547477115050040},
url={https://doi.org/10.1134/S1547477115050040}
}

@book{Drake2023,
author="Drake, G W F",
editor="Drake, G. W. F.",
chapter="High Precision Calculations for Helium",
title="Springer Handbook of Atomic, Molecular, and Optical Physics",
year="2023",
publisher="Springer International Publishing",
address="Cham",
pages="199--216",
isbn="978-3-030-73893-8",
doi="10.1007/978-3-030-73893-8_12",
url="https://doi.org/10.1007/978-3-030-73893-8_12"
}

@article{Drake1972,
  title = {Radiative Transition Rates from the $2p3p$ {${}^{3}$P} and $2p3d$ {${}^{1,3}$D} States of the Helium Isoelectronic Sequence},
  author = {Drake, G W F},
  journal = {Phys. Rev. A},
  volume = {5},
  issue = {2},
  pages = {614--619},
  numpages = {0},
  year = {1972},
  month = {Feb},
  publisher = {American Physical Society},
  doi = {10.1103/PhysRevA.5.614},
  url = {https://link.aps.org/doi/10.1103/PhysRevA.5.614}
}

@article{Kar2008,
author = {Kar, S and Ho, Y K},
title = {Unnatural parity states of helium with screened {Coulomb} potentials},
journal = {Int. J. Quant. Chem.},
volume = {108},
number = {9},
pages = {1491-1504},
doi = {https://doi.org/10.1002/qua.21661},
url = {https://onlinelibrary.wiley.com/doi/abs/10.1002/qua.21661},
year = {2008}
}

@article{Hesse2001,
doi = {10.1088/0953-4075/34/8/308},
url = {https://dx.doi.org/10.1088/0953-4075/34/8/308},
year = {2001},
month = {apr},
publisher = {},
volume = {34},
number = {8},
pages = {1425},
author = {Hesse, M and Baye, D},
title = {Lagrange-mesh calculations of excited states of three-body atoms and molecules},
journal = {J. Phys. B: At. Mol. Opt. Phys.}
}

@article{Nikitin1985a,
doi = {10.1088/0022-3700/18/22/006},
url = {https://dx.doi.org/10.1088/0022-3700/18/22/006},
year = {1985},
month = {nov},
publisher = {},
volume = {18},
number = {22},
pages = {4349},
author = {Nikitin, S I and Ostrovsky, V N},
title = {Vibro-rotational states of the two-electron atom. {I}. {Euler} angles coordinate basis},
journal = {J. Phys. B: At. Mol. Opt. Phys.}
}

@article{Nikitin1985b,
doi = {10.1088/0022-3700/18/22/007},
url = {https://dx.doi.org/10.1088/0022-3700/18/22/007},
year = {1985},
month = {nov},
publisher = {},
volume = {18},
number = {22},
pages = {4371},
author = {Nikitin, S I and Ostrovsky, V N},
title = {Vibro-rotational states of the two-electron atom. {II}. {Two} interacting particles on the sphere},
journal = {J. Phys. B: At. Mol. Opt. Phys.}
}

@article{Hilger1996,
title = {Accurate nonrelativistic energies for some doubly excited {P}${}^{\text{e}}$ states of {He}},
journal = {Chem. Phys. Lett.},
volume = {262},
number = {3},
pages = {400-404},
year = {1996},
issn = {0009-2614},
doi = {https://doi.org/10.1016/0009-2614(96)01071-8},
author = {R Hilger and H-P Merckens and H Kleindienst}
}

@book{Gradshteyn2015,
author = {I S Gradshteyn and I M Ryzhik},
editor = {Daniel Zwillinger and Victor Moll},
title = {Table of Integrals, Series, and Products},
publisher = {Academic Press},
year = {2015},
doi = {},
address = {},
edition   = {Eighth},
url = {https://www.sciencedirect.com/book/9780123849335/table-of-integrals-series-and-products#editors}
}

@book{Brualdi2010,
  title={Introductory Combinatorics},
  author={R A Brualdi},
  edition={5th},
  year={2010},
  publisher={Pearson/Prentice Hall},
  isbn={978-0136020400}
}

@misc{Note,
  author = {Note},
  title = {},
  howpublished = {},
  year = {2026},
  note = {It should be noted that we have used a similar transformation to the (internal + Euler angles) coordinate system $(r_{1},r_{2},\theta,\alpha,\beta,\gamma)$ in the original paper \textbf{1}. However, in that case the transformation is mapped as
  $\mathbf{r}_{i} \rightarrow\mathbb{R}_{\alpha\beta\gamma}\left[ r_i\cos{\frac{\theta}{2}},(-1)^i r_i\sin{\frac{\theta}{2}},0\right]^\mathsf{T}$. For details please consult the text accompanying Eqs.~(12--16) of \textbf{1}.}
}

@article{Drake1979,
  title = {Unified relativistic theory for $1s2p\,^{3}\text{{P}}_{1}\ensuremath{-}1{s}^{2}\,{}^{1}\text{{S}}_{0}$ and $1s2p\,^{1}\text{{P}}_{1}\ensuremath{-}1{s}^{2}\,{}^{1}\text{{S}}_{0}$ frequencies and transition rates in heliumlike ions},
  author = {Drake, G W F},
  journal = {Phys. Rev. A},
  volume = {19},
  issue = {4},
  pages = {1387--1397},
  numpages = {0},
  year = {1979},
  month = {Apr},
  publisher = {American Physical Society},
  doi = {10.1103/PhysRevA.19.1387},
  url = {https://link.aps.org/doi/10.1103/PhysRevA.19.1387}
}

@article{Drake1988a,
title = {High precision variational calculations for the $1s^2$ ${}^1${S} state of {H}$^{-}$ and the $1s^2$ ${}^1${S}, $1s2s$ ${}^1${S} and $1s2s$ ${}^3${S} states of helium},
journal = {Nucl. Instrum. Methods Phys. Res., Sect. B},
volume = {31},
number = {1},
pages = {7-13},
year = {1988},
issn = {0168-583X},
doi = {https://doi.org/10.1016/0168-583X(88)90387-4},
url = {https://www.sciencedirect.com/science/article/pii/0168583X88903874},
author = {Drake, G W F}
}

@article{Yan1996,
title = {On the evaluation of two-electron integrals in {Hylleraas} coordinates},
journal = {Chem. Phys. Lett.},
volume = {259},
number = {1},
pages = {96-102},
year = {1996},
issn = {0009-2614},
doi = {https://doi.org/10.1016/0009-2614(96)00706-3},
url = {https://www.sciencedirect.com/science/article/pii/0009261496007063},
author = {Z-C Yan and G W F Drake}
}

@article{Goldman1997,
  title = {Accurate Modified Configuration Interaction Calculations for Many Electron Systems Made Easy},
  author = {Goldman, S P},
  journal = {Phys. Rev. Lett.},
  volume = {78},
  issue = {12},
  pages = {2325--2328},
  numpages = {0},
  year = {1997},
  month = {Mar},
  publisher = {American Physical Society},
  doi = {10.1103/PhysRevLett.78.2325},
  url = {https://link.aps.org/doi/10.1103/PhysRevLett.78.2325}
}

@article{Yan1997,
doi = {10.1088/0953-4075/30/21/012},
url = {https://doi.org/10.1088/0953-4075/30/21/012},
year = {1997},
month = {nov},
publisher = {},
volume = {30},
number = {21},
pages = {4723},
author = {Z-C Yan and G W F Drake},
title = {Computational methods for three-electron atomic systems in {Hylleraas} coordinates},
journal = {J. Phys. B: At. Mol. Opt. Phys.}
}

@article{Liu1997,
  title = {High-order nonlinear susceptibilities of helium},
  author = {Liu, W-C},
  journal = {Phys. Rev. A},
  volume = {56},
  issue = {6},
  pages = {4938--4945},
  numpages = {0},
  year = {1997},
  month = {Dec},
  publisher = {American Physical Society},
  doi = {10.1103/PhysRevA.56.4938},
  url = {https://link.aps.org/doi/10.1103/PhysRevA.56.4938}
}

@article{Lee2003,
  title = {Core-polarization potentials for {Si} and {Ti}},
  author = {Lee, Y and Needs, R J},
  journal = {Phys. Rev. B},
  volume = {67},
  issue = {3},
  pages = {035121},
  numpages = {10},
  year = {2003},
  month = {Jan},
  publisher = {American Physical Society},
  doi = {10.1103/PhysRevB.67.035121},
  url = {https://link.aps.org/doi/10.1103/PhysRevB.67.035121}
}

@article{Harris2005,
  title = {Singular integrals and their application to a hypervirial theorem},
  author = {Harris, F E and Frolov, A M and Smith, V H},
  journal = {Phys. Rev. A},
  volume = {72},
  issue = {1},
  pages = {012511},
  numpages = {6},
  year = {2005},
  month = {Jul},
  publisher = {American Physical Society},
  doi = {10.1103/PhysRevA.72.012511},
  url = {https://link.aps.org/doi/10.1103/PhysRevA.72.012511}
}

@article{Harris2005a,
author = {Harris, F E and Smith, V H and Frolov, A M},
title = {Correlated exponential-basis integrals with logarithmic integrands},
journal = {Mol. Phys.},
volume = {103},
number = {15-16},
pages = {2047--2054},
year = {2005},
publisher = {Taylor \& Francis},
doi = {10.1080/00268970500084034},
URL = {https://doi.org/10.1080/00268970500084034}
}

@article{Drake2005,
doi = {10.1088/0953-4075/38/18/009},
url = {https://doi.org/10.1088/0953-4075/38/18/009},
year = {2005},
month = {sep},
publisher = {},
volume = {38},
number = {18},
pages = {3377},
author = {Drake, G W F and Grigorescu, M},
title = {Binding energy of the positronium negative ion: relativistic and {QED} energy shifts},
journal = {J. Phys. B: At. Mol. Opt. Phys.}
}

@article{Laughlin2009,
doi = {10.1088/1751-8113/42/26/265004},
url = {https://doi.org/10.1088/1751-8113/42/26/265004},
year = {2009},
month = {jun},
publisher = {},
volume = {42},
number = {26},
pages = {265004},
author = {Laughlin, C and Chu, S-I},
title = {A highly accurate study of a helium atom under pressure},
journal = {J. Phys. A: Math. Theor.}
}

@article{Kar2012,
  title = {Dynamic dipole polarizability of the helium atom with {Debye-H\"uckel} potentials},
  author = {Kar, S},
  journal = {Phys. Rev. A},
  volume = {86},
  issue = {6},
  pages = {062516},
  numpages = {5},
  year = {2012},
  month = {Dec},
  publisher = {American Physical Society},
  doi = {10.1103/PhysRevA.86.062516},
  url = {https://link.aps.org/doi/10.1103/PhysRevA.86.062516}
}

@article{Kar2013,
    author = {Kar, S and Li, H W and Jiang, P},
    title = {Dynamic dipole polarizabilities of {H}$^{-}$ and {Ps}$^{-}$ in weakly coupled plasmas},
    journal = {Phys. Plasmas},
    volume = {20},
    number = {8},
    pages = {083302},
    year = {2013},
    month = {08},
    issn = {1070-664X},
    doi = {10.1063/1.4818600},
    url = {https://doi.org/10.1063/1.4818600}
}

@article{Zhong2013,
  title = {Pseudostate calculation of the Bethe logarithm for {H}${{}_{2}}^{+}$},
  author = {Zhong, Z-X and Yan, Z-C and Shi, T-Y},
  journal = {Phys. Rev. A},
  volume = {88},
  issue = {5},
  pages = {052520},
  numpages = {4},
  year = {2013},
  month = {Nov},
  publisher = {American Physical Society},
  doi = {10.1103/PhysRevA.88.052520},
  url = {https://link.aps.org/doi/10.1103/PhysRevA.88.052520}
}

@article{Kruppa2014,
  title = {Nuclear three-body problem in the complex energy plane: Complex-scaling {Slater} method},
  author = {Kruppa, A T and Papadimitriou, G and Nazarewicz, W and Michel, N},
  journal = {Phys. Rev. C},
  volume = {89},
  issue = {1},
  pages = {014330},
  numpages = {15},
  year = {2014},
  month = {Jan},
  publisher = {American Physical Society},
  doi = {10.1103/PhysRevC.89.014330},
  url = {https://link.aps.org/doi/10.1103/PhysRevC.89.014330}
}

@Inbook{Jiao2014,
author={Jiao, L G and Ho, Y K},
editor={Sen, K D},
chapter={Bound and Resonant States in Confined Atoms},
title={Electronic Structure of Quantum Confined Atoms and Molecules},
year={2014},
publisher={Springer International Publishing},
pages={145--168},
isbn={978-3-319-09982-8},
doi={10.1007/978-3-319-09982-8_6},
url={https://doi.org/10.1007/978-3-319-09982-8_6}
}

@article{Frolov2015,
author={Frolov, A M and Wardlaw, D M},
title={On the products of bipolar harmonics},
journal={J. Math. Chem.},
year={2015},
month={Apr},
day={01},
volume={53},
number={4},
pages={1068-1079},
issn={1572-8897},
doi={10.1007/s10910-014-0466-2},
url={https://doi.org/10.1007/s10910-014-0466-2}
}

@article{Jiao2015,
title = {Computation of two-electron screened {Coulomb} potential integrals in {Hylleraas} basis sets},
journal = {Comput. Phys. Commun.},
volume = {188},
pages = {140-147},
year = {2015},
issn = {0010-4655},
doi = {https://doi.org/10.1016/j.cpc.2014.11.019},
url = {https://www.sciencedirect.com/science/article/pii/S0010465514004020},
author = {L G Jiao and Y K Ho}
}

@article{Kar2015,
title = {Energies and transition wavelengths for two-electron atoms under {Debye} screening},
journal = {At. Data Nucl. Data Tables},
volume = {102},
pages = {42-63},
year = {2015},
issn = {0092-640X},
doi = {https://doi.org/10.1016/j.adt.2014.12.003},
url = {https://www.sciencedirect.com/science/article/pii/S0092640X14000813},
author = {S Kar and Z Jiang}
}

@article{Jiang2015,
author = {Jiang, Z and Zhang, Y-Z and Kar, S},
title = {Resonances in positron-hydrogen scattering in dense quantum plasmas},
journal = {Phys. Plasmas},
volume = {22},
number = {5},
pages = {052105},
year = {2015},
month = {05},
issn = {1070-664X},
doi = {10.1063/1.4919933},
url = {https://doi.org/10.1063/1.4919933}
}

@article{Kar2015a,
author={Kar, S and Wang, Y and Li, W-Q and Sun, X-D},
title={Doubly Excited Nonautoionizing {F} and {G} States of Two-Electron Ions},
journal={Few-Body Syst.},
year={2015},
month={Oct},
day={01},
volume={56},
number={10},
pages={651-657},
issn={1432-5411},
doi={10.1007/s00601-015-1011-y},
url={https://doi.org/10.1007/s00601-015-1011-y}
}

@article{Zalialiutdinov2016,
  title = {Spin-statistic selection rules for multiphoton transitions: Application to helium atoms},
  author = {Zalialiutdinov, T and Solovyev, D and Labzowsky, L and Plunien, G},
  journal = {Phys. Rev. A},
  volume = {93},
  issue = {1},
  pages = {012510},
  numpages = {7},
  year = {2016},
  month = {Jan},
  publisher = {American Physical Society},
  doi = {10.1103/PhysRevA.93.012510},
  url = {https://link.aps.org/doi/10.1103/PhysRevA.93.012510}
}

@article{Zalialiutdinov2017,
doi = {10.1088/1361-6455/aa8e6e},
url = {https://doi.org/10.1088/1361-6455/aa8e6e},
year = {2017},
month = {nov},
publisher = {IOP Publishing},
volume = {51},
number = {1},
pages = {015003},
author = {Zalialiutdinov, T and Solovyev, D and Labzowsky, L},
title = {{BBR}-induced {Stark} shifts and level broadening in a helium atom},
journal = {J. Phys. B: At. Mol. Opt. Phys.}
}

@article{Jiao2019,
  title = {High-precision calculation of the geometric quantities of two-electron atoms based on the {Hylleraas} configuration-interaction basis},
  author = {Jiao, L G and Zan, L R and Zhu, L and Zhang, Y Z and Ho, Y K},
  journal = {Phys. Rev. A},
  volume = {100},
  issue = {2},
  pages = {022509},
  numpages = {16},
  year = {2019},
  month = {Aug},
  publisher = {American Physical Society},
  doi = {10.1103/PhysRevA.100.022509},
  url = {https://link.aps.org/doi/10.1103/PhysRevA.100.022509}
}

@article{Zhang2019,
author = {Zhang, Y Z and Gao, Y C and Jiao, L G and Liu, F and Ho, Y K},
title = {Linear dependence in {Hylleraas} configuration-interaction calculations of {He} atom},
journal = {Int. J. Quant. Chem.},
volume = {120},
number = {7},
pages = {e26136},
doi = {https://doi.org/10.1002/qua.26136},
url = {https://onlinelibrary.wiley.com/doi/abs/10.1002/qua.26136},
year = {2019}
}

@article{Korobov2020,
  title = {Hyperfine structure in the {${\mathrm{H}}_{2}^{+}$} and {${\mathrm{HD}}^{+}$} molecular ions at order $m{\ensuremath{\alpha}}^{6}$},
  author = {Korobov, V I and Karr, J-P and Haidar, M and Zhong, Z-X},
  journal = {Phys. Rev. A},
  volume = {102},
  issue = {2},
  pages = {022804},
  numpages = {14},
  year = {2020},
  month = {Aug},
  publisher = {American Physical Society},
  doi = {10.1103/PhysRevA.102.022804},
  url = {https://link.aps.org/doi/10.1103/PhysRevA.102.022804}
}

@article{Sims2020,
author = {Sims, J S and Padhy, B and Ruiz, M B},
title = {Exponentially correlated {Hylleraas}-configuration interaction nonrelativistic energy of the ${}^{1}${S} ground state of the helium atom},
journal = {Int. J. Quant. Chem.},
volume = {121},
number = {4},
pages = {e26470},
doi = {https://doi.org/10.1002/qua.26470},
url = {https://onlinelibrary.wiley.com/doi/abs/10.1002/qua.26470},
year = {2020}
}

@article{Korobov2022,
author={Korobov, V I},
title={Variational Methods in the Quantum Mechanical Three-Body Problem with a {Coulomb} Interaction},
journal={Phys. Part. Nucl.},
year={2022},
month={Feb},
day={01},
volume={53},
number={1},
pages={1-20},
issn={1531-8559},
doi={10.1134/S1063779622010038},
url={https://doi.org/10.1134/S1063779622010038}
}

@article{Jiao2022,
  title = {Bound and resonance states near the critical charge region in two-electron atoms},
  author = {Jiao, L G and Zheng, R Y and Liu, A and Montgomery, H E and Ho, Y K},
  journal = {Phys. Rev. A},
  volume = {105},
  issue = {5},
  pages = {052806},
  numpages = {11},
  year = {2022},
  month = {May},
  publisher = {American Physical Society},
  doi = {10.1103/PhysRevA.105.052806},
  url = {https://link.aps.org/doi/10.1103/PhysRevA.105.052806}
}

@article{Qi2024,
author = {Qi, X-Q and Zhong, Z-X and Zhang, P-P},
title = {Singular integrals in {QED} corrections of three-body systems},
journal = {Mod. Phys. Lett. B},
volume = {38},
number = {07},
pages = {2450025},
year = {2024},
doi = {10.1142/S0217984924500258},
URL = {https://doi.org/10.1142/S0217984924500258}
}

@article{Davis1983,
year = {1983},
month = {dec},
publisher = {},
volume = {16},
number = {18},
pages = {4237},
author = {C L Davis and E N Maslen},
title = {On exact analytical solutions for the few-particle {Schr\"{o}dinger} equation. {III}. {Spatially} symmetric {S} states of two identical particles in the field of a massive third particle},
journal = {J. Phys. A: Math. Gen.},
doi = {10.1088/0305-4470/16/18/025},
url = {https://doi.org/10.1088/0305-4470/16/18/025}
}

@article{Davis1982,
    author = {Davis, C L and Maslen, E N and Varghese, J N},
    title = {On exact analytical solutions for the few-particle {Schr\"{o}dinger} equation. {I}. {A} perturbation study},
    journal = {Proc. R. Soc. Lond. A Math. Phys. Sci.},
    volume = {384},
    number = {1786},
    pages = {57-88},
    year = {1982},
    month = {11},
    issn = {0080-4630},
    doi = {10.1098/rspa.1982.0148},
    url = {https://doi.org/10.1098/rspa.1982.0148}
}

@article{Davis1982a,
    author = {Davis, C L and Maslen, E N},
    title = {On exact analytical solutions for the few-particle {Schr\"{o}dinger} equation. {II}. {The} ground state of helium},
    journal = {Proc. R. Soc. Lond. A Math. Phys. Sci.},
    volume = {384},
    number = {1786},
    pages = {89-105},
    year = {1982},
    month = {11},
    issn = {0080-4630},
    doi = {10.1098/rspa.1982.0149},
    url = {https://doi.org/10.1098/rspa.1982.0149}
}

@article{Davis1983a,
doi = {10.1088/0305-4470/16/18/026},
url = {https://doi.org/10.1088/0305-4470/16/18/026},
year = {1983},
month = {12},
publisher = {},
volume = {16},
number = {18},
pages = {4255},
author = {E L Davis and E N Maslen},
title = {On exact analytical solutions for the few-particle {Schr\"{o}dinger} equation. {IV}. {The} asymptotic form and normalisability of the wavefunction},
journal = {J. Phys. A: Math. Gen.}
}

@article{Davis1983b,
author = {Davis, C. L. and Maslen, E. N.},
title = {Series wave functions for the helium atom},
journal = {Int. J. of Quant. Chem.},
volume = {24},
number = {S17},
pages = {217-225},
doi = {https://doi.org/10.1002/qua.560240828},
url = {https://onlinelibrary.wiley.com/doi/abs/10.1002/qua.560240828},
year = {1983}
}

@article{Abbott1986,
doi = {10.1088/0022-3700/19/11/014},
url = {https://doi.org/10.1088/0022-3700/19/11/014},
year = {1986},
month = {jun},
publisher = {},
volume = {19},
number = {11},
pages = {1595},
author = {P C Abbott and E N Maslen},
title = {A model wavefunction including electron correlation for the ground state of the helium isoelectronic sequence},
journal = {J. Phys. B: At. Mol. Phys.}
}

@article{Abbott1984,
doi = {10.1088/0022-3700/17/15/001},
url = {https://doi.org/10.1088/0022-3700/17/15/001},
year = {1984},
month = {aug},
publisher = {},
volume = {17},
number = {15},
pages = {L489},
author = {P C Abbott and E N Maslen},
title = {Expansion of two-body potentials in hyperspherical harmonics},
journal = {J. Phys. B: At. Mol. Phys.}
}

@article{McIsaac1987,
author = {McIsaac, K and Maslen, E N},
title = {Exact wave functions for few-particle systems: {The} choice of expansion for {Coulomb} potentials},
journal = {Int. J. of Quant. Chem.},
volume = {31},
number = {3},
pages = {361-368},
doi = {https://doi.org/10.1002/qua.560310307},
url = {https://onlinelibrary.wiley.com/doi/abs/10.1002/qua.560310307},
year = {1987}
}

@article{Gottschalk1985,
doi = {10.1088/0305-4470/18/10/023},
url = {https://doi.org/10.1088/0305-4470/18/10/023},
year = {1985},
month = {jul},
publisher = {},
volume = {18},
number = {10},
pages = {1687},
author = {J E Gottschalk and E N Maslen},
title = {Three-body {S}-state wavefunctions: {Symmetry} and degrees of freedom associated with normalisation of the exact wavefunction},
journal = {J. Phys. A: Math. Gen.}
}

@article{Yerokhin2023,
  title = {{QED} $m{\ensuremath{\alpha}}^{7}$ effects for triplet states of heliumlike ions},
  author = {Yerokhin, V A and Patk\'o\ifmmode \check{s}\else \v{s}\fi{}, V and Pachucki, K},
  journal = {Phys. Rev. A},
  volume = {107},
  issue = {1},
  pages = {012810},
  numpages = {7},
  year = {2023},
  month = {Jan},
  publisher = {American Physical Society},
  doi = {10.1103/PhysRevA.107.012810},
  url = {https://link.aps.org/doi/10.1103/PhysRevA.107.012810}
}

@article{Wienczek2019,
  title = {Quantum-electrodynamic corrections to the $1s3d$ states of the helium atom},
  author = {Wienczek, A and Pachucki, K and Puchalski, M and Patk\'o\ifmmode \check{s}\else \v{s}\fi{}, V and Yerokhin, V A},
  journal = {Phys. Rev. A},
  volume = {99},
  issue = {5},
  pages = {052505},
  numpages = {12},
  year = {2019},
  month = {May},
  publisher = {American Physical Society},
  doi = {10.1103/PhysRevA.99.052505},
  url = {https://link.aps.org/doi/10.1103/PhysRevA.99.052505}
}

@article{Yerokhin2020,
  title = {{QED} calculation of ionization energies of $1snd$ states in helium},
  author = {Yerokhin, V A and Patk\'o\ifmmode \check{s}\else \v{s}\fi{}, V and Puchalski, M and Pachucki, K},
  journal = {Phys. Rev. A},
  volume = {102},
  issue = {1},
  pages = {012807},
  numpages = {5},
  year = {2020},
  month = {Jul},
  publisher = {American Physical Society},
  doi = {10.1103/PhysRevA.102.012807},
  url = {https://link.aps.org/doi/10.1103/PhysRevA.102.012807}
}

@article{Pachucki2017,
  title = {Testing fundamental interactions on the helium atom},
  author = {Pachucki, K and Patk\'o\ifmmode \check{s}\else \v{s}\fi{}, V and Yerokhin, V A},
  journal = {Phys. Rev. A},
  volume = {95},
  issue = {6},
  pages = {062510},
  numpages = {8},
  year = {2017},
  month = {Jun},
  publisher = {American Physical Society},
  doi = {10.1103/PhysRevA.95.062510},
  url = {https://link.aps.org/doi/10.1103/PhysRevA.95.062510}
}

@article{Patkos2016,
  title = {Higher-order recoil corrections for triplet states of the helium atom},
  author = {Patk\'o\ifmmode \check{s}\else \v{s}\fi{}, V and Yerokhin, V A and Pachucki, K},
  journal = {Phys. Rev. A},
  volume = {94},
  issue = {5},
  pages = {052508},
  numpages = {24},
  year = {2016},
  month = {Nov},
  publisher = {American Physical Society},
  doi = {10.1103/PhysRevA.94.052508},
  url = {https://link.aps.org/doi/10.1103/PhysRevA.94.052508}
}

@article{Pachucki2015,
    author = {Pachucki, K and Yerokhin, V A},
    title = {Theory of the Helium Isotope Shift},
    journal = {J. Phys. Chem. Ref. Data},
    volume = {44},
    number = {3},
    pages = {031206},
    year = {2015},
    month = {06},
    doi = {10.1063/1.4921428},
    url = {https://doi.org/10.1063/1.4921428},
}

@article{Pachucki2011,
doi = {10.1088/1742-6596/264/1/012007},
url = {https://doi.org/10.1088/1742-6596/264/1/012007},
year = {2011},
month = {jan},
publisher = {},
volume = {264},
number = {1},
pages = {012007},
author = {Pachucki, K and Yerokhin, V A},
title = {Helium fine structure theory for determination of $\alpha$},
journal = {J. Phys. Conf. Ser.}
}

@article{Pachucki2011a,
author = {Pachucki, K and Yerokhin, V A},
title = {Fine structure of helium and light helium-like ions},
journal = {Can. J. Phys.},
volume = {89},
number = {1},
pages = {95-101},
year = {2011},
doi = {10.1139/P10-050},
URL = {https://doi.org/10.1139/P10-050}
}

@article{Pachucki2006,
  title = {Helium energy levels including $m{\ensuremath{\alpha}}^{6}$ corrections},
  author = {Pachucki, K},
  journal = {Phys. Rev. A},
  volume = {74},
  issue = {6},
  pages = {062510},
  numpages = {5},
  year = {2006},
  month = {Dec},
  publisher = {American Physical Society},
  doi = {10.1103/PhysRevA.74.062510},
  url = {https://link.aps.org/doi/10.1103/PhysRevA.74.062510}
}

@article{Yerokhin2010,
  title = {Theoretical energies of low-lying states of light helium-like ions},
  author = {Yerokhin, V A and Pachucki, K},
  journal = {Phys. Rev. A},
  volume = {81},
  issue = {2},
  pages = {022507},
  numpages = {15},
  year = {2010},
  month = {Feb},
  publisher = {American Physical Society},
  doi = {10.1103/PhysRevA.81.022507},
  url = {https://link.aps.org/doi/10.1103/PhysRevA.81.022507}
}

@article{Drake1970a,
  title = {Second Bound State for the Hydrogen Negative Ion},
  author = {Drake, G W F},
  journal = {Phys. Rev. Lett.},
  volume = {24},
  issue = {4},
  pages = {126--127},
  numpages = {0},
  year = {1970},
  month = {Jan},
  publisher = {American Physical Society},
  doi = {10.1103/PhysRevLett.24.126},
  url = {https://link.aps.org/doi/10.1103/PhysRevLett.24.126}
}

@article{Becker1964,
  title = {Double Excitation of Helium by Electron Impact},
  author = {Becker, P M and Dahler, J S},
  journal = {Phys. Rev.},
  volume = {136},
  issue = {1A},
  pages = {A73--A86},
  numpages = {0},
  year = {1964},
  month = {Oct},
  publisher = {American Physical Society},
  doi = {10.1103/PhysRev.136.A73},
  url = {https://link.aps.org/doi/10.1103/PhysRev.136.A73}
}

@article{Kar2011,
  title = {Doubly excited ${}^{1,3}${P}$^{\text{e}}$ resonance states of helium and the hydrogen negative ion interacting with {Coulomb} and screened {Coulomb} potentials},
  author = {Kar, S and Ho, Y K},
  journal = {Phys. Rev. A},
  volume = {83},
  issue = {4},
  pages = {042506},
  numpages = {8},
  year = {2011},
  month = {Apr},
  publisher = {American Physical Society},
  doi = {10.1103/PhysRevA.83.042506},
  url = {https://link.aps.org/doi/10.1103/PhysRevA.83.042506}
}

@article{Kar2010a,
  title = {Comment on ``Doubly excited bound and resonance (${}^{3}${P}$^{\text{e}}$) states of helium''},
  author = {Kar, S and Ho, Y K},
  journal = {Phys. Rev. A},
  volume = {82},
  issue = {3},
  pages = {036501},
  numpages = {4},
  year = {2010},
  month = {Sep},
  publisher = {American Physical Society},
  doi = {10.1103/PhysRevA.82.036501},
  url = {https://link.aps.org/doi/10.1103/PhysRevA.82.036501}
}

@article{Eiglsperger2010,
  title = {Spectral representation of the three-body {Coulomb} problem. {II}. {Autoionizing} doubly excited states of unnatural parity in helium},
  author = {Eiglsperger, J and Piraux, B and Madro\~nero, J},
  journal = {Phys. Rev. A},
  volume = {81},
  issue = {4},
  pages = {042528},
  numpages = {9},
  year = {2010},
  month = {Apr},
  publisher = {American Physical Society},
  doi = {10.1103/PhysRevA.81.042528},
  url = {https://link.aps.org/doi/10.1103/PhysRevA.81.042528}
}

@Article{Duan2002,
author={Duan, B and Gu, X-Y and Ma, Z-Q},
title={Numerical calculation of energies of some excited states in a helium atom},
journal={Eur. Phys. J. D},
year={2002},
month={Apr},
day={01},
volume={19},
number={1},
pages={9-12},
issn={1434-6079},
doi={10.1140/epjd/e20020049},
url={https://doi.org/10.1140/epjd/e20020049}
}

@article{Duan2001,
title = {Precise calculation for energy levels of a helium atom in {P} states},
journal = {Phys. Lett. A},
volume = {283},
number = {3},
pages = {229-236},
year = {2001},
issn = {0375-9601},
doi = {https://doi.org/10.1016/S0375-9601(01)00222-5},
url = {https://www.sciencedirect.com/science/article/pii/S0375960101002225},
author = {Duan, B and Gu, X-Y and Ma, Z-Q}
}

@article{Banyard1992,
doi = {10.1088/0953-4075/25/16/003},
url = {https://doi.org/10.1088/0953-4075/25/16/003},
year = {1992},
month = {aug},
publisher = {},
volume = {25},
number = {16},
pages = {3405},
author = {K E Banyard and D R T Keeble and G W F Drake},
title = {The doubly-excited state $2p^2$ ${}^3${P} for $1\leq {Z} \leq 4$: {Coulomb} holes derived from explicitly correlated wavefunctions},
journal = {J. Phys. B: At. Mol. Opt. Phys.}
}

@article{dePrunele1992,
  title = {Schr\"{o}dinger equation for two-electron atoms expressed in terms of symmetric sparse matrices involving only {O}(4,2) representations},
  author = {de Prunel\'e, E.},
  journal = {Phys. Rev. A},
  volume = {45},
  issue = {5},
  pages = {2757--2762},
  numpages = {0},
  year = {1992},
  month = {Mar},
  publisher = {American Physical Society},
  doi = {10.1103/PhysRevA.45.2757},
  url = {https://link.aps.org/doi/10.1103/PhysRevA.45.2757}
}

@article{Goodson1991,
  title = {Energies of doubly excited two-electron atoms from interdimensional degeneracies},
  author = {Goodson, D Z and Watson, D K and Loeser, J G and Herschbach, D R},
  journal = {Phys. Rev. A},
  volume = {44},
  issue = {1},
  pages = {97--102},
  numpages = {0},
  year = {1991},
  month = {Jul},
  publisher = {American Physical Society},
  doi = {10.1103/PhysRevA.44.97},
  url = {https://link.aps.org/doi/10.1103/PhysRevA.44.97}
}

@article{Bishop1989,
    author = {Bishop, D M and R\'{e}rat, M},
    title = {Higher-order polarizabilities for the helium isoelectronic series},
    journal = {J. Chem. Phys.},
    volume = {91},
    number = {9},
    pages = {5489-5491},
    year = {1989},
    month = {11},
    doi = {10.1063/1.457550},
    url = {https://doi.org/10.1063/1.457550}
}

@article{Ho1986,
doi = {10.1088/0031-8949/34/6B/009},
url = {https://doi.org/10.1088/0031-8949/34/6B/009},
year = {1986},
month = {dec},
publisher = {},
volume = {34},
number = {6B},
pages = {766},
author = {Ho, Y K },
title = {Transition Wavelengths Among Doubly Excited States of {He} {I}, {Li} {II}, and {Be} {III}},
journal = {Phys. Scr.}
}

@article{Conneely1978,
doi = {10.1088/0022-3700/11/24/008},
url = {https://doi.org/10.1088/0022-3700/11/24/008},
year = {1978},
month = {dec},
publisher = {},
volume = {11},
number = {24},
pages = {4135},
author = {M J Conneely and L Lipsky},
title = {Widths and configuration mixings of two-electron systems below the {$N=2$} threshold},
journal = {J. Phys. B: At. Mol. Phys.}
}

@article{Wu1944,
  title = {Auto-Ionization in Doubly Excited Helium and the $\ensuremath{\lambda}320.4$ and $\ensuremath{\lambda}357.5$ Lines},
  author = {Wu, T-Y},
  journal = {Phys. Rev.},
  volume = {66},
  issue = {11-12},
  pages = {291--294},
  numpages = {0},
  year = {1944},
  month = {Dec},
  publisher = {American Physical Society},
  doi = {10.1103/PhysRev.66.291},
  url = {https://link.aps.org/doi/10.1103/PhysRev.66.291}
}

@article{Eiglsperger2010a,
  title = {Spectral representation of the three-body {Coulomb} problem. {I}. {Nonautoionizing} doubly excited states of high angular momentum in helium},
  author = {Eiglsperger, J and Piraux, B and Madro\~nero, J},
  journal = {Phys. Rev. A},
  volume = {81},
  issue = {4},
  pages = {042527},
  numpages = {6},
  year = {2010},
  month = {Apr},
  publisher = {American Physical Society},
  doi = {10.1103/PhysRevA.81.042527},
  url = {https://link.aps.org/doi/10.1103/PhysRevA.81.042527}
}
